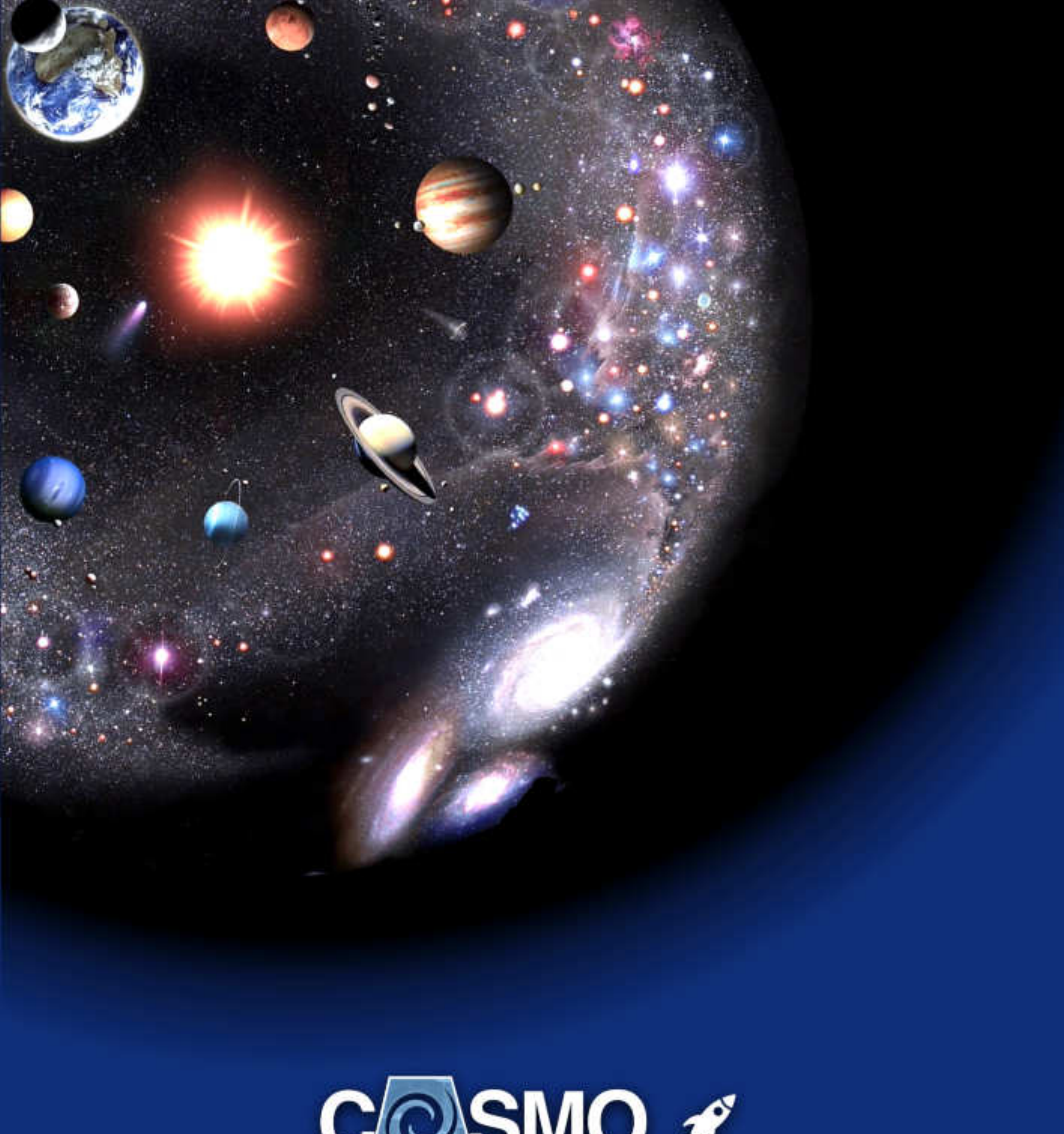

COSMO
Amautas

# Astrofísica en el Aula

CosmoAmautas 2022

Guía de contenido y actividades en temas selectos de astrofísica

**Segunda edición**

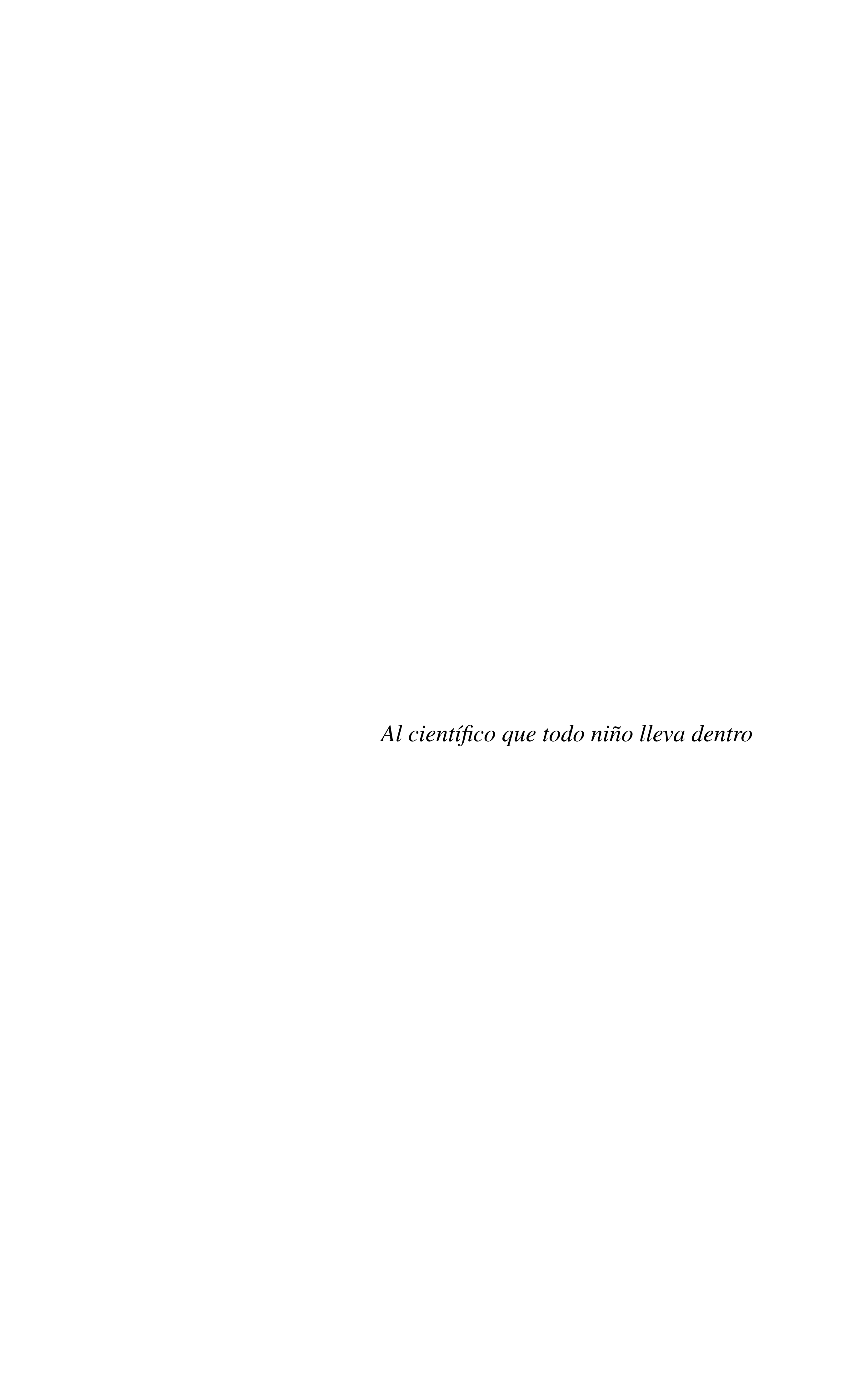

*Al científico que todo niño lleva dentro*

*Autores:* El Equipo CosmoAmautas

*Edición de contenido:*
Gabriela Calistro Rivera, Ph.D., Daniella Bardalez Gagliuffi, Ph.D.
*Edición de formato y diseño:*
Gabriela Calistro Rivera, Ph.D., Daniella Bardalez Gagliuffi, Ph.D., Erika Torre Ramirez, Diego Alvarado Urrunaga
*Capítulo 1:*
José Ricra, Adita Quispe Quispe, Lisseth Gonzales Quevedo.
*Capítulo 2:*
Erika Torre Ramirez, Diego Alvarado Urrunaga.
*Capítulo 3:*
Lisseth Gonzales Quevedo, Erika Torre Ramirez.
*Capítulo 4:*
Bruno Rodríguez Marquina, Esly Calcina.
*Capítulo 5:*
Jenny Margot Ramos Lázaro, Bruno Rodriguez Marquina.
*Capítulo 6:*
Esly Calcina, Diego Alvarado Urrunaga, Santiago Casas, Ph.D., Jenny Margot Ramos Lázaro.
*Actividades Aprendizaje por indagación (API):*
Gabriela Calistro Rivera, Ph.D., Daniella Bardalez Gagliuffi, Ph.D.

*Agradecimientos para la segunda edición:*
*Expresamos nuestra gratitud a* Ramiro Moro, Ph.D., *y* Vanessa Navarrete *por la detallada lectura, sus comentarios y sugerencias nos ayudaron a mejorar esta edición.*

Segunda edición, 2022
Publicado por CosmoAmautas
WWW.COSMOAMAUTAS.ORG
Primera impresión, marzo 2022.

*Acknowledgement:* CosmoAmautas is a project funded by the IAU Office of Astronomy for Development (OAD). The OAD is a joint project of the International Astronomical Union (IAU) and the South African National Research Foundation (NRF) with the support of the Department of Science and Innovation (DSI).

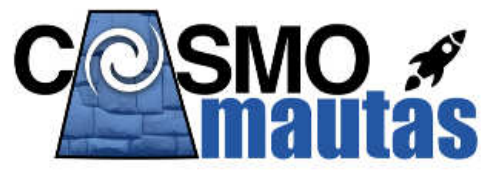

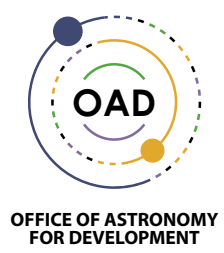

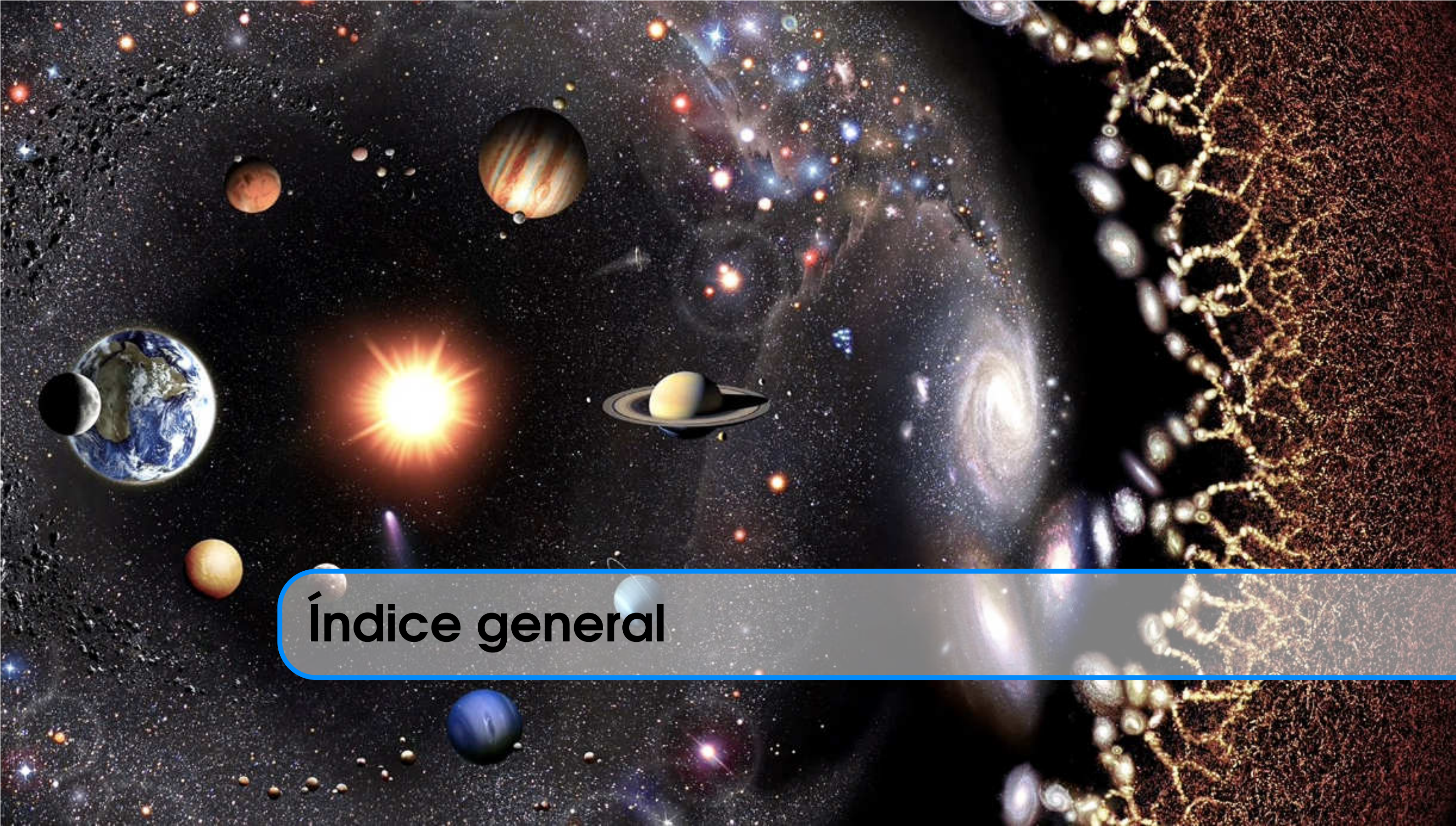

# Índice general

## I Conociendo el universo

# I Conociendo el universo

# 1. La Tierra, el Sol y la Luna

Desde el momento en que los primeros seres humanos empezaron a observar el cielo, fue necesario establecer mecanismos para determinar la posición y ciclos de movimiento de los objetos celestes. Es por esto que en la primera parte de esta sección repasamos algunos conceptos básicos de la astronomía de posición. Asimismo, todas las observaciones hechas desde la Tierra dependen de la forma en la que nuestra atmósfera interactúa con la radiación electromagnética proveniente del espacio exterior. Por ello es fundamental conocer nuestra atmósfera y cómo esta influye en nuestras observaciones. Finalmente, tratamos conceptos básicos sobre nuestro satélite, la Luna, sus fases y su relación dinámica con el Sol y la Tierra a través de los eclipses.

## 1.1 Ubicándonos en el cielo

Al levantar la mirada al cielo, los primeros objetos que llamarán nuestra atención (debido a su brillo), son el Sol, la Luna y las estrellas. Luego de algunas horas notaremos que estos objetos se mueven aproximadamente de la zona del este al oeste; pero no son los objetos los que se mueven, somos nosotros. Este movimiento aparente de los objetos celestes es causado por el movimiento de rotación de la Tierra entorno a un eje imaginario llamado **eje de rotación**. La rotación de la Tierra también da origen al día y la noche. Dependiendo de nuestra posición en el planeta, notaremos que la trayectoria este-oeste de estos objetos estará inclinada ligeramente hacia el norte si nos ubicamos en el hemisferio sur, y al sur, si nos ubicamos en el hemisferio norte (ver Figura 1.1).

### 1.1.1 Sistemas de coordenadas

Para determinar la posición de un objeto en el cielo, no es necesario conocer la distancia a la cual este se encuentra, sino solo su dirección en el cielo. Esta dirección

---

Imagen de encabezado: Analema de la Luna, curva que describe la posición y fases de la Luna. Crédito: Fotografía de Giorgia Hofer

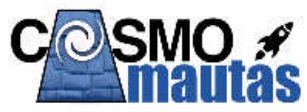

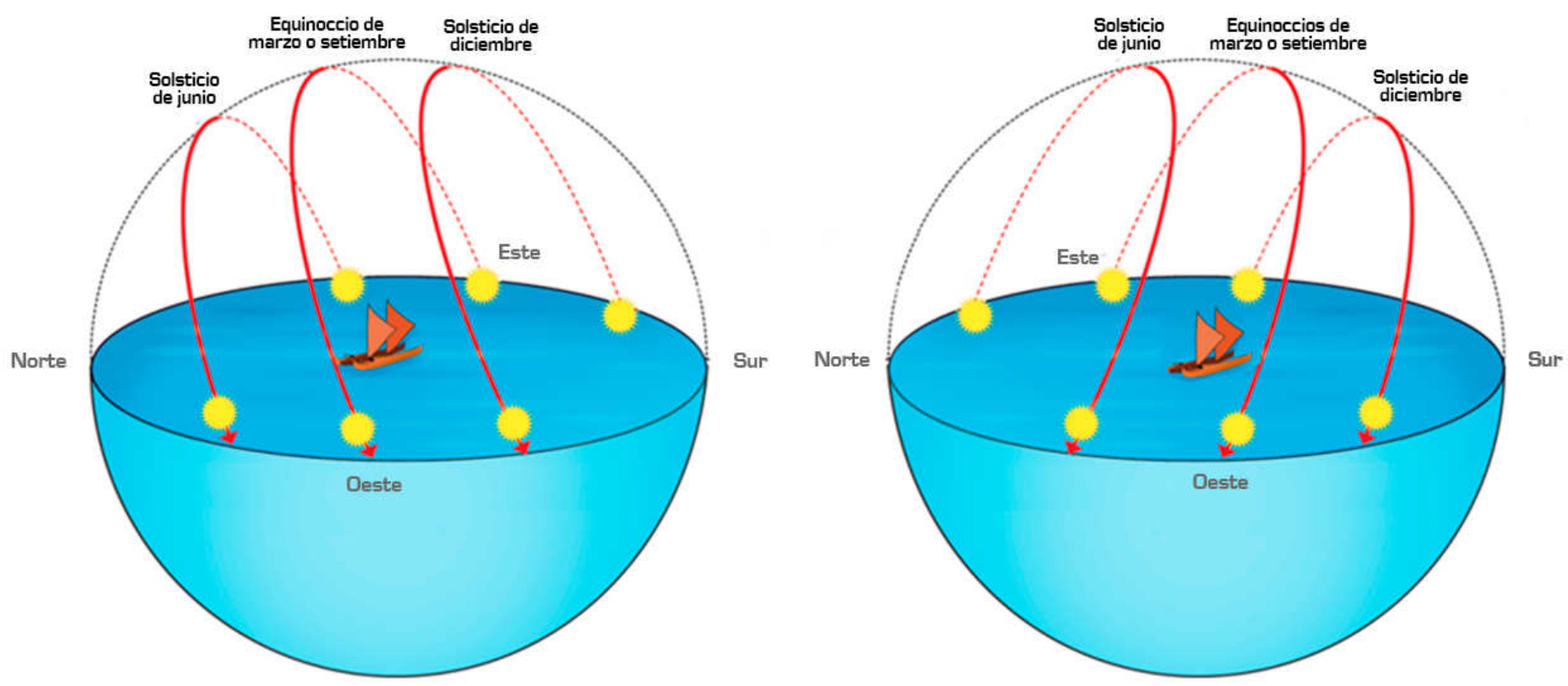

Figura 1.1: El movimiento aparente del Sol en el cielo visto desde el hemisferio sur *(izquierda)* y el hemisferio norte *(derecha)* [Crédito: imagen adaptada de University of Waikato].

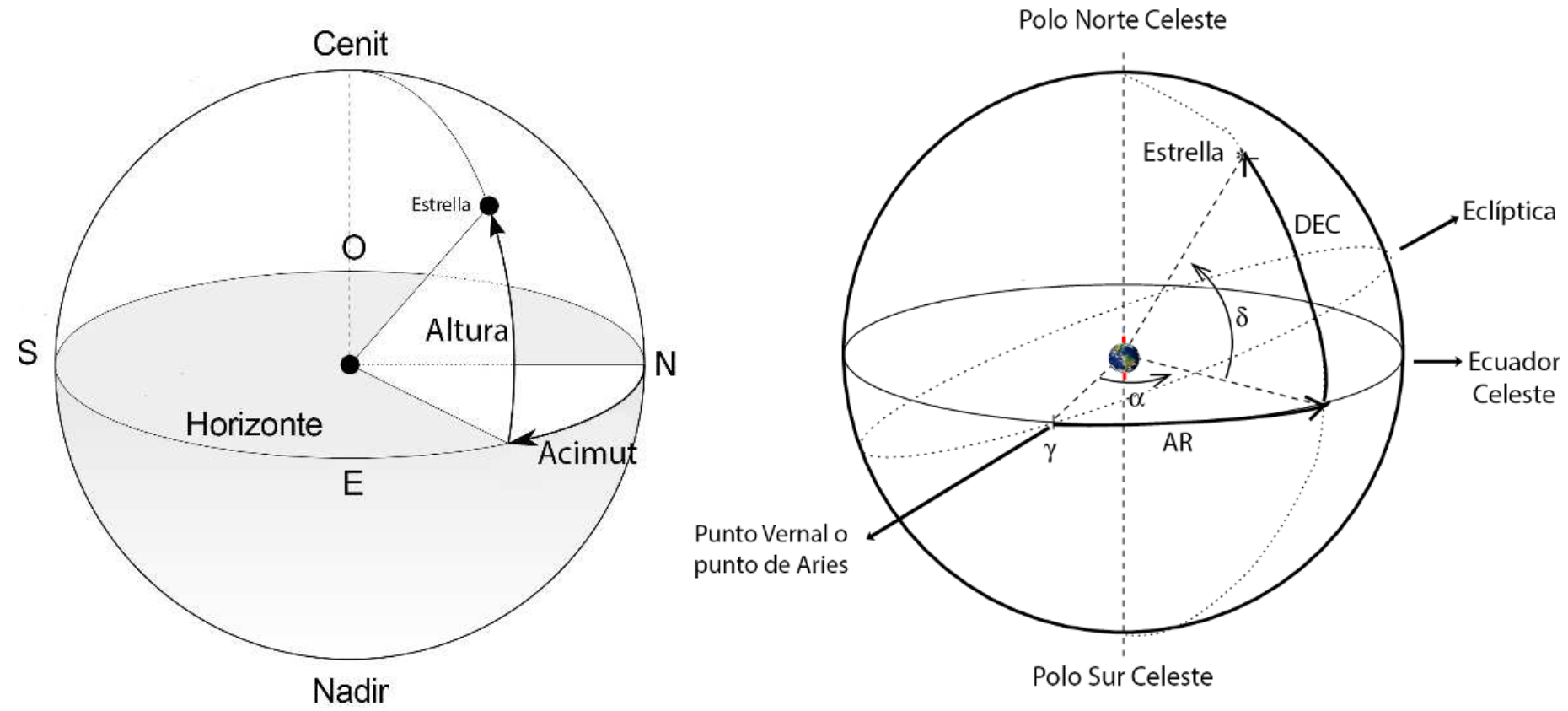

Figura 1.2: *(Izquierda)* Sistema de coordenadas horizontal [Crédito: imagen adaptada de TW Carlson]. *(Derecha)* Sistema de coordenadas ecuatorial [Crédito: imagen adaptada de Creative Commons].

se puede obtener empleando un modelo denominado **esfera celeste**, que es una esfera hipotética de radio arbitrario centrada en el observador, y en cuya superficie están ubicados todos los objetos del cielo (planetas, estrellas, galaxias, etc.). La esfera celeste puede ser utilizada para establecer diversos sistemas de coordenadas, entre ellos, el sistema de coordenadas horizontal y el ecuatorial[**1**, **2**].

De acuerdo al panel izquierdo de la Figura 1.2, para un observador en Tierra notará que las salidas y puestas de Sol no se dan exactamente por los puntos cardinal este y oeste respectivamente, pues notará que a lo largo del año estas van cambiando. Por ejemplo, para un observador en el hemisferio sur o el hemisferio norte las únicas dos ocasiones en las que el Sol sale exactamente por el Este y se oculta por el Oeste sucede en los equinoccios y el resto del año varía.

### Sistema de coordenadas horizontal (altacimutal)

Se trata de un sistema de coordenadas local que posee dos coordenadas:

- **Altura (h)**. Es un ángulo vertical contenido en un plano perpendicular al plano horizontal del observador, y varía en el intervalo de -90° a 90°. Es medido a partir del plano horizontal del observador y toma valores positivos cuando el objeto está por encima del horizonte y negativos cuando está por debajo. Adquiere sus valores extremos cuando el objeto se encuentra en el cenit (90°) o en el nadir (-90°).

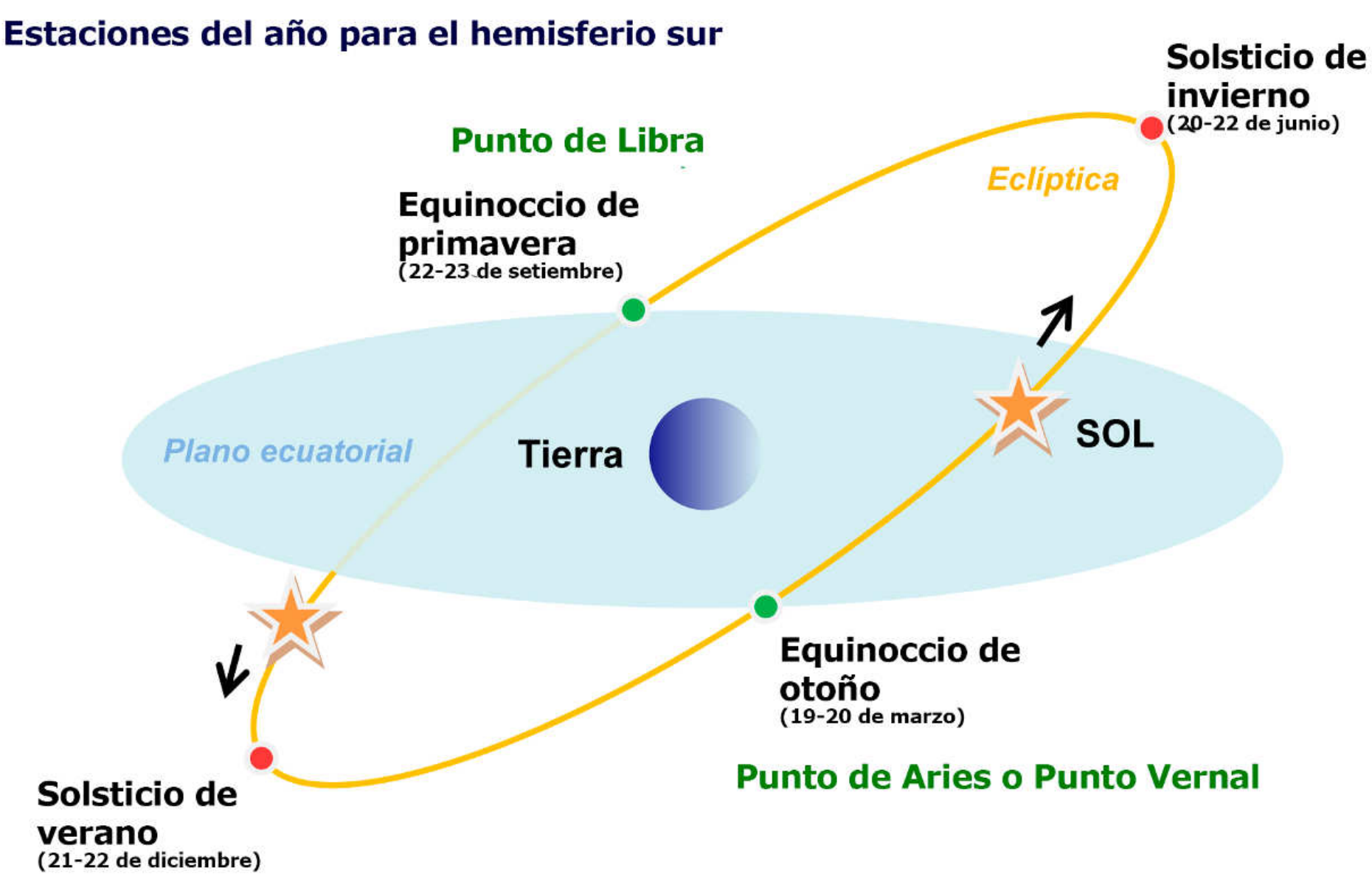

Figura 1.3: Trayectoria del Sol a lo largo del año considerando un sistema de coordenadas ecuatorial [Crédito: Wikipedia/Divad].

- **Acimut (A)**. Es un ángulo medido en el plano horizontal del observador y varía entre 0° y 360°. Generalmente es medido a partir del punto cardinal norte y toma valores positivos cuando es medido en dirección al este. En el hemisferio Sur es medido desde el punto cardinal Sur hacia el oeste.

### Sistema de coordenadas ecuatorial

Para comprender cómo funciona el sistema de coordenadas ecuatorial es necesario considerar un sistema de referencia geocéntrico. De esta manera, el centro de la Tierra coincidirá con el origen de coordenadas y la prolongación del eje de rotación terrestre determinará la posición del Polo Norte y Sur celeste en el modelo de la esfera celeste. Asimismo, la prolongación de la línea ecuatorial dará origen al plano del Ecuador celeste (ver Figura 1.2, derecha).

**La eclíptica** es la trayectoria que sigue el Sol en la esfera celeste considerando un sistema de referencia geocéntrico (ver Figura 1.3). Otro componente importante en este modelo es el Punto de Vernal o **Punto de Aries**. Como se puede ver en la Figura 1.3, el Punto de Aries corresponde al nodo ascendente de la intersección del plano del Ecuador celeste y el plano de la eclíptica, es decir, el punto en el que el Sol pasa del hemisferio sur celeste al hemisferio norte celeste). El Punto de Aries permanece fijo en una escala corta de tiempo (años o décadas). Sin embargo, en una escala larga (cientos de años) su posición varía respecto al fondo de estrellas debido a la precesión del eje de rotación terrestre.

Con estas definiciones, ya podemos establecer las coordenadas que componen el sistema de coordenadas ecuatorial. Estas coordenadas son:

- **Ascensión Recta** ($\alpha$). Es un ángulo, medido en el plano del Ecuador celeste,

partiendo desde el Punto Vernal hasta el meridiano que corresponde al objeto observado. Su valor puede ser expresado en horas (de 0h a 24h), de manera que 24h corresponde a 360º.

- **Declinación** ($\delta$). Es un ángulo contenido en un plano perpendicular al plano del Ecuador celeste, que parte del Ecuador celeste hasta el paralelo correspondiente al objeto observado. Su valor varía entre -90° y 90°.

### 1.1.2 Constelaciones

Otra forma de ubicarse en el cielo es usando las posiciones de las estrellas. Para esto, en la antigüedad se asociaron conjuntos de las estrellas más brillantes de la bóveda celeste para crear figuras imaginarias de criaturas o seres mitológicos. Además de ser una herramienta para la ubicación, las constelaciones reflejan las creencias, tradición y cultura de los pueblos que las crearon.

**Historia 1.1.1: Constelaciones incas**

La cultura Inca alcanzó un gran desarrollo en distintas áreas del conocimiento, gran parte de estos conocimientos fueron heredados de las culturas previas que se desarrollaron en el territorio peruano. En el aspecto astronómico los incas fueron grandes observadores del cielo, lograron medir el paso del tiempo empleando para ello los movimientos cíclicos del Sol, la Luna y las estrellas. Existen dos tipos de constelaciones Incas: las oscuras y las brillantes. Sus observaciones de las salidas y puestas del Sol les permitieron conocer las fechas de los solsticios y equinoccios [**3**, **4**, **5**], y con ello, predecir las épocas de llegada de lluvia y sequía. Las observaciones solares también fueron importantes para establecer el calendario inca. Cronistas como Juan Betanzos y Pedro Cieza de León sugieren que el año (guata) estuvo basado en el ciclo solar y lunar, con una duración de 360 días debido al ciclo solar y doce meses de 30 días, debido al ciclo lunar [**3**, **6**]. Dado que el año no tiene 360 días, es probable que el año inca fuera reajustado al final de cada ciclo agregando cierto número de días para compensar la diferencia. En la imagen se muestran posibles marcadores de horizonte conocidas como *saywas* o *sukanqas*.

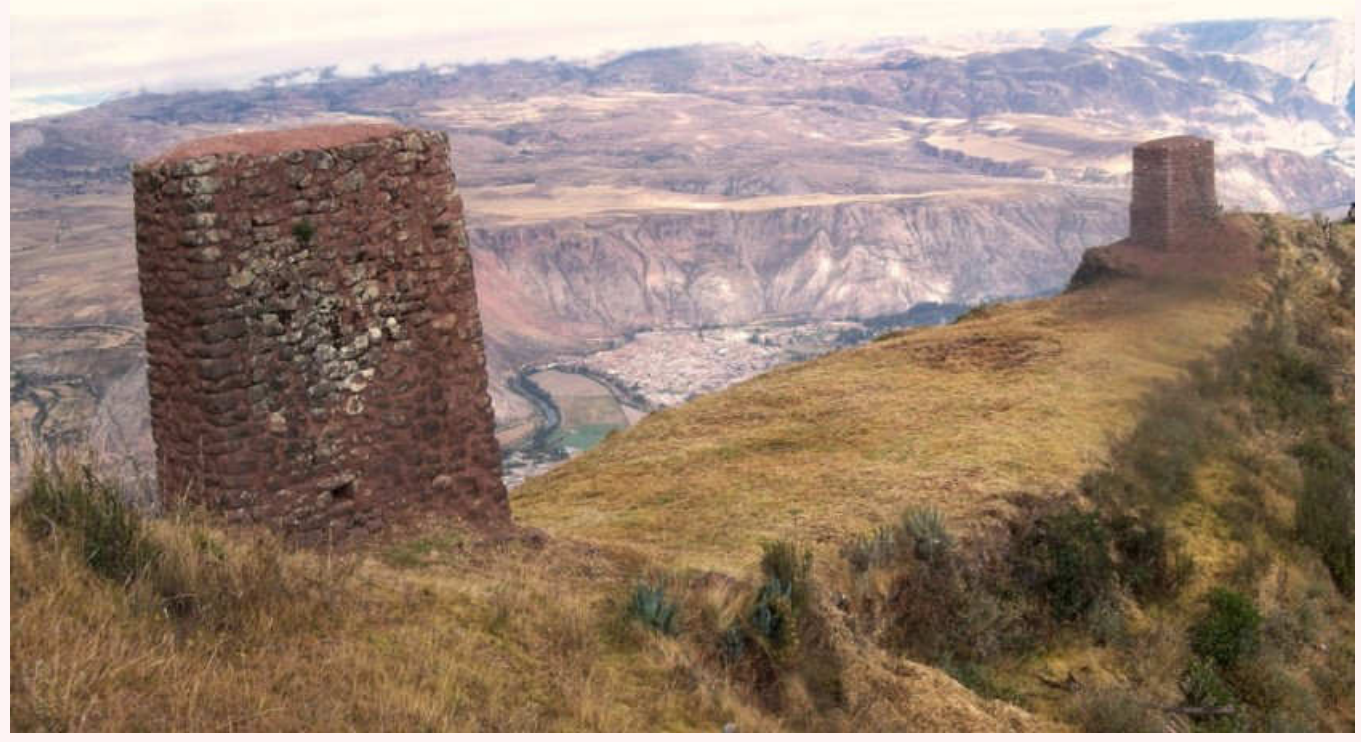

Figura 1.4: Posibles marcadores de horizonte conocidas como *saywas* o *sukanqas* [Crédito: imagen extraída de Salazar [**7**]]

- **Constelaciones incas oscuras.** Se ubican en las regiones oscuras de la Vía Láctea, regiones donde el polvo y el gas molecular bloquean la luz de las estrellas de fondo. De esta manera se produce un contraste entre estas regiones oscuras y con el resto de estrellas que conforman nuestra galaxia. Entre las principales constelaciones oscuras tenemos a: *Yakana* (llama negra), Uña Llama (llama bebe), *Atoq* (zorro), *Michiq* (pastor), *Lluthu* (perdiz andina), *Han'patu* (sapo) y *Mach'aqway*(serpiente).

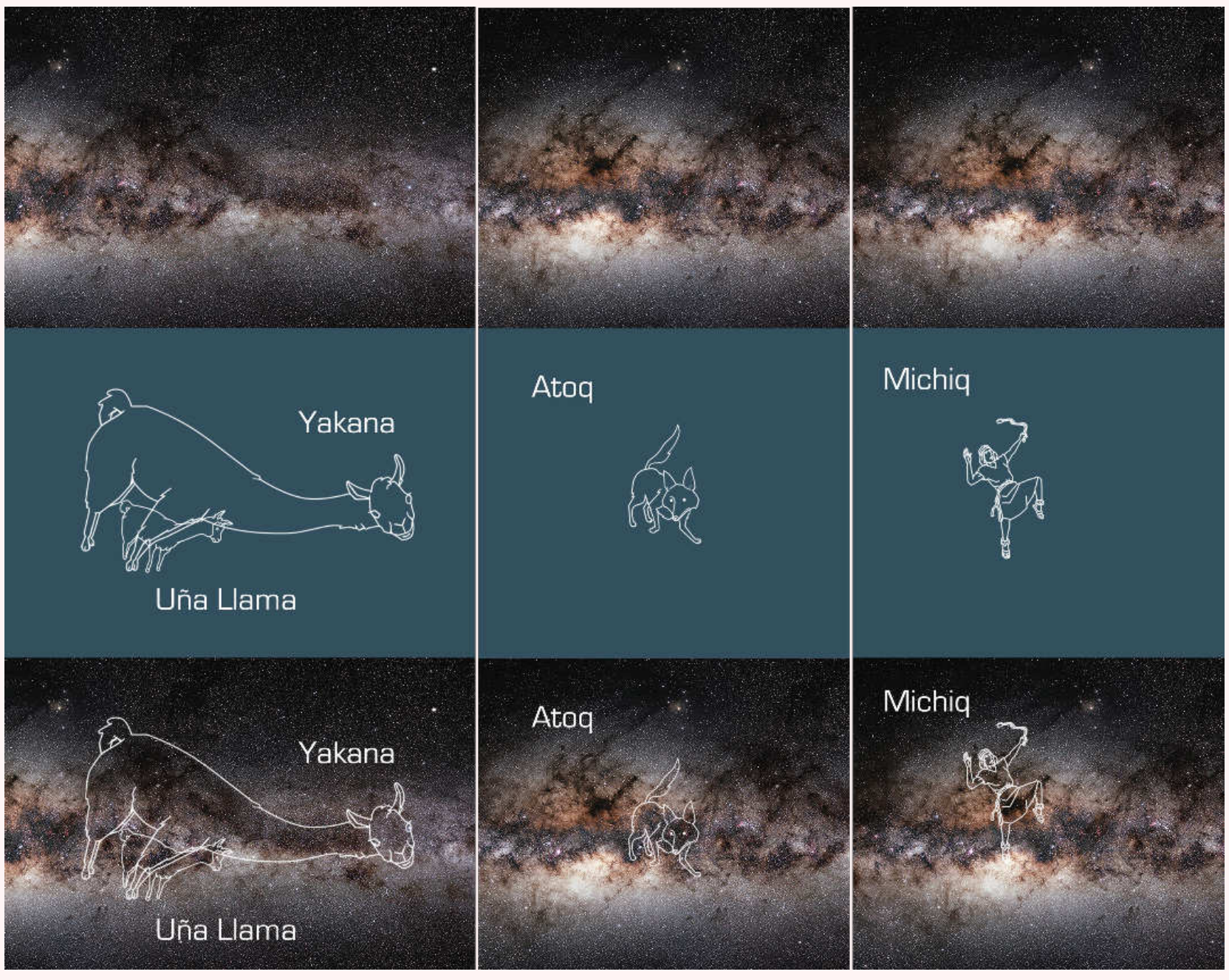

Figura 1.5: ***Yakana* (Llama negra)**, es una de las constelaciones más importantes del mundo inca. Existen diversos escritos que hablan de su existencia, por ejemplo el Manuscrito de Huarochirí. Es una constelación grande que se puede apreciar bajo cielos oscuros. Los ojos de la llama están representados por las estrellas Alfa y Beta Centauri. Según Salazar [7]. ***Atoq* (Zorro)**, es una zona oscura que se encuentra ubicado entre las dos llamas (madre y cría) y el *Michiq* (el pastor). Existen diversas historias sobre este intrépido animal. Una de ellas la encontramos en Salazar [7] y nos explica por qué el zorro tiene la cola de color oscuro. "El zorro se convirtió, mediante un encantamiento, en un tronco de huarango por causa de una venganza del gallinazo, su viejo enemigo. Un campesino lo encontró en su chacra y se lo llevó a casa para usarlo como tranquera; cada noche, en la oscuridad, el zorro rompía su encantamiento y asolaba la vecindad hasta que al amanecer volvía a adoptar la forma de tronco. El labriego al darse cuenta del engaño arrojó el tronco al fuego que al empezar a arder rompió el encanto y el zorro huyó despavorido con la cola chamuscada." Por historias como ésta, el zorro es un personaje reconocido del ande que terminó siendo perennizado en el cielo. ***Michiq* (Pastor)**, esta constelación se encuentra encima de la constelación de Sagitario muy cerca del Centro Galáctico. Tiene la forma alargada y representa un pequeño niño pastor que va arrojando piedras con su *warak'a*, tratando de alejar al *Atoq* de la cría de la llama (Salazar [7]). En otros lugares del Perú, como en el lago Titicaca, los campesinos ven en esta región del cielo a un balsero, que lleva en la mano un palo muy largo con el cual guía a su embarcación. [Crédito: José Ricra & Josué Maguiña.]

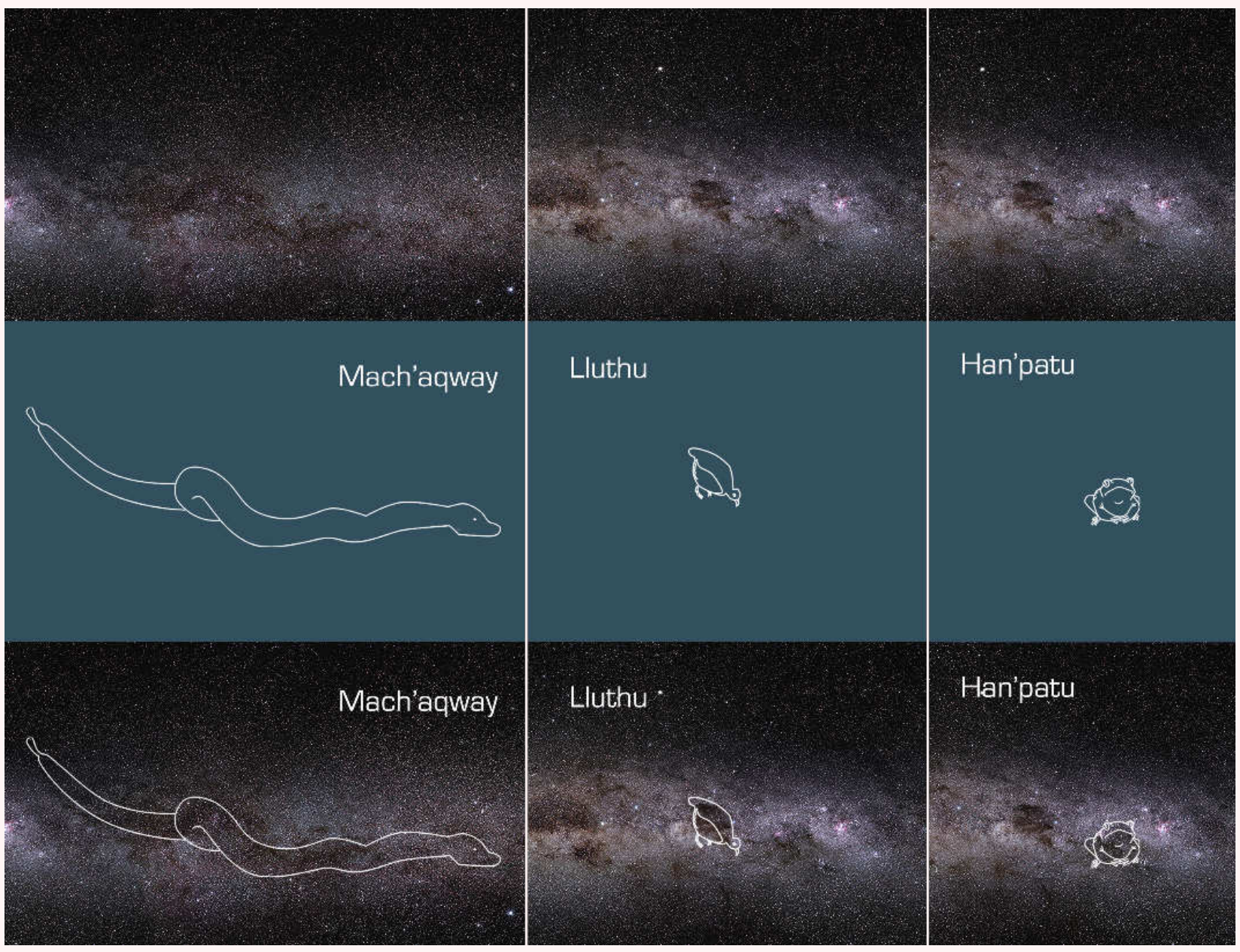

Figura 1.6: ***Mach'aqway*** **(Culebra)**, esta constelación oscura en forma de serpiente, un poco más tenue que las demás. Se encuentra entre las constelaciones de Vela y Puppis. Es un poco más difícil de observar; según el cronista Fray Martín de Morúa, *Mach'aqway* es aquella constelación que tiene a su cargo las serpientes, y era venerada para evitar que estos animales hicieran daño a las personas. ***Lluthu*** **(Perdiz)**, es un ave que representaría la perdiz andina (*Nothoprocta Pentladii*), denominada en algunas regiones como *Chawa* o *P'isaqa* [**7**]. Se puede ubicar fácilmente junto a la Cruz del Sur, en el espacio conocido como Saco de Carbón. Se consideraba a *Lluthu* como un ave sagrada, y por ello, era muy utilizada en los sacrificios. ***Han'patu*** **(Sapo)**, esta constelación se ubica muy cerca de la constelación de la Cruz del Sur en la constelación de Carina. Su orto helíaco se produce a mediados del mes de octubre, por lo cual, era tomado como un anunciante de las primeras descargas de lluvia [**7**]. ***Michiq*** **(Pastor)**, esta constelación se encuentra encima de la constelación de Sagitario muy cerca del centro galáctico. Tiene la forma alargada y representa un pequeño niño pastor que va arrojando piedras con su *warak'a*, tratando de alejar al *Atoq* de la cría de la llama [**7**]. En otros lugares del Perú, como en el lago Titicaca, los campesinos ven en esta región del cielo a un balsero, que lleva en la mano un palo muy largo con el cual guía a su embarcación. [Crédito: José Ricra & Josué Maguiña.]

- **Constelaciones incas brillantes.** Estaban conformadas por estrellas. Entre ellas tenemos a la *Qollqa* (almacén o racimo de estrellas) y *Urkuchillay* (la llamita de plata).

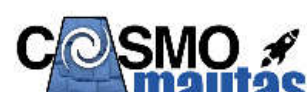

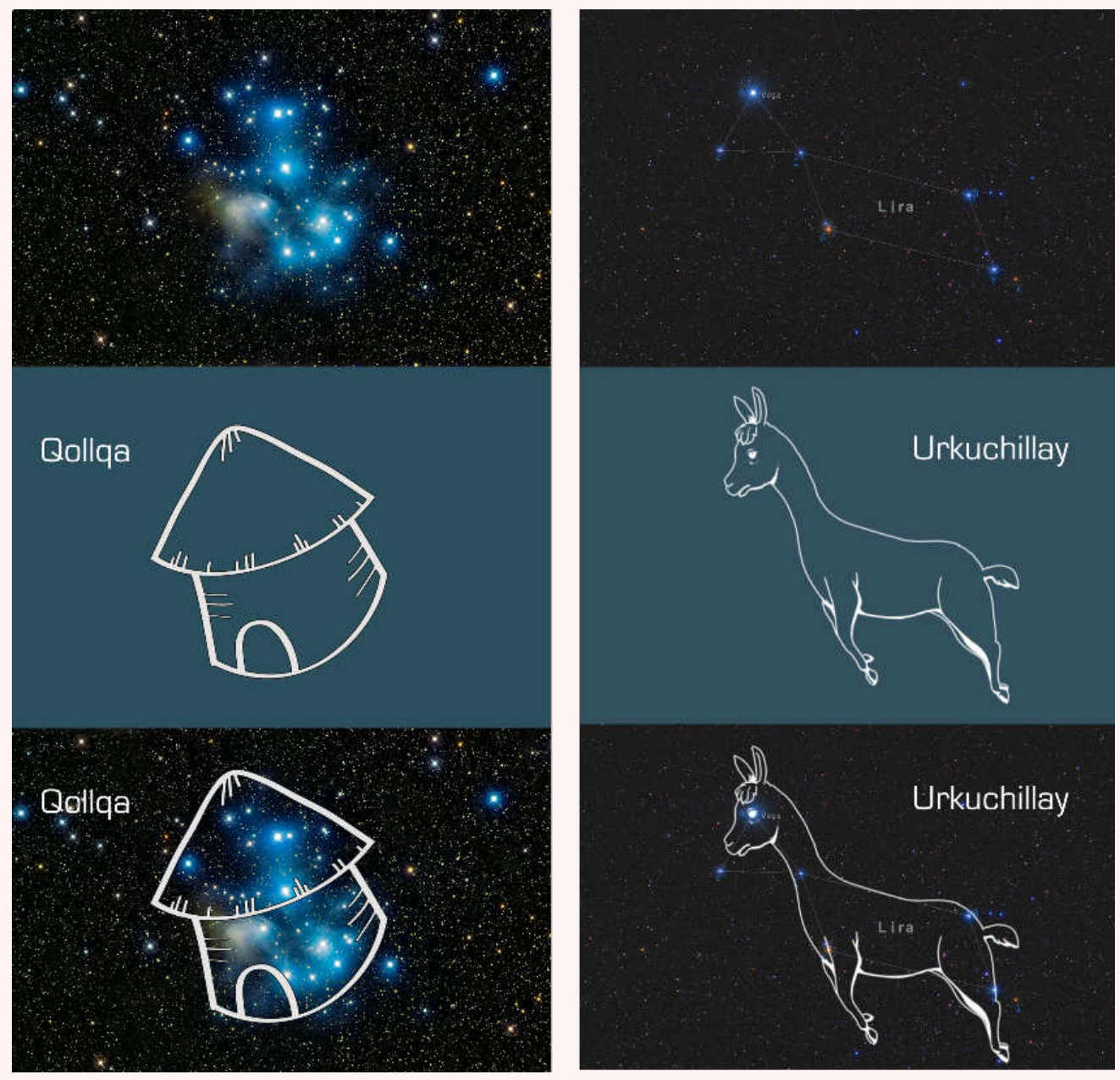

Figura 1.7: ***La Qollqa*** **(Las Pléyades)**, es una agrupación de estrellas que forman un cúmulo abierto. En occidente es denominado como las Pléyades o M45 (objeto 45 del catálogo de Messier). En países de habla hispana también es conocido como “las 7 cabrillas”. La *Qollqa* era considerada por los incas como la madre de todas las estrellas, era una constelación tan importante que se le consideraba una *wak’a* del cielo; es decir, una entidad sagrada, como también lo eran los templos, las lagunas o las montañas. Se piensa que su nombre está relacionado con la agricultura, pues en el idioma quechua la palabra *Qollqa* hace referencia a un almacén, granero o depósito. En otros lugares la conocen como Qoto, que significa puñado, manojo o racimo de tubérculos, debido al número de estrellas que la conforman. ***Urkuchillay*** **(Lira)**, esta constelación ocupa el lugar en el cielo que en occidente es conocido como la constelación de la Lira (un instrumento musical). Tiene como característica interesante que una de sus estrellas es Vega, una de las estrellas más brillantes del cielo. Se le consideraba ente protector de los auquénidos bebés y fue de gran veneración para los pastores y criadores de llamas. [Crédito: José Ricra & Josué Maguiña.]

Hoy en día sabemos que las constelaciones son solo una percepción óptica debido a que percibimos el cielo en dos dimensiones. En realidad, las estrellas que la conforman no necesariamente están relacionadas espacialmente si consideramos la profundidad. Es decir, en una constelación, las estrellas pueden encontrarse por ejemplo a millones de años luz una de la otra (ver Figura 1.8). En 1929 la Unión Astronómica Internacional (IAU, por sus siglas en inglés) definió a las constelaciones como mapas o secciones de la esfera celeste [**8**]. En esta asamblea se dividió el cielo en 88 sectores o constelaciones, en donde en su mayoría se conservaron los nombres provenientes de la cultura griega.

En el sistema de coordenadas ecuatorial, el Sol recorre el plano de la eclíptica y en su recorrido anual el astro rey tiene de fondo a las **constelaciones del Zodiaco** (Ver Figura 1.9). En nuestros días las constelaciones del Zodiaco son muy conocidas y utilizadas en la Astrología (pseudo-ciencia cuyo origen se remonta a los inicios de la Astronomía, pero no cuenta con validación científica ni poder explicativo), para la cual estas son 12. Sin embargo, el recorrido del Sol no es el mismo que en el 3000 o 2000 A.C. y actualmente no solo recorre estas 12 secciones o constelaciones, sino que su recorrido es a través de 13 secciones o constelaciones, adicionando la la constelación de Ofiuco como nueva constelación. Por lo que podemos decir que para la Astronomía existen 13 constelaciones Zodiacales (Figura 1.9).

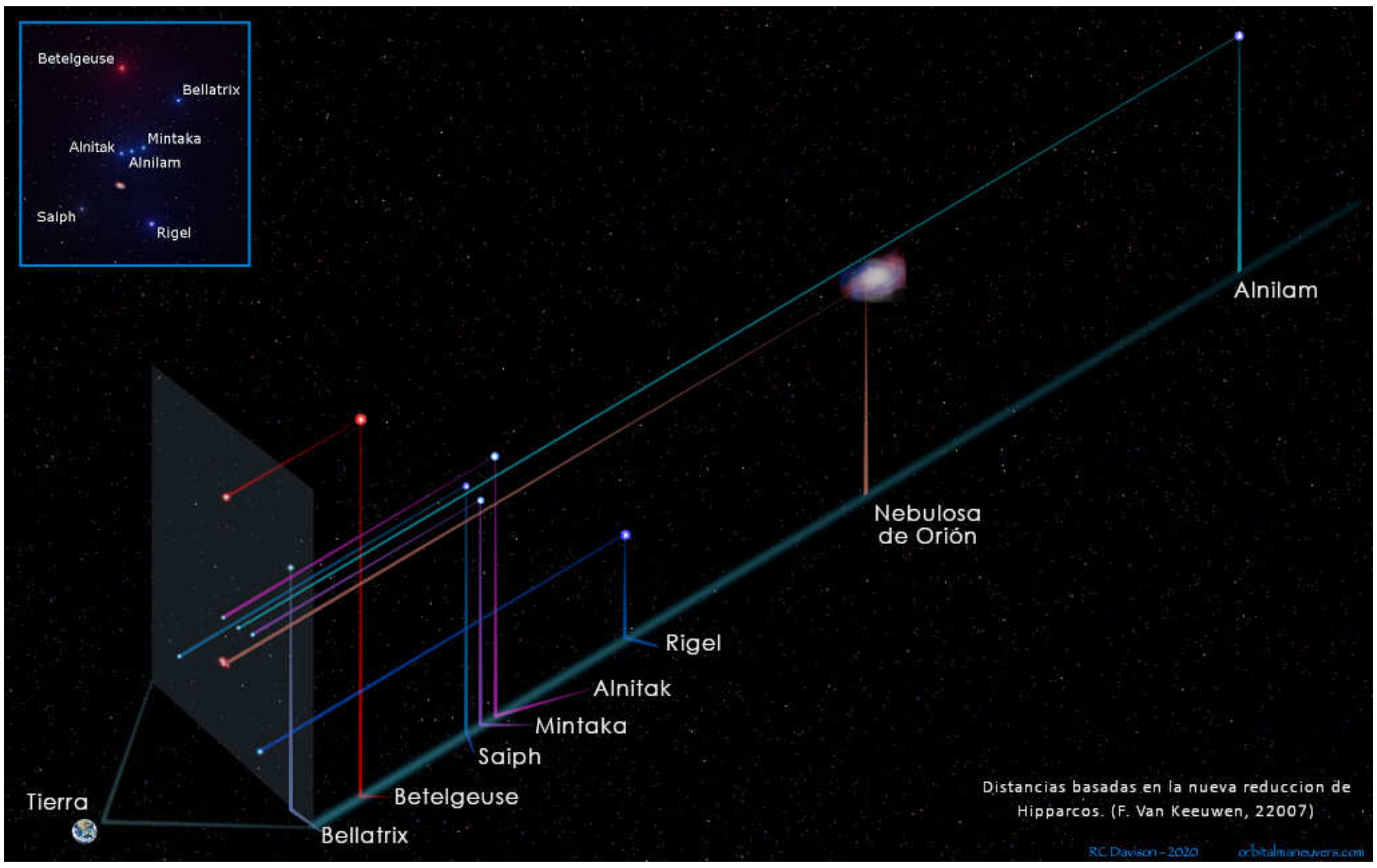

Figura 1.8: Se observa que las estrellas de la constelación de Orión se encuentran muy distantes unas de las otras [Crédito: adapatado de Ronald Davison, `orbitalmaneuvers.com`].

**Recurso TIC 1.1.1:**

*Stellarium* es un software astronómico que muestra los objetos visibles del cielo desde tu ubicación y de acuerdo a la hora de observación. Ingresa a su versión web `https://stellarium-web.org/` y descubre qué constelaciones son observables en estas épocas del año.

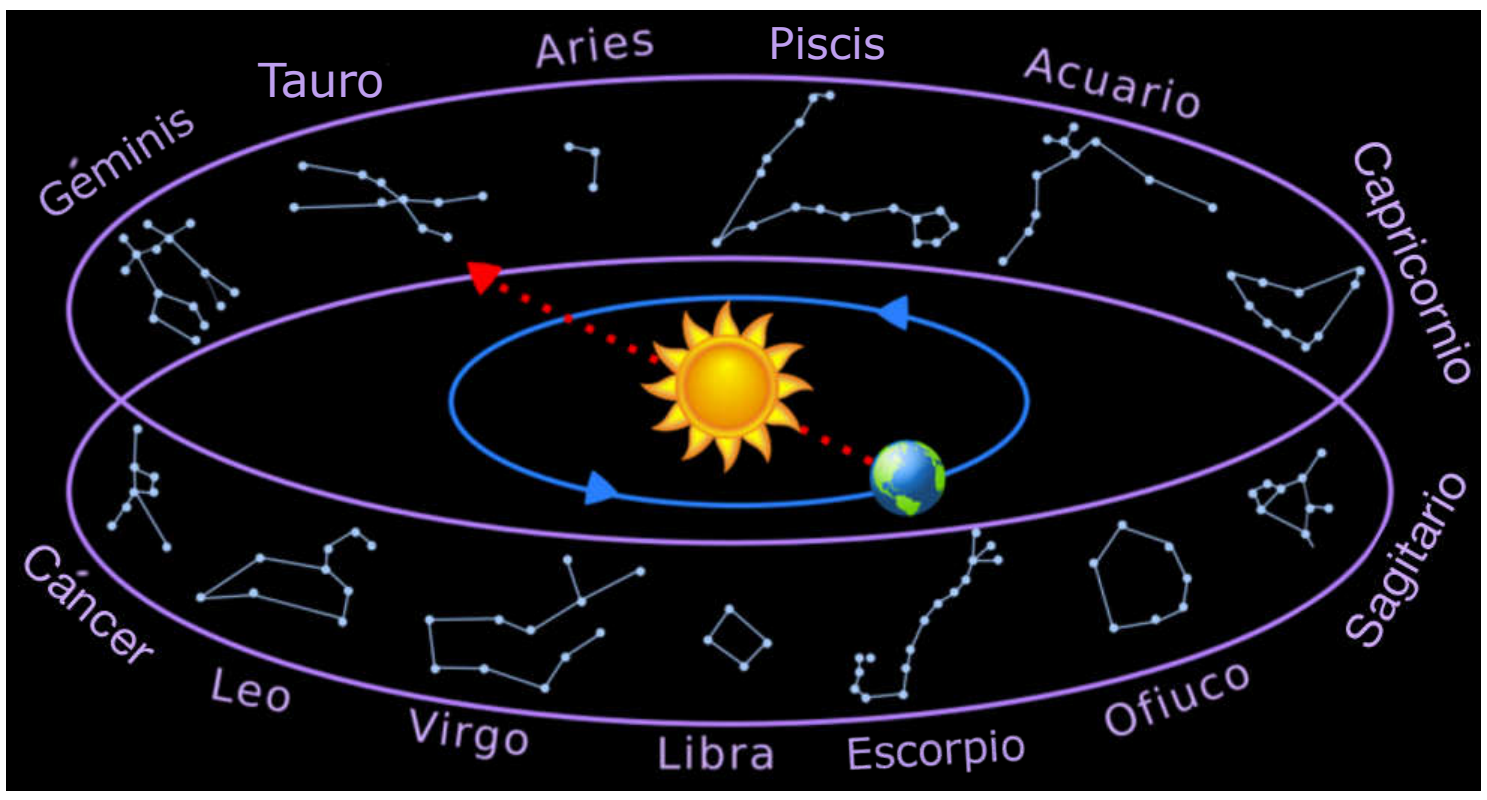

Figura 1.9: Las 13 constelaciones zodiacales. A diferencia de las 12 constelaciones zodiacales antiguas, actualmente el recorrido del Sol incluyen también a la constelación de Ofiuco [Crédito: imagen adaptada de Germán Martínez Gordillo].

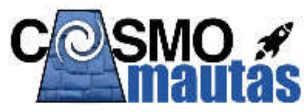

## 1.2 Los movimientos de la Tierra

La Tierra tiene cuatro diferentes movimientos (ver Figura 1.10), que ocurren simultáneamente. Los movimientos principales y de mayor impacto en nuestras vidas son el de traslación, en su órbita alrededor del Sol, y el de rotación, alrededor de su propio eje. Además de éstos, la Tierra presenta también movimientos más sutiles que son el de precesión y nutación, los cuales producen cambios en escalas de tiempo muy largas, imperceptibles para la vida humana.

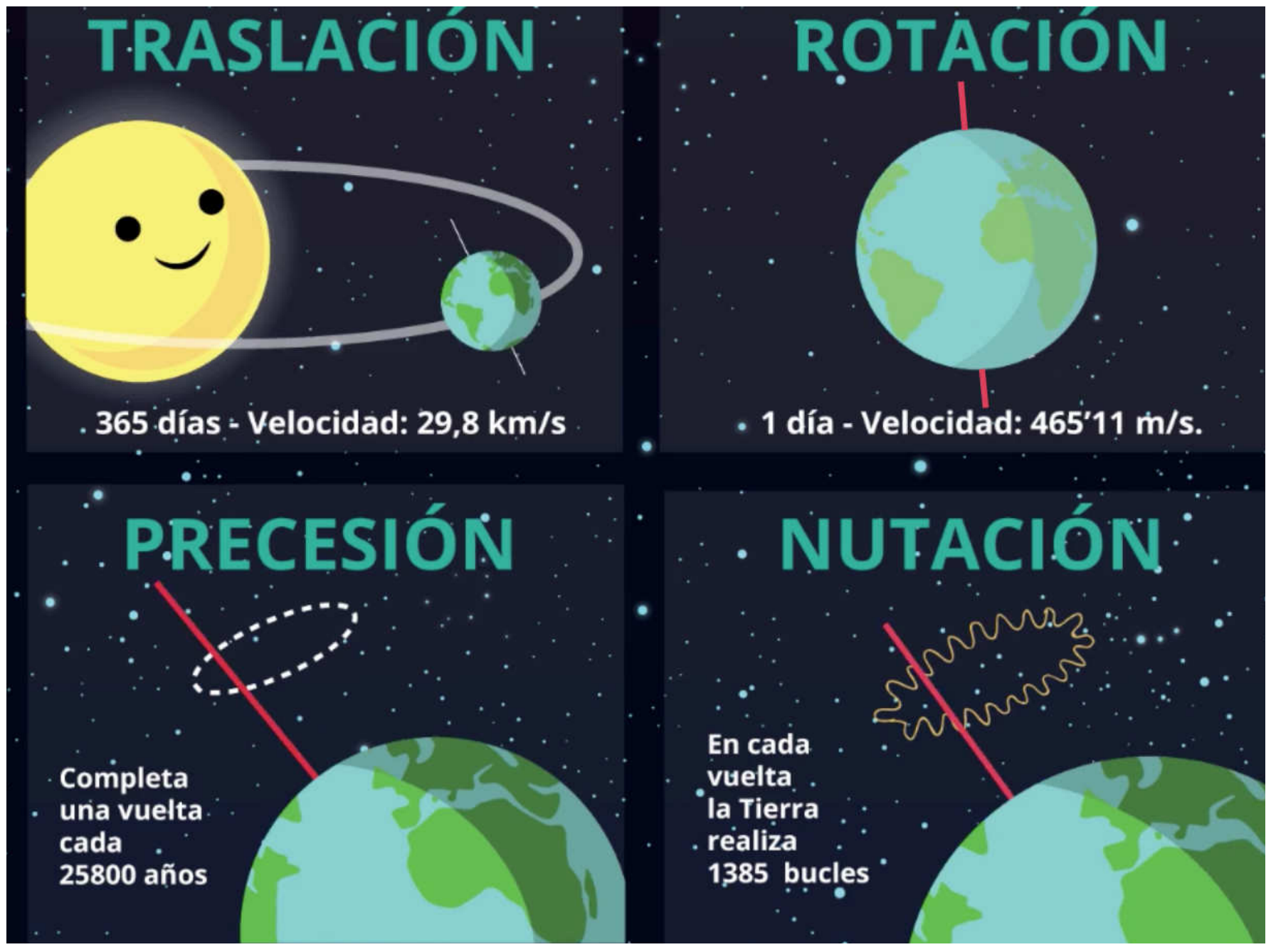

Figura 1.10: Los cuatro movimientos de la Tierra [Crédito: imagen adaptada de `Astroaficion.com`]

**El movimiento de rotación** origina el día y la noche, y el movimiento aparente de todos los objetos celestes. La Tierra completa una rotación sobre su eje en 23 horas, 56 minutos y 4 segundos.

**El movimiento de traslación** es el movimiento de la Tierra en su órbita alrededor del Sol. Esta órbita posee una excentricidad de 0.0167, lo que indica que presenta una órbita casi circular [**16**]. Esto quiere decir que la distancia entre el Sol y la Tierra prácticamente no cambia durante el año. Esta distancia es de aproximadamente 150 millones de kilómetros, y es usada también como una unidad básica de distancia en la astronomía denominada una Unidad Astronómica (UA). La órbita de la Tierra tiene un periodo de un año, y es este movimiento lo que define nuestro concepto de año.

Durante su traslación alrededor del Sol, la Tierra mantiene su rotación entorno a un eje inclinado (Figura 1.11). Esta inclinación forma un ángulo de 23°26' con el plano de la eclíptica. Esta inclinación del eje terrestre, en combinación con la traslación alrededor del Sol, es lo que da el origen a **las estaciones del año**: verano, otoño, invierno y primavera. Es muy importante resaltar que las estaciones del año no tienen ninguna relación con la distancia de la Tierra al Sol, ya que la órbita de la Tierra es casi circular, la distancia de la Tierra al Sol *no* varía significativamente a lo largo del año.

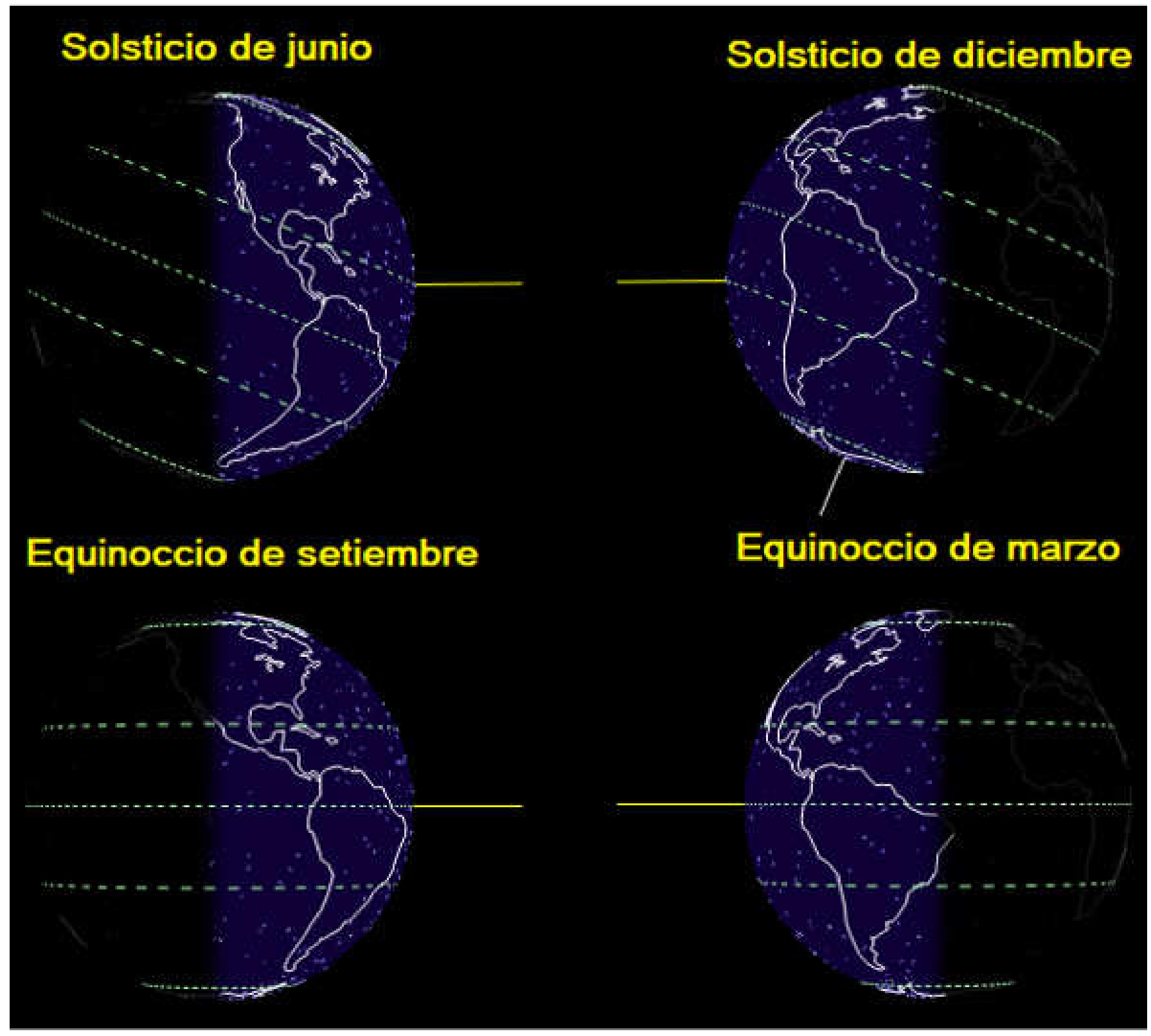

Figura 1.11: Movimiento de traslación de la Tierra en torno al Sol y las estaciones del año, las cuales se originan debido a la inclinación (23°26') del eje de rotación de la Tierra. Se observa que durante los solsticios las regiones polares están o completamente iluminadas o en la oscuridad. [Crédito: Ramiro Moro]

A su vez, esta inclinación ocasiona que, en el sistema de coordenadas ecuatoriales, la intersección del plano de la eclíptica con el plano del ecuador de la esfera celeste no coincidan, sino que formen el ángulo de 23°26' (ver Figura 1.3). En los puntos de intersección de estos dos planos (punto de Aries y punto de Libra) ocurren los **equinoccios** (otoño y primavera). En estas fechas la duración del día y la noche son iguales. Por otro lado, los **solsticios** (invierno y verano) son fechas en las que el Sol tiene su máxima separación del plano del ecuador, haciendo que en verano la duración del día sea más largo y las noches más cortas, mientras que en invierno los días son más cortos y las noches son más largas. Este efecto es especialmente notorio en localidades lejanas al Ecuador y cercanas a los polos. En el Perú, por encontrarse cercano al Ecuador, esta diferencia no es tan marcada.

## 1.3 La atmósfera terrestre

Toda la radiación electromagnética que nos llega desde el espacio exterior (la luz del Sol, la Luna, las estrellas, las galaxias, etc.) interactúa con la atmósfera terrestre antes de llegar a nosotros, los observadores. Por ello, conocer nuestra atmósfera y sus propiedades es fundamental para la astronomía. La atmósfera es la capa más externa de la Tierra, la cual está constituida principalmente por nitrógeno (78 %), oxígeno (21 %) y una pequeña cantidad de gases como el argón, el dióxido de carbono, el helio y el neón. El vapor de agua y el polvo también son parte de la atmósfera terrestre [**9**].

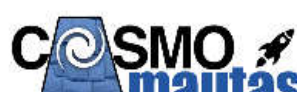

### 1.3.1 Capas de la atmósfera

10000 KM
Exósfera
SATELITE JASON-3
1280 KM
SATELITE SUOMI NPP
800 KM
Termósfera
700 KM
ESTACION ESPACIAL INTERNACIONAL
TRANSBORDADOR ESPACIAL
AURORA
320 KM
LINEA DE KARMAN
100 KM
Mesósfera
80 KM
METEOROS
Estratósfera
50 KM
AVIONES JET COMERCIALES
Tropósfera
11 KM
NOAA Satellite and Information Service
www.nesdis.noaa.gov

Figura 1.12: Capas de la atmósfera [Crédito: imagen modificada de NOAO].

**Tropósfera.** La tropósfera es la capa más baja de la atmósfera (hasta 6-10 km) donde se desarrollan los diferentes sistemas climáticos debido a que contiene casi todo el vapor de agua de la atmósfera. Con el incremento de la altitud la densidad del aire disminuye y con ello la temperatura. Es por eso que las cimas de las montañas suelen ser mucho más frías que las zonas costeras.

**Estratósfera.** Debido a que en esta capa los vientos son horizontales, la poca turbulencia permite que los aviones puedan volar por esta capa de la atmósfera. Dentro de la estratosfera existe una capa denominada **capa de ozono**, la cual posee una considerable concentración de ozono. Esta capa es importante para la vida porque absorbe la radiación ultravioleta proveniente del Sol y evita que llegue a la superficie de la Tierra. Al mismo tiempo, este proceso de absorción ocasiona un incremento de la temperatura con la altitud en la estratosfera.

**Mesósfera.** Los meteoros que atraviesan la atmósfera terrestre son visibles en la mesósfera. Éstos son partículas del tamaño de un grano de arena que despiden luz debido a la fricción que se produce con los gases y partículas que componen la mesosfera. La mayoría de estos objetos se queman y no llegan a alcanzar la superficie sólida de la Tierra. Sin embargo, si se trata de objetos con mayor densidad y tamaño, pueden sobrevivir su paso por la atmósfera y alcanzar la superficie (adquiriendo así el nombre de meteorito).

**Termósfera.** En esta capa de la atmósfera también encontramos objetos artificiales en órbita como el *Telescopio Espacial Hubble* (*HST*, por sus siglas en inglés) y la Estación Espacial Internacional (ISS, por sus siglas en inglés). Debido a la altitud de su órbita, se dice que estos objetos se encuentran en la "órbita terrestre baja".

**Exósfera.** La exósfera es la región más alta de la atmósfera cuya densidad es muy baja y se confunde con la densidad del espacio exterior por la mayor presencia de átomos de hidrógeno y helio.

### 1.3.2 Ventanas atmosféricas

Una de las características más importantes de la atmósfera terrestre es que esta interactúa tanto con la radiación electromagnética proveniente del espacio exterior, así como con la radiación emitida por la superficie terrestre. Un tipo de interacción se denomina **absorción**, y consiste básicamente en que gases como el vapor de agua ($H_2O$), el dióxido de carbono ($CO_2$) y el ozono ($O_3$) logran absorber la radiación correspondiente a ciertas regiones del espectro electromagnético, impidiendo que esta llegue a la superficie terrestre. En contraste, existen otras regiones del **espectro electromagnético** (ver Caja 3.1.3, para más detalles) donde la atmósfera es transparente (poca o ninguna absorción de radiación). Estas bandas de longitud de onda se conocen como "ventanas atmosféricas", ya que permiten que la radiación pase fácilmente a través de la atmósfera hasta la superficie de la Tierra (o salgan al espacio exterior). Como podemos apreciar en la Figura 1.13, las principales regiones del espectro electromagnético que pueden atravesar la atmósfera son la ventana óptica, la ventana de radio y varias ventanas infrarrojas estrechas [**10**, **11**, **12**]. Por esta razón, existen muchos telescopios ópticos y telescopios de radio terrestres, mientras que los telescopios que observan en bandas ultra-violetas y de rayos-X se encuentran principalmente en el espacio, fuera de nuestra atmósfera. Ciertas longitudes de onda infrarrojas se pueden detectar desde la Tierra, pero para observar en el infrarrojo lejano se necesita ir al espacio.

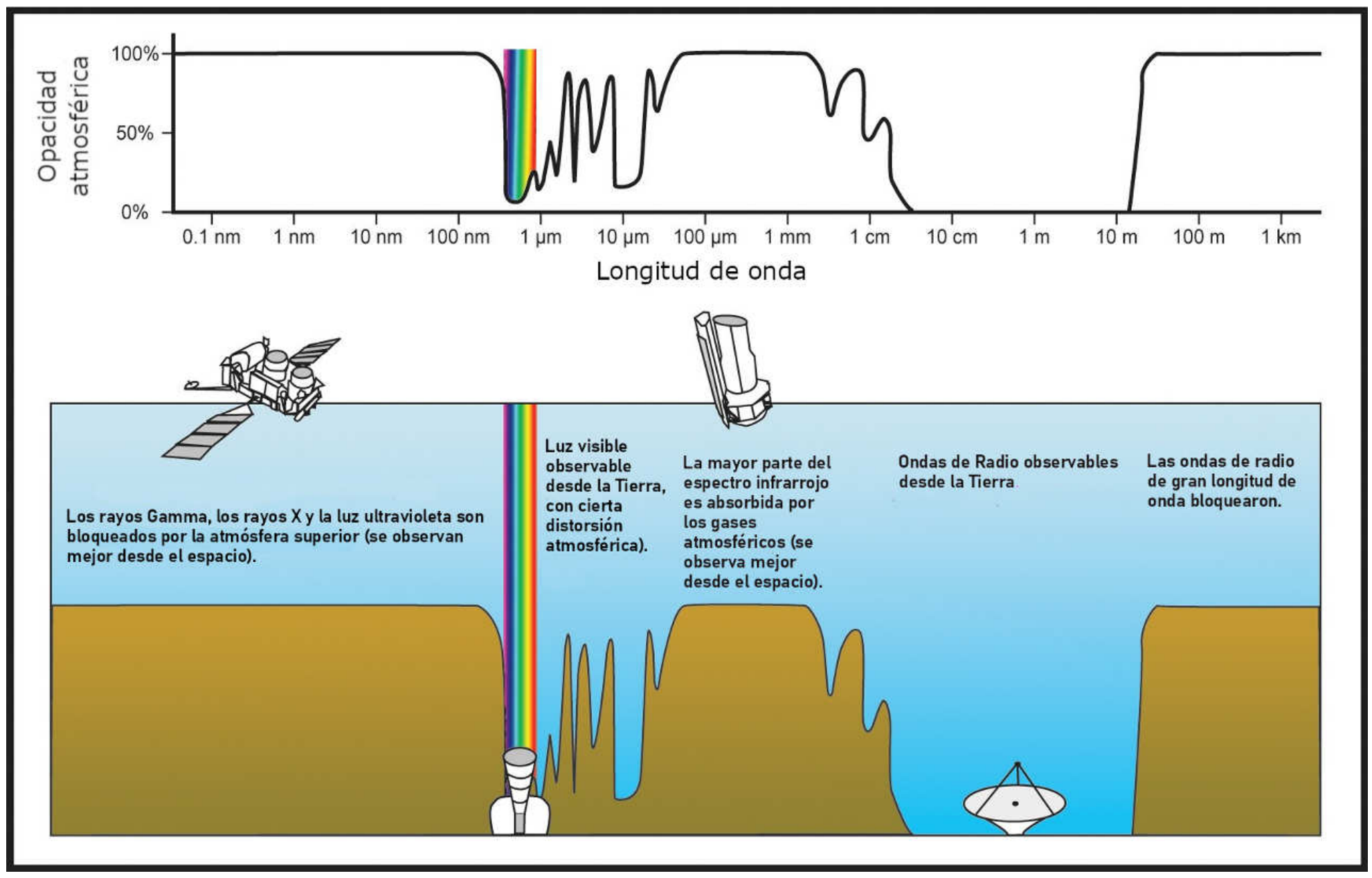

Figura 1.13: Ventanas atmosféricas [Crédito: NASA].

### 1.3.3 El color del cielo

Otro tipo de interacción que tiene la luz con la atmósfera es la **dispersión**. Al incidir la luz solar en la atmósfera, esta interactúa con las partículas y moléculas que la componen, ocasionando que la luz se disperse en todas direcciones. El grado de dispersión está ligado a la longitud de onda del rayo incidente: la luz de longitud de onda más corta se dispersa más que la radiación de longitud de onda larga [**13**, **14**].

**El color azul del cielo.** Cuando el Sol está ubicado en lo alto del cielo, el haz de luz que incide en la atmósfera recorre una distancia más corta para llegar al observador si lo comparamos con la posición del Sol al amanecer o atardecer (ver Figura 1.14). Al interactuar con las moléculas de nitrógeno y oxígeno, la luz es dispersada principalmente en la longitud de onda corta correspondiente al azul y en menor medida en la longitud de onda larga correspondiente al rojo. De esta manera, la luz azul es dispersada por todo el cielo, dando origen a su color característico.

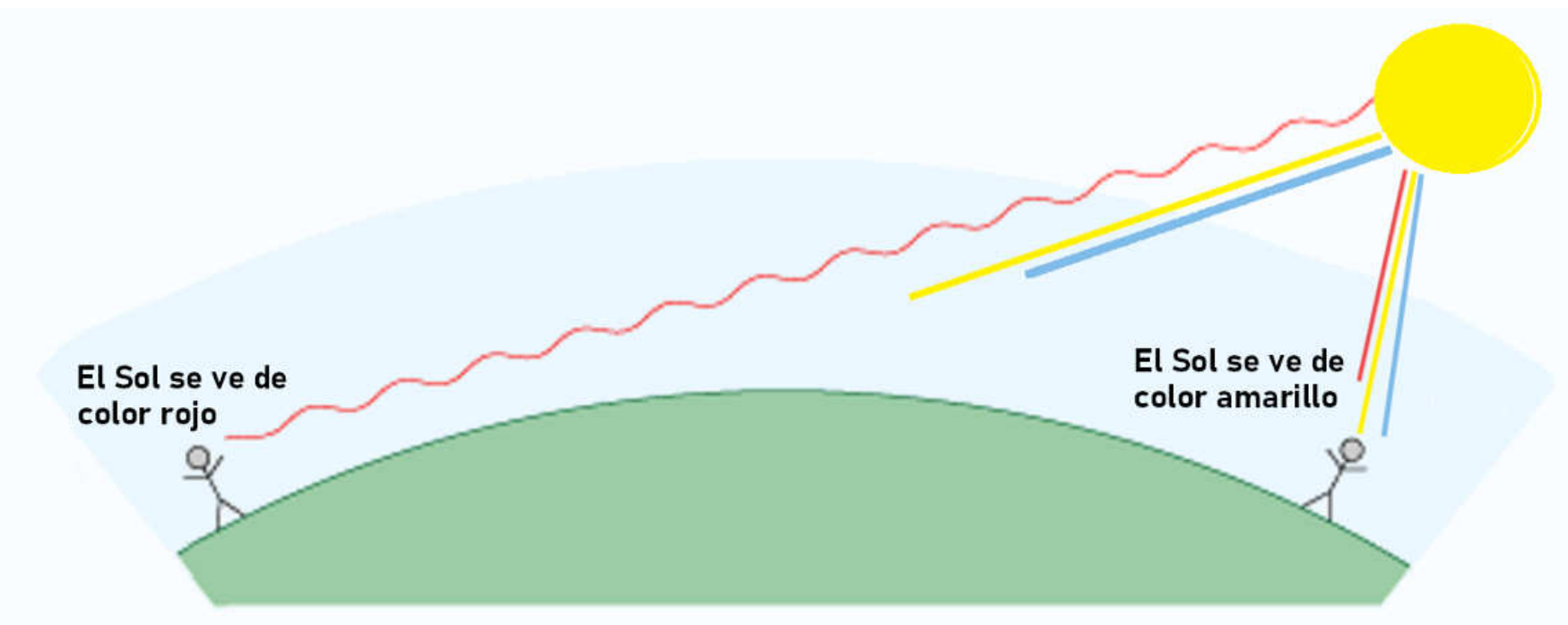

Figura 1.14: Los distintos grados de dispersión dan origen a los distintos colores del cielo. [Crédito: `explainingscience.org`].

**El rojo de los atardeceres.** Al caer el día el Sol se encuentra muy abajo en el cielo, entonces el camino que tiene que recorrer su luz antes de alcanzar al observador es claramente más largo que al mediodía. A lo largo de todo su recorrido, la luz azul habrá sido dispersada en un mayor grado si la comparamos con el mediodía. Dado que la porción roja de la luz es menos dispersada que la azul, la mayor parte de la luz que alcanzará al observador corresponderá a la longitud de onda correspondiente al rojo o naranja.

**Gravedad 1.3.1: Ley de gravitación universal de Newton**

Según la física clásica o newtoniana, la gravedad es la fuerza con la que un objeto atrae hacia su centro todo objeto con masa. La ley de gravitación universal dice que toda partícula de materia en el universo atrae a todas las demás partículas con una fuerza directamente proporcional a la masa de las partículas e inversamente proporcional al cuadrado de la distancia que los separa. De manera cuantitativa tenemos:

$$F_g = \frac{GM_1M_2}{r^2} \tag{1.1}$$

Dónde $M_1$ y $M_2$ son las masas de las partículas expresadas en kilogramos (kg), la distancia entre ellas, $r^2$, en metros (m) y el valor numérico de la fuerza, su módulo, es expresado en Newtons (N). Si combinamos esta ecuación con la segunda ley de Newton, podremos hallar la aceleración de la gravedad (g) en la superficie de cualquier planeta o luna, y esto a su vez nos permitirá hallar el **peso** de un objeto en dicha superficie con la siguiente ecuación:

$$P = mg \tag{1.2}$$

Donde $P$ es **el peso** expresado en Newtons (N), $m$ es la masa del objeto en kg y $g$ es la aceleración de la gravedad en $m/s^2$. La masa es una cantidad invariante: no importa en qué parte del universo nos encontremos, siempre vamos a tener la misma masa. El peso, en cambio, depende del campo gravitacional en el que nos encontremos. Por ejemplo, para una persona con una masa de 60 kg:

| **Planeta** | **g** ($m/s^2$) | **Peso (N)** |
|---|---|---|
| Tierra | 9,81 | 588,6 |
| Luna | 1,72 | 103,2 |
| Marte | 3,71 | 222,6 |

## 1.4 La Luna

La Luna es el satélite natural de nuestro planeta. La hipótesis más aceptada sobre su origen es la llamada “hipótesis del gran impacto”. Esta hipótesis sostiene que hace 4 400 millones de años un planeta más o menos del tamaño de Marte, llamado Theia (o Tea), chocó con la Tierra, y de esta manera, parte del material lanzado al espacio se terminó reagrupando por efecto de la gravedad, dando origen a nuestro satélite. Esto explicaría por qué la Tierra y la Luna son geoquímicamente similares [**17**]. Sin embargo, recientes investigaciones sugieren que el origen de nuestro satélite corresponde a una variedad de colisiones más pequeñas [**18**]. Se calcula que la Luna tiene más de 300 mil cráteres con tamaño mayor a 1 km, producto del impacto de asteroides y meteoroides sobre su superficie. Su radio es de 1 737,5 km y nos separa de ella una distancia de 384 400 km, es decir unas 30 Tierras juntas aproximadamente.

La Luna, tiene un movimiento de rotación en torno a su eje y otro de traslación alrededor de la Tierra. Para la Luna, el periodo de rotación es igual al periodo de traslación, entonces un año lunar es igual a un día lunar. Esto se conoce como **rotación sincrónica** y explica porqué la Luna siempre nos muestra la misma cara. La Tierra y la Luna se encuentran unidas por la gravedad y esto las obliga a moverse alrededor de un punto común denominado **baricentro** (ver Caja 4.2.1). En otras palabras, tenemos una danza de dos cuerpos que se mueven alrededor de su baricentro. Sin embargo, debido a que la masa de la Tierra es casi 100 veces mayor a la masa de la Luna, el baricentro se encuentra muy pegado al centro de la Tierra, no en una posición fija, sino, describiendo una trayectoria dentro de la Tierra. De forma particular, cuando la Luna está justo sobre nosotros, el baricentro está a 1700 km bajo nuestros pies, en otros momentos la distancia puede ser mayor. Es por esto que, de forma simplificada, podemos afirmar que la Luna orbita alrededor de la Tierra, en un periodo de 27,3 días, también llamado *mes sidéreo* [**19**].

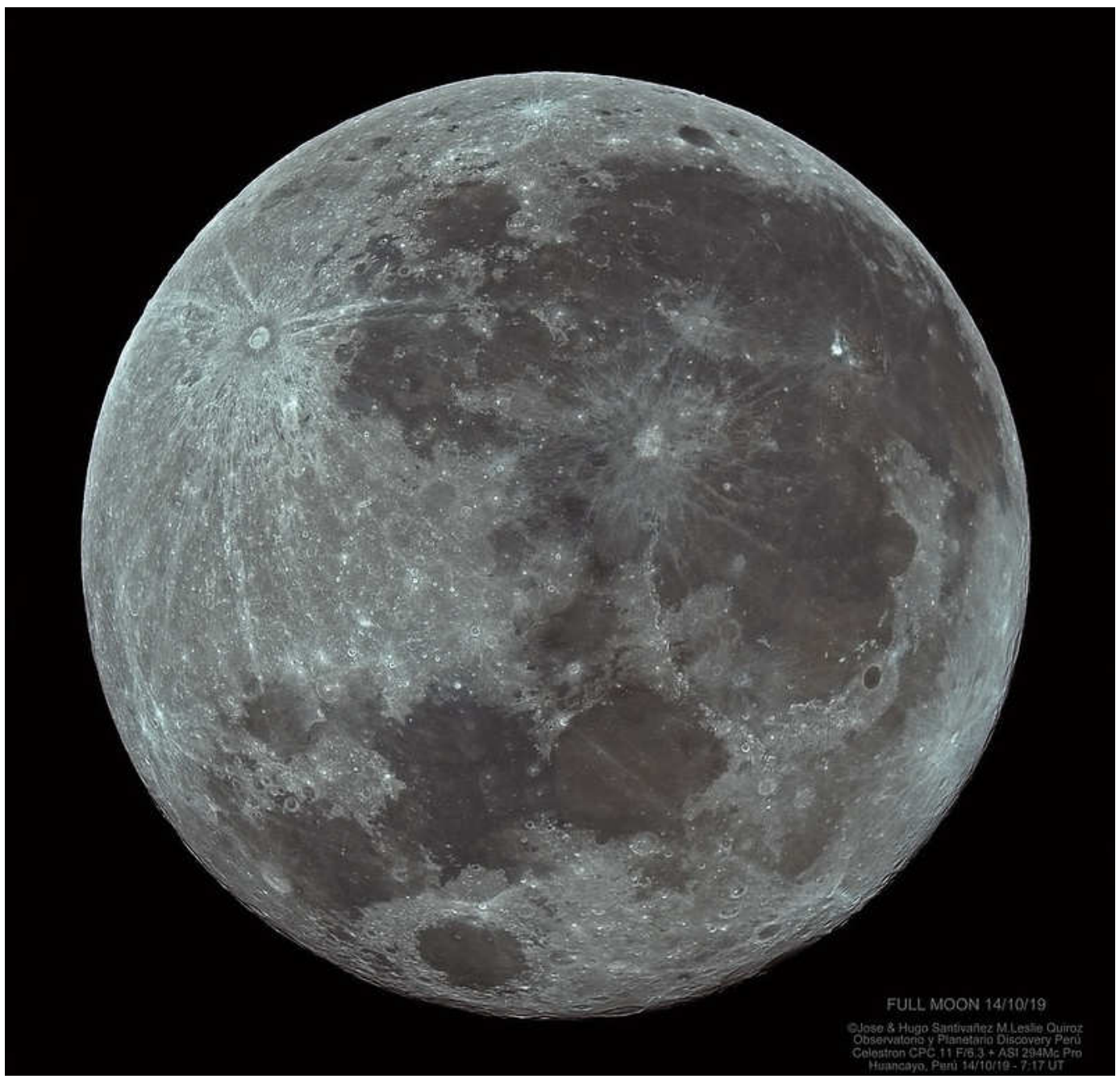

Figura 1.15: Luna en fase de luna llena. [Crédito: José y Hugo Santivañez, M. Leslie Quiroz, Observatorio y planetario Discovery, Perú.]

### 1.4.1 ¿Cómo influye la Luna sobre la Tierra?

Seguramente hemos escuchado hablar de los efectos de la Luna sobre nuestro planeta. Muchos de estos están basados en evidencia científica, pero muchos otros son parte de mitos y leyendas. Vamos a revisar este punto brevemente.

El principal efecto de la Luna es debido a la atracción gravitatoria con nuestro planeta. Este efecto es mayor en masas mayores, por ejemplo, sobre las concentraciones de masa ubicadas en la superficie terrestre, los océanos. Por la posición de la Luna en su órbita, los océanos son sometidos a cambios periódicos en el nivel del mar conocidos como **mareas** (ver Figura 1.16).

Teniendo en cuenta que la atracción gravitatoria entre dos cuerpos es proporcional a las masas e inversamente proporcional a la distancia entre los cuerpos, podemos entender que, aunque la influencia de la Luna sea significativa para cuerpos masivos como los océanos, no lo es para cuerpos como los seres humanos debido a su poca masa.

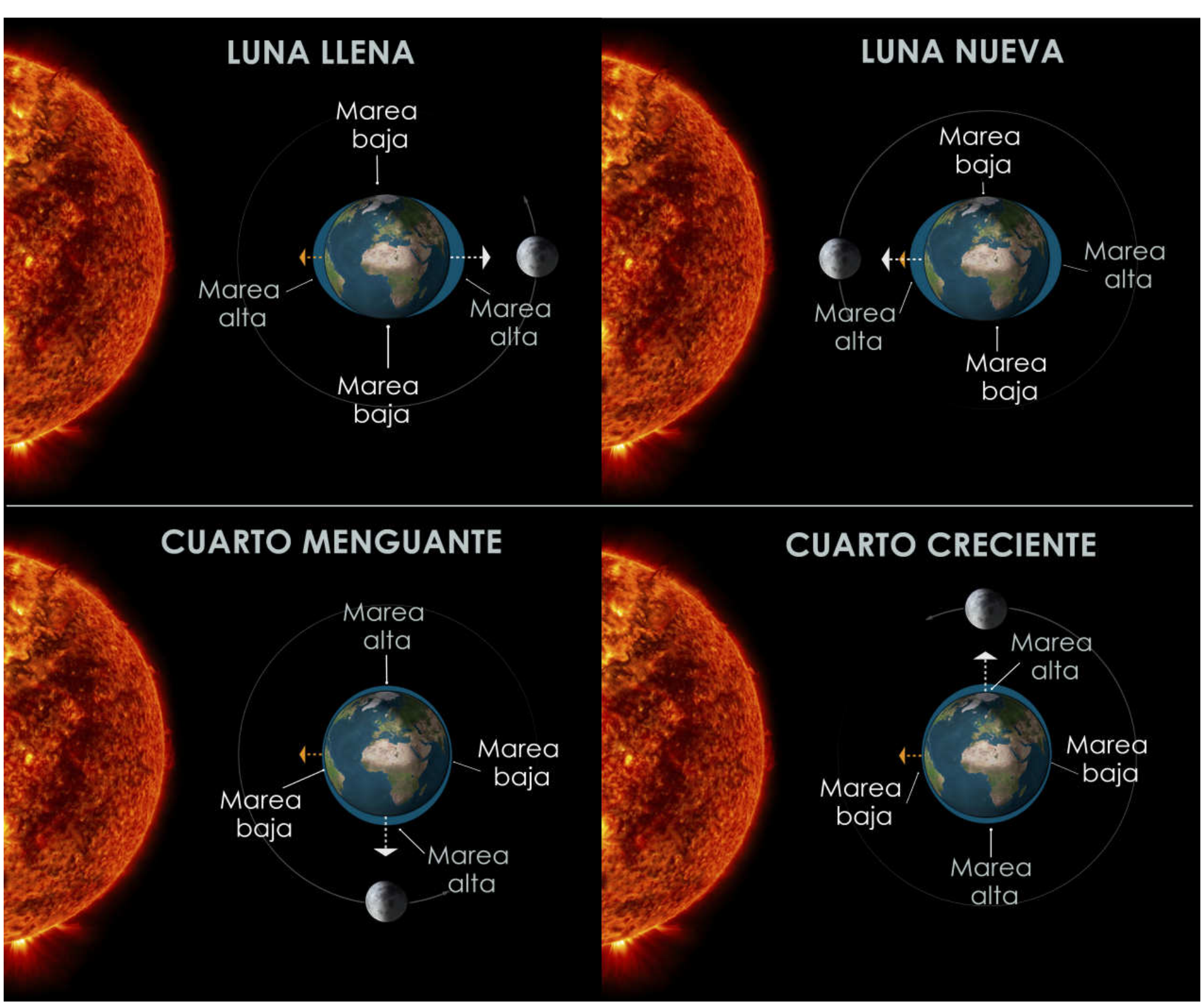

Figura 1.16: Las mareas en el mar y océanos son originadas principalmente por la atracción gravitatoria de la Luna sobre la Tierra, sin embargo, el Sol también produce mareas pero de mayor intensidad. El efecto máximo de las fuerzas de marea se origina cuando la Luna, la Tierra y el Sol están alineados. [Crédito: adaptado de `https://tablademareas.com/mareas/tipos-mareas`]

### 1.4.2 Fases de la Luna

Llamamos fases lunares a las variaciones en la iluminación de la Luna percibidas desde la Tierra. Estos cambios son dados por la ubicación relativa del Sol y la Tierra con respecto a la órbita lunar. El tiempo que transcurre en cada ciclo de las fases lunares se denomina **lunación** y tiene una duración de 29.5 días. Podemos dividir una lunación en cuatro fases principales:

Llamamos fases lunares a las variaciones en la iluminación de la Luna percibidas desde la Tierra. Estos cambios son dados por la ubicación relativa del Sol y la Tierra con respecto a la órbita lunar.

- Durante una **luna nueva**, la Luna se ubica entre el Sol y la Tierra. El porcentaje de iluminación proyectado a la Tierra es nulo, por eso no podemos observar a la Luna en esta fase.
- En los **cuartos creciente y menguante** el hemisferio lunar que está orientado a la Tierra posee una iluminación del 50 %. En el cuarto creciente, un observador ubicado en el hemisferio sur observará que la mitad izquierda iluminada y la derecha a oscuras. En la fase de cuarto menguante, el efecto es el contrario.

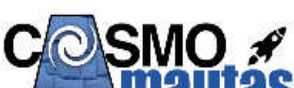

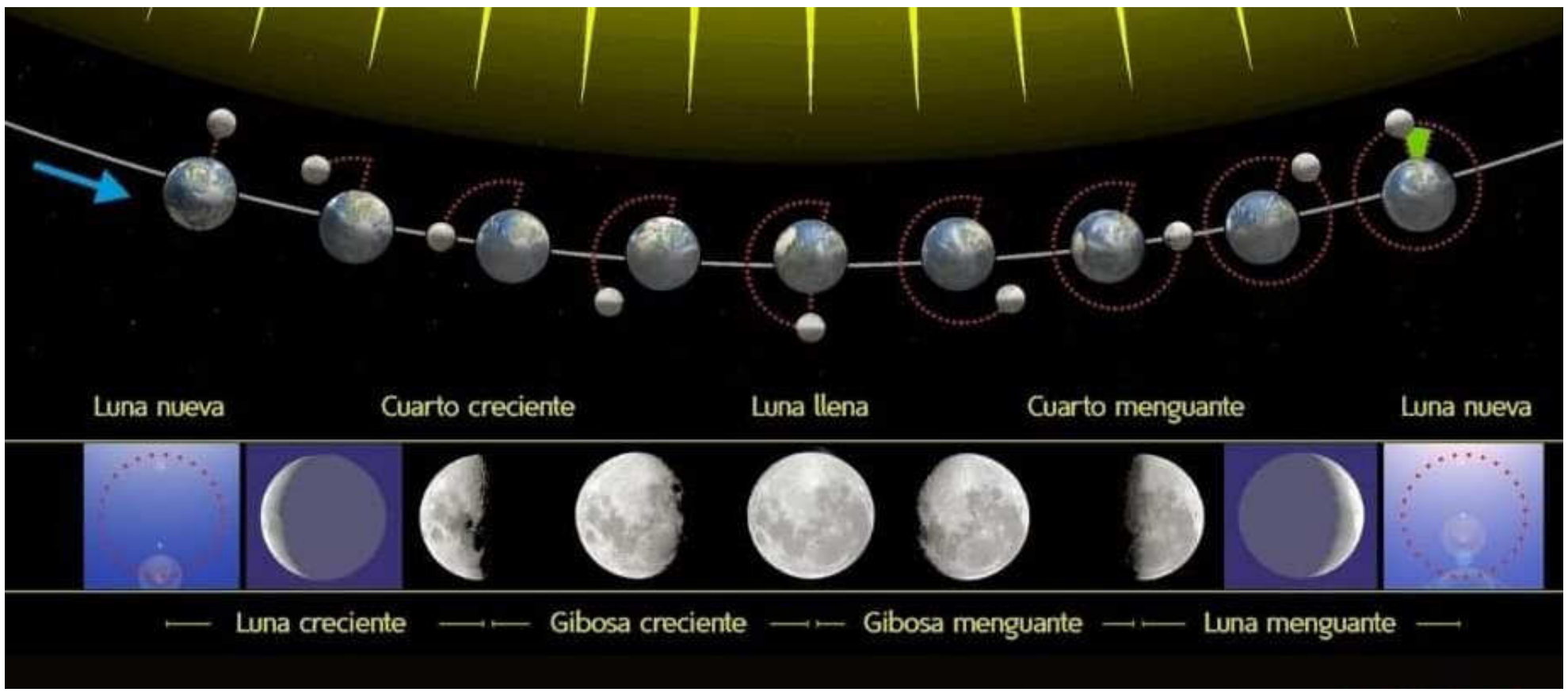

Figura 1.17: Fases lunares observadas desde el hemisferio sur [Crédito: imagen adaptada de Orion 8 (Creative Commons)].

- Tenemos una **luna llena** cuando la Tierra se ubica entre el Sol y la Luna, de esta manera el hemisferio lunar que está orientado a la Tierra se encuentra completamente iluminado.

### 1.4.3 Eclipses

#### Eclipses solares

Los eclipses solares ocurren cuando la Luna se ubica entre el Sol y la Tierra, formando una línea, justo cuando intercepta el plano de la eclíptica. En esta posición, la Luna proyecta, en la superficie terrestre, un cono de sombra que presentará dos zonas bien marcadas: la umbra que se caracteriza por ser la zona más oscura y la penumbra como la menos oscura [**20**]. Los tipos de eclipses solares son:

- Un **eclipse total de Sol** ocurre cuando la zona umbral alcanza la superficie terrestre. Los habitantes que estén ubicados en esta zona lograrán observar a la Luna cubriendo completamente al disco solar. Este hecho ocurre por una singular circunstancia, el Sol es 400 veces más grande que la Luna, pero al mismo tiempo, está 400 veces más alejado. Es por esto que ambos objetos aparentan tener las mismas dimensiones en el cielo.
- Un **eclipse parcial de Sol** ocurre cuando el observador es alcanzado por la zona de la penumbra. En este caso la Luna cubre parcialmente el disco del Sol.
- Un **eclipse anular** ocurre cuando la Luna se ubica en el punto de su órbita más alejado de la Tierra (apogeo). En este caso el observador en Tierra es alcanzado por el cono de sombra opuesto de la umbra (la antumbra). De esta manera, se podrá apreciar un anillo luminoso durante la fase central del eclipse debido a que el tamaño aparente de la Luna es menor que la del Sol.
- Los **eclipses solares híbridos** ocurren cuando durante un mismo evento un eclipse empieza siendo un eclipse anular, pasa a ser un eclipse total y finaliza como un eclipse anular.

#### Eclipses lunares

Para que sea posible un eclipse lunar, la Luna debe encontrarse en fase de luna llena y estar ubicada muy cerca al plano de la eclíptica. En este caso, el cono de sombra es proyectado por la Tierra, y al igual que en los eclipses solares, este también se divide en una zona de umbra y otra de penumbra. Los tipos de eclipses lunares son:

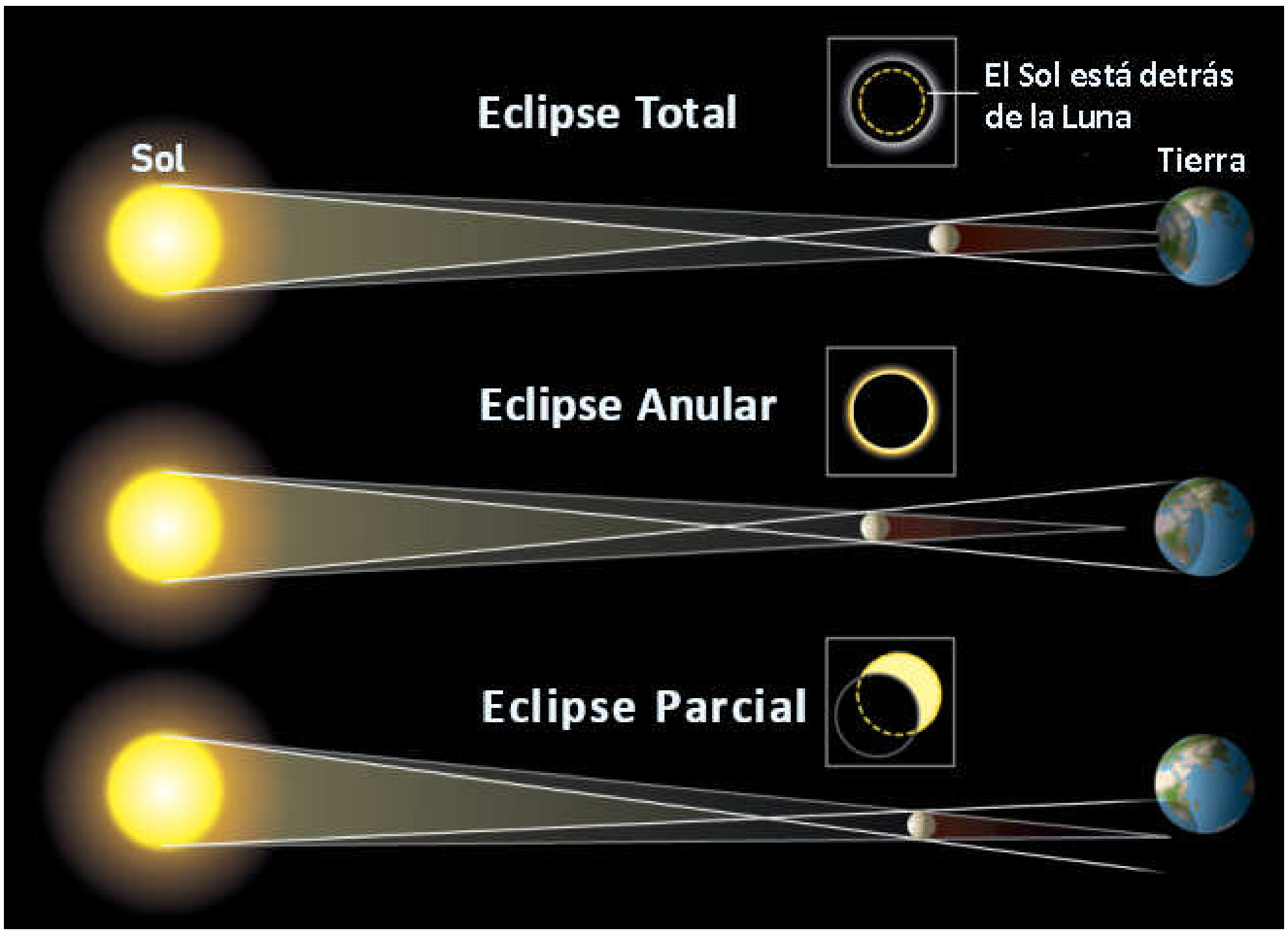

Figura 1.18: Tipos de eclipses solares [Crédito: `https://www.drishtiias.com`].

- Un **eclipse total de Luna** ocurre cuando la Luna ingresa por completo en la zona de la umbra. Durante la fase central de este tipo de eclipse, se observa que la Luna adquiere un color rojo-naranja. Esto se debe a que los rayos solares que atraviesan la atmósfera de la Tierra son refractados y dispersados por las partículas y moléculas que componen la atmósfera terrestre. El grado de dispersión es mayor en el azul, es por esto que solo la radiación correspondiente al rojo logra llegar a la superficie lunar.
- Un **eclipse parcial de Luna** ocurre cuando la Luna solo entra parcialmente en la umbra.
- Un **eclipse penumbral de Luna** ocurre cuando la Luna pasa a través de la penumbra de la sombra terrestre. Un tipo especial de eclipse penumbral es un eclipse lunar penumbral total, durante el cual la Luna se encuentra exclusivamente dentro de la penumbra de la Tierra.

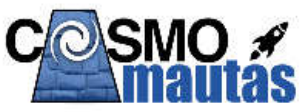

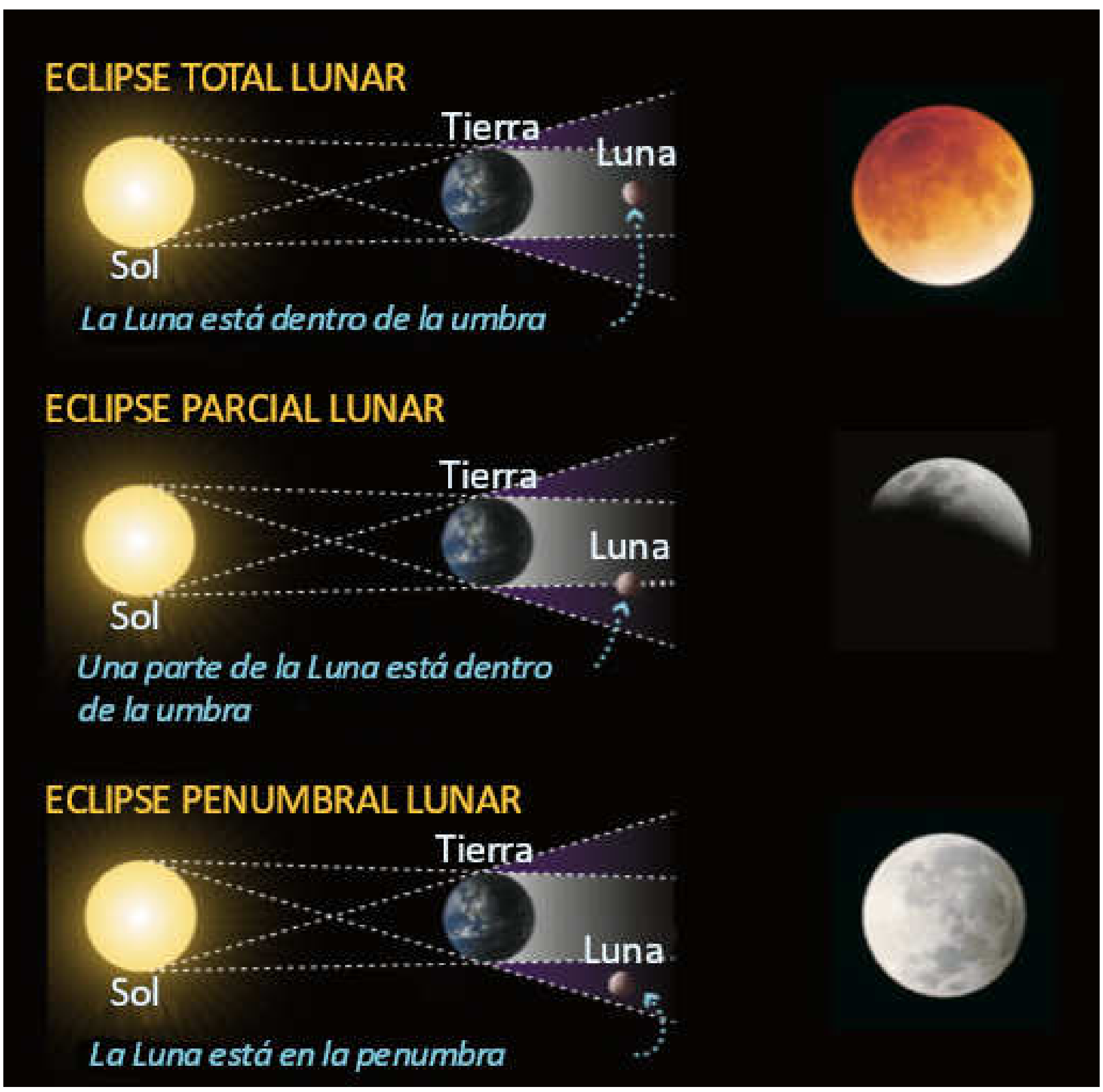

Figura 1.19: Tipos de eclipses lunares [Crédito: Imagen traducida de `nineplanets.org`].

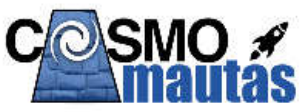

# 2. El sistema solar

Durante el día podemos observar al Sol, y a medida que este se va ocultando por el oeste aparecen más objetos brillantes aunque de menor tamaño. Seguido del Sol, la Luna es el objeto más brillante, y después de ella, se observa una infinidad de puntos brillantes esparcidos por todo el cielo. A simple vista podemos ver algunos de los planetas del sistema solar: Mercurio y Venus al amanecer y al atardecer por estar ubicados entre la Tierra y el Sol; Marte, de un característico color rojo. Júpiter y Saturno son visibles a simple vista, y con un telescopio pequeño podemos observar los satélites galileanos de Júpiter y los anillos de Saturno. Debido a que Urano y Neptuno están mucho más lejos de nosotros, para observarlos necesitamos un cielo despejado y sin polución luminosa, además de un ojo muy bien entrenado. ¿Qué otros objetos forman parte de nuestro sistema solar? ¿Conocemos sus límites? ¿Qué tan lejos han llegado las exploraciones espaciales?

Imagen de encabezado: Impresión artística del sistema solar. Crédito: IAU/Kornmesser

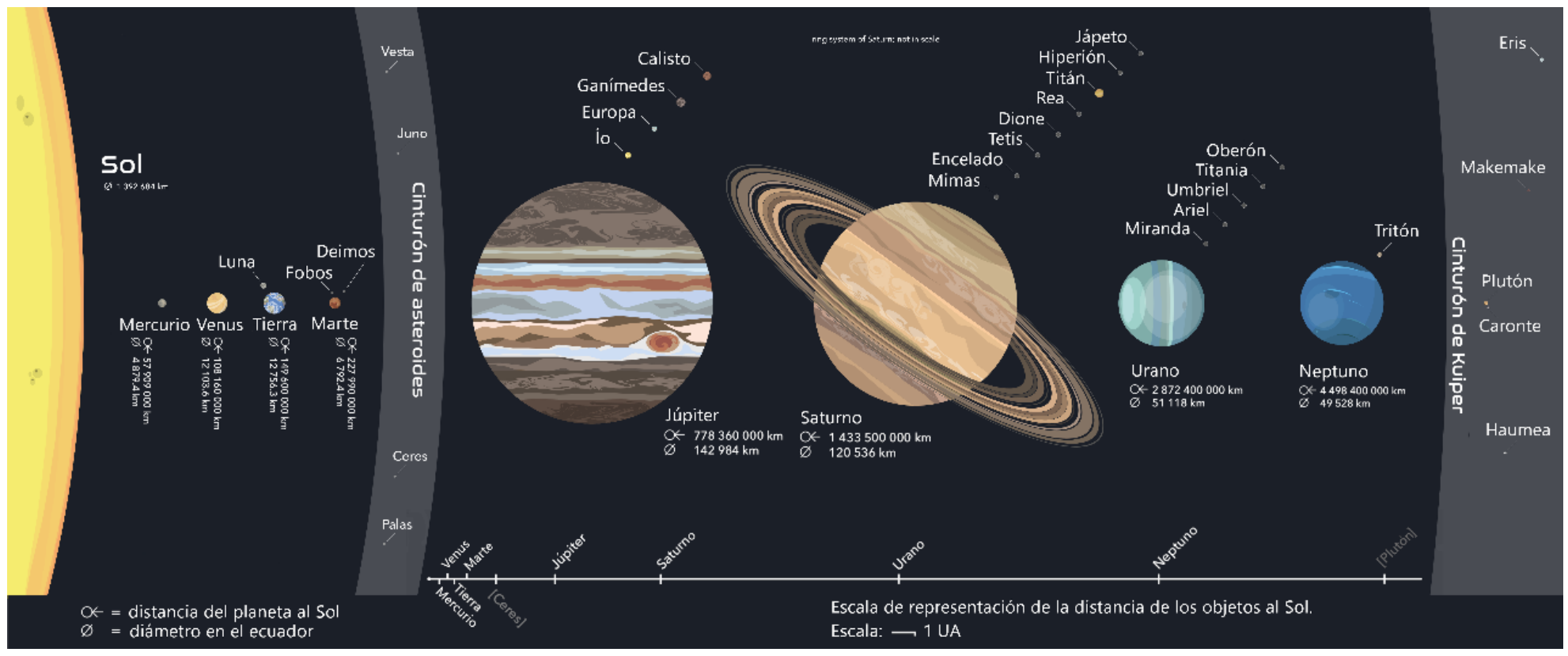

Figura 2.1: El Sol es el objeto más grande de nuestro sistema y contiene el 99,8 % de la masa total. Distancias no a escala [Crédito: Julian Graeve, Beinahegut (Wikipedia)].

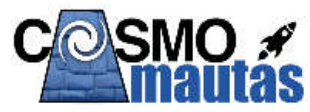

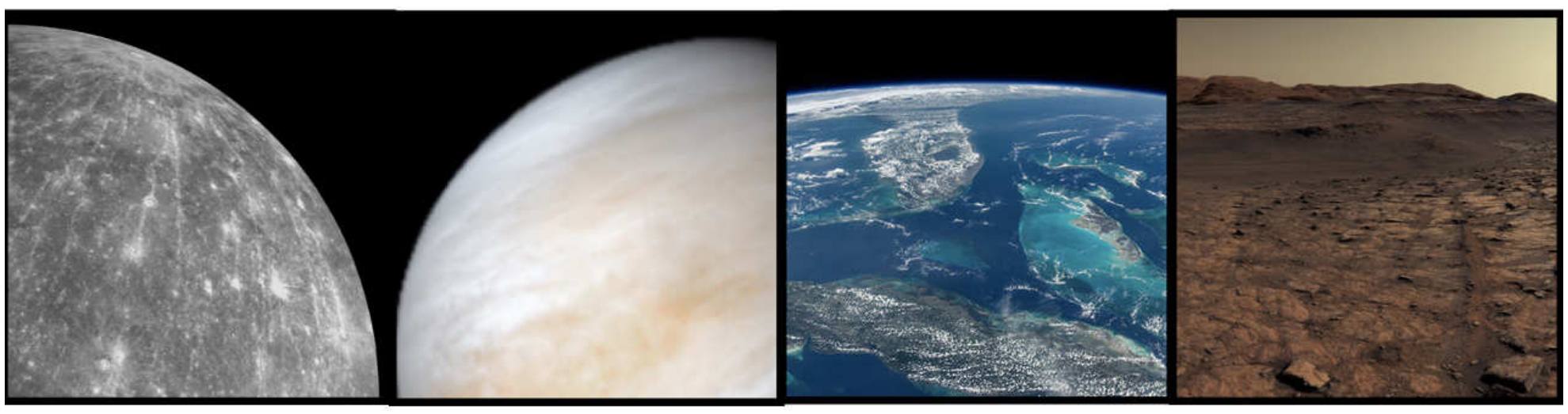

Figura 2.2: Planetas rocosos. De izquierda a derecha: Mercurio, Venus, la Tierra y Marte. [Crédito: NASA]

## 2.1 Componentes del sistema solar

Nuestro sistema solar está formado por una sola estrella, el Sol, y por todos los objetos celestes que están unidos a él por gravedad. Esto abarca ocho planetas, cinco planetas enanos, decenas de satélites naturales o también llamadas lunas, millones de asteroides (tanto de roca como de hielo), cometas y meteoroides.

#### El Sol: nuestra estrella

Es una estrella enana amarilla, una bola caliente de gas incandescente en el corazón de nuestro sistema solar. El Sol es la principal fuente de energía de la Tierra, la cual recibimos como luz y que es producto de la fusión de núcleos de hidrógeno en helio. El Sol se compone de 73 % de hidrógeno, 25 % de helio y pequeñas cantidades de elementos más pesados. Aunque es especial para nosotros, el Sol es una de aproximadamente $10^{11}$ estrellas esparcidas por la Vía Láctea, en un universo con $10^{11}$ galaxias. El Sol es nuestra principal fuente de energía, la misma que recibimos como luz y que es producto de la fusión de núcleos de Hidrógeno en Helio

#### Mercurio: una roca sin atmósfera

Es el planeta más pequeño del sistema, con un tercio del radio de la Tierra, su tamaño es similar a nuestro satélite natural, la Luna. La corta distancia entre Mercurio y el Sol (0.3 UA) hace que ambos objetos se vean .$^{ex}$tendidos"(al analizar la atracción gravitatoria entre ellos, no podemos considerarlos objetos puntuales, ver cajita de gravedad), de tal forma que su órbita está anclada por fuerzas de mareas, de la misma forma en que el efecto gravitatorio de la Luna produce la subida y bajada de mareas en los océanos de la Tierra, el efecto gravitatorio del Sol hace que Mercurio tenga una forma esferoidal y esto afecta su periodo de rotación, haciéndolo más lento. Como consecuencia, 58 días terrestres en Mercurio equivalen a 1 día y 88 días terrestres a un año, lo que se conoce como una resonancia 3:2 entre rotación-órbita.

Mercurio está tan cerca del Sol que recibe entre 5 y 11 veces la irradiación que recibe la Tierra (Tierra: 1.366 $W/m^2$). Esta alta radiación ha hecho que la atmósfera de Mercurio se evapore al punto que solo queda una delgada capa llamada exósfera y está compuesta principalmente de oxígeno, sodio, hidrógeno, helio y potasio. Esta resonancia y la falta de atmósfera en Mercurio hace que la temperatura entre día y noche varíe más que en cualquier otro planeta del sistema solar, yendo desde 100 K (-173° C) en la noche, hasta 700 K (427° C) durante el día en las regiones ecuatoriales. Las regiones polares siempre están a menos de 180 K (-93° C). La inclinación axial de Mercurio (el ángulo entre su eje de rotación y su plano orbital) es la menor de todos los planetas del sistema solar (0.01°), lo que nos haría pensar que no tiene cuatro estaciones como nuestro planeta. Sin embargo, debido a su gran excentricidad, podemos decir que tiene invierno y verano, según la cantidad de irradiacion que recibe.

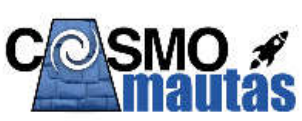

**Gravedad 2.1.1: Aproximaciones**

En Física, hacemos aproximaciones para generar modelos que nos permiten explicar el universo. Una aproximación que usamos frecuentemente es la de objetos puntuales. La fuerza de gravedad es una **fuerza central** que se manifiesta desde el centro de masa de un objeto al centro de masa de otro, de tal forma que podemos considerar a cada objeto como un punto que acarrea toda su masa. Cuando no podemos usar la aproximación de objetos puntuales, tenemos que considerarlos como objetos extendidos, y aparecen fuerzas de superficie, como la fricción y las fuerzas de mareas.

### Venus: el gemelo de la Tierra

Venus es el segundo planeta desde el Sol (0.7 UA) y el planeta más cercano a la Tierra. Venus, al igual que Urano, son los planetas que rotan en sentido opuesto a los demás planetas del sistema solar. A este movimiento se le conoce como rotación retrógrada y como consecuencia de esto, si un habitante en la superficie de este planeta observase el movimiento aparente del Sol, se daría cuenta que sale por el oeste y se oculta por el este. La razón del movimiento retrógrado no se conoce con claridad, pero se piensa que puede haber sido producto de una colisión durante el periodo de formación de Venus con otro cuerpo del sistema. Después de la colisión, Venus habría desacelerado y cambiado su dirección de rotación. Un día en Venus equivalen a 243 días terrestres y un año, a 225 días, lo que significa que en Venus, un día es más largo que un año.

Venus, al igual que Urano, son los planetas que rotan en sentido opuesto a los demás planetas.

Venus ha sido llamado el gemelo de la Tierra por la similitud en sus dimensiones y masa, a pesar que su atmósfera es totalmente distinta a la nuestra. El 98 % de la atmósfera de Venus es dióxido de carbono ($CO_2$), con capas de nubes de ácido sulfúrico ($H_2SO_4$) y dióxido de azufre ($SO_2$). El alto porcentaje de $CO_2$ y la presencia de ácido sulfúrico son consecuencia de su intensa actividad volcánica. Estos gases crean condiciones para un intenso efecto invernadero que se manifiesta en la constante y elevada temperatura de la superficie del planeta, 735 K (462 °C), convirtiéndolo en el planeta más caliente del sistema solar. Los compuestos de las nubes son altamente reflectantes, por ello, Venus es el planeta más brillante visto de la Tierra, y es el tercer objeto celeste más brillante del cielo visto desde nuestro planeta. Simulaciones climatológicas estiman que hace 3 mil millones de años Venus puede haber tenido agua líquida y condiciones habitables por 700-750 millones de años [**21**].

**Unidades 2.1.1: Billón**

En el mundo de habla hispana, un **billón** equivale a $10^{12}$. En el mundo de habla inglesa, un billón equivale a $10^9$, lo que en español equivale a mil millones o un *millardo*. Esta diferencia la debemos tomar en cuenta al leer literatura científica en inglés. El prefijo *giga* siempre significa $10^9$.

### Tierra: el planeta azul

Nuestro planeta Tierra es el único planeta que conocemos con agua líquida y vida en su superficie. Por definición, la Tierra se encuentra a 1 UA del Sol, dentro de la zona habitable de esta estrella. Su radio ecuatorial es de 6 374 km, y en los polos es 22 km menor. Esta diferencia se debe a la rotación de la Tierra, pues un punto en el Ecuador se mueve más distancia que un punto cerca al polo durante el mismo periodo de rotación, haciendo que la velocidad y la fuerza centrífuga sean mayores en el Ecuador. La Tierra completa una rotación alrededor de su eje en 23 horas y 56 minutos. La órbita de la Tierra es casi circular, con una excentricidad de 0,0167 y un periodo orbital de 365,25 días. Ese cuarto de día extra es la causa de que cada cuatro años se añada un día más al calendario, lo que da origen a los años bisiestos. La inclinación de 23,45° del eje de rotación de la Tierra con respecto al plano de su órbita, produce las estaciones. Por la segunda ley de Kepler y la órbita circular de la Tierra, las cuatro estaciones tienen casi la misma duración. Con una masa de $5{,}972 \times 10^{24}$ kg, la Tierra es el planeta rocoso más masivo del sistema solar. Posee un único satélite natural, la Luna, que posiblemente se formó como consecuencia de varios impactos.

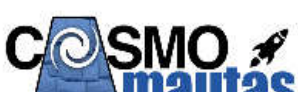

La atmósfera de la Tierra está compuesta principalmente por nitrógeno (78 %), oxígeno (21 %) y otros gases como argón, dióxido de carbono y neón. Cuando la luz solar llega a la atmósfera y atraviesa estas moléculas, ellas absorben ciertas longitudes de luz y reflejan con mayor intensidad la luz de color azul, por ello vemos al cielo de ese color. La atmósfera también afecta el clima global a largo plazo y el clima local a corto plazo y nos protege de gran parte de la radiación dañina proveniente del Sol.

**Unidades 2.1.2: Unidad Astronómica (UA)**

Una Unidad Astronómica (UA) es la distancia promedio entre la Tierra y el Sol y equivale a $1{,}5 \times 10^{11}$ m.

### Marte: el planeta rojo

El radio de Marte es aproximadamente la mitad del de la Tierra y su eje de inclinación es de 25° respecto a su plano orbital, por ello, Marte presenta estaciones al igual que en la Tierra. Sin embargo, la órbita de Marte es mucho más excéntrica ($e = 0{,}0935$) que la terrestre, lo cual lleva al planeta bastante más lejos del Sol durante el verano en el polo sur, haciendo que el polo sur esté más frío que el polo norte durante esa estación. Marte posee casquetes polares formados por hielo seco ($CO_2$ en estado sólido) y hielo ordinario, los cuales varían en área dependiendo de la estación del planeta. Mediciones del *Rover Spirit* de la temperatura en Marte registran extremos de $-153$°C (120 K) a 20°C (293 K) en su superficie, además de fuertes vientos de más de 100 m/s que arrastran grandes cantidades de polvo.

Marte tiene un décimo de la masa de la Tierra ($6{,}39 \times 10^{23}$ kg) y 62,5 % menos gravedad que la Tierra, entonces si una balanza marca 60 kg para una persona en la Tierra, marcaría 22.5 kg en Marte. La baja masa y gravedad superficial de Marte ocasiona que su atmósfera sea poco densa y relativamente delgada (30 km aproximadamente). A pesar de estar compuesta principalmente por $CO_2$, el efecto invernadero no es tan significativo en la atmósfera de Marte. En la superficie se observan matices de colores entre marrones, dorados y rojos, pero es conocido como el planeta rojo debido a la oxidación del hierro en las rocas y la superficie. Marte tiene dos satélites naturales, Fobos y Deimos, los cuales tienen características más similares a asteroides que a otros satélites naturales del sistema solar, indicando que probablemente fueron capturados por el campo gravitatorio del planeta [**22**].

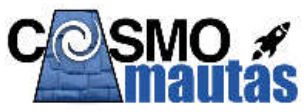

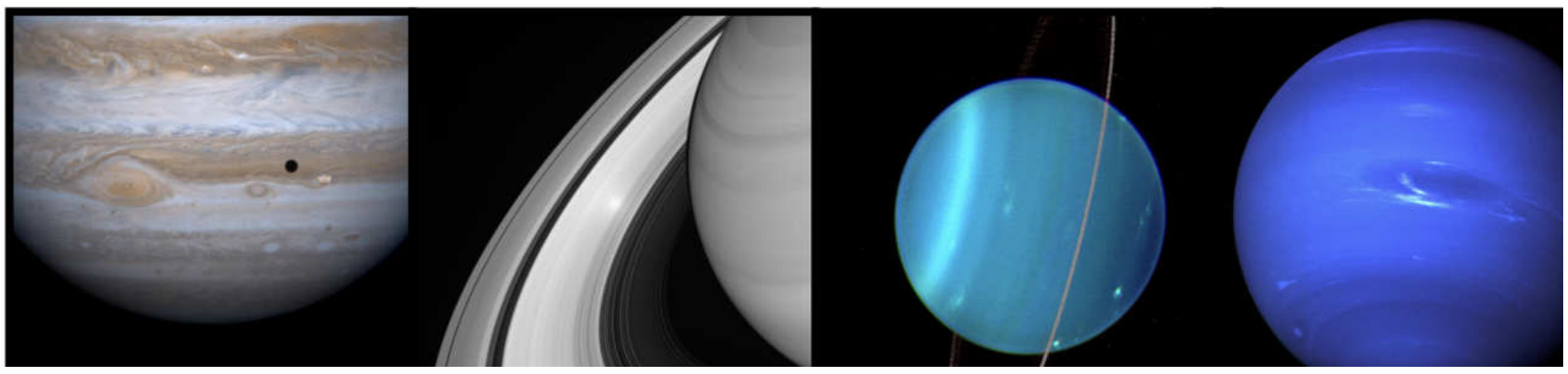

Figura 2.3: Planetas gaseosos. De izquierda a derecha: Júpiter, Saturno, Urano y Neptuno [Crédito: NASA].

**Unidades 2.1.3: Unidades solares ($\odot$)**

Las cantidades con las que nos encontramos en astronomía estelar son tan grandes que es conveniente definir unidades con respecto al Sol. Por ejemplo, una masa solar equivale a $1\,M_\odot = 1{,}989 \times 10^{30}$ kg, y un radio solar equivale a $1\,R_\odot = 6{,}957 \times 10^{8}$ m.

### Júpiter: el gigante gaseoso

Júpiter es el planeta más grande del sistema solar, siendo su radio 11 veces el radio terrestre. También es el más masivo, con una masa de $1{,}898 \times 10^{27}$ kg, que equivale a 0,1 % de la masa del Sol y a 2,5 veces la masa de todos los demás planetas del sistema solar *juntos*. Júpiter es tan masivo, que el centro de gravedad de Júpiter y el Sol se encuentra fuera del Sol, a una distancia de $1{,}068\,R_\odot$ desde el centro del Sol.

Su periodo de rotación es el más corto de los planetas, con días que duran solo 9 horas y 55 minutos. Esta rotación también es responsable de crear fuertes vientos en su atmósfera que separan los gases causando las bandas alternantes de color paralelas a su ecuador, las cuales se mueven en direcciones opuestas creando vórtices y remolinos turbulentos. La Gran Mancha Roja, uno de estos vórtices, es una tormenta persistente por más de 300 años aproximadamente del tamaño de la Tierra.

La atmósfera de Júpiter contiene hidrógeno (86 %) y helio (13,6 %) principalmente y es más similar a la atmósfera del Sol en cuanto a composición que a las de los planetas rocosos. Además también contiene agua, metano ($CH_4$), sulfuro de hidrógeno ($H_2S$), amoniaco ($NH_3$) y fosfano ($PH_3$). Sin embargo, aún no existe una teoría que describa la dinámica atmosférica de Júpiter.

El planeta parece no tener una superficie sólida, causando una convección vertical a través de todas sus capas. Es posible que exista un núcleo sólido de hierro y níquel del tamaño de la Tierra, que se vuelve difuso hacia capas más superficiales. Este núcleo estaría rodeado por una capa de hidrógeno líquido metálico, donde la temperatura superaría los 10 000 K y la presión estaría a 3 millones de atmósferas. Estas condiciones de presión provocan que el gas hidrógeno se configure en una estructura metálica, reticular, o cristalina de protones con electrones libres.

Júpiter irradia el doble de energía que recibe del Sol, lo que parece indicar que el planeta continúa en su proceso de contracción gravitacional como consecuencia de su formación. Cuenta con un anillo tenue y más de 60 satélites naturales, de los cuales Io, Ganímedes, Calisto y Europa son los más grandes, habiendo sido observados por primera vez por Galileo Galilei con la invención del telescopio.

### Saturno: el señor de los anillos

Saturno es el segundo planeta más grande del sistema solar con un radio 9 veces el terrestre y una masa 95 veces la de la Tierra ($5{,}68 \times 10^{26}$ kg). Esto hace que su densidad sea 0.687 g/cm$^3$, lo que quiere decir que Saturno podría flotar en el agua si tuviéramos un contenedor suficientemente grande. Su periodo de rotación dura 10 horas y 45 minutos, lo cual ocasiona una forma esferoide aplanada en los polos al igual que en Júpiter. Su eje de rotación está inclinado 27° respecto a su plano orbital por lo que los polos norte y sur pueden ser observados cada 15 años.

La característica que más resalta de este planeta es el sistema de anillos que lo rodea en su plano ecuatorial. Los anillos están hechos de hielo de agua y el tamaño de las partículas oscila entre micras y trozos del tamaño de un camión. La razón de su origen es aún desconocido: se creía que era partículas residuales de la formación del planeta, pero datos recientes enviados por la sonda Cassini podrían indicar que estos anillos se formaron después que el planeta.

La atmósfera de Saturno tiene aproximadamente un 75 % de hidrógeno y un 25 % de helio con trazas de otras sustancias como metano, amoniaco y hielo de agua formando neblinas. Su estructura interna es similar a la de Júpiter, aunque la capa de hidrógeno metálico es de menor grosor. Saturno irradia 2,8 veces más calor del que recibe del Sol en longitudes de onda infrarrojas, pero ya que tiene una masa menor a la de Júpiter, se cree que el origen de su energía interna es distinto.

#### Urano: el planeta lateral

Urano tiene un radio sólo 4 veces el radio terrestre y una masa 14,5 veces la masa de la Tierra, lo que le confiere la segunda densidad más baja del sistema solar, después de Saturno. La inclinación de su eje de rotación es de 98° respecto a su plano orbital, posiblemente a causa de una colisión. Debido a esta característica, los polos están iluminados o en oscuridad durante décadas: cuando el Sol brilla directamente sobre un polo, sumerge a la otra mitad del planeta en un oscuro invierno de 21 años aproximadamente. El período de rotación, confirmado por las mediciones del Voyager 2 en 1986, es de 17,3 horas.

#### Neptuno: el planeta más lejano

Su existencia fue predicha por el astrónomo francés Urbain Le Verrier en 1845 dadas irregularidades observadas en la órbita de Urano con respecto a predicciones de la ley gravitacional de Newton. Al año siguiente, el planeta fue observado a solo 1° de la posición predicha. A Neptuno le demora 165 años en dar una vuelta alrededor del Sol, y según los datos del Voyager 1, su periodo de rotación es de 16 horas. Su eje de rotación está inclinado 29° con respecto al plano orbital por lo que experimenta estaciones como los otros planetas de similar oblicuidad. Sin embargo, cada estación dura más de 40 años debido a su órbita de 30 UA.

### 2.1.1 Los planetas enanos

Desde la década de 1990, se descubrieron varios objetos más allá de la órbita de Neptuno (objetos trans-Neptunianos), algunos de los cuales eran incluso más grandes que Plutón. Plutón siempre fue un objeto peculiar, con una órbita altamente excéntrica ($e = 0{,}2488$) que hace que su distancia al Sol varíe entre 30-49 UA, a veces moviéndose dentro de la órbita de Neptuno. Plutón, a diferencia de los gigantes gaseosos, es un planeta relativamente pequeño y principalmente rocoso, con una masa de 0.002 veces la masa de la Tierra y 0.19 veces su radio. Es decir, Plutón es más parecido a los objetos trans-Neptunianos descubiertos recientemente que a los planetas gaseosos del sistema solar que se encuentran cerca a él. Por esta razón, la IAU declaró en el 2006 una nueva definición de **planeta**, suplementándola con una definición de **planeta enano**. Para llamarse planeta, un objeto debe cumplir con 3 requisitos:

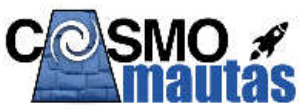

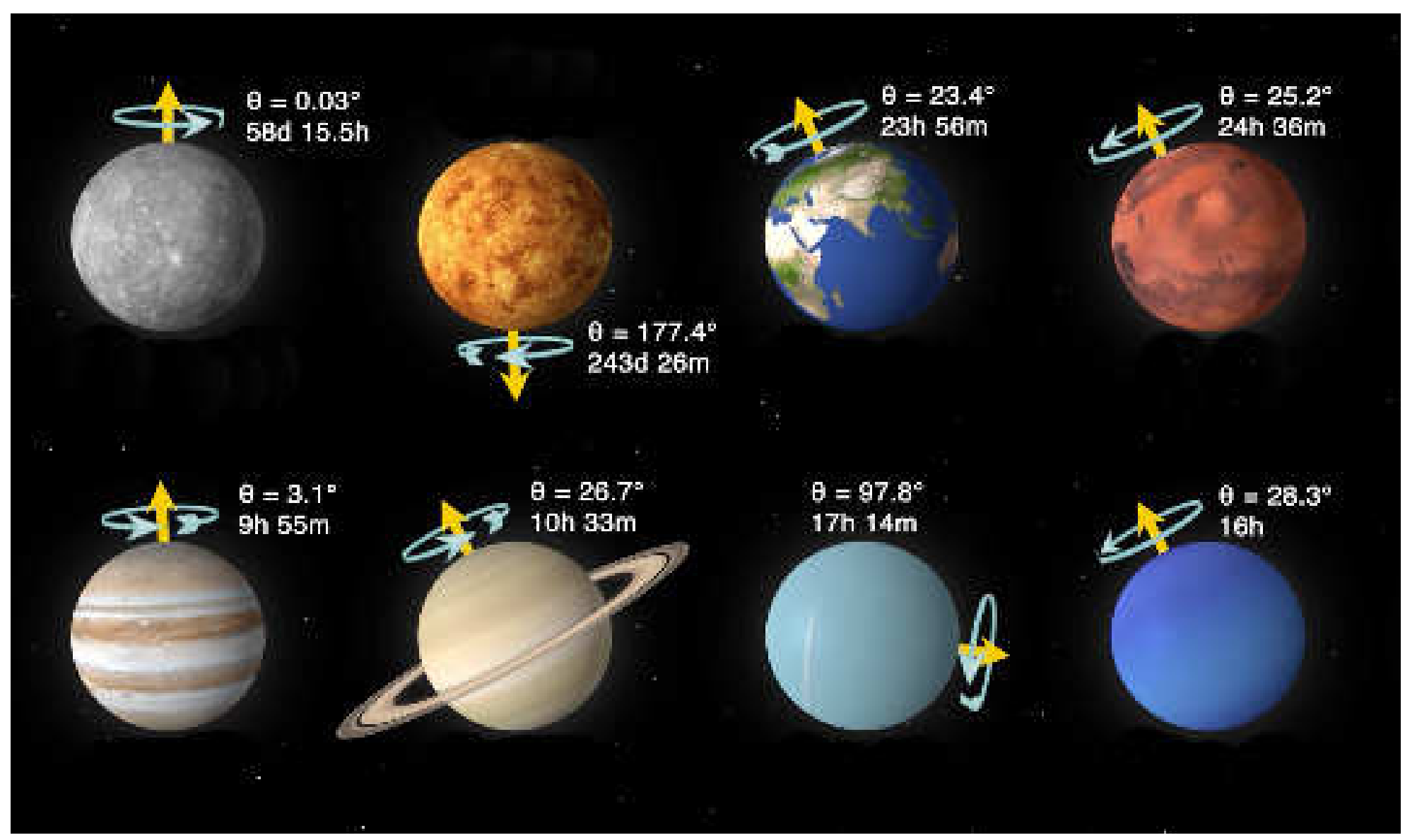

Figura 2.4: La imagen muestra los diferentes ángulos de inclinación del eje de rotación de los planetas. [Crédito: James O'Donoghue/JHUAPLNASA]

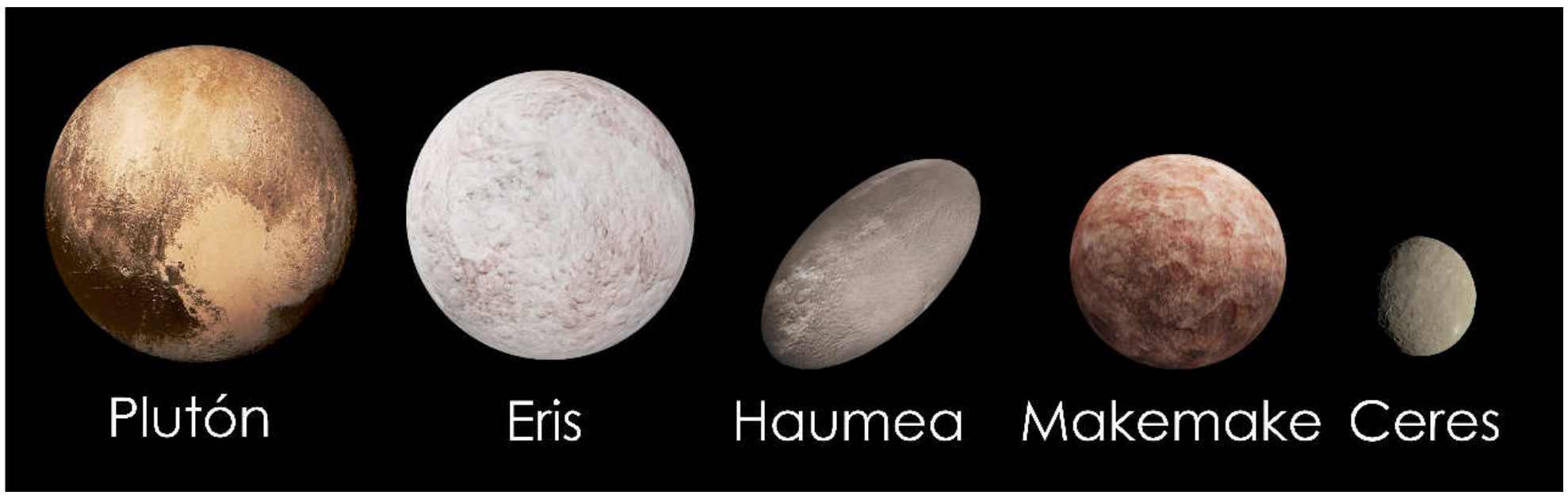

Figura 2.5: La imagen representa la escala aproximada del tamaño de los planetas enanos, sin considerar el orden en las distancias al Sol. Solo las imágenes de Plutón y Ceres corresponden a fotografías, mientras que Eris Haumea y Makemake son modelos hechos con predicciones matemáticas. [Crédito: NASA]

1. Debe orbitar alrededor del Sol (no alrededor de cualquier otra estrella).
2. Debe contar con suficiente masa para estar en equilibrio hidrostático y adoptar una forma casi esférica.
3. Debe ser lo suficientemente grande para haber "limpiado" la vecindad de su órbita, es decir, debe ser el objeto más prominente en toda su trayectoria.

Bajo esta definición, si movemos a la Tierra más allá de la órbita de Neptuno, ¡no la llamaríamos planeta! En el caso de Plutón, no cumple con la tercera condición. Los planetas enanos son aquellos que cumplen con los dos primeros requisitos. La IAU reconoce a cinco planetas enanos en nuestro sistema solar: Ceres, Plutón, Eris, Haumea y Makemake.

### 2.1.2 Cometas, asteroides y meteoroides

Los **cometas** son "bolas de nieve sucias" que vienen a visitarnos desde las partes más alejadas de nuestro sistema solar: el cinturón de Kuiper (30-50 UA) y la nube de Oort (2 000-200 000 UA). Estos son reservorios de rocas sobrantes de cuando se formaron los planetas del sistema solar. Los cometas están formados por rocas, polvo y hielo, con un núcleo de hielo de unos pocos kilómetros de diámetro. A medida que se acercan al Sol, el calor hace que el hielo se sublime, creando una formando una cola larga y brillante llamada "coma".

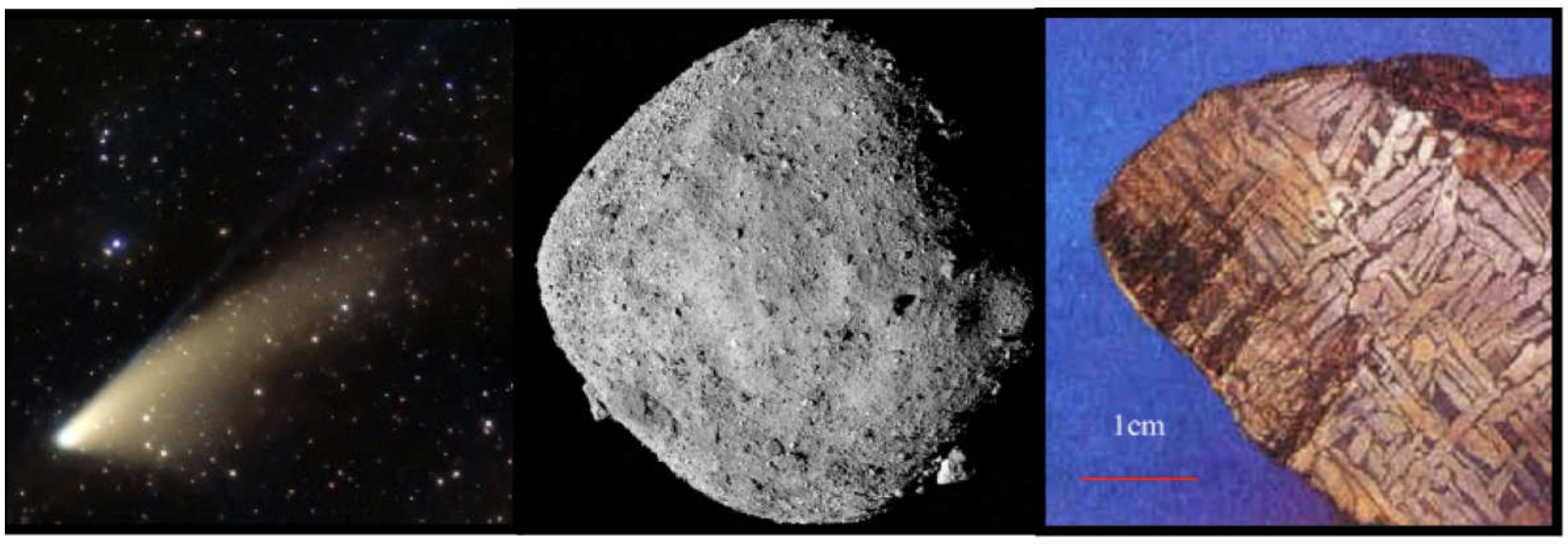

Figura 2.6: *(Izquierda)* Cometa NEOWISE, pasando por cielo peruano [Crédito: José y Hugo Santivañez, Observatorio y Planetario Discovery, Perú]. *(Centro)* Asteroide Bennu [Crédito: NASA]. *(Derecha)* Meteorito caído en Tambo Quemado, Ayacucho, Perú; muestra del Museo Nacional de Historia Natural [Crédito: INGENMET].

A diferencia de los cometas, los **asteroides** vienen del cinturón de asteroides que se encuentra entre las órbitas de Marte y Júpiter (2,2 – 3,2 UA). Los asteroides están compuestos principalmente por roca y polvo y no contienen hielo por encontrarse más cerca al Sol. El enorme campo gravitacional de Júpiter en cercanía al de Marte evita que los fragmentos de roca en el cinturón de asteroides se coagulen para formar un planeta nuevo, aunque la masa total de todos los asteroides combinados es menor que la de la Luna. La mayoría de éstos no representan ningún peligro, siempre y cuando pasen a una distancia mayor a 0,5 UA de la Tierra. Aún no queda claro si fue un asteroide o cometa (10 – 80 km de diámetro) el que impactó la Tierra hace 65 millones de años y extinguió a los dinosaurios.

Los **meteoroides** son fragmentos pequeños de roca, provenientes de colisiones de cometas y asteroides. Pueden ser rocosos, metálicos o una combinación de ambos. Se le llama **meteoro** a un meteoroide que ingresa a la atmósfera. La fricción del aire erosiona al meteoro por la alta velocidad a la que cae y se puede ver como una estrella fugaz. Si el meteoro logra impactar con la Tierra, recibe el nombre de **meteorito**.

## 2.2 ¿Cómo se formó el sistema solar?

La composición y ubicación de los componentes del sistema solar nos ayuda a entender su formación como una consecuencia de la formación del Sol. Notamos que la estrella es el objeto con mayor masa del sistema y por lo tanto, el que ejerce la mayor influencia gravitacional sobre los demás objetos, además de ser la fuente principal de energía para todo el Sistema Solar. Los planetas rocosos son de menor tamaño que los gaseosos y se encuentran entre el Sol y Júpiter. A partir de Júpiter, todos los planetas son gaseosos.

Luego de la formación del Sol hace 4 600 millones de años, quedó una nube de polvo y gas que giraba a su alrededor. El calor del Sol evaporó la mayoría de los compuestos volátiles en su cercanía, dejando rocas y polvo que luego fueron chocando en colisiones inelásticas formando fragmentos cada vez más grandes (planetesimales) que eventualmente formaron los planetas rocosos o terrestres. Se cree que el mismo proceso ocurrió con los planetas gaseosos para formar un núcleo rocoso y luego adquirir una atmósfera gaseosa recolectando gases durante su órbita gracias a su propia gravedad. Estos planetas, al estar más alejados del Sol, se encuentran en una zona de menor temperatura y con una gran reserva de gases. Hoy en día, las órbitas de los planetas están alineadas ($\pm 3^{\circ}$) con el plano eclíptico definido por la órbita de la Tierra alrededor del Sol. Esto nos indica que los planetas se originaron de un mismo disco.

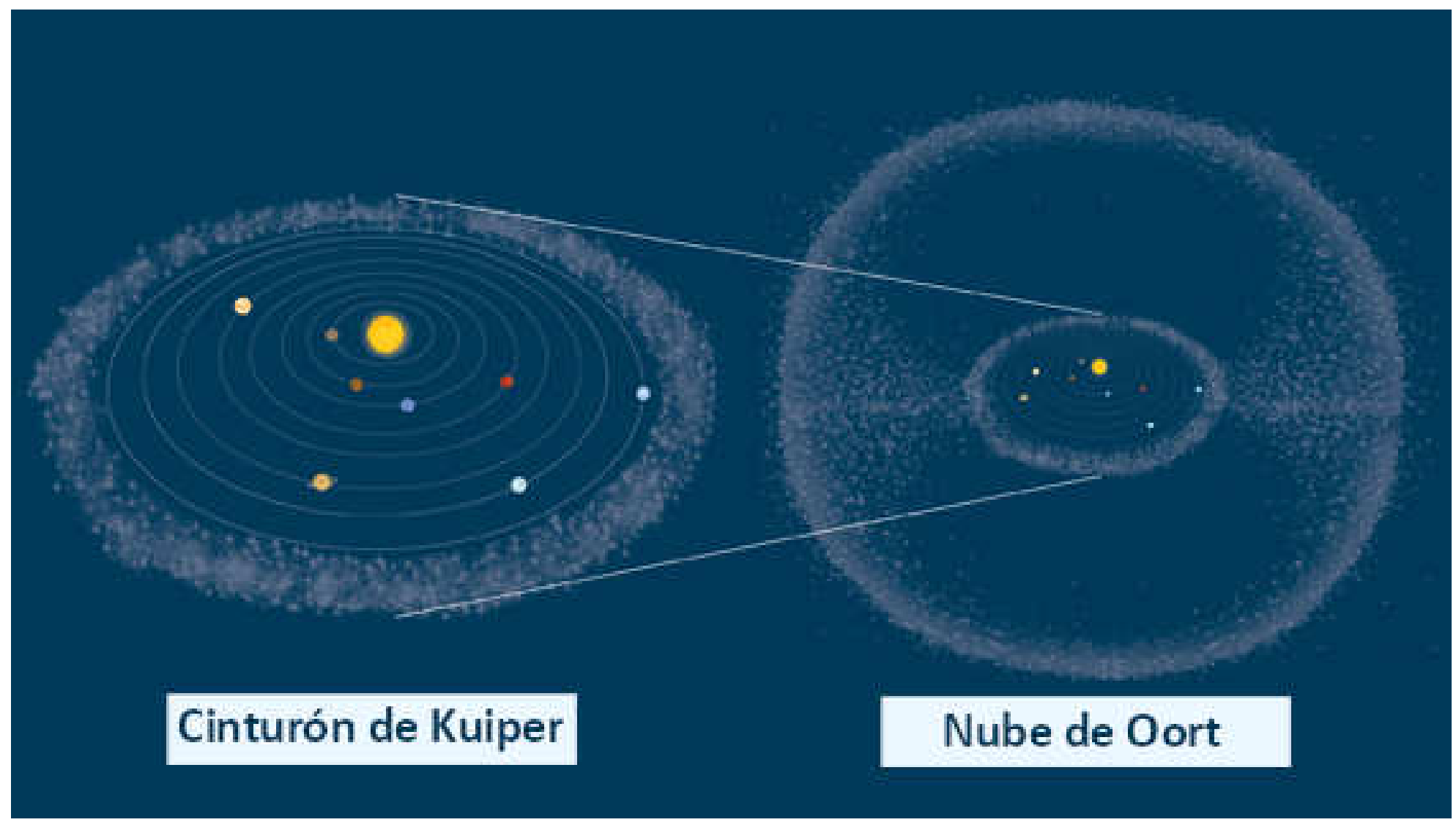

Figura 2.7: Ilustración que muestra los dos reservorios principales de cometas en el Sistema Solar, el Cinturón de Kuiper, a una distancia de 30 a 50 UA, y la Nube de Oort, que puede extenderse hasta 200 000 UA del Sol. Se piensa que el cometa Halley, uno de los cometas más famosos, proviene de la Nube de Oort [Crédito: ESA].

y polvo que luego fueron chocando en colisiones inelásticas formando fragmentos cada vez más grandes (planetesimales) que eventualmente formaron los planetas. Los cometas, asteroides y meteoritos son restos de rocas que no lograron formar parte de un planeta, y ofrecen pistas sobre el origen del agua y las materias primas que hicieron posible la vida en la Tierra.

**Recurso TIC 2.2.1:**

En el siguiente enlace: `https://phet.colorado.edu/sims/html/gravity-and-orbits/latest/gravity-and-orbits_en.html` puede encontrar una simulación interactiva para investigar los parámetros que definen una órbita estable.

## 2.3 Órbitas

Todos los objetos del sistema solar siguen una trayectoria alrededor del Sol, enteramente definida por su gravedad. Los antiguos griegos y egipcios creían que la Tierra era el centro del universo y que todos los planetas y estrellas giraban en torno a ella. En el siglo XII, astrónomos árabes fueron los primeros en sugerir que los planetas se movían como consecuencia de un intercambio de energía. En 1543, el modelo de Nicolaus Copernicus postulaba que los planetas giraban en torno al Sol. Sin embargo, entre 1609 y 1616, Johannes Kepler le dio sentido a los datos de posiciones planetarias de Tycho Brahe, resumiendo sus movimientos en tres leyes.

En el siglo XVII, al postular Isaac Newton su ley de gravitación universal, logró explicar las leyes empíricas de Kepler como manifestaciones de la gravedad:

1. Ley de órbitas: los planetas se mueven en órbitas elípticas como resultado de la proporcionalidad inversa de la distancia al cuadrado. Podemos pensar que el planeta está "cayendo" hacia el Sol de la misma forma que una pelota cae hacia el piso, pero como el planeta tiene una velocidad tangencial en su órbita, entonces continúa avanzando.
2. Ley de áreas: los planetas cubren áreas iguales en tiempos iguales como consecuencia de la conservación de momento angular: cuando el planeta está más cerca al Sol, se mueve más rápido y cuando se aleja del Sol se mueve más lento.
3. Ley de periodos: el cuadrado del periodo de traslación de un planeta ($P$) es proporcional al cubo del semieje mayor de su órbita ($a$). La constante de proporcionalidad en la ley de periodos es la constante de gravitación universal, $G$.

$$P^2 = \left(\frac{4\pi}{MG}\right) a^3 \tag{2.1}$$

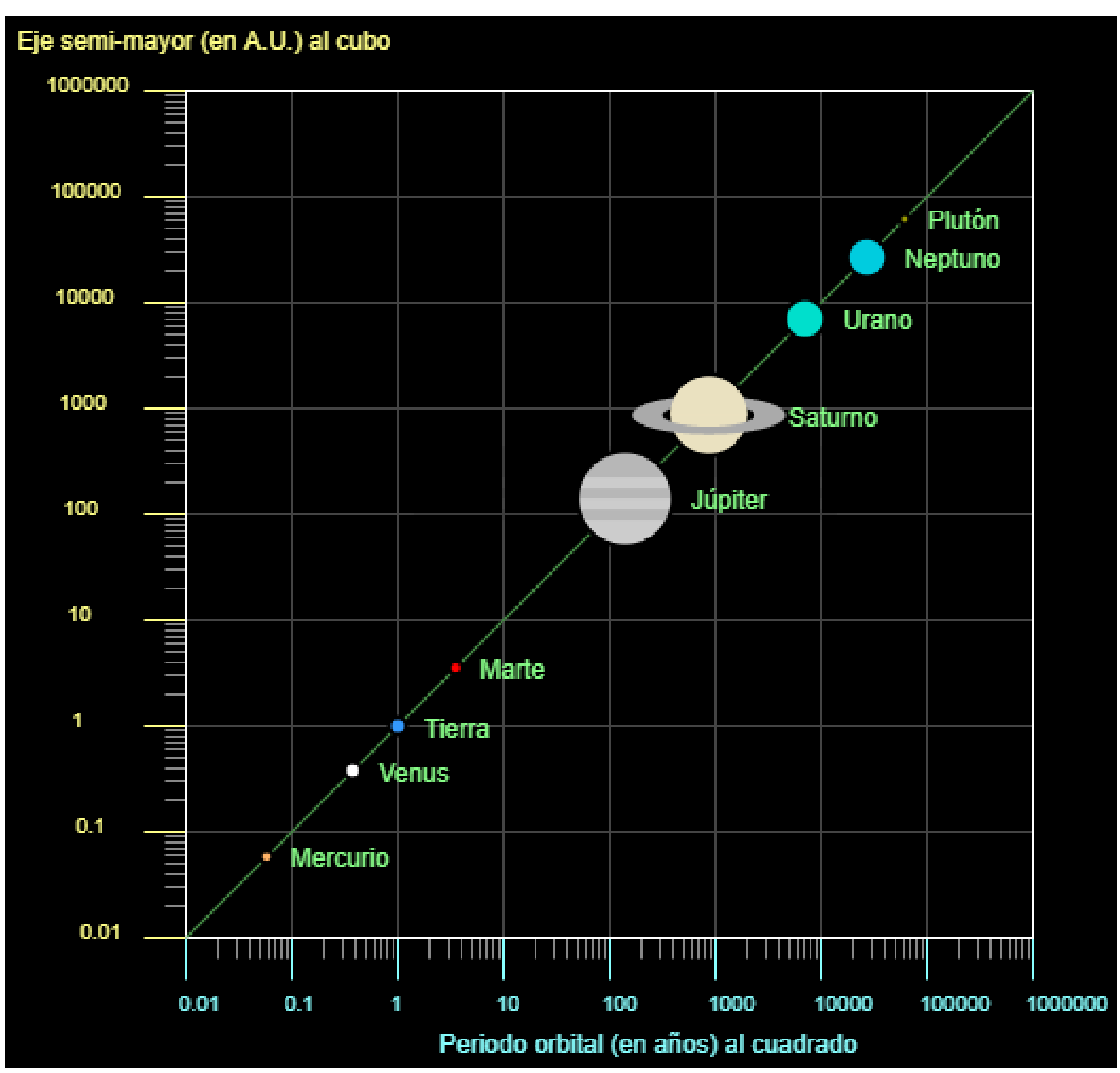

Figura 2.8: La gráfica muestra la proporcionalidad entre el cuadrado del periodo orbital y el cubo del semieje mayor comprendida en la tercera ley de Kepler [Crédito: Ramiro Moro].

## 2.4 Exploración del sistema solar

Durante la guerra fría entre Estados Unidos y la Unión Soviética en la década de 1960, ambas partes buscaban mejorar su tecnología en vuelos espaciales. A paritr de esta época fueron lanzadas misiones tripuladas y no tripuladas para explorar el sistema solar. Aquí detallamos algunas de las misiones más importantes:

1. **Sputnik-1 (1957):** Es el primer satélite artificial del mundo puesto en órbita, tenía aproximadamente el tamaño de una pelota de playa (58 cm de diámetro) y su masa era solo de 83,6 kg. Este lanzamiento marcó el comienzo de nuevos desarrollos políticos, militares, tecnológicos y científicos.
2. **Apolo XI (1969):** Apolo fue un programa diseñado para la exploración lunar que constó de 11 misiones, de las cuales la primera en pisar suelo lunar fue el Apolo 11. Los astronautas lograron traer muestras de rocas lunares para su estudio, con las cuales se pudo determinar, por ejemplo, que la edad de la Luna es similar al de la Tierra.
3. **Pioneer (1972):** Fue la primera nave espacial en viajar a través del cinturón de asteroides y la primera nave espacial en realizar observaciones directas y obtener imágenes de cerca de Júpiter. Dejo de enviar señales en 2003.
4. **Voyager I y II (1977):** Esta misión cuenta con dos sondas gemelas lanzados casi al mismo tiempo, la primera en agosto y la segunda en setiembre. Voyager I tenía como misión principal la exploración de las atmósferas de Júpiter y Saturno. Gracias a sus observaciones se descubrieron los volcanes en Io y además, nos brindó detalles de la complejidad de los anillos de Saturno. Ambas sondas ya se encuentran en el espacio interestelar, es decir, fuera del sistema solar.
5. **Cassini (1997):** Esta nave espacial no tripulada se dedicó a explorar principalmente a Saturno y satélites naturales. Con los datos enviados se pudo observar el hexágono polar de nubes que tiene este planeta, y puso a Encelado, una de sus lunas, como candidata a albergar vida. La misión culminó el 2017.
6. **Juno (2011):** Es una sonda enviada para explorar Júpiter. El principal objetivo es comprender su origen y evolución. Con los datos enviados se espera describir mejor la atmósfera y magnetósfera de este planeta gigante. La misión terminará en el 2025.

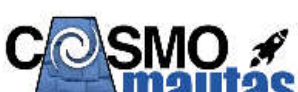

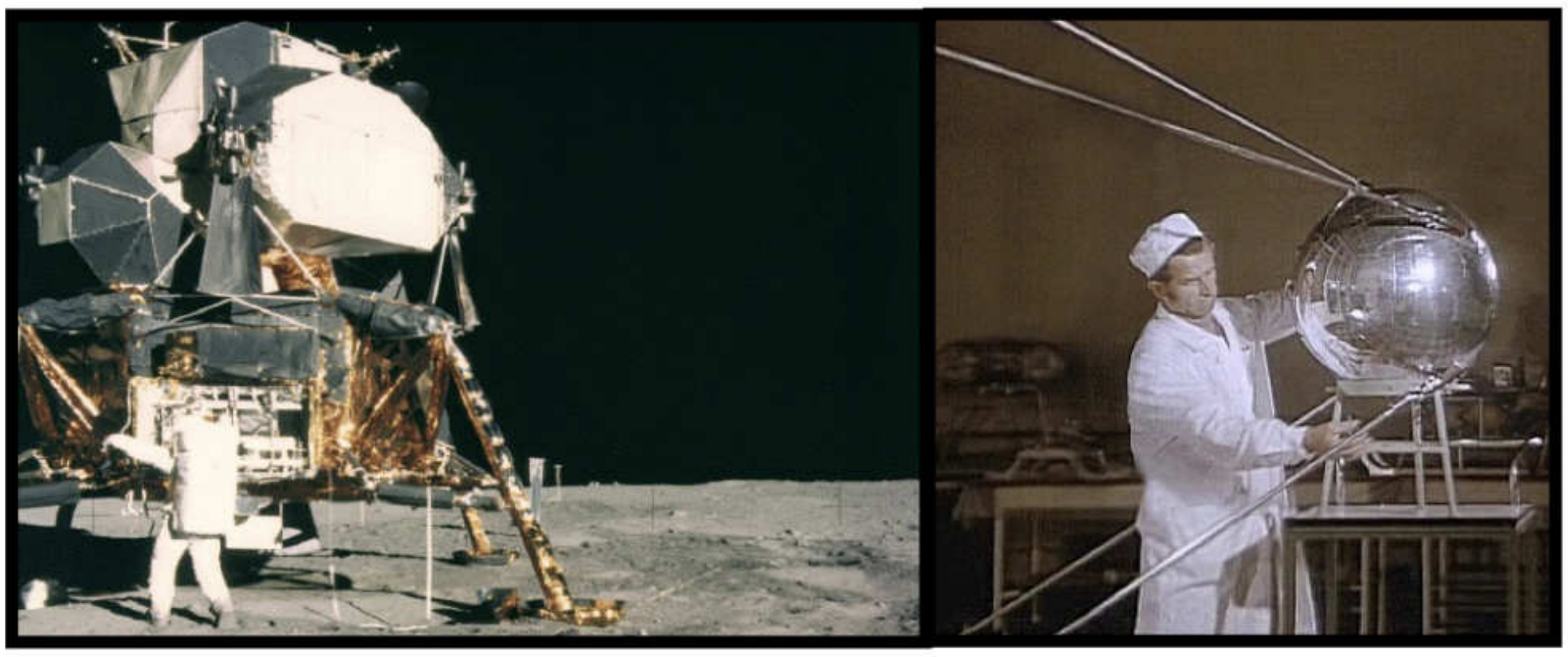

Figura 2.9: *(Izquierda)* El módulo lunar con el que descendieron a la superficie lunar los tripulantes del Apolo 11 [Crédito: NASA]. *(Derecha)* El satélite Sputnik durante pruebas técnicas [Crédito: ROSCOSMOS].

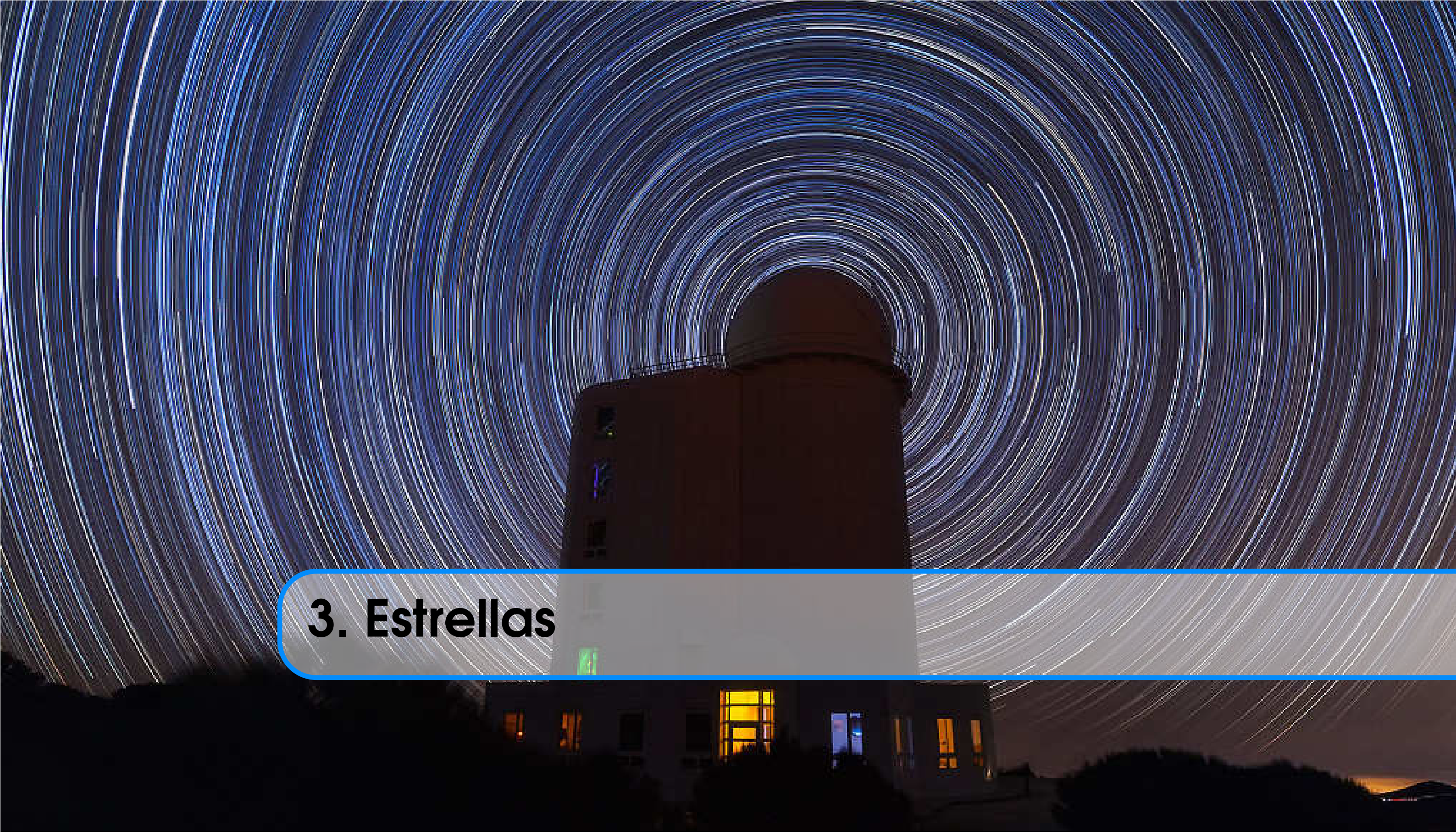

# 3. Estrellas

Las luces brillantes que la humanidad ha observado con asombro desde tiempos antiquísimos en una noche despejada son lo que hoy en día conocemos como estrellas. Las estrellas son objetos similares a nuestro Sol, con propiedades físicas casi enteramente determinadas por su masa. En este capítulo vamos a hablar sobre las estrellas, brindando una perspectiva general sobre el estudio de estas, mostrando sus características generales y las leyes que las gobiernan.

## 3.1 ¿Cómo funcionan las estrellas?

Las estrellas son gigantes esferas de plasma de hidrógeno y helio que se mantienen unidas gracias a su propia gravedad. El plasma es un estado de la materia que ocurre cuando un gas se calienta a temperaturas tan altas que sus átomos se han ionizado, de tal forma que los electrones, núcleos atómicos (o elementos altamente ionizados) y protones que se encuentran libres. La gravedad de la estrella genera altas presiones en su núcleo, produciendo un incremento en la temperatura. Las altas temperaturas y presiones en el núcleo de una estrella crean las condiciones idóneas para la fusión de átomos de hidrógeno en helio y esta reacción nuclear expulsa radiación que se dirige del núcleo hacia la superficie. El continuo balance de la fuerza de gravedad (que apunta hacia adentro) y la presión termal causada por la fusión de hidrógeno (que apunta hacia afuera) hace que las estrellas se mantengan del mismo tamaño emitiendo una cantidad constante de energía por millones de años. A esta lucha constante entre la fuerza de gravedad y la presión termal de la estrella, se le conoce como la **ley de equilibrio hidrostático** (Figura 3.1).

---

Imagen de encabezado: Por la rotación de la Tierra las estrellas parecen girar en torno al polo norte celeste. Fotografía de larga exposición tomada en el Observatorio El Teide, Islas Canarias, España [Crédito y copyright: Gabriel Funes].

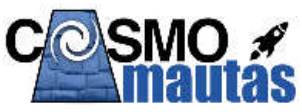

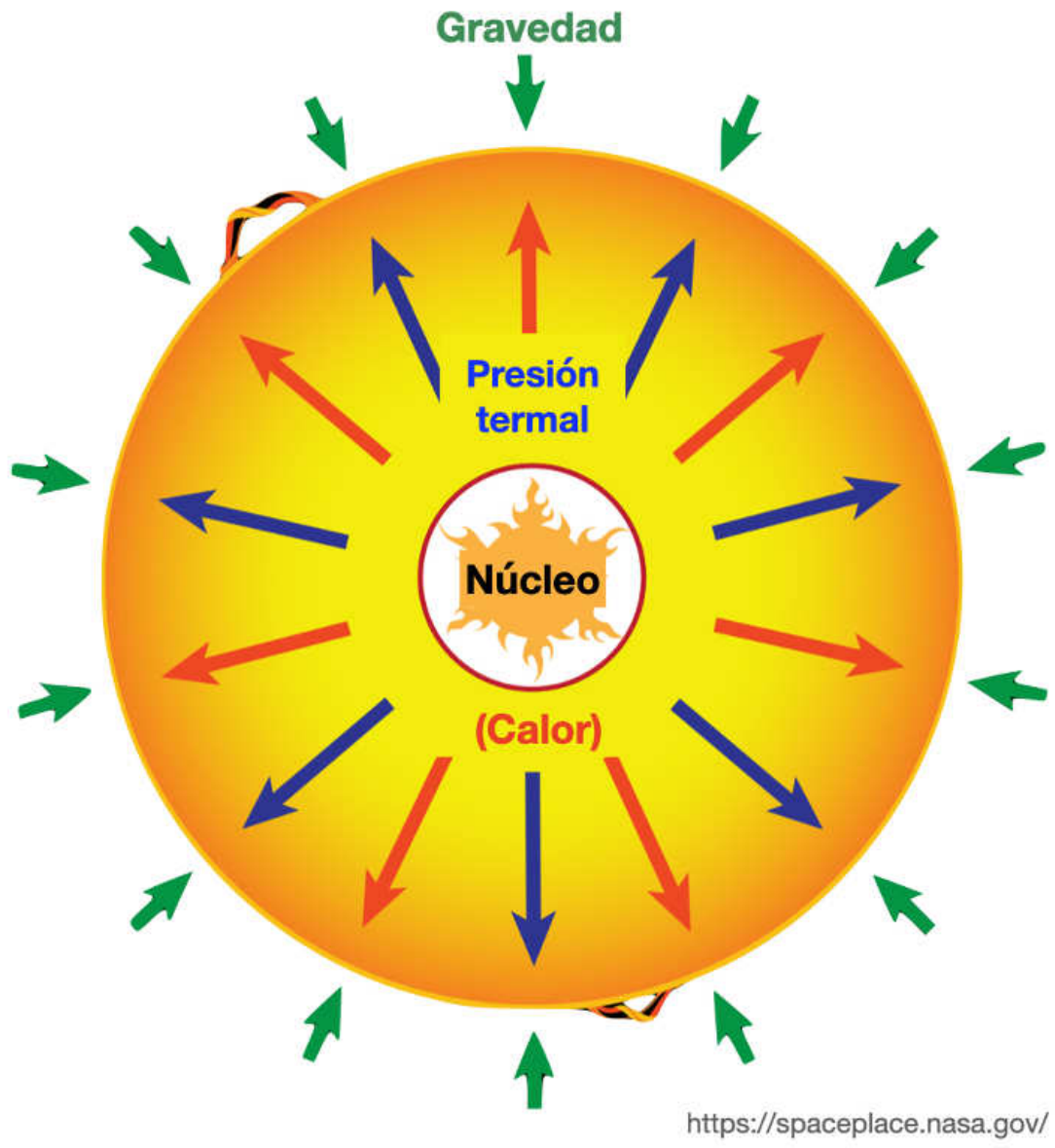

Figura 3.1: El balance entre la gravedad que trata de comprimir a la estrella (flechas verdes) y la presión termal o calor producido por las reacciones nucleares en el núcleo de la estrella (flechas azules y rojas) se conoce como equilibrio hidrostático [Crédito: NASA Science Space Place, `https://spaceplace.nasa.gov/`].

### 3.1.1 El motor: la fusión de hidrógeno

Las estrellas generan energía a través de la fusión de átomos de hidrógeno en helio. En general, están compuestas de $\sim 90\%$ hidrógeno, $\sim 10\%$ helio y trazas de otros elementos. Los astrónomos llamamos "metales" a todos los elementos con números atómicos mayores a 2 que corresponde a helio por la gran abundancia de hidrógeno y helio en el universo. Todos los elementos químicos que constituyen el universo fueron sintetizados en el núcleo de las estrellas (nucleosíntesis estelar) o en eventos altamente energéticos (supernovas, fusión de estrellas de neutrones, etc.), a diferencia de los elementos más ligeros (hidrógeno, helio y litio) que se formaron durante el *Big Bang*.

**Física 3.1.1: Introducción a la física de partículas**

Recordemos que en el universo hay 3 partículas principales que forman los átomos: **protones, neutrones y electrones.**

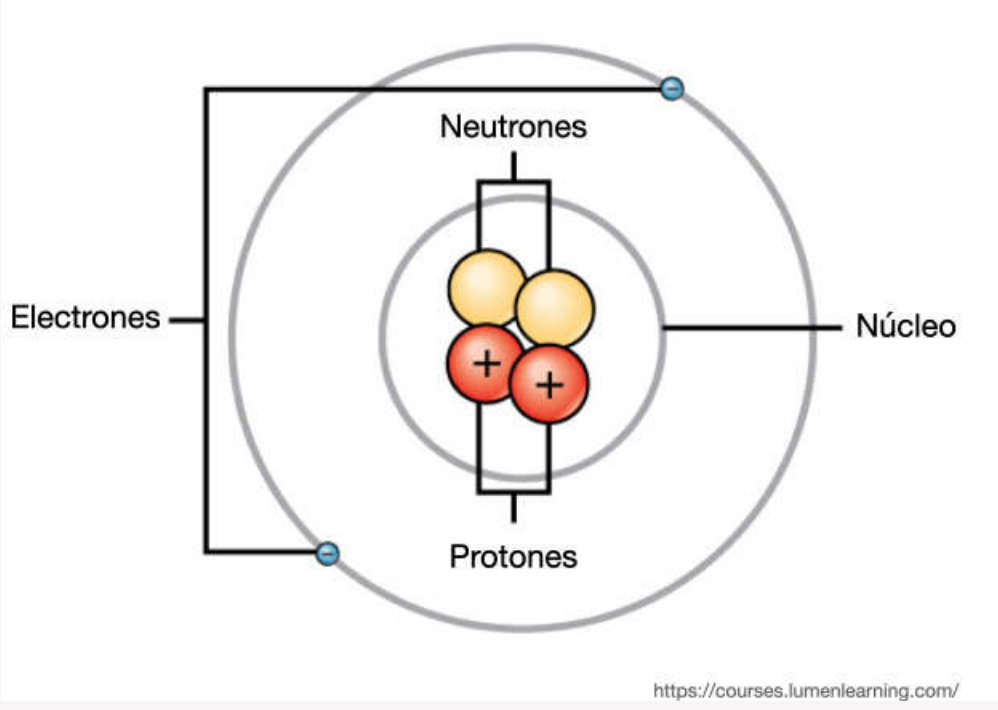

El protón y el neutrón tienen tienen aproximadamente la misma masa, sin embargo, el protón presenta carga positiva y el neutrón no presenta carga. Por otro lado, el electrón presenta cara negativa y una masa que es dos mil veces menor que el protón. Lo que dicta la identidad de cada elemento químico es el número de protones (también llamado número atómico), por ejemplo, el hidrógeno (H) tiene 1 protón, el helio (He) tiene 2 protones y el hierro (Fe) tiene 26 protones.

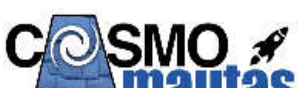

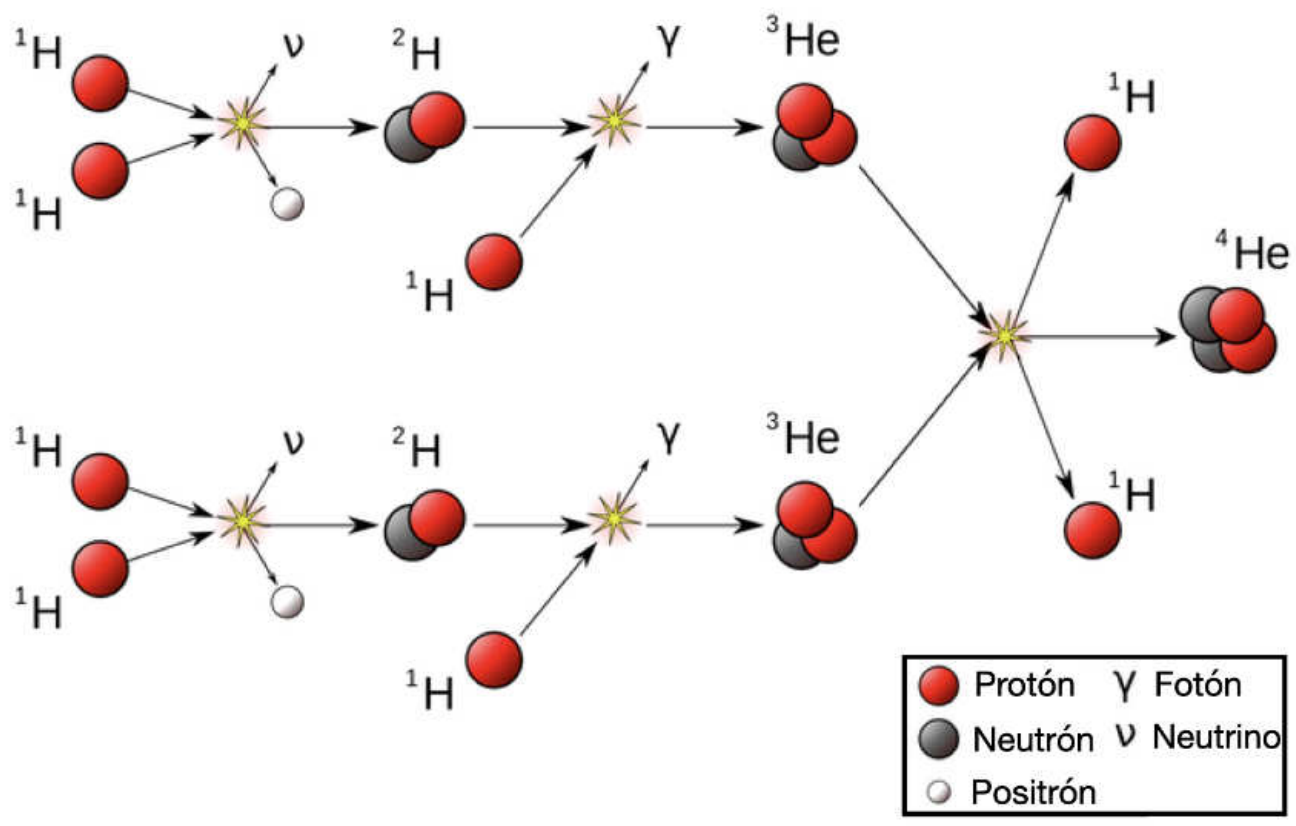

Figura 3.2: La cadena protón-protón [Crédito: imagen adaptada de CESAR booklet.]

La principal vía de producción de energía para estrellas como el Sol es la **cadena protón-protón** en donde la fusión de cuatro átomos de hidrógeno produce un átomo de helio y libera energía. En el ambiente ionizado de las estrellas, un átomo de hidrógeno tiene solo un protón. Este complejo proceso se resume en tres pasos principales:

1. Dos protones fusionan generando deuterio o hidrógeno pesado, liberando un "positrón" (una partícula con la misma masa de un electrón pero con carga positiva) y un neutrino (una partícula muy ligera). Al liberarse un positrón, uno de los protones cambia a neutrón quedando el núcleo como un protón y un neutrón. Este núcleo tiene la misma carga que el hidrógeno, pero su masa es el doble, es el hidrógeno pesado o deuterio.
2. Un protón colisiona con un deuterio generando helio-3 y liberando energía. El helio-3 es un isótopo de helio con dos protones y un neutrón. Este helio tiene el mismo número atómico que el helio común, pero es más ligero en masa.
3. Al repetir los pasos 1 y 2 otra vez, tenemos 2 átomos de helio-3, los cuales se fusionan para crear un átomo de helio común y dos átomos de hidrógeno.

En esta cadena, cuatro núcleos de hidrógeno han formado un núcleo de helio. Se necesitan entre 100 mil y 1 millón de años para que la energía generada en el núcleo de la estrella sea irradiada al espacio porque este es el tiempo le toma a un fotón en llegar a la superficie de la estrella.

### 3.1.2 El color de luz: depende de su temperatura

Las estrellas se distinguen por su color, el mismo que está relacionado a su temperatura superficial, por ello, las estrellas más calientes presentan un color azulado y las estrellas más frías son mas bien rojizas. La energía emitida por las estrellas puede ser representada mediante un modelo teórico llamado radiación de **cuerpo negro**, donde el cuerpo absorbe toda la energía incidente en él y la re-emite toda la radiación, sin reflejarla ni dispersarla, en todas las longitudes de onda. Sin embargo, lo que caracteriza la radiación de un cuerpo negro es su temperatura (independientemente de su forma, material y constitución interna, ¡por que es un modelo teórico!). La distribución de intensidad de radiación de un cuerpo negro en longitudes de onda sigue la **ley de Planck** (Figura 3.3). De hecho el ejemplo más cercano y real que tenemos de un cuerpo negro son las estrellas.

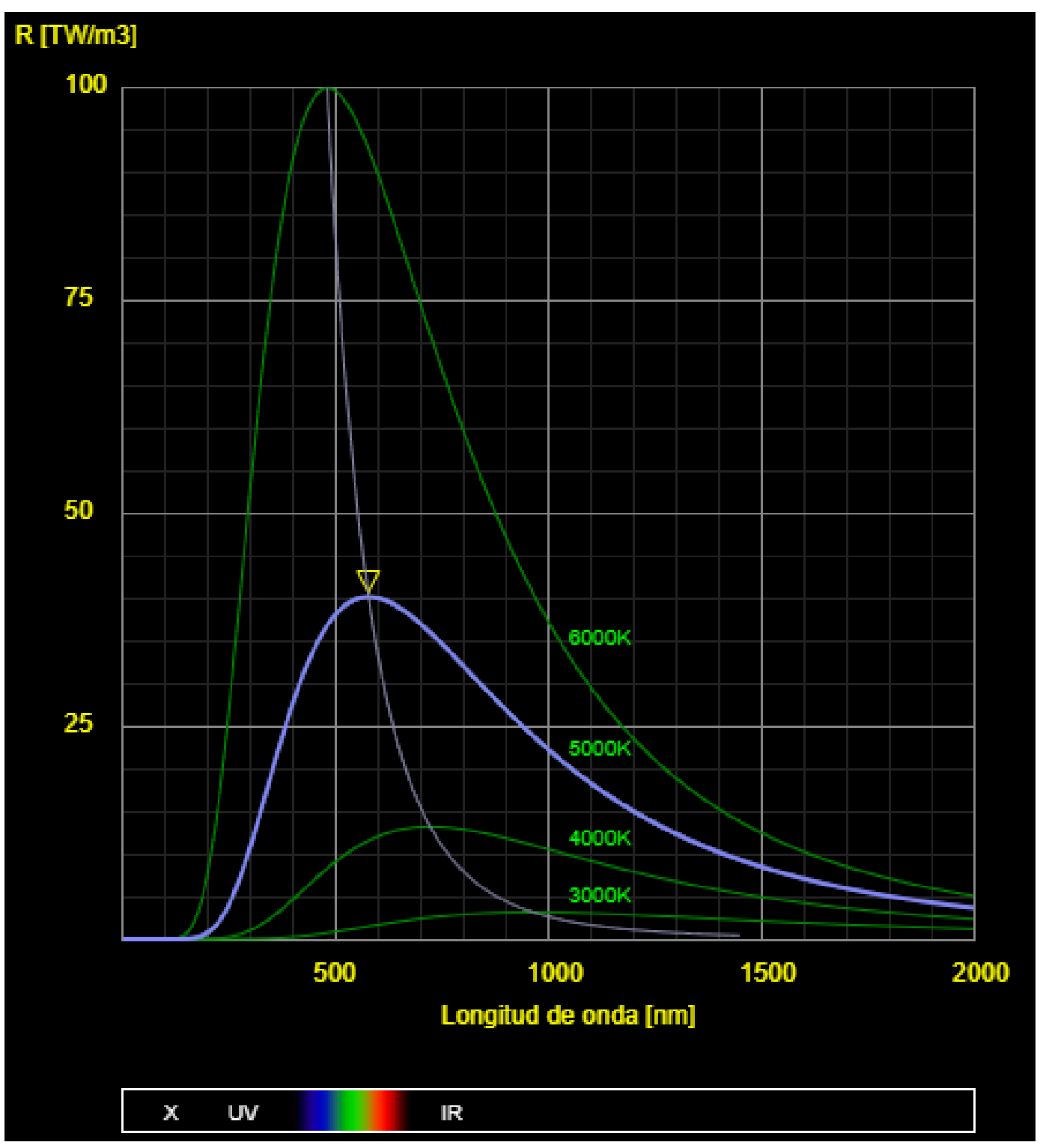

Figura 3.3: Esta imagen muestra la distribución de intensidad de radiación de un cuerpo negro, conocida como la ley de Planck. A menores temperaturas, el pico de intensidad de radiación se encuentra a longitudes de onda más largas y menos energéticas. A esta relación entre temperatura y longitud de onda para el pico de intensidad se le conoce como ley de Wien [Crédito: Ramiro Moro, `http://rok.50webs.com/courseware/modern1/blackbody_sim.html`].

Basados en la Figura 3.3, podemos encontrar la longitud de onda en la cual se alcanza el pico máximo de intensidad de la ley de Planck (en matématicas, a esa operación se le llama derivación). A esta relación se le conoce como la ley de Wien:

$$\lambda_{max} = \frac{2{,}898 \times 10^{-3}\, mK}{T} \tag{3.1}$$

De esta manera demostramos lo que dijimos al inicio: que las estrellas más calientes producen radiación que alcanza su punto máximo en longitudes de onda más cortas, más energéticas y más “azules”, mientras que las estrellas más frías producen radiación que alcanza su punto máximo en longitudes de onda más largas, menos energéticas y mas “rojas”, como se muestra en la Figura 3.3.

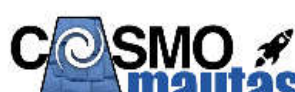

### 3.1.3 La luminosidad: depende de su temperatura

Al sumar la intensidad de la ley de Planck sobre todas las longitudes de onda emitidas por el cuerpo negro (en matemáticas, a esta operación se le llama integración), obtenemos la ley de Stefan-Boltzmann, la cual establece que un cuerpo negro emite radiación proporcional a la cuarta potencia de su temperatura:

$$F = \sigma T^4, \qquad \sigma = 5{,}670 \times 10^{-8} W m^{-2} K^{-4} \tag{3.2}$$

Donde $F$ es el flujo del cuerpo negro (en unidades de $W/m^2$), es decir, la energía total irradiada por unidad de área y por unidad de tiempo, $T$ es la temperatura del cuerpo negro (en K), y $\sigma$ es la constante de Stefan-Boltzmann. Cuando hablamos de estrellas, usamos el término "temperatura efectiva" ($T_{eff}$, en inglés) para referirnos a la temperatura de cuerpo negro que mejor reproduce la temperatura de la estrella. Por ejemplo, el Sol tiene una temperatura efectiva de $\sim 6000\,$K. Una estrella roja como Próxima Centauri, la estrella más cercana al Sol, tiene una temperatura efectiva de $\sim 3000\,$K. Por la ley de Stefan-Boltzmann, el flujo del Sol ($F_\odot$) vendría a ser 16 veces el flujo de Próxima Centauri ($F_{PC}$):

$$\frac{F_\odot}{F_{PC}} = \frac{\sigma(6000\,K)^4}{\sigma(3000\,K)^4} = \left(\frac{6000}{3000}\right)^4 = 2^4 = 16 \tag{3.3}$$

A partir del flujo, podemos calcular la luminosidad de una estrella, que viene a ser la *energía total* irradiada por la estrella en una unidad de tiempo. Mientras que el flujo proveniente de la estrella depende de la distancia a la que hacemos la medición (Figura 3.4), la luminosidad es una propiedad intrínseca de la estrella. Para obtenerla, tenemos que multiplicar el flujo (que es energía por unidad de área) por el área de la superficie de la estrella, $4\pi R^2$, donde $R$ es el radio de la estrella:

$$L = AF = (4\pi R^2)(\sigma T^4) \tag{3.4}$$

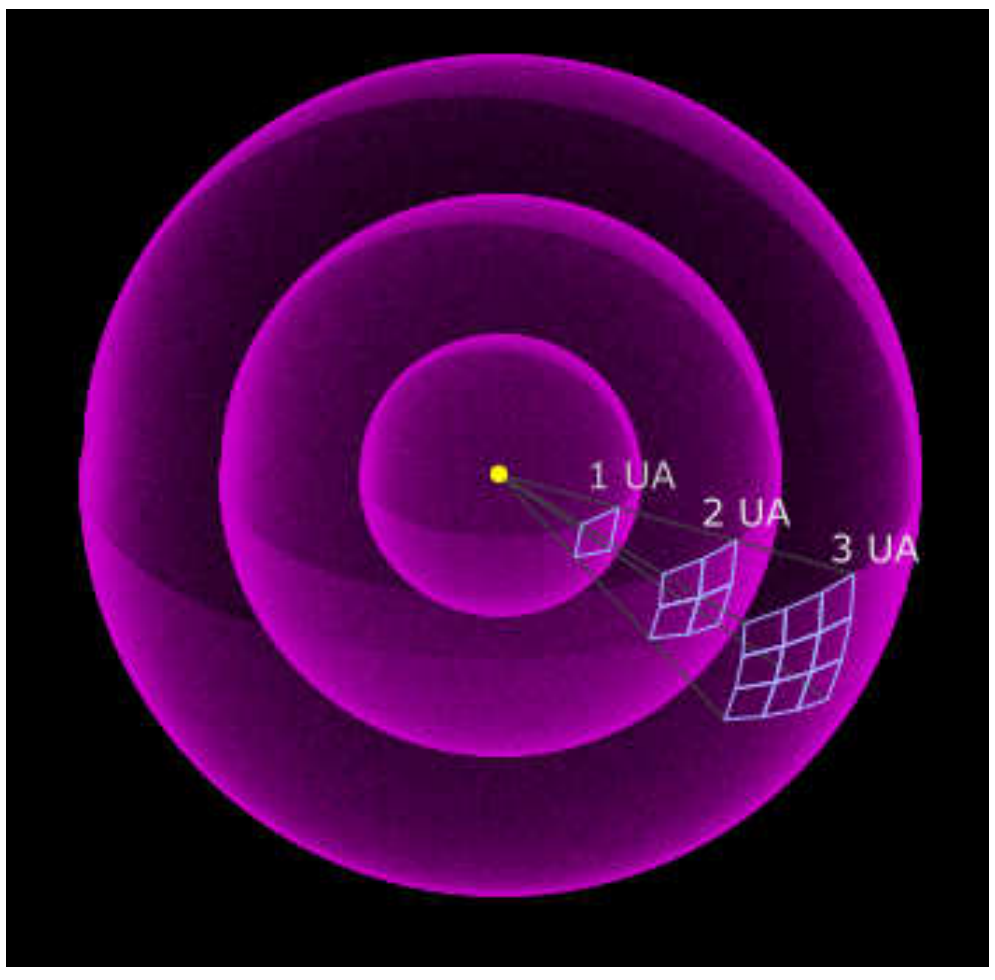

Figura 3.4: La luminosidad es la energía total que emite la en todas las direcciones, mientras que el flujo depende de la distancia a la que se encuentre la unidad de área donde se realice la medición. Conforme se incrementa la distancia desde la estrella, crece el radio de la esfera y se reduce el flujo, o la luminosidad por unidad de área [Crédito: Ramiro Moro].

### Luz 3.1.1: El espectro electromagnético

La luz visible es la radiación electromagnética que puede ser detectada por el ojo humano, se encuentra en el rango aproximado de los 400 nm para la luz violeta hasta 700 nm para la luz roja. Sin embargo, la luz abarca un rango mucho más amplio de longitudes de onda, llamado el espectro electromagnético. Desde ondas de radio de baja energía con longitudes de onda que se miden en metros, hasta rayos gamma de alta energía con longitudes de onda inferiores a $10^{-12}$ metros, la radiación electromagnética describe las fluctuaciones de los campos eléctricos y magnéticos que transportan energía a la velocidad de la luz (que es de 300000 km/s en el vacío). La luz visible no es inherentemente diferente de otras partes del espectro electromagnético, con la excepción de que el ojo humano puede detectar ondas visibles.
La luz también se puede describir en términos de una corriente de **fotones**: paquetes de energía sin masa, cada uno de los cuales viaja a la velocidad de la luz. Un fotón es la cantidad más pequeña (cuántica) de energía que puede ser transportada y equivale a:

$$E = h\nu \tag{3.5}$$

donde $E$ es la energía de un fotón de frecuencia $\nu$ y $h = 6{,}626 \times 10^{-34}$ J s es la constante de Planck. Cabe recordar que la frecuencia es la cantidad inversa a la longitud de onda $\lambda$ y se relacionan a través de la velocidad de la luz, $c$:

$$c = \lambda\nu \tag{3.6}$$

Comprender que la luz viajaba en cuantos discretos significó el origen de la teoría cuántica. La luz es algo que escapa de nuestro entendimiento, por eso usamos modelos de ondas y partículas como convenga [Crédito: imagen adaptada de Wikipedia/Creative Commons].

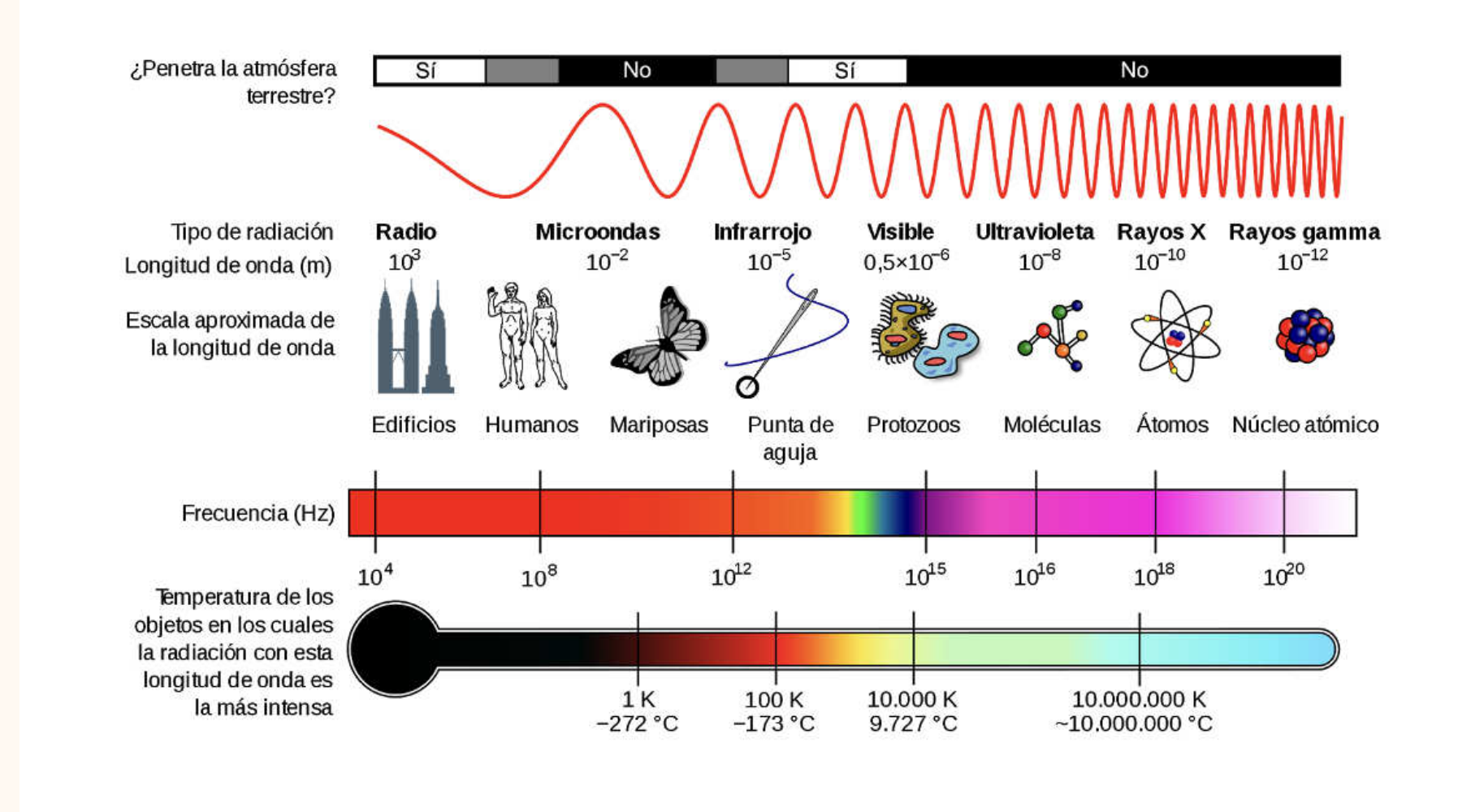

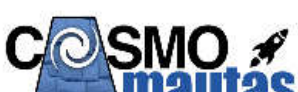

## 3.2 Magnitudes estelares

Casi toda la información que los astrónomos han recibido sobre el universo más allá de nuestro sistema solar proviene del estudio cuidadoso de la luz emitida por estrellas, galaxias y nubes interestelares de gas y polvo. Nuestra comprensión moderna del universo ha sido posible por la medición cuantitativa de la intensidad de la luz en cada parte del espectro electromagnético. Sin embargo, antes del inicio de la era del telescopio, los astrónomos ancestrales observaban estrellas a simple vista. El ojo humano ha evolucionado para identificar cantidades en escalas logarítmicas (a esto se le conoce como la ley de Weber–Fechner, para los curiosos) para asegurar su supervivencia. Por eso, a simple vista podemos diferenciar entre uno y varios pumas que nos amenazan, pero nos cuesta ver si hay 30 o 32 alpacas en un rebaño. De la misma forma, al ver estrellas de diferente brillo, estamos haciendo cálculos logarítmicos sin darnos cuenta.

En la sección anterior aprendimos sobre flujo y luminosidad de forma teórica. En la práctica, medir cantidades como el radio, la temperatura, o incluso la longitud de onda en el pico de intensidad de una estrella es difícil o imposible. Por esta razón, los astrónomos se valen de cantidades más accesibles para poder inferir las propiedades físicas de las estrellas.

### 3.2.1 La magnitud aparente de una estrella es análoga al flujo

El astrónomo griego Hiparco fue uno de los primeros observadores en catalogar las estrellas que vio. Hiparco inventó un sistema numérico en escala para describir cuán brillante es una estrella en el cielo. Él asignó una **magnitud aparente** $m = 1$ a la estrella más brillante en el cielo, y asignó a la estrella más brillante la magnitud 1 y a la de brillo más débil, apenas distinguible a simple vista, magnitud 6. Desde los tiempos de Hiparco hasta nuestros día, los astrónomos han ampliado y mejorado la escala de la magnitud manteniendo el sentido de la escala, es decir, a medida que la estrella sea más brillante, el valor de la magnitud será un número más pequeño y si la estrella es más débil el valor de la magnitud será mayor. Esta magnitud la llamaremos magnitud aparente y la podemos expresar en función al flujo de dos estrellas:

$$m_1 - m_2 = -2{,}5 log_{10}\left(\frac{F_1}{F_2}\right) \tag{3.7}$$

### 3.2.2 La magnitud absoluta de una estrella es análoga a la luminosidad

Si sólo sabemos la magnitud aparente de una estrella, no podemos determinar si estamos viendo a una estrella lejana muy brillante y caliente o una estrella cercana débil y fría. Podríamos resolver nuestro problema si pudiéramos traer a todas las estrellas a la misma distancia para comparar sus brillos. La **magnitud absoluta (M)** se define como la magnitud aparente de una estrella a una distancia de 10 *parsec* ( pc).

$$m - M = 5\, log_{10}\left(\frac{d}{10\, pc}\right) \tag{3.8}$$

donde $m$ es la magnitud aparente de la estrella, $M$ es la magnitud absoluta, y $d$ es la distancia de la estrella desde la Tierra.

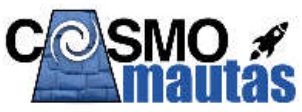

**Unidades 3.2.1: Paralaje trigonométrico**

El método que usamos para medir distancias a estrellas cercanas se conoce como **paralaje trigonométrico**, que viene a ser el cambio de posición *aparente* de una estrella con respecto al fondo distante de estrellas “fijas” debido al movimiento anual de la Tierra. Si estiramos un brazo frente a nosotros, levantamos un dedo y abrimos y cerramos un ojo a la vez, vemos como si nuestro dedo se moviera de derecha a izquierda. De la misma forma, a medida que la Tierra orbita al Sol, podemos medir la posición de una estrella con respecto al fondo de estrellas más distantes con 6 meses de diferencia. El movimiento aparente de la estrella cercana forma una línea que viene a ser la base de un triángulo. El ángulo opuesto a esta “base” es el doble del **paralaje**. Por la propiedad de triángulos semejantes, ese ángulo es el vértice de otro triángulo cuya base es el diámetro de la órbita de la Tierra. Como muestra la Figura 3.2.2, podemos encontrar la distancia a la estrella en parsecs midiendo la tangente del paralaje, $p$, en segundos de arco y sabiendo que el radio de la órbita de la Tierra es 1 UA:

$$tanp \approx p = \frac{1\,UA}{d} \tag{3.9}$$

Un *parsec* es el “paralaje de un arcosegundo”. Al hacer cambios de unidades correspondientes de radianes a arcosegundos ($1'' = 1/3600°$) y de unidades astronómicas a metros, encontramos que 1 pc = 206265 UA. Incluso para las estrellas más cercanas, los ángulos de paralajes son minúsculos. Para Próxima Centauri, la estrella más cercana al Sol, el paralaje es de tan solo $0''\!.7687 \pm 0''\!.0003$. Actualmente, la misión *Gaia* está midiendo el paralaje de $1,5 \cdot 10^9$ estrellas con precisiones de microarcosegundos [Crédito: imagen modificada del libro Fundamental Astronomy].

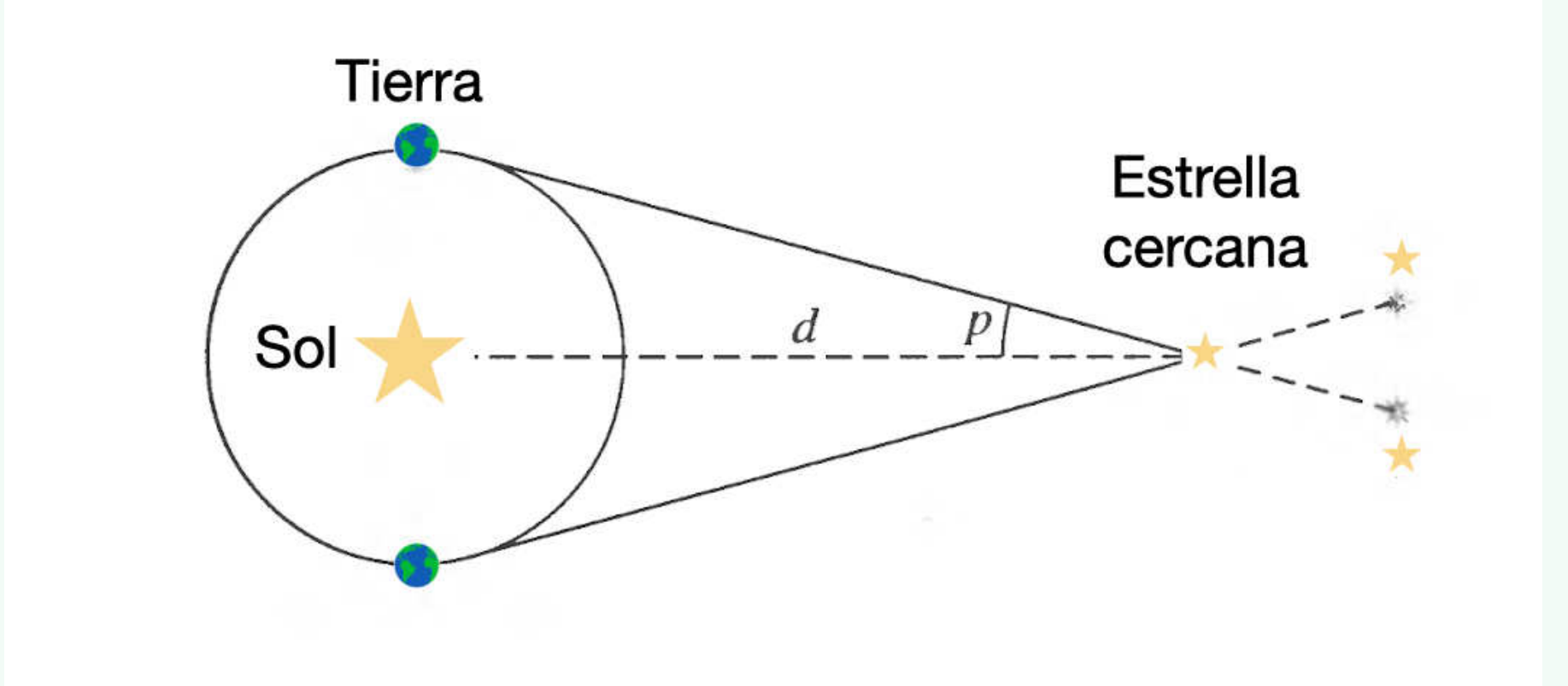

## 3.3 La huella “dactilar” de una estrella

En 1814, Joseph Fraunhofer, astrónomo, ingeniero y vidriero alemán, puso un prisma frente a un telescopio y vio que la luz de la estrella Sirius mostraba líneas oscuras en un orden diferente al de las líneas oscuras del Sol. Así se dio cuenta que la razón no podía ser la atmósfera de la Tierra y que eran una especie de firma o “huella dactilar” de cada estrella. Más aún, al comparar los espectros de las estrellas con experimentos de gases calientes en laboratorios de química, los científicos se dieron cuenta que la luz emitida por ciertos gases correspondían en longitud de onda a las líneas oscuras. De esta forma se descubrió que las líneas oscuras indicaban los componentes de la atmósfera de las estrellas.

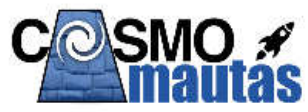

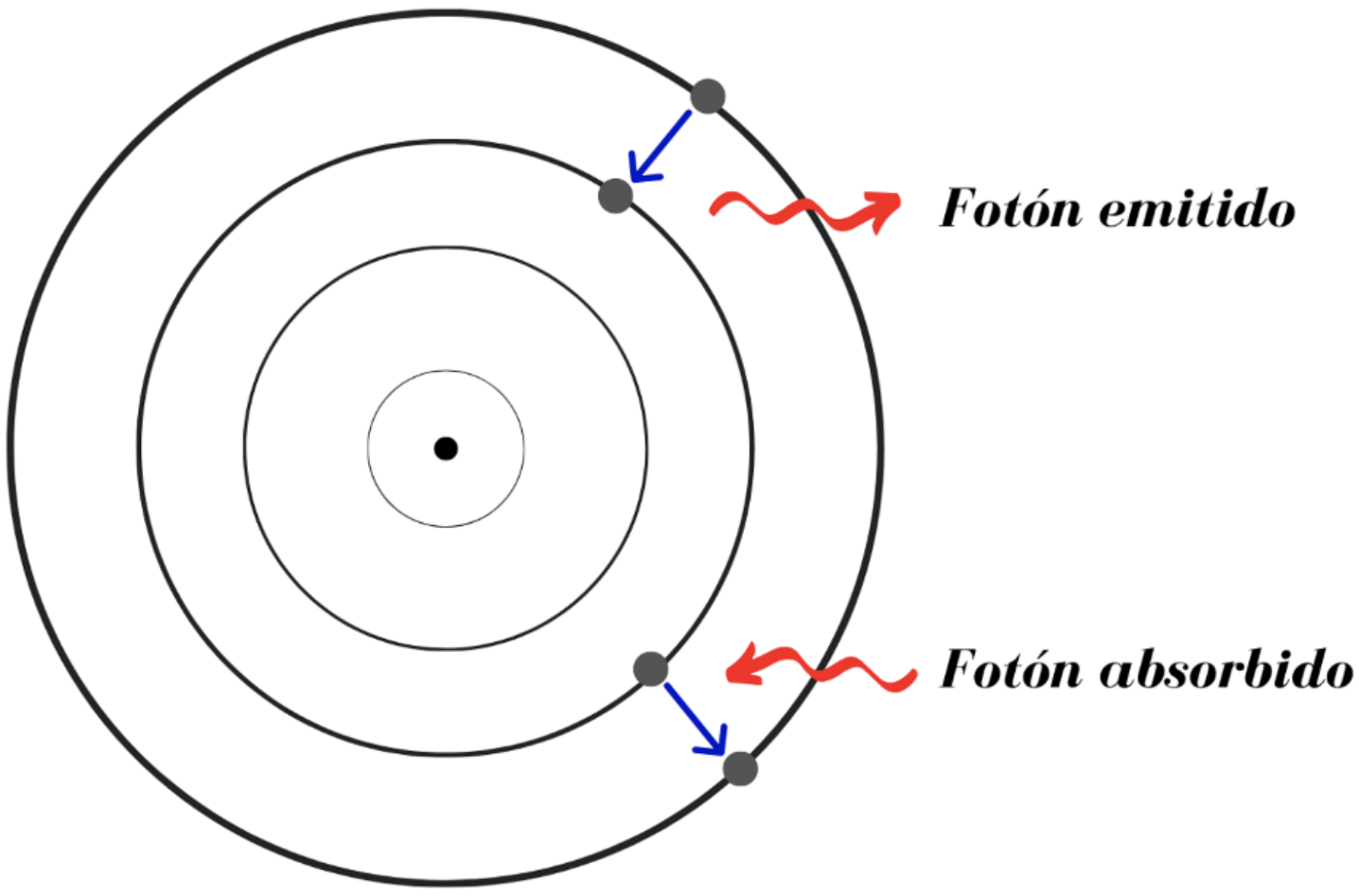

Figura 3.5: Saltos de nivel de energía dentro de un átomo. Cuando un electrón emite un fotón, baja de nivel de energía y viceversa [Crédito: Lisseth Gonzales].

Para entender este proceso, tenemos que visualizar la interacción entre los átomos y la luz (Figura 3.5). Los átomos están compuestos por un núcleo denso que contiene protones y neutrones y rodeados por “orbitales” de electrones. Los electrones ocupan ciertos niveles de energía que siguen las reglas de mecánica cuántica. Cuando un electrón absorbe un fotón de luz, su energía total se incrementa y el electrón tiene que “saltar” a un nivel más energético. De la misma forma, si un electrón emite un fotón, su energía se reduce y tiene que saltar a un nivel menos energético. La energía perdida por el electrón es llevada por un solo fotón.

Como vimos en la Sección 3.1.2, un cuerpo negro produce un espectro continuo (Figura 3.3), como un arco iris sin las líneas espectrales oscuras que vio Fraunhofer. Un gas denso y caliente o un objeto sólido caliente también produce un espectro continuo (Figura 3.6), en cambio un gas difuso y caliente produce líneas de emisión brillantes. Las líneas de emisión se producen cuando un electrón hace una transición descendiente de una órbita de más energía a una órbita de menos energía. Por otro lado, un gas frío y difuso frente a una fuente de espectro continuo produce líneas de absorción oscuras en el espectro continuo. Las líneas de absorción se producen cuando un electrón hace una transición de una órbita de menor a una de mayor energía.

El espectro del Sol que observó Fraunhofer nos permite identificar los elementos que componen la atmósfera solar. El Sol es un cuerpo negro que produce un espectro continuo y también es un gas caliente que produce líneas de emisión. Sin embargo, cuando la luz producida por el Sol atraviesa su atmósfera, que es relativamente más fría, los gases absorben fotones “quitando” ciertas líneas del espectro (las líneas de Fraunhofer). A la Tierra nos llega un espectro casi continuo al cual le faltan líneas en longitudes de onda específicas, que sabemos que corresponden a hidrógeno, helio y otros gases trazas.

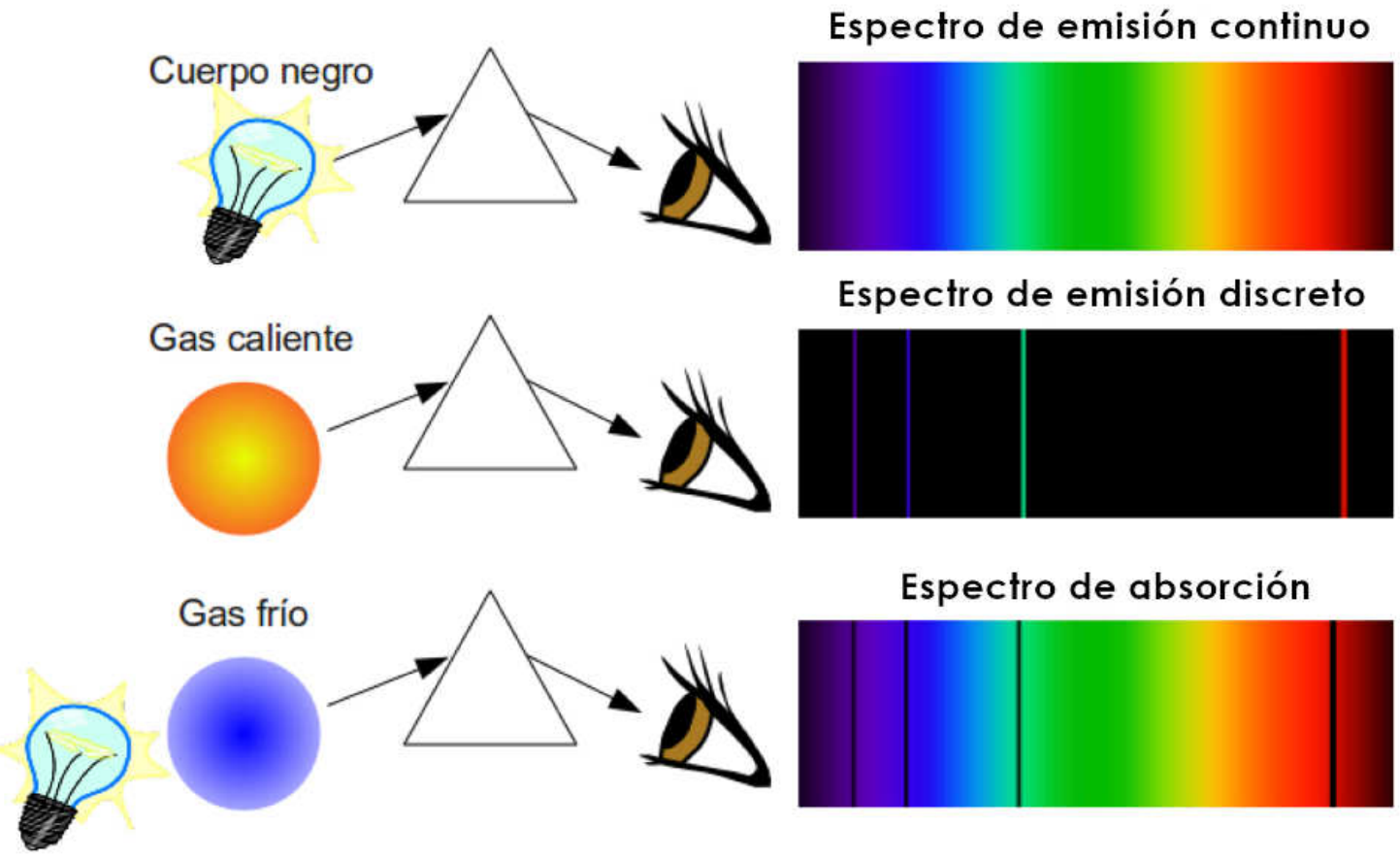

Figura 3.6: En la imagen se observa el espectro de emisión continuo (A), el espectro de emisión discreto (B) y el espectro de absorción (C) de una fuente [Crédito: Carlos Rudamás, El Salvador Ciencia y Tecnología (2007)].

## 3.4 Clasificación espectral

Dado que podemos identificar a las estrellas por su espectro, los astrónomos diseñaron un sistema de clasificación estelar basado en espectros. El esquema de clasificación espectral de las estrellas usado actualmente fue desarrollado por el Observatorio de Harvard a inicios del siglo XX. El trabajo fue iniciado por Henry Draper, quien en 1872 tomó la primera fotografía del espectro de la estrella Vega. Más adelante, Mary Anna Draper, viuda de Draper, donó el el equipo de observación junto una a una suma de dinero al Observatorio para continuar con el trabajo de clasificación.

La mayor parte de la clasificación fue realizada por Annie Jump Cannon , junto con un grupo de astrónomas, usando espectros de prisma tomados desde Perú. El catálogo Henry Draper (HD) fue publicado en 1918 - 1924. Este contenía 225 estrellas hasta una magnitud de 9. En total 390 estrellas fueron clasificadas por las “computadoras” de Harvard. La clasificación espectral estaba denotada por letras mayúsculas (O–B–A–F–G–K–M), de acuerdo a su temperatura. La secuencia abarca desde la más caliente (tipo O) hasta la más fría (tipo M) como se muestra en la Figura 3.7. En la actualidad, se han añadido las letras L, T e Y continuando la secuencia más allá de M, las cuales representan a las enanas marrones [**29**, **30**]. El Sol es una estrella de tipo G2.

**Historia 3.4.1: Las computadoras de Harvard**

Se le conoce como “las computadoras de Harvard” a un grupo de mujeres que trabajaron calculando, clasificando e investigando placas fotográficas de estrellas de todo tipo espectral tomadas en los observatorios de Harvard en Cambridge, Massachusetts (E.E.U.U.) y Carmen Alto en Arequipa, Perú. Entre las figuras mas representativas de este grupo tenemos: Williamina Fleming, Annie Jump Cannon, Henrietta Swan Leavitt, Cecilia Payne-Gaposchkin y Antonia Maury.

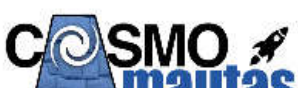

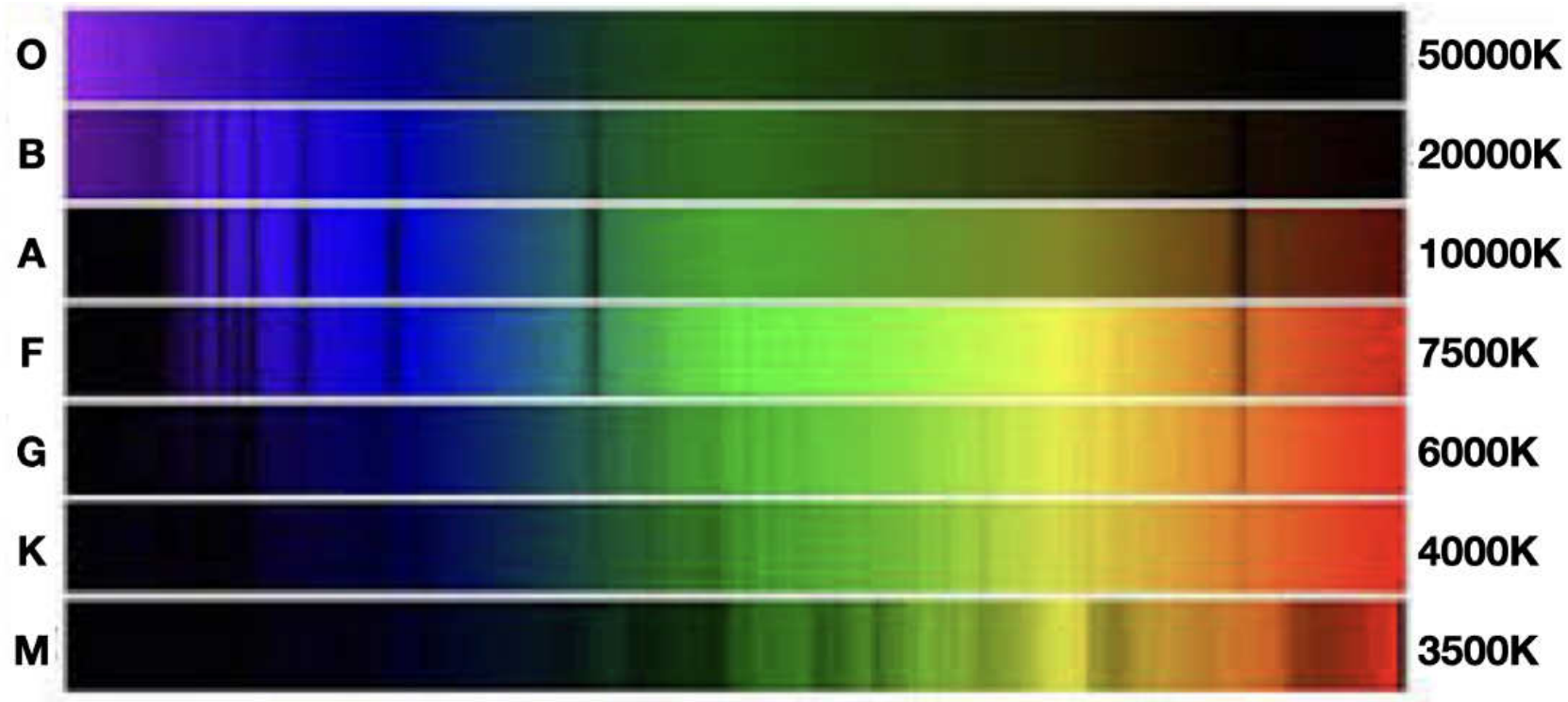

Figura 3.7: Espectros ópticos de acuerdo al tipo espectral. Las estrellas más calientes emiten luz en longitudes de onda más cortas y viceversa, tal como descrito por la ley de Wien [Crédito: imagen adaptada de KPNO 0.9-m Telescope, AURA, NOAO, NSF].

## 3.5 La tabla periódica de las estrellas

Alrededor de 1910, Ejnar Hertzsprung y Henry Norris Russell estudiaron la relación entre el tipo espectral de las estrellas y su magnitud absoluta. Ellos descubrieron que graficando estos parámetros de las estrellas, estas se organizaban en algunas regiones determinadas. A este resultado se le llamó el diagrama de Hertzsprung-Russell (HR) (Figura 3.8). En el diagrama el tipo espectral usa el color de las estrellas en el eje $x$ y la luminosidad o magnitud absoluta en el eje $y$. A partir de la ubicación de una estrella en el diagrama de HR se puede conocer su luminosidad, tipo espectral, color, temperatura, masa, radio, edad e historia evolutiva. El lugar en el diagrama HR en donde las estrellas se ubican la mayor parte de su vida se denomina "secuencia principal". A continuación describimos algunas propiedades físicas de las estrellas organizadas en el diagrama HR:

1. **Color/Temperatura/Tipo espectral**: dado que las estrellas más calientes emiten luz en longitudes de ondas más energéticas (ley de Wien), su color tiende al azul y se ubican en el lado izquierdo del diagrama, mientras que las estrellas más frías se ubican al lado derecho. De acuerdo a la morfología de su espectro, las estrellas se clasifican en tipos espectrales O–B–A–F–G–K–M, que también indican una clasificación en temperatura. De este modo, las estrellas tipo O se encuentran en el lado izquierdo del diagrama y las estrellas tipo M en el lado derecho.
2. **Brillo/Luminosidad/Magnitud absoluta**: por la ley de Stefan-Boltzmann, la temperatura de una estrella influye en su producción total de energía. Por otro lado, una estrella gigante roja o supergigante roja, a pesar de ser estrellas frías, también son muy brillantes debido a la gran superficie de emisión de energía que poseen. Con todo ello, podemos afirmar que las estrellas más brillantes se ubican en la parte superior del diagrama HR y las menos brillantes en la parte inferior.
3. **Masa:** la propiedad fundamental de las estrellas es su masa, a partir de la cual se determina cómo será su evolución estelar y su temperatura. Mientras mayor sea la masa de una estrella, mayor será su temperatura efectiva. Por lo tanto, en la secuencia principal del diagrama HR, las estrellas de mayor masa se encuentran hacia el lado izquierdo y las estrellas de menor masa, al lado derecho.
4. **Radio:** para las estrellas en la secuencia principal, a mayor masa, mayor radio, de tal forma que la estrellas más grandes se ubican en la parte superior y las estrellas más pequeñas, en la parte inferior de la secuencia principal.

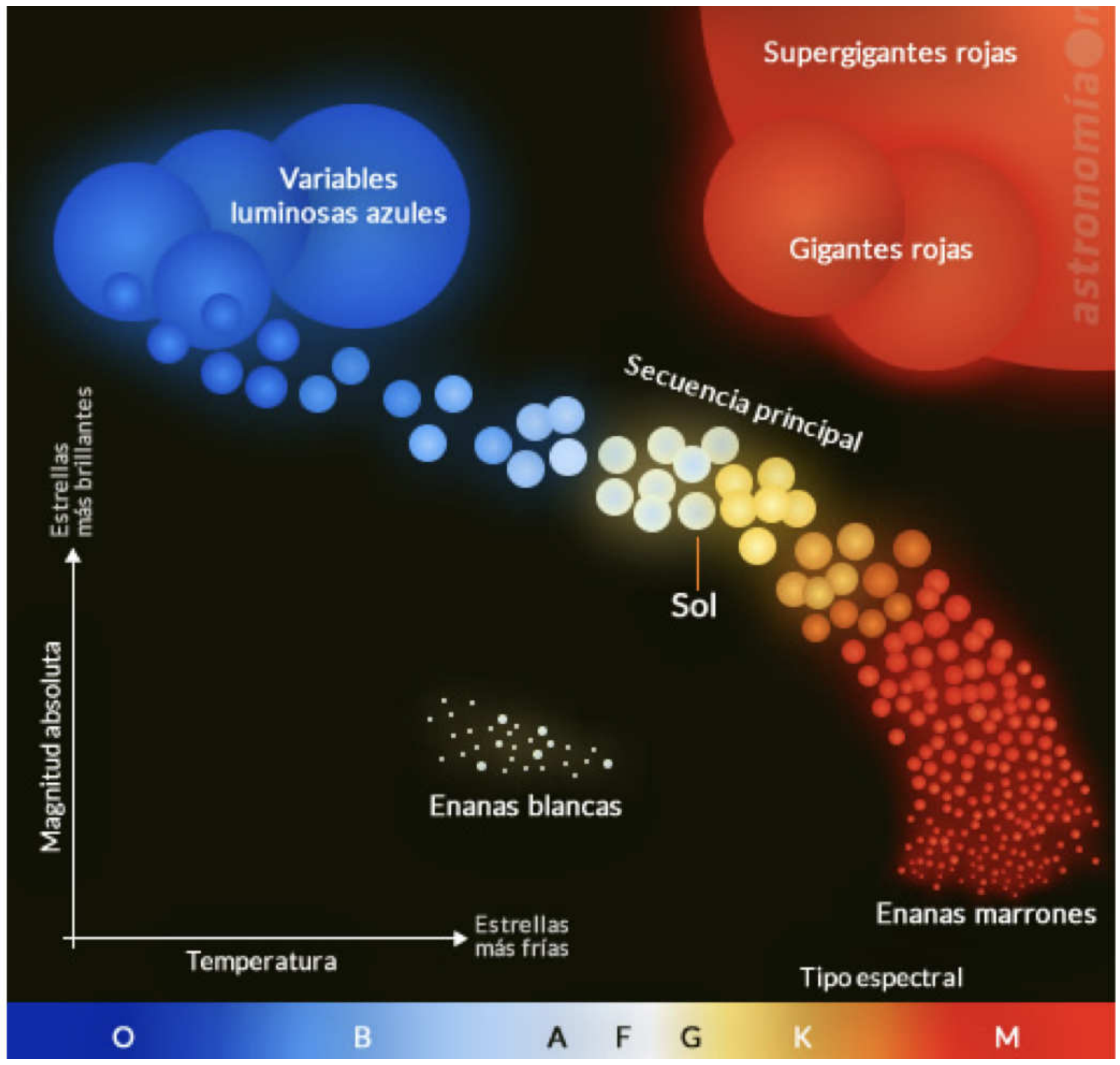

Figura 3.8: El diagrama Hertzsprung-Russell muestra la secuencia principal donde se ubican estrellas durante la parte de su vida donde fusionan hidrógeno en su núcleo. Las estrellas más masivas son más luminosas y de mayor temperatura, mientras que las estrellas menos masivas presentan temperaturas más bajas y son menos luminosas [Crédito: Ricardo J. Tohmé].

Las estrellas “ingresan”a la secuencia principal cuando logran estabilizar la tasa de fusión de hidrógeno, dependiendo de su masa inicial. Una vez que encuentran su lugar en la secuencia principal, se quedan en el mismo punto la mayor parte de su vida, hasta acabar el hidrógeno en su núcleo, luego pasa por otras etapas más cortas con las cuales sale de la secuencia principal mientras sintetiza elementos más pesados en el núcleo y sus capas externas van creciendo.

**Recurso TIC 3.5.1:**

Ingrese al siguiente enlace: `http://astro.unl.edu/mobile/HRdiagram/HRdiagramStable.html` y varíe el radio, color y temperatura de una estrella para observar en que región del diagrama de HR se encuentra.

## 3.6 La vida de las estrellas

### 3.6.1 ¿Cómo nacen las estrellas?

Las estrellas se forman en nubes de gas y polvo, llamadas nebulosas o nubes moleculares (Figura 3.9). Las nubes moleculares se extienden por 20-200 pc, a una temperatura promedio de 10 K y están compuestas por varios millones de masas solares de gas y polvo. Las nubes moleculares se estabilizan compensando su propia gravedad con el movimiento termal de las partículas que la componen. Sin embargo, perturbaciones en la nube pueden ocasionar una reacción en cadena donde la gravedad “gana” en escalas locales y partes de la nube colapsan en fragmentos individuales.

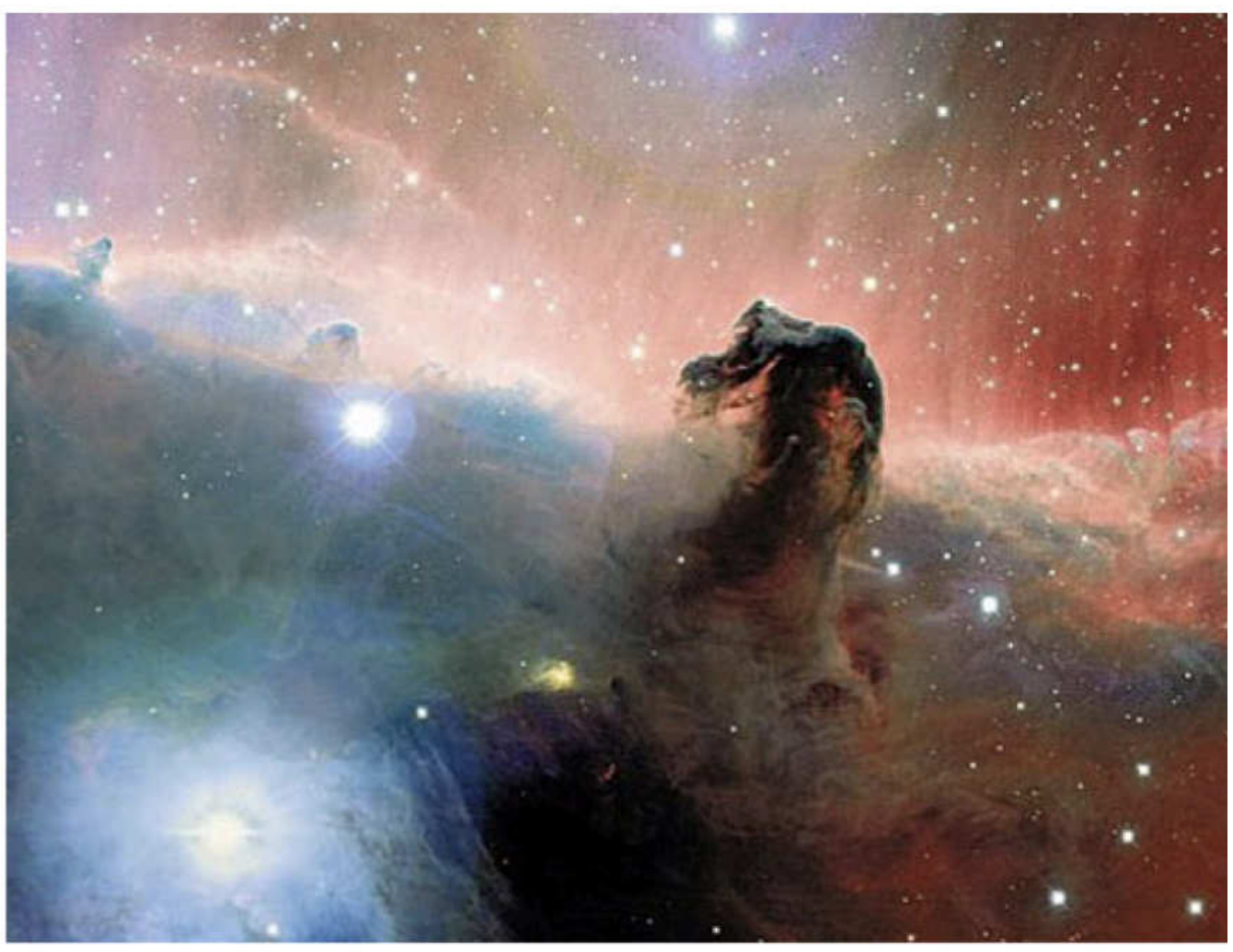

Figura 3.9: La nebulosa cabeza de caballo es un “semillero” de estrellas [Crédito: NASA].

Estos fragmentos o “núcleos” alcanzan densidades más altas que el material que los rodea y por su propia gravedad logran atraer material que incrementa su masa total y la temperatura de su interior. Toda la nube gira en su conjunto, pero las zonas en donde se incrementa la masa, por conservación del momento angular, giran más rápido cada vez, este movimiento también produce la formación de un disco en torno al objeto masivo principal, iniciando así la formación de una protoestrella en el centro y un disco protoplanetario alrededor de ella. La protoestrella continua atrayendo más materia del disco y reduciendo su tamaño. Cuando el interior de la protoestrella alcanza unos 13 millones de Kelvin, se inicia la fusión de hidrógeno y ha nacido una estrella. Todo este proceso dura alrededor de un millón de años.

#### ¿Qué pasa si no hay suficiente masa para iniciar la fusión de hidrógeno?

Si una estrella en formación tiene una masa de 8 % la masa del Sol ($0.08\,M_\odot$) o menos, no logrará llegar a las temperaturas y presiones necesarias para iniciar y mantener la fusión de hidrógeno en su núcleo. En este caso, su propia gravedad comprime el objeto hasta más o menos el radio de Júpiter, que es alrededor de 10 % del radio del Sol. Estos objetos se llaman **enanas marrones**.

### 3.6.2 ¿Cómo mueren las estrellas?

La vida y evolución de una estrella depende enteramente de su masa. Estrellas más masivas requieren una alta tasa de fusión nuclear para mantener una alta presión termal interna y poder contrarrestar su propia fuerza de gravedad. Por esta razón, utilizan su combustible más rápidamente que una estrella de baja masa y viven vidas mucho más cortas. Las estrellas tipo O viven en promedio 3 millones de años, mientras que las estrellas tipo M pueden vivir por $10^{13}$ años, que es $\sim$ 7700 veces la edad del universo. Podemos clasificar el destino final de las estrellas dependiendo de su masa en tres posibles finales.

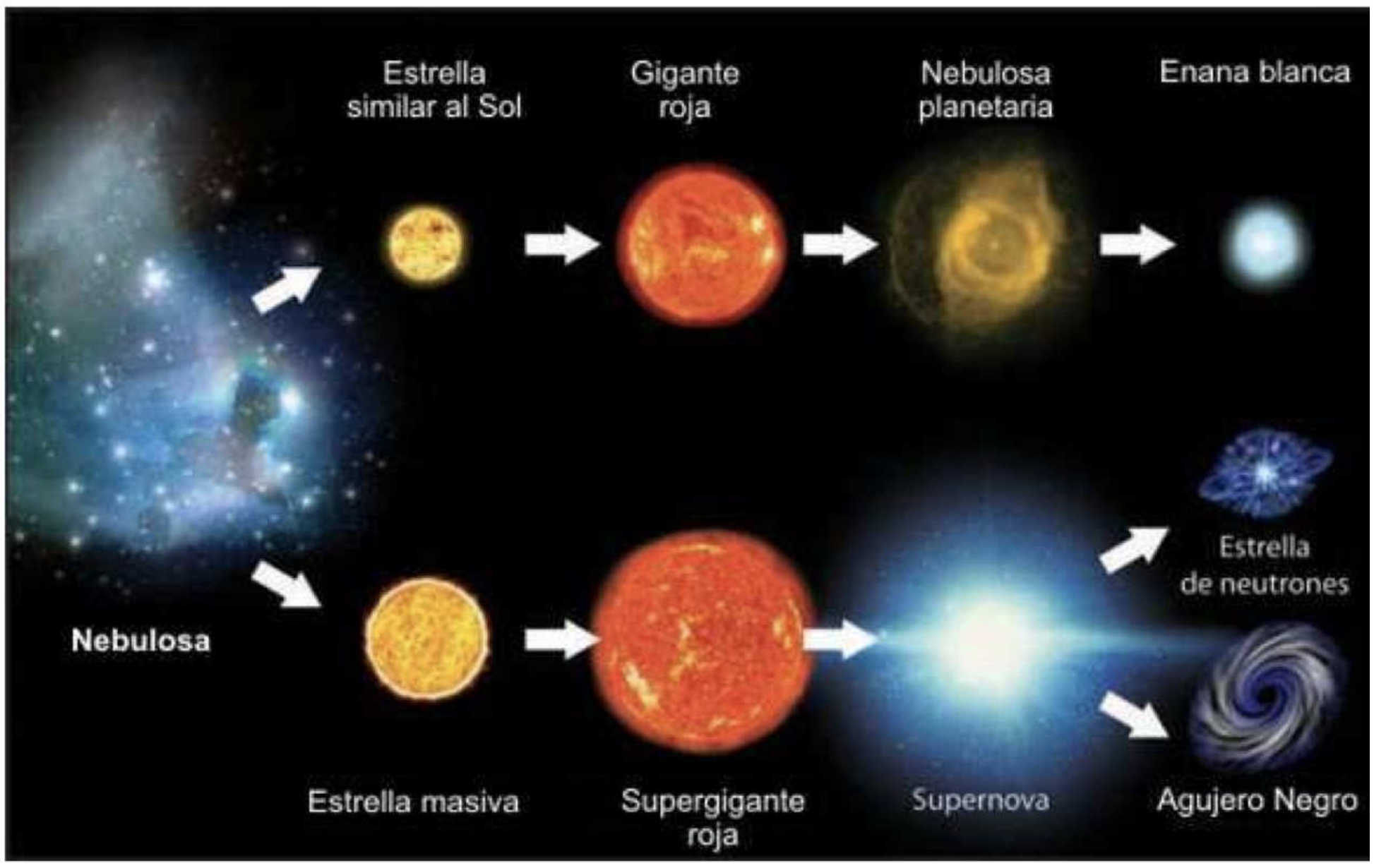

Figura 3.10: Evolución estelar desde la formación de una estrella en una nebulosa o nube molecular, hasta su colapso en enana blanca, estrella de neutrones o agujero negro, dependiendo de su masa inicial [Crédito: Wikipedia, Creative Commons].

#### Enanas blancas

Las estrellas de masa intermedia (0.5-1.5 $M_\odot$) fusionan hidrógeno en su núcleo durante millones de años. Al acabarse el hidrógeno en el núcleo de la estrella, se rompe el equilibrio hidrostático. La fuerza de gravedad de las capas superiores presiona al núcleo que ya no genera reacciones nucleares y la estrella se empieza a contraer. Esto hace que aumente la densidad y temperatura en el núcleo de la estrella.. Con el incremento de temperatura, empieza una nueva etapa de fusión de hidrógeno en las capas superiores de la estrella, haciendo que se expandan y la estrella entra en la fase de gigante roja. En este momento, la estrella abandona la secuencia principal del diagrama de HR y se mueve hacia la esquina superior derecha donde residen estrellas muy grandes y muy rojas. Cuando el Sol se vuelva una gigante roja dentro de 5 mil millones de años, su radio se expandirá hasta la órbita de la Tierra. En el núcleo, las altas temperaturas y presiones hacen que sea factible la fusión de átomos de helio en carbono y oxígeno.

Si la estrella es muy masiva, al acabarse el helio, otra vez más se comprime el núcleo y se inicia fusión de helio en las capas superiores, mientras que en el núcleo, se inicia la fusión de carbono en silicio. De lo contrario, la estrella expulsa sus capas superiores formando una **nebulosa planetaria** y dejando a su núcleo comprimido en el centro, **una enana blanca**. Un ejemplo famoso de nebulosa planetaria es la nebulosa Ojo de gato. El nombre de nebulosa planetaria tiene su origen en el siglo XIX, debido a que en ese entonces algunas de ellas visualmente parecían ser planetas, como Urano. Sin embargo, no tienen nada que ver con estos.

Las enanas blancas generalmente están compuestas de carbono y oxígeno y llegan a temperaturas de 10 000 K. Tienen masas de aproximadamente 1.4 $M_\odot$, pero por estar comprimidas en un tamaño similar al de la Tierra, son extremadamente densas. Alrededor del 97 % de todas las estrellas en la Vía Láctea terminarán su vida como enanas blancas [**31**].

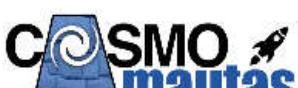

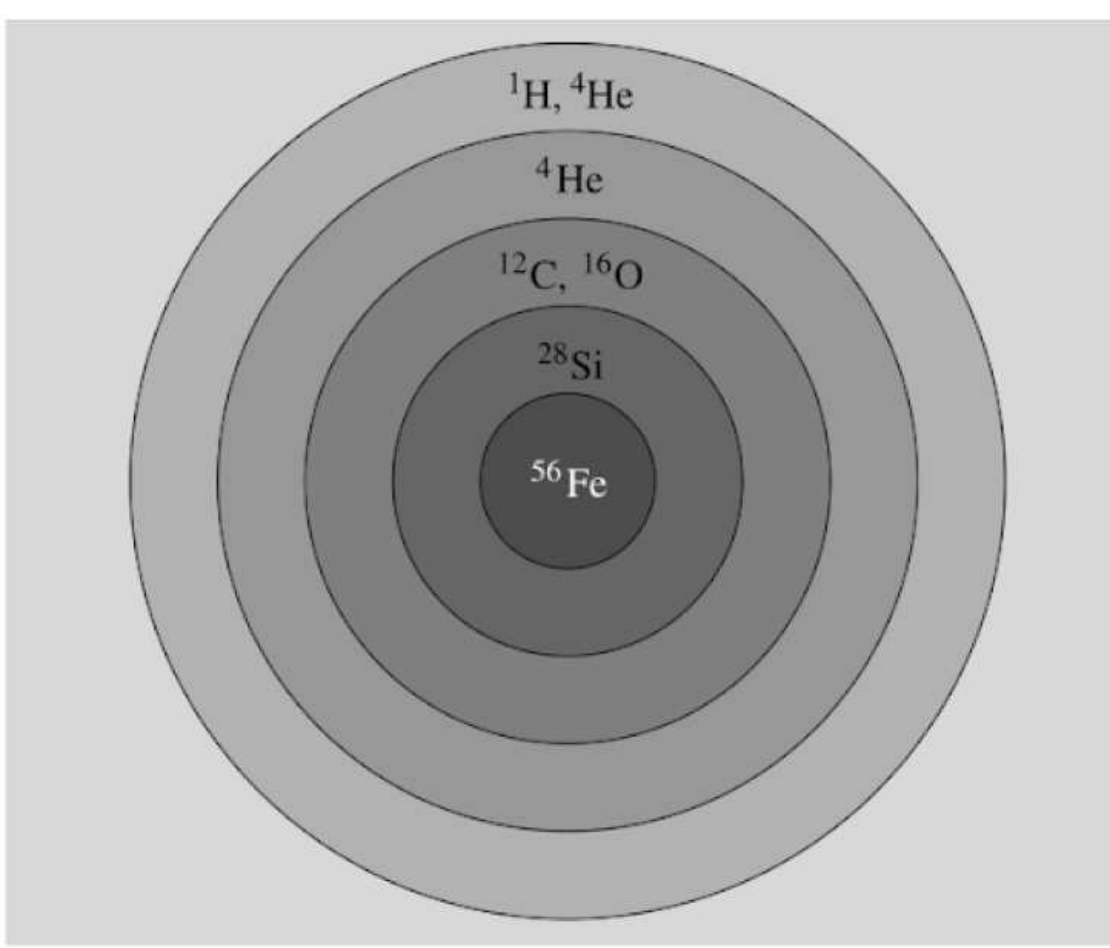

Figura 3.11: Las reacciones nucleares en una estrella evolucionada ocurren en capas de acuerdo a la presión y temperatura de cada una de ellas. En la capa superior, se alcanza la temperatura necesaria para quemar hidrógeno y helio, mientras que en el núcleo de la estrella, las presiones y temperaturas son suficientemente altas para obtener hierro [Crédito: Fundamental Astronomy].

### Estrellas de neutrones

En estrellas masivas de 10-25 $M_\odot$ ocurren los mismos pasos previos que para las estrellas de masa intermedia: entran a la fase de gigantes rojas al acabarse el hidrógeno y se calienta el núcleo al punto donde es factible fusionar helio en carbono y oxígeno y carbono en silicio. De la misma forma, al acabarse el carbono, se elevan las temperaturas otra vez para fusionar silicio, oxígeno, magnesio y otros metales hasta el hierro. En las capas superiores de la estrella también ocurren reacciones nucleares de acuerdo a su presión y temperatura (Figura 3.11). El hierro es el átomo más estable de toda la tabla periódica y fusionar hierro no libera energía. Una vez llegado a este punto, no se puede continuar con la fusión en el núcleo, haciendo que este se comprima por acción de la gravedad hasta colapsar en una explosión extremadamente luminosa llamada **supernova.** La cantidad de energía liberada por una supernova equivale a la luminosidad de 10 mil millones de soles, por lo que una sola estrella al término de su vida puede ser tan brillante como una galaxia entera. En promedio, ocurre una supernova cada 50 años en nuestra Vía Láctea.

Una supernova deja atrás al núcleo comprimido de la estrella, que viene a ser **una estrella de neutrones** con una masa máxima de 2-3 $M_\odot$y de unos 10 km de radio. La propia gravedad de la estrella de neutrones es tan alta que hace que protones y electrones se fusionen en neutrones. La densidad de una estrella de neutrones es tan alta que una cucharadita de su material en la Tierra “pesaría” más de 10 millones de toneladas.

### Agujeros negros

En las estrellas más masivas, después de una supernova, no queda residuo. La fuerza de gravedad comprime toda la masa del núcleo de la estrella a un punto, que virtualmente tiene una densidad y gravedad infinitas y del que ni siquiera la luz puede escapar. Aprenderemos más sobre agujeros negros en la sección 5.3.

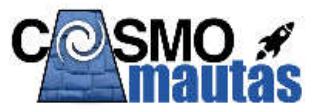

# 4. Exoplanetas

## 4.1 Mundos distantes

En los capítulos anteriores hemos estudiado algunos de los objetos más conocidos que podemos observar en el cielo nocturno, como los planetas de nuestro sistema solar y las estrellas que habitan nuestra galaxia. Con los conocimientos que tenemos hasta ahora podemos preguntarnos: ¿podrían haber planetas alrededor de otras estrellas? Y de ser así, ¿cómo podríamos estudiarlos?

Hasta inicios de la década de 1990, los astrónomos únicamente conocían de la existencia de los 8 planetas que orbitan el Sol, y por lo tanto, todas nuestras teorías acerca de la formación de sistemas planetarios se basaban en observaciones de nuestro sistema solar. Sin embargo, hoy sabemos que los planetas son muy comunes en nuestra galaxia, al menos tan abundantes como las estrellas mismas. En promedio, hay 1.6 planetas por cada estrella, y alrededor de cien mil millones de estrellas ($10^{11}$, [**32**]) en nuestra galaxia. Aquellos planetas que orbitan otras estrellas que no son el Sol reciben el nombre de planetas extrasolares o **exoplanetas**.

El estudio de los exoplanetas es un campo relativamente reciente de la astronomía, que ha crecido gracias a avances tecnológicos y nuevas técnicas de análisis desarrolladas en las últimas décadas. Observar exoplanetas es una tarea inherentemente difícil, ya que los planetas son muy pequeños en comparación con las estrellas que orbitan, y son mucho menos brillantes que ellas, por lo que la luz de una estrella lejana domina por completo cualquier imagen tomada de esta y es muy complejo distinguir la emisión proveniente de un planeta. Cabe resaltar que la única luz visible que podría detectarse de la mayoría de planetas es en realidad luz proveniente de su estrella que es reflejada por la superficie del planeta. Sin embargo, existen varias técnicas para poder

Imagen de encabezado: Impresión artística de la estrella TRAPPIST-1 rodeada por sus siete planetas de tamaño parecido a nuestra Tierra. Crédito:NASA/JPL-Caltech detectar exoplanetas, de manera directa o indirecta, y obtener información importante sobre ellos.

## 4.2 Métodos de detección de exoplanetas

### 4.2.1 Detección directa

La detección directa, como su nombre lo indica, consiste en obtener la imagen de un planeta con la cámara de un telescopio. Para poder observar exoplanetas de este modo, éstos deben ser jóvenes, es decir, que han sido formados hace solo unos pocos millones de años y por lo tanto continúan calientes ($T_{eff} \gtrsim 1\,000$ K, [**33**]), de tal forma que su temperatura los hace lo suficientemente brillantes para ser detectados directamente. También deben tener un gran tamaño, mayor incluso que el de Júpiter, para que la luz que emiten sea suficientemente intensa para ser detectada. Finalmente, deben orbitar a su estrella a una gran distancia de ella, de modo que la luz proveniente de los exoplanetas pueda distinguirse claramente de aquella de la estrella. Este método de detección permite restringir la órbita del exoplaneta. Además, una ventaja de realizar este tipo de observaciones es que, idealmente, uno podría obtener un espectro directamente del planeta de interés, y así deducir información crucial acerca de su composición atmosférica y temperatura. Captar imágenes de un planeta es más viable en luz infrarroja que en luz visible, ya que los planetas emiten luz infrarroja debido a su temperatura [**34**]. No obstante, esta técnica solamente ha podido usarse para detectar alrededor de 50 exoplanetas, debido a las restricciones que mencionamos anteriormente. Telescopios más avanzados en el futuro podrán alcanzar contrastes más altos entre estrellas y exoplanetas para poder observar exoplanetas más fríos, más pequeños y en órbitas más cercanas, pero es probable que el número de detecciones directas posibles continúe siendo muy limitado. Los siguientes métodos de detección que describiremos son indirectos, es decir, consisten en analizar el comportamiento de la luz proveniente de una estrella para deducir la presencia de exoplanetas en órbita.

**Gravedad 4.2.1: Centro de Masa**

Comúnmente se piensa que los planetas orbitan alrededor de sus respectivas estrellas mientras que éstas se quedan quietas, pero lo que realmente ocurre es que tanto los planetas como la estrella misma orbitan alrededor de su centro de masa común, también llamado **baricentro**. Este es el centro exacto de toda la materia del sistema, y se ubica más cerca del cuerpo más masivo. Por ejemplo, el centro de masa del sistema Sol-Júpiter se encuentra muy cerca de la superficie del Sol, pues el Sol contiene la mayor parte de la masa del sistema. Ambos se mueven en torno a este punto, pero el movimiento del Sol es mucho menos notable, ya que su distancia al baricentro es mucho menor (ver Figura 4.2). Así, se dice que los planetas tienen un efecto gravitacional sobre sus estrellas que causa que estas oscilen periódicamente, lo cual es clave para comprender dos de los métodos de detección indirecta de exoplanetas.

### 4.2.2 Monitoreo astrométrico

La astrometría consiste en la medición precisa de la posición de las estrellas en el cielo, así como del cambio de sus posiciones con el paso del tiempo. Existen varios telescopios y misiones espaciales dedicadas a esta tarea (como *Gaia*, por ejemplo) con el objetivo de obtener mapas precisos de la ubicación de las estrellas en nuestra galaxia, pero también han permitido descubrir nuevos planetas extrasolares. Varias mediciones astrométricas han revelado estrellas que no permanecen estáticas en una única posición ni se mueven en líneas rectas en el cielo, sino que oscilan ligeramente en torno a un punto invisible. Esta oscilación se debe al movimiento sutil de la estrella causado por la influencia gravitacional de un planeta, el cual no puede verse pues la estrella es mucho más brillante. Tanto la estrella como el planeta se mueven alrededor del baricentro del sistema, y los astrónomos pueden observar el movimiento de la estrella [**35**]. En imágenes de una estrella solitaria, el "centro de luz" la estrella (la parte del disco con la máxima intensidad de luz) siempre coincide con el centro de la imagen. En cambio, si la estrella tiene un planeta en órbita, cuando se comparan imágenes de ella tomadas en épocas diferentes, se puede ver que el centro de la estrella se desplaza trazando una pequeña órbita, que es proporcional a la órbita del planeta alrededor de la estrella.

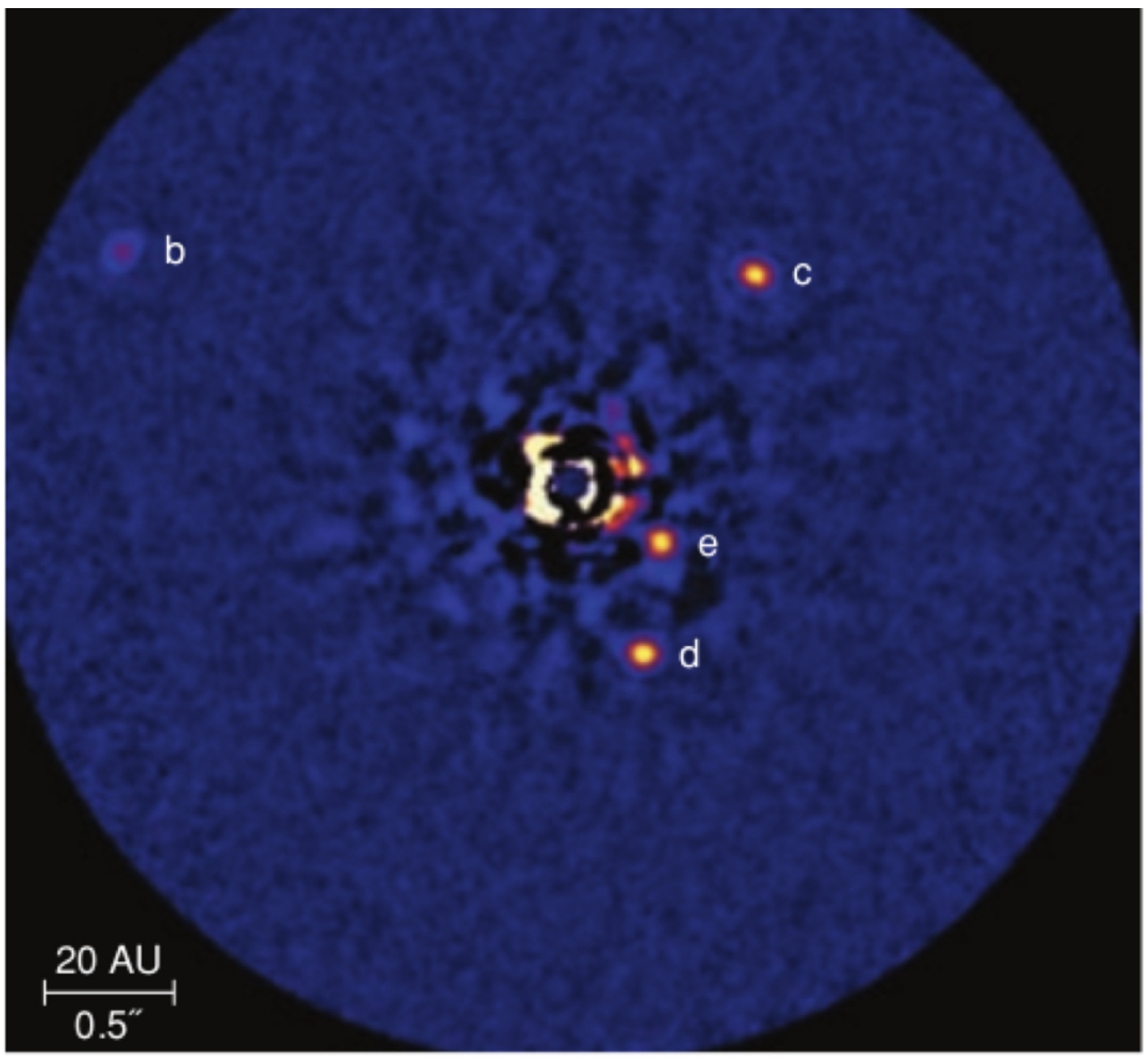

Figura 4.1: Esta imagen infrarroja tomada desde el Telescopio Keck muestra la detección directa de un sistema de cuatro exoplanetas orbitando la estrella HR 8799. La luz de la estrella fue bloqueada durante esta observación. Estos exoplanetas son mucho más grandes y brillantes que los planetas del sistema solar [Crédito: Jason Wang (Caltech) / Christian Marois (NRC Herzberg)].

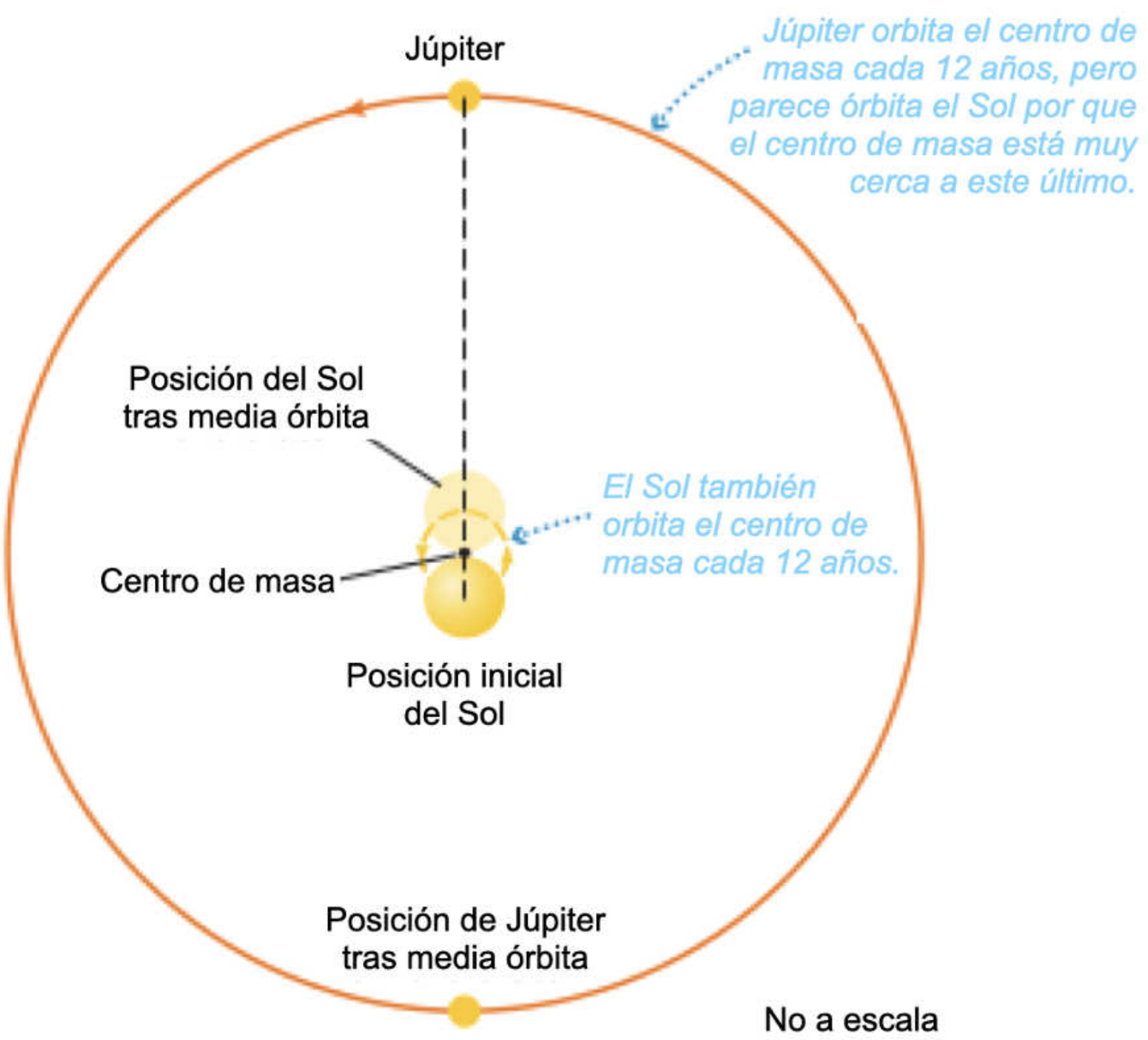

Figura 4.2: Este diagrama muestra cómo el Sol y Júpiter orbitan en torno a su centro de masa común, el cual se encuentra muy cerca al Sol. El diagrama no está a escala, y el tamaño de la órbita del Sol ha sido exagerado para que sea más notable [Crédito: imagen adaptada de Pearson Education].

**Recurso TIC 4.2.1:**

En la siguiente página web podrás visualizar los métodos de detección de exoplanetas, con explicaciones de los instrumentos utilizados en la práctica: `https://exoplanets.nasa.gov/alien-worlds/ways-to-find-a-planet/#/5`.

El método astrométrico es eficiente en detectar exoplanetas de gran masa y separados una distancia intermedia de su estrella. Esto se debe a que los exoplanetas de masa muy grande tienen mayores efectos gravitacionales sobre sus estrellas, lo cual produce mayores oscilaciones que son más fáciles de observar. Además, es un método complementario a la detección directa y velocidad radial en cuanto a la separación entre estrella y exoplaneta. Mientras que es más factible utilizar la detección directa para un exoplaneta a una separación larga, y más efectivo utilizar el método de Doppler para una separación muy corta (ver Sección 4.2.3), el método astrométrico funciona mejor en separaciones intermedias. Esto se debe a que mientras más lejos esté el exoplaneta de la estrella, el centro de masa del sistema también se aleja más del centro de masa de la estrella, haciendo que la oscilación de la estrella con respecto al centro de masa del sistema sea más notable.

Este método permite medir el periodo orbital del planeta, correspondiente al periodo de oscilación de la posición de la estrella. Si también se conoce la masa de la estrella, $M_\star$, y la separación entre el planeta y la estrella, $a_p$, se puede calcular la masa del planeta, $M_p$, usando la fórmula general de la Tercera Ley de Kepler[**34**]:

$$4\pi^2 a_p^3 = G M_{total} P_p^2 \tag{4.1}$$

donde $G$ es la constante universal de gravitación, $M_{total} = M_\star + M_p$ es la masa total del sistema y $P_p$ es el periodo orbital del planeta, el cual equivale al periodo de la estrella alrededor del centro de masa del sistema. Este método es difícil de aplicar ya que requiere mediciones de posiciones extremadamente precisas y por lo tanto, solo un planeta 28 veces tan masivo como Júpiter ha sido descubierto de esta forma [**36**], aunque con este método se ha confirmado la presencia de varios exoplanetas tras haber sido descubiertos con otros métodos.

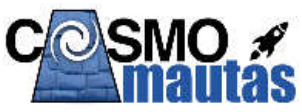

## Luz 4.2.1: Efecto Doppler

El efecto Doppler es un fenómeno que describe la relación entre la frecuencia de una onda y la velocidad relativa entre la fuente de dicha onda y un observador, lo que causa que este último perciba una mayor frecuencia de la onda cuando la fuente se acerca hacia él, y una menor frecuencia cuando ésta se aleja. Este efecto se observa en situaciones cotidianas con ondas de sonido, como cuando uno oye el sonido de un tren al pasar. Cuando el tren se acerca, los frentes de la onda sonora se encontrarán más cerca entre sí que si el tren estuviera estático, por lo que el observador percibirá una mayor frecuencia y oirá el sonido más agudo. Análogamente, si el tren se aleja, los frentes de onda estarán más espaciados entre sí, causando que se perciba una menor frecuencia y un sonido más grave. Sin embargo, este fenómeno también se observa en otros tipos de onda, como la luz, lo cual permite obtener información sobre la velocidad relativa de un cuerpo. Así, cuando uno observa el espectro de una fuente de luz con un espectrómetro, verá que este espectro se encuentra desplazado hacia frecuencias más altas o más bajas si esta fuente está acercándose o alejándose, respectivamente, del observador. La imagen inferior ilustra cómo el efecto Doppler actúa en ondas de sonido y de luz. Los círculos representan los frentes de onda expandiéndose en todas direcciones [Crédito: Pearson Education].

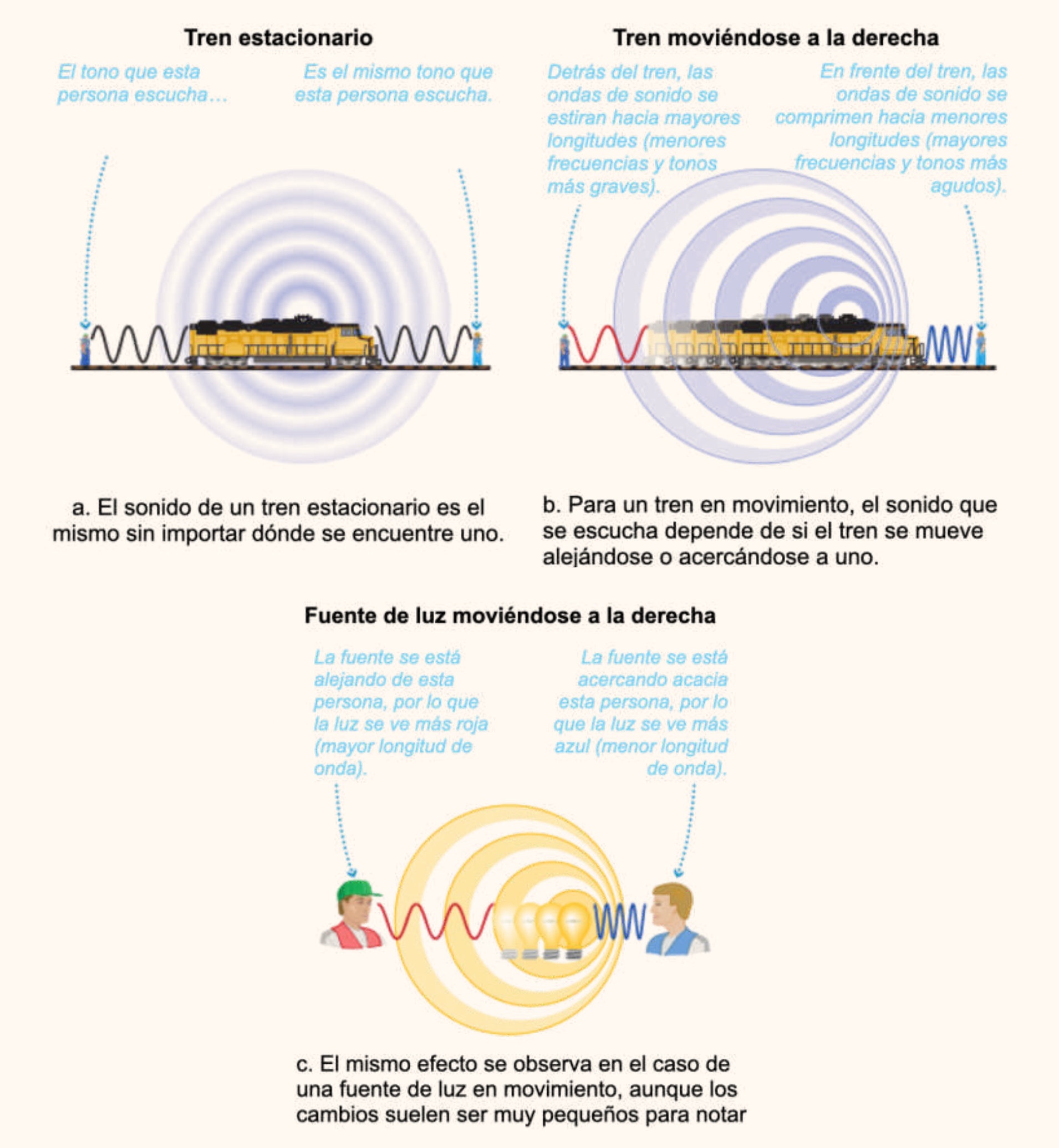

a. El sonido de un tren estacionario es el mismo sin importar dónde se encuentre uno.

b. Para un tren en movimiento, el sonido que se escucha depende de si el tren se mueve alejándose o acercándose a uno.

c. El mismo efecto se observa en el caso de una fuente de luz en movimiento, aunque los cambios suelen ser muy pequeños para notar

### 4.2.3 Método de Doppler

El método de Doppler o de velocidad radial también aprovecha la influencia gravitacional de un planeta sobre su estrella, la cual causa que ambos orbiten alrededor del baricentro de su sistema. Desde nuestro punto de vista fijo en la Tierra, a veces la estrella se mueve hacia nosotros y a veces se aleja de nosotros. La componente de la velocidad de la estrella en la dirección de nuestra línea de visión se llama **velocidad radial**. La otra componente es la velocidad tangencial, que es la proyección del vector de velocidad en el plano del cielo. Conforme el exoplaneta y su estrella orbitan el baricentro del sistema, la velocidad radial de la estrella relativa a la Tierra irá variando, y podemos medir esa variación haciendo uso del efecto Doppler. En un momento dado, la estrella se moverá acercándose hacia la Tierra, lo que hará que su espectro de emisión se desplace hacia el azul (equivalente a menores longitudes de onda). Notamos la diferencia al comparar las longitudes de onda de líneas de emisión o absorción atómica en el espectro de la estrella con respecto a las longitudes de onda de líneas de emisión de los mismos gases medidos en condiciones de laboratorio en la Tierra. Cuando la estrella se aleje de la Tierra mientras sigue su órbita, su espectro se desplazará hacia el rojo (a mayores longitudes de onda). Al observar desplazamientos periódicos en el espectro de una estrella, alternando entre el rojo y el azul, podemos deducir la presencia de un exoplaneta que ejerce un efecto gravitacional sobre ella [**35**] (ver Figura 4.3).

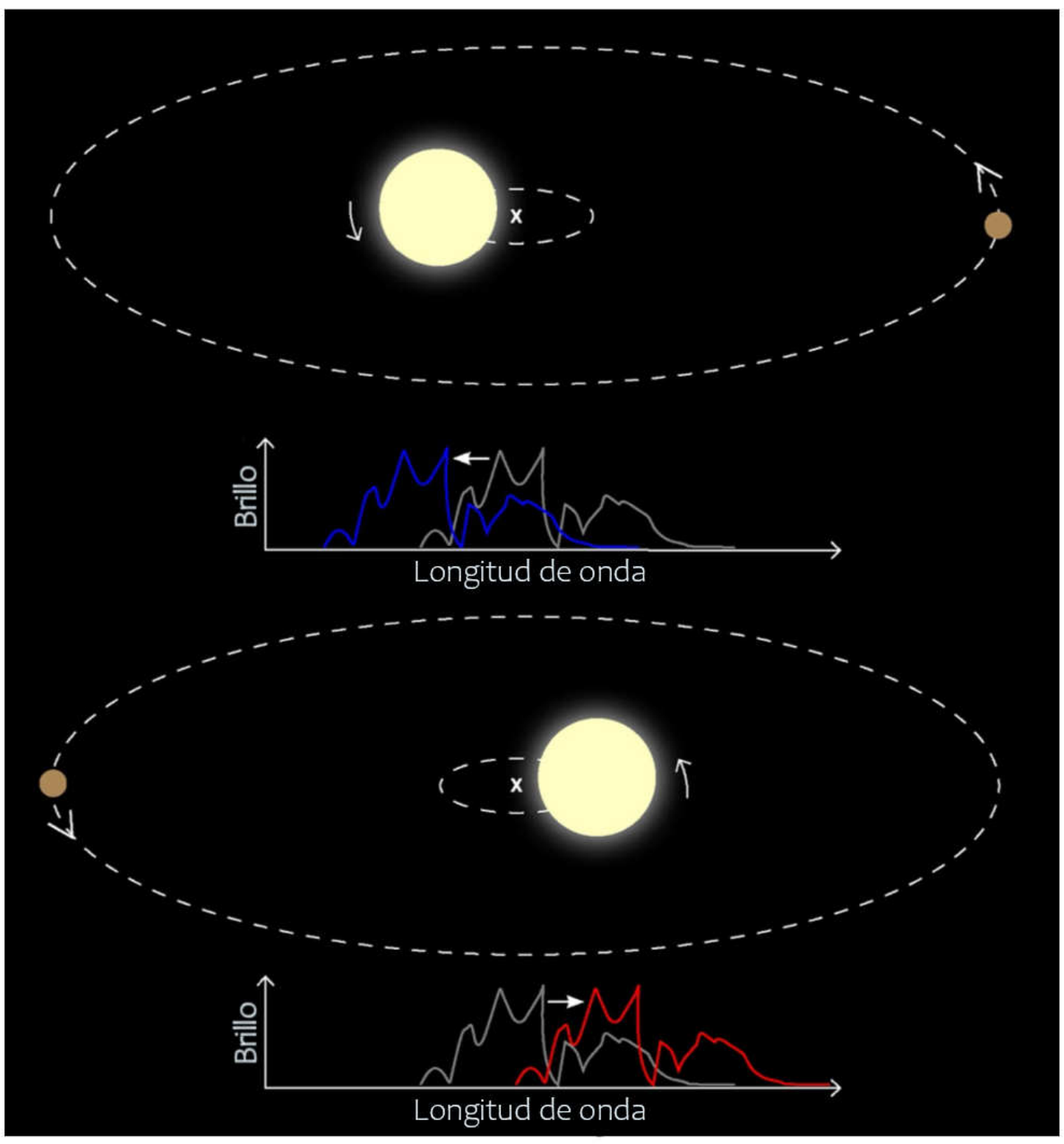

Figura 4.3: Esquema de detección de un exoplaneta mediante el método de Doppler. Cuando la estrella se mueve hacia la Tierra (*arriba*), su espectro se desplaza hacia el azul. Cuando la estrella se aleja (*abajo*), el espectro se desplaza hacia el rojo. Es importante mencionar que, los desplazamientos en la imagen han sido exagerados para distinguirlos, si se realizan las mediciones, el desplazamientos hacia el azul (*arriba*) será menor que el desplazamiento hacia el rojo (*abajo*), debido a que estos son proporcionales a la longitud de onda. [Crédito: ESA].

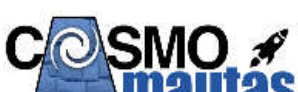

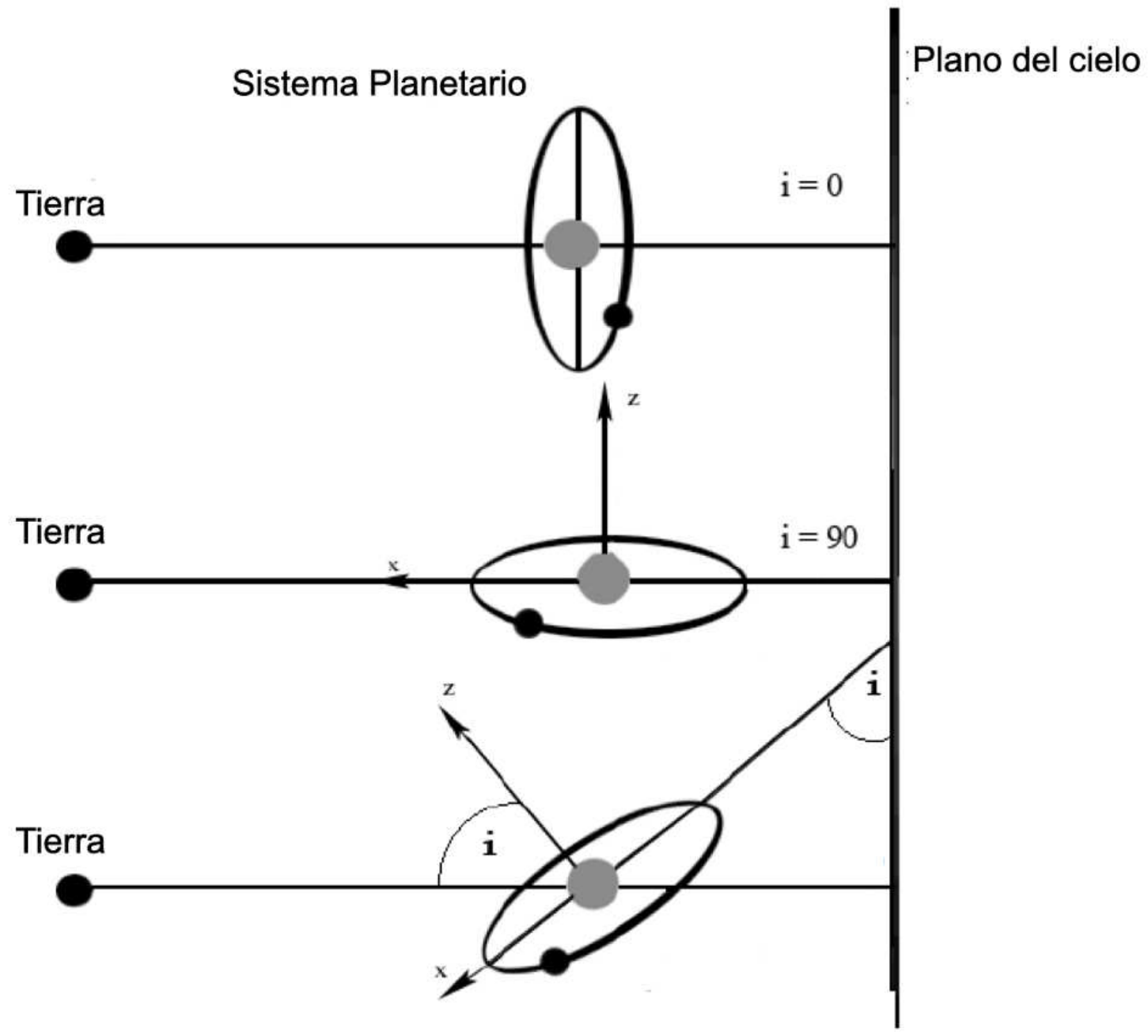

Figura 4.4: Este diagrama muestra la convención para el ángulo de inclinación $i$ de la órbita de un exoplaneta visto desde la Tierra. El ángulo $i = 0°$ corresponde a un sistema orientado "de cara", mientras que $i = 90°$ corresponde a una orientación "de borde" [Crédito: Thomson Learning].

El método de Doppler es muy apto para detectar exoplanetas de gran masa y exoplanetas que orbitan muy cerca de su estrella, pues así el efecto gravitacional que tengan sobre la estrella será mayor [**37**]. Por otro lado, si el ángulo de inclinación de la órbita (denotado por $i$) es definido tal como se muestra en la Figura 4.4, entonces el componente radial de la velocidad de la estrella está dado por $\mathrm{v}\,sen(i)$, donde v es la magnitud de la velocidad orbital. Esto quiere decir que solo es posible observar exoplanetas con este método si el plano de su órbita no está orientado "de cara" con respecto a la Tierra, pues en ese caso la velocidad de la estrella no tiene un componente radial [**36**]. Estudiar la velocidad radial de la estrella en función del tiempo permite calcular ciertas propiedades del exoplaneta. El periodo de oscilación de la velocidad radial corresponde al periodo orbital, a partir del cual se puede calcular la separación entre el exoplaneta y la estrella usando la Tercera Ley de Kepler. Si se conoce la masa de la estrella, también es posible determinar la masa del exoplaneta a partir de la relación de conservación de momento lineal (en la dirección radial):

$$M_{\star}v_{\star} = M_p v_p, \tag{4.2}$$

donde $v_{\star}$ es la velocidad radial de la estrella obtenida con este método, y la velocidad del exoplaneta $v_p$ se calcula a partir de su periodo y distancia orbital[**34**]. Si la órbita del exoplaneta no tiene una orientación "de borde" con respecto a la Tierra, la masa que se calcula con este método es solo un límite inferior, equivalente a $\mathrm{M}_p\,sen(i)$; por ello, para conocer el valor exacto de la masa del exoplaneta es necesario conocer la inclinación de su órbita, lo cual no siempre se puede observar.

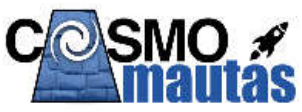

**Historia 4.2.1: La primera detección de un exoplaneta**

En el año 1995, los astrónomos suizos Michel Mayor y Didier Queloz anunciaron la primera detección de un exoplaneta orbitando una estrella similar al Sol, en este caso la estrella 51 Pegasi. Este descubrimiento se realizó utilizando el método de Doppler, y le valió a estos científicos el Premio Nobel de Física en el 2019 debido a su gran aporte a la ciencia de planetas extrasolares.

### 4.2.4 Tránsitos

El método del tránsito consiste en medir continuamente el brillo aparente de una estrella por un tiempo prolongado, y observar disminuciones en el brillo que se repiten regularmente [**37**]. Esto se debe al tránsito de un exoplaneta, el cual bloquea parte de la luz de la estrella al pasar frente a ella, al igual que la Luna bloquea la luz del Sol durante un eclipse solar. Para poder detectar este tipo de señal, es necesario que el plano de la órbita del planeta esté orientado "de borde" con respecto a la Tierra, o que tenga un ángulo de inclinación relativamente alto ($i \geq 70°$), pues si estuviera orientado de otra forma jamás se le observaría transitar frente a su estrella.

Este método también nos permite conocer el periodo orbital de un exoplaneta, el cual corresponde al intervalo de tiempo entre dos mínimos en el brillo aparente de la estrella. Por otro lado, al medir la fracción de la luz de la estrella que es bloqueada, podemos calcular el tamaño del exoplaneta, ya que planetas más grandes bloquearán más luz que exoplanetas más pequeños. En algunos casos, también es posible comparar la cantidad de luz infrarroja que se observa antes de y durante un eclipse (cuando el exoplaneta es bloqueado por la estrella), para obtener la cantidad que es emitida solo por el exoplaneta y así calcular su temperatura superficial.

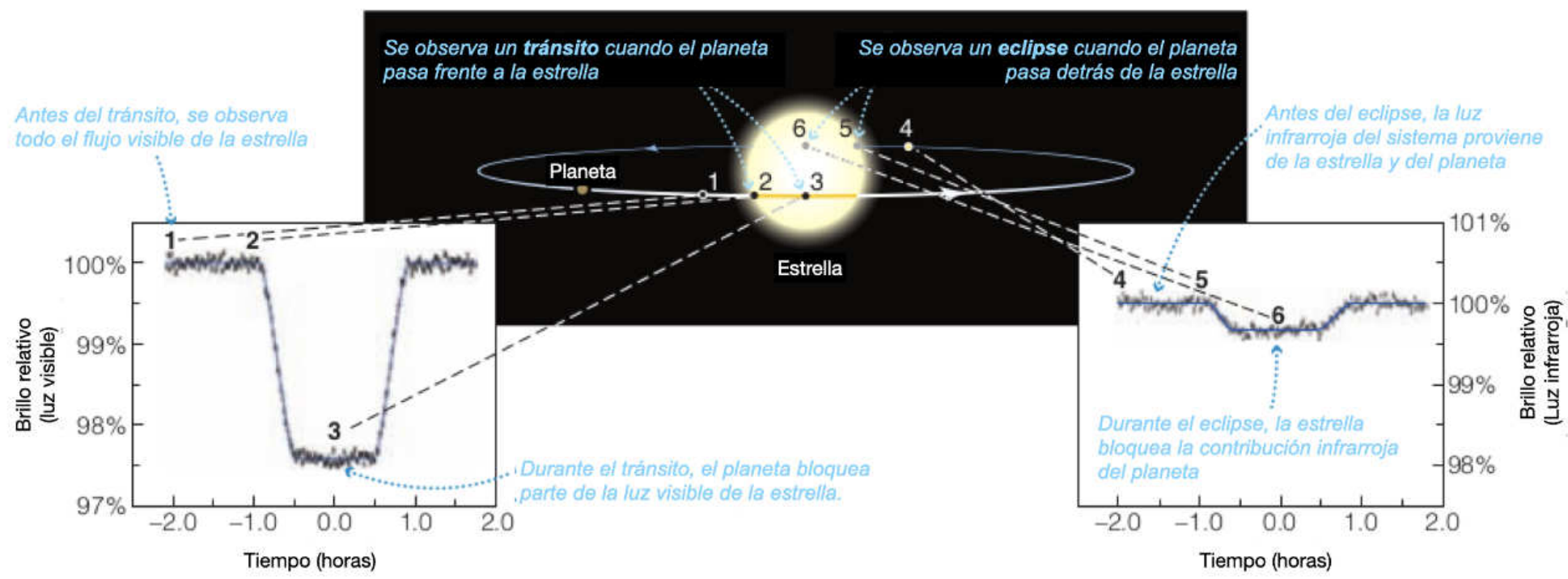

Figura 4.5: Este diagrama representa cómo se observan los tránsitos y eclipses de exoplanetas, usando como ejemplo el exoplaneta que orbita la estrella HD 189733. Las curvas de luz muestran que cada tránsito dura 2 horas, en las cuales el brillo de la estrella disminuye en 2.5 %. Los eclipses son observables en el infrarrojo, pues la estrella bloquea la luz infrarroja del exoplaneta. Los tránsitos y eclipses ocurren una vez por cada órbita del exoplaneta [Crédito: imagen adaptada de Pearson Education].

## 4.3 Formación de exoplanetas

Una de las mayores preguntas sin responder sobre los exoplanetas es cómo se forman. El gran número de planetas extrasolares hasta ahora detectados da fundamento a la hipótesis de que la formación de planetas es un fenómeno muy común en el universo, y probablemente sea una consecuencia directa del proceso de formación estelar. De hecho, el mecanismo más aceptado de formación de planetas, conocido como "acreción de núcleo" [**38**], describe cómo los planetas se forman naturalmente a partir del material que se encuentra en los discos protoplanetarios que se encuentran alrededor de estrellas. Conforme estos discos se enfrían, las moléculas más ligeras y volátiles se evaporan, dejando a las más pesadas en el disco. Estas moléculas, mayormente metales, pueden condensarse para formar partículas de polvo, las cuales colisionan unas con otras y se unen entre sí, formando partículas cada vez más grandes que con el tiempo formarán rocas, asteroides y cometas. Si estos llegan a acumular suficiente masa, su propia gravedad los hará adoptar una forma esférica, y formarán planetas rocosos como la Tierra. Además, su gravedad les permitirá capturar moléculas más ligeras que formarán su atmósfera, mientras que aquellos de mayor masa podrán acumular atmósferas muy extensas y densas, dando así lugar a los exoplanetas gigantes y gaseosos como Júpiter. Esta teoría es apoyada por simulaciones computacionales, así como por observaciones directas de discos protoplanetarios [**39**] que han detectado regiones en las que las partículas de polvo se acumulan, y que posteriormente pueden convertirse en exoplanetas que orbiten las estrellas recientemente formadas [**40**].

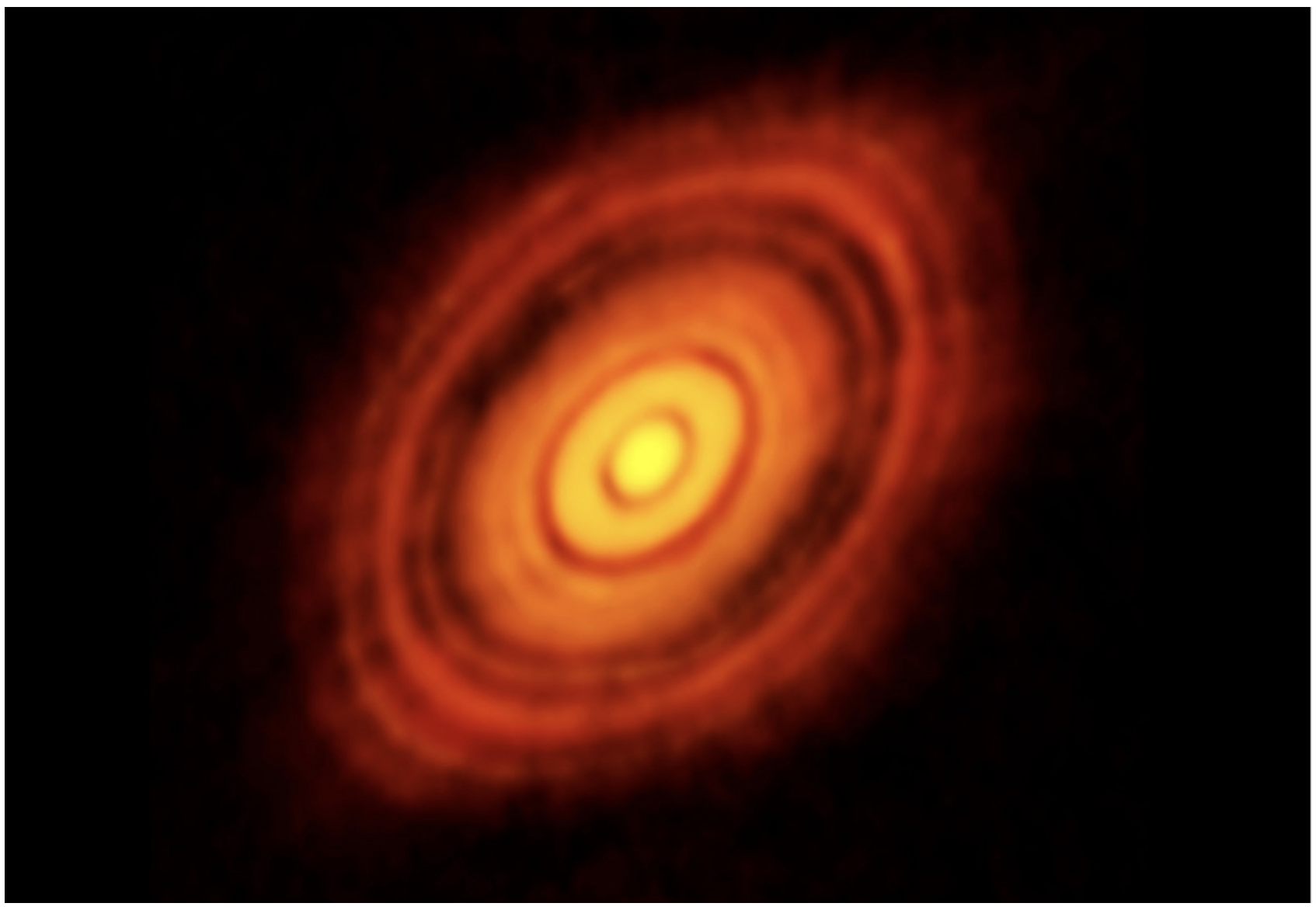

Figura 4.6: Imagen del disco protoplanetario alrededor de la estrella HL Tauri, tomada con el sistema de antenas ALMA. Los anillos oscuros son regiones en las que el contenido de gas y polvo ha disminuido en el proceso de formación de planetas [Crédito: ALMA (NRAO/ESO/NAOJ)].

## 4.4 Poblaciones de exoplanetas

### 4.4.1 Número de descubrimientos por año

En tan solo 30 años, el número de planetas que conocemos ha aumentado de 8 a más de 4 mil [**41**], y nuevos planetas son descubiertos cada año. Al principio, la mayoría de estos descubrimientos fueron realizados con el método de Doppler, pero durante la década de los 2000 empezaron a aumentar las detecciones por tránsitos, y el número de estas creció incluso más con el éxito de la misión *Kepler*. Este telescopio espacial de la NASA fue puesto en órbita en el año 2009, con el objetivo de medir continuamente el brillo de las estrellas en una pequeña región del cielo, y así buscar tránsitos de planetas con periodos orbitales similares al de la Tierra. Con los miles de exoplanetas confirmados en los datos de *Kepler*, los tránsitos se han convertido en el método de detección más exitoso hasta ahora, y se proyecta que muchos planetas más sean descubiertos por misiones como el *Satélite de Sondeo de Exoplanetas en Tránsito* (*TESS*, por sus siglas en inglés), el sucesor de *Kepler*, que empezó a funcionar en el 2018 y ya ha revelado más de 100 planetas nuevos. Nuevos instrumentos, como aquellos incluidos en el *Telescopio Espacial James Webb* o el Telescopio Europeo Extremadamente Grande (E-ELT, por sus siglas en inglés), ayudarán a detectar y caracterizar aún más exoplanetas en los próximos años.

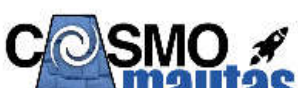

**Recurso TIC 4.4.1:**

Descubre las características de otros sistemas planetarios en el enlace `https://exoplanets.nasa.gov/eyes-on-exoplanets/`. Además, podrás ver las estrellas, los planetas que los orbitan y compararlos con nuestro sistema solar.

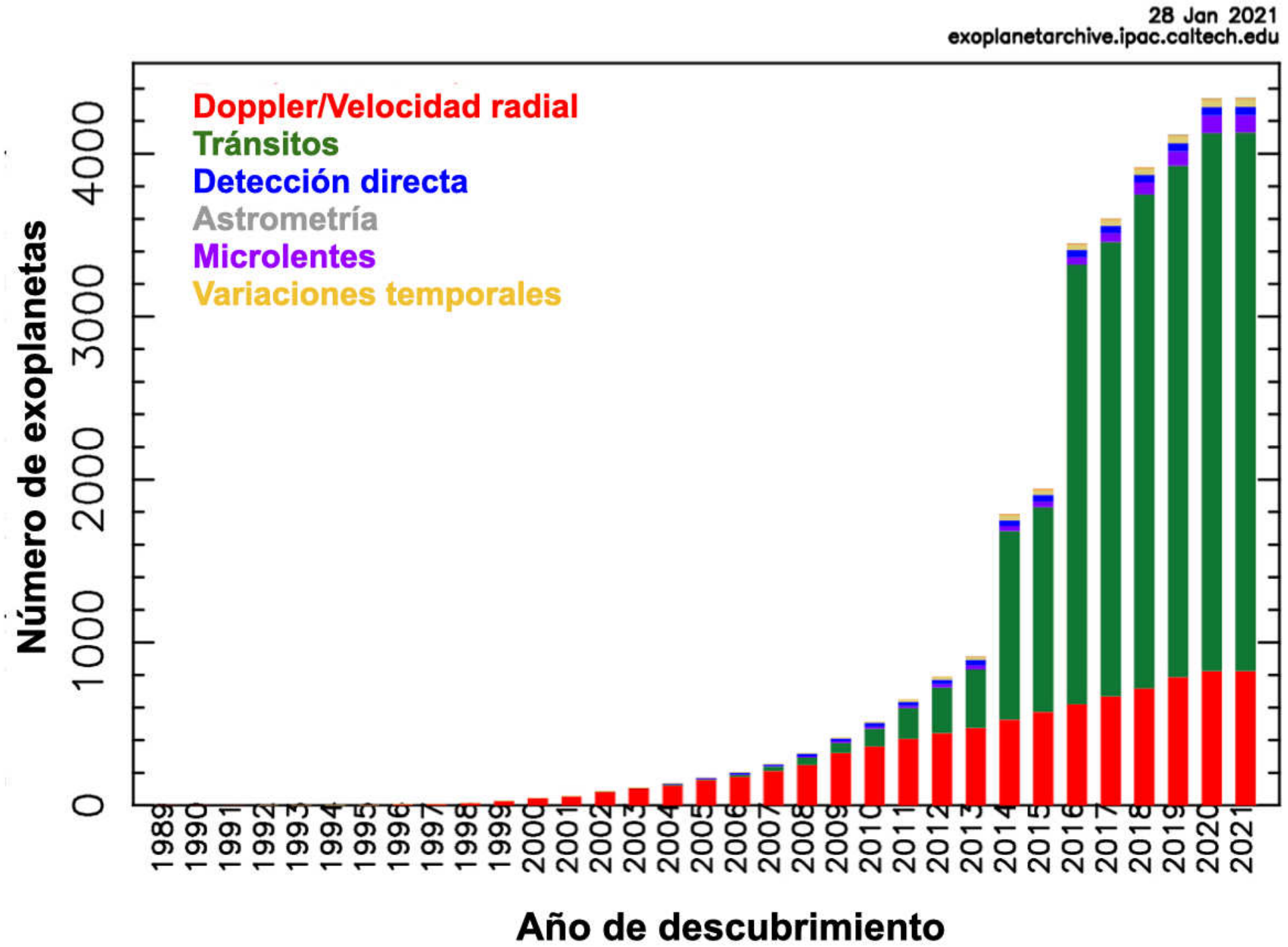

Figura 4.7: Este gráfico muestra el número acumulado de exoplanetas detectados por año desde 1989. Los colores indican el método de detección que fue usado, incluyendo algunos métodos adicionales a los que hemos descrito en este capítulo [Crédito: NASA Exoplanet Archive].

### 4.4.2 Tipos de exoplanetas detectados

Los exoplanetas hasta ahora encontrados son extremadamente variados en sus tamaños, masas y distancias orbitales. Basándose en estas propiedades, los astrónomos han diseñado distintas categorías para clasificar la mayoría de planetas descubiertos, algunos de los cuales son muy diferentes a cualquier planeta de nuestro Sistema Solar

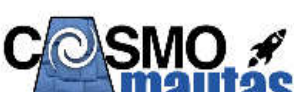

[**42**]:

- Los **gigantes gaseosos** o **mundos jovianos** tienen tamaños similares al de Júpiter y atmósferas muy densas, pero algunos de ellos, conocidos como "Júpiter calientes" tienen temperaturas superficiales extremadamente altas debido a que orbitan muy cerca de sus estrellas.
- Los **planetas "neptunianos"** son similares en tamaño a Neptuno y Urano, aunque algunos son mucho más pequeños, y se caracterizan por tener atmósferas compuestas principalmente de hidrógeno y helio.
- Los **súper-Tierras** son más masivos que la Tierra pero menos masivos que Neptuno, están típicamente compuestos de roca y pueden poseer atmósfera.
- Los **planetas "terrestres"**, son del tamaño de la Tierra o más pequeños, están principalmente compuestos de rocas y metales y podrían tener atmósferas, pero aún no se sabe si contienen océanos u otros ambientes potencialmente habitables.

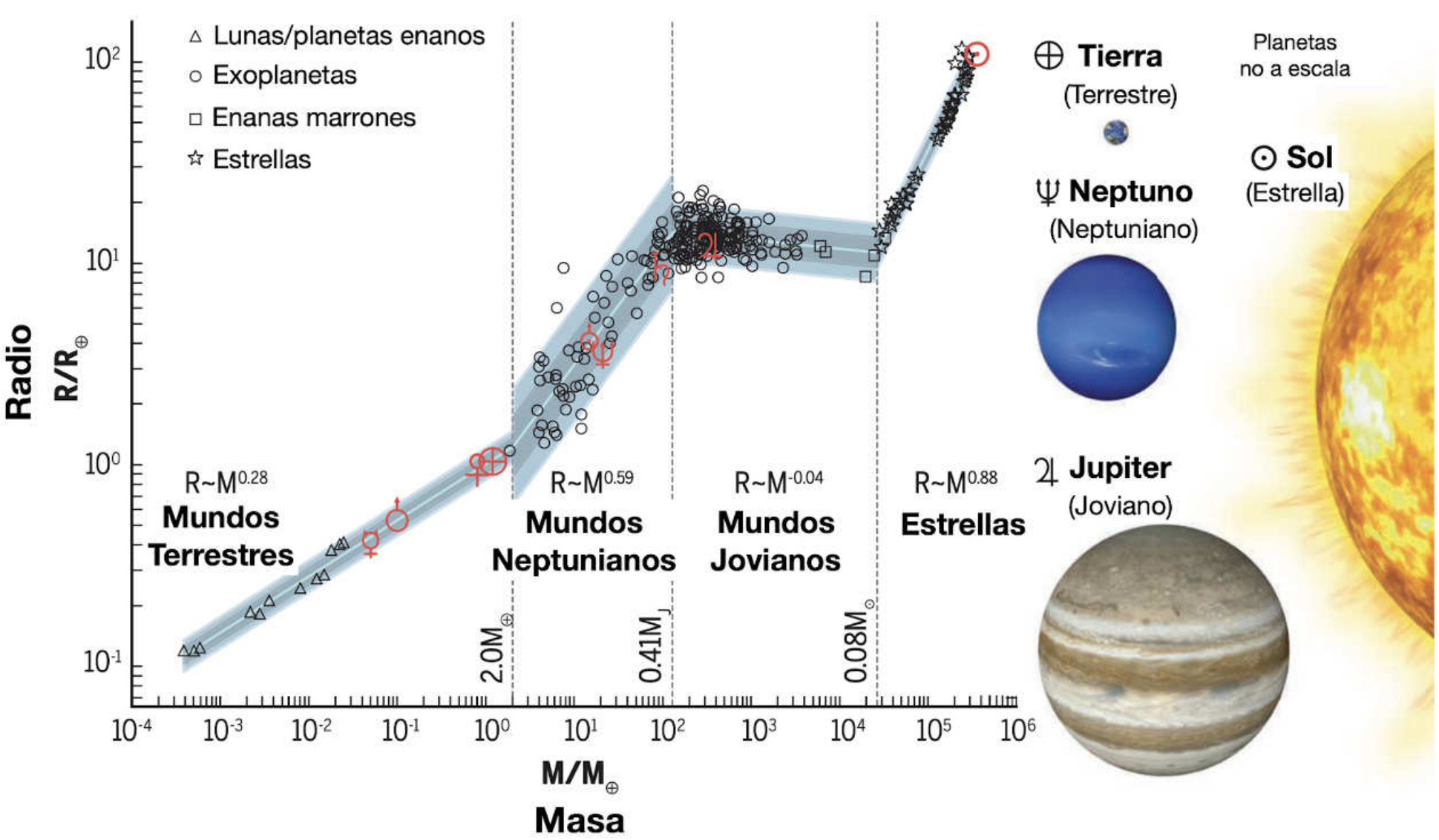

Figura 4.8: Este gráfico muestra la relación entre la masa y el radio para planetas, determinada con datos de exoplanetas detectados. Las distintas regiones en el gráfico con pendientes diferentes corresponden a diferentes tipos de planetas. También se muestra la relación correspondiente a estrellas [Crédito: adaptado de Chen & Kipping (2017)].

### 4.4.3 Habitabilidad

La principal motivación para querer detectar y observar exoplanetas es la búsqueda de vida fuera de la Tierra, para saber si los seres humanos estamos solos en el universo. Para ello, los astrónomos buscan encontrar exoplanetas que orbiten dentro de la zona habitable de una estrella, la región alrededor de la misma que se encuentra a una distancia ideal, tal que podría existir agua líquida en la superficie de un exoplaneta. De estar muy cerca a la estrella, la alta irradiación en el exoplaneta incrementa la temperatura en su superficie, haciendo que el agua se evapore, pero de estar muy lejos, el agua se congelará. Ya que estrellas de distintos tipos emiten radiación con diferente intensidad, la delimitación exacta de la zona habitable depende del tipo de estrella. La presencia de agua es un indicador importante debido a que este es un ingrediente crucial para el desarrollo de formas de vida en la Tierra, pero incluso si un exoplaneta se encuentra en la zona habitable de su estrella, esto no implica necesariamente que haya agua líquida en su superficie, ni mucho menos que exista vida en él. De hecho, se han encontrado varios exoplanetas que se encuentran en la zona habitable, pero la mayoría de ellos orbitan enanas rojas, estrellas que emiten grandes cantidades de radiación ultravioleta que podrían resultar muy peligrosas para los seres vivos. Idealmente, se esperaría encontrar planetas orbitando la zona habitable de estrellas similares al Sol, pero estos sistemas son mucho menos comunes y más difíciles de identificar [**43**]. Aunque se conocen exoplanetas de masa y tamaño similar al de la Tierra, y otros que orbitan en la zona habitable, aún no se ha encontrado un planeta realmente análogo al nuestro, que tenga las condiciones perfectas para albergar vida como la conocemos, por lo que la búsqueda todavía continúa.

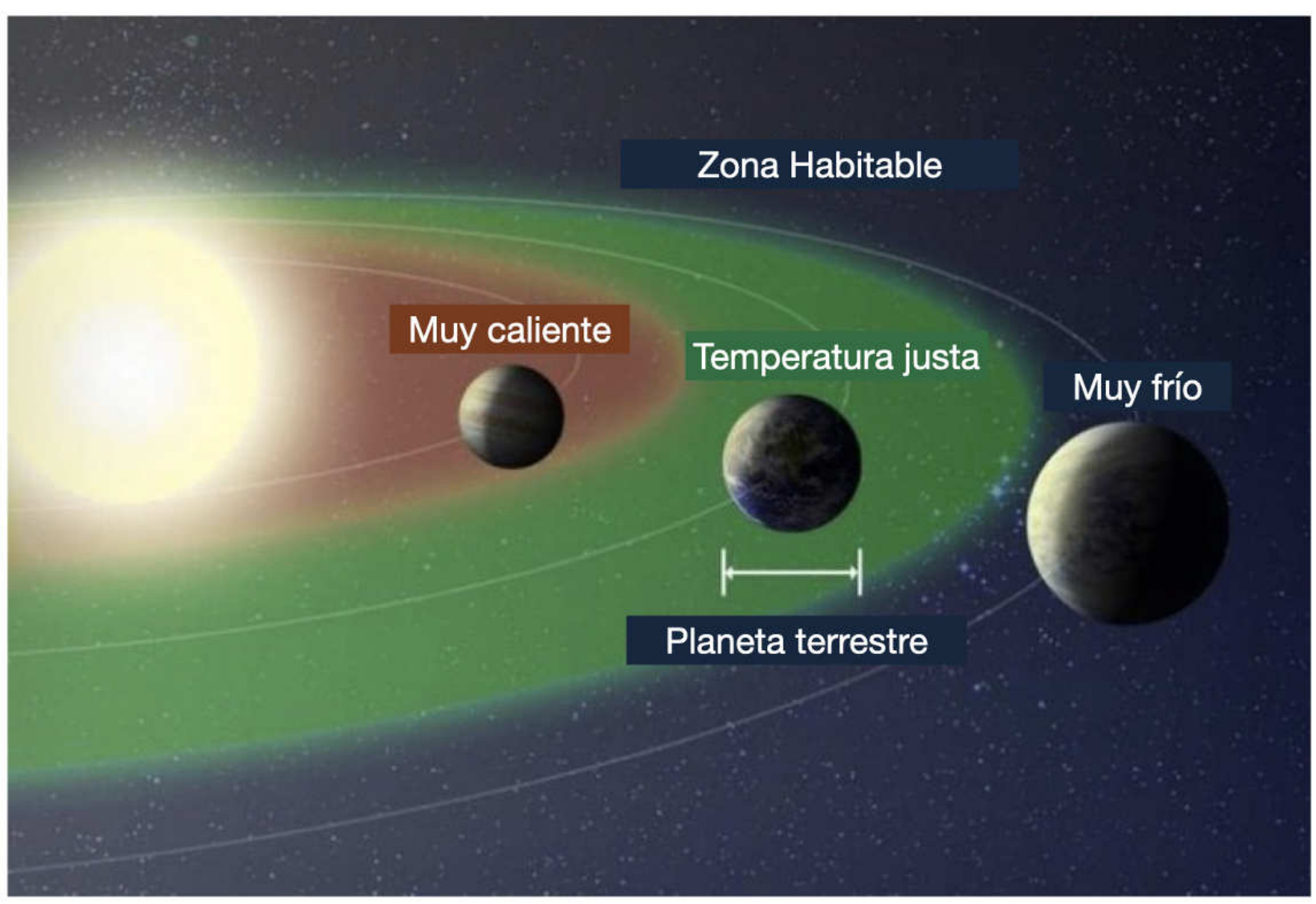

Figura 4.9: Esquema que muestra de la zona habitable alrededor de una estrella, donde la temperatura es justa para que puede haber agua líquida en la superficie de un planeta [Crédito: NASA].

## 4.5 Crisis ambiental

La búsqueda de exoplanetas nos ha llevado a descubrir mundos distantes muy variados, pero nuestro planeta es el único que sabemos que es capaz de albergar vida, por lo que es importante que cuidemos de él y de los ecosistemas que alberga. Actualmente, la Tierra y las especies que viven en ella se ven amenazadas por una severa crisis ambiental, que es causada en gran parte por el incremento de las actividades industriales.

### 4.5.1 El efecto invernadero

De todos los factores que determinan el clima a nivel global, el más importante es el efecto invernadero, un fenómeno natural que ocurre en los planetas que tienen una densa atmósfera, como la Tierra o Venus, mediante el cual parte de la energía que el planeta recibe del Sol se queda “atrapada” en la atmósfera. Las superficies de estos planetas, calentadas con la luz solar, emiten radiación infrarroja, pero gran parte de esta radiación es absorbida por las moléculas de algunos gases que se encuentran en la atmósfera, conocidos como gases de efecto invernadero. Luego de absorber los fotones infrarrojos, estos son emitidos en direcciones aleatorias, algunos hacia el espacio y otros de vuelta hacia la superficie. Muchos de ellos son absorbidos y re-emitidos repetidas veces por varias moléculas, lo cual ralentiza su escape hacia el espacio, y permite que la energía se disperse en el aire de la atmósfera baja y aumente la temperatura (ver Figura 4.10). En condiciones normales, este proceso mantiene la temperatura superficial de la Tierra a niveles adecuados para los seres vivos que la habitan. Sin embargo, cuando la concentración de gases de efecto invernadero es muy alta, la energía se acumula gradualmente en la atmósfera y causa el incremento de la temperatura promedio en la superficie, al cual se le conoce como cambio climático o calentamiento global. Desde hace varias décadas, los científicos han mantenido registros de la temperatura en la atmósfera, y lo que han observado es un incremento sostenido de la temperatura promedio. El consenso de la comunidad científica es que esta tendencia es causada por las emisiones de gases de efecto invernadero por las actividades humanas, principalmente la quema de combustibles fósiles que liberan dióxido de carbono, un potente gas de efecto invernadero. Las mediciones de la concentración de dióxido de carbono en la atmósfera también muestran un sostenido aumento que data desde la revolución industrial, cuando se inició el uso de combustibles fósiles como fuente de energía, y este incremento coincide con el alza de las temperaturas promedio (ver Figura 4.11).

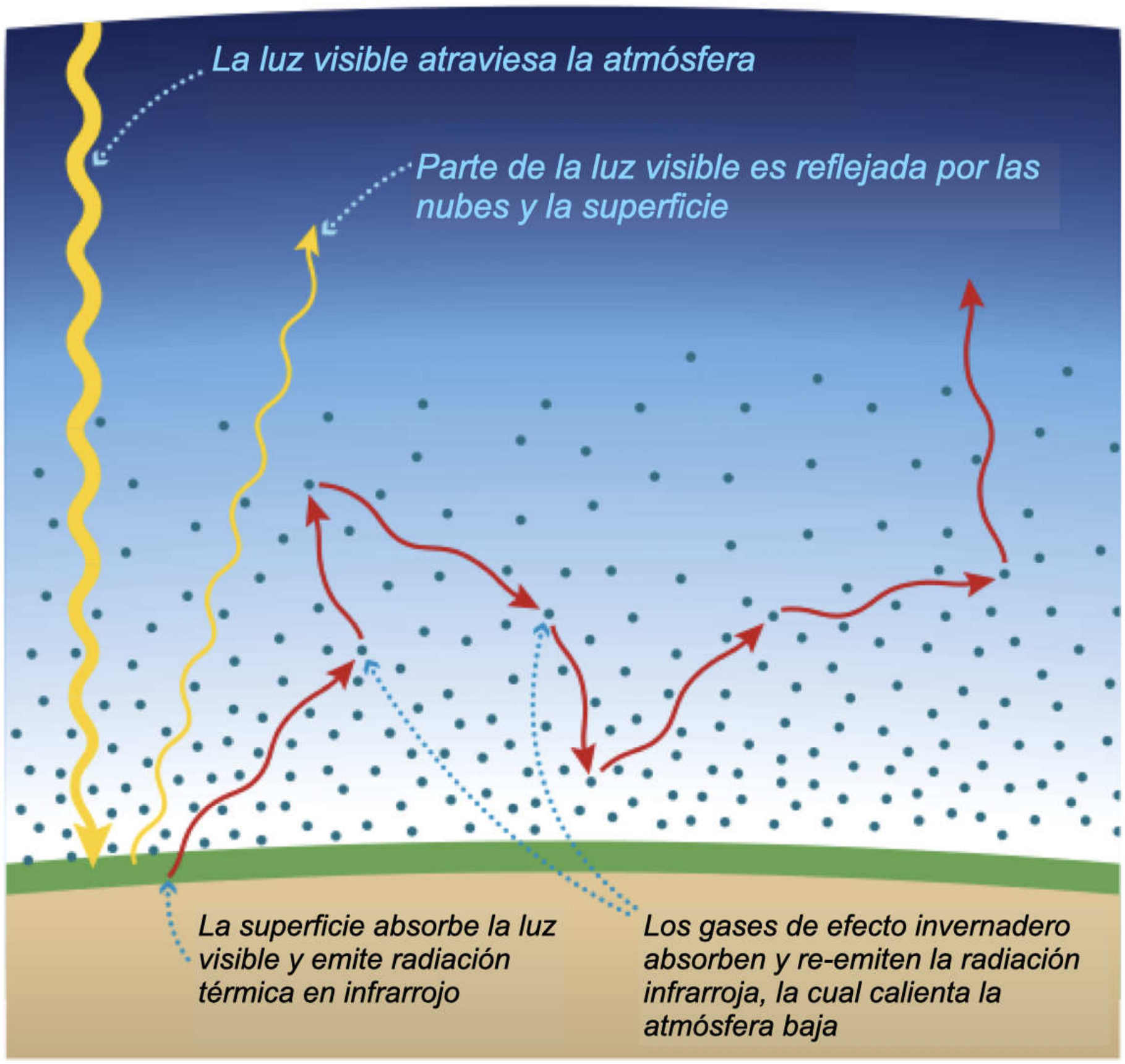

Figura 4.10: Este diagrama muestra cómo los gases de efecto invernadero en la atmósfera hacen que la radiación infrarroja quede atrapada, por lo que la temperatura en la superficie es mayor que si estos gases no estuvieran presentes [Crédito: Pearson Education].

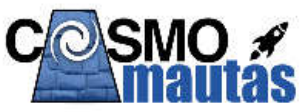

### 4.5.2 Consecuencias del calentamiento global

El incremento descontrolado de la temperatura promedio, incluso solo un par de grados centígrados, puede tener devastadoras consecuencias para el ambiente, incluyendo el derretimiento de glaciares, el alza del nivel del mar, y el aumento en la severidad de desastres naturales como tormentas, incendios forestales y sequías. Estos efectos no solo amenazan los ecosistemas naturales y las especies de plantas y animales salvajes, sino que también ponen en peligro a los humanos, ya que el alza del nivel del mar implica que muchas comunidades costeras se verán inundadas dentro de pocas décadas, y las sequías hacen más difícil cultivar y cosechar alimentos para sustentar a la población. Todavía estamos a tiempo de evitar las peores consecuencias del calentamiento global, pero eso requerirá que hagamos un esfuerzo conjunto como sociedad para detener el calentamiento global. Debemos adoptar medidas para disminuir las emisiones de gases de efecto invernadero, abandonando los combustibles fósiles como principal fuente de energía en favor de alternativas ecoamigables como la energía solar, hidroeléctrica, geotérmica o eólica. También es importante que prioricemos iniciativas para reforestar y preservar los bosques, ya que estos son uno de los principales mecanismos naturales para absorber el dióxido de carbono de la atmósfera, pero actualmente no son capaces de hacerlo a un ritmo que contrarreste las continuas emisiones de este gas.

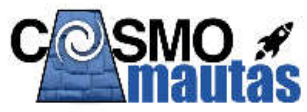

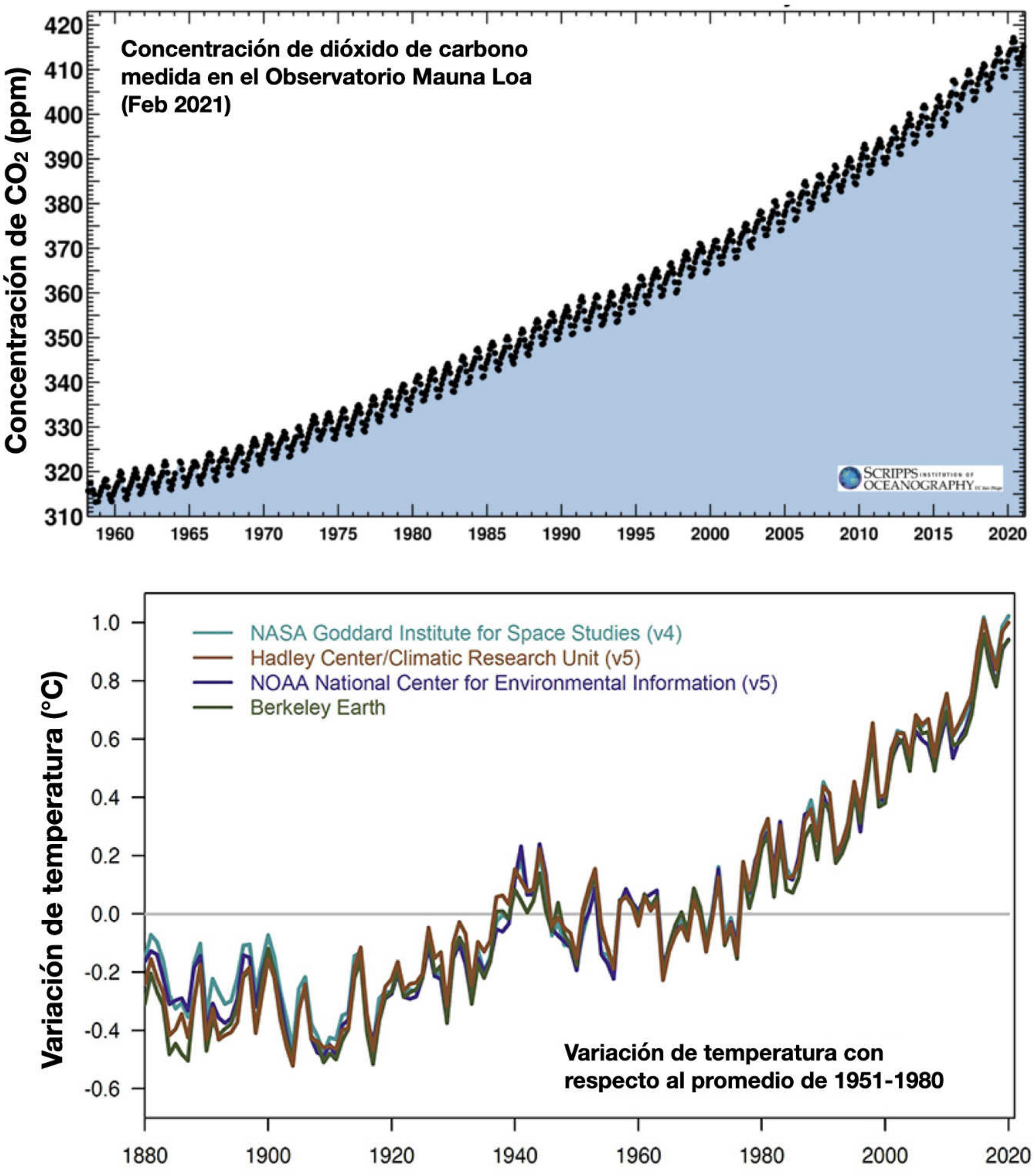

Figura 4.11: *(Arriba)* Incremento de la concentración de dióxido de carbono en la atmósfera en los últimos 50 años [Crédito: Scripps CO2 Program]. *(Abajo)* Variación de la temperatura promedio en la superficie terrestre desde 1880 [Crédito: NASA's Goddard Institute for Space Studies]. Los aumentos en la concentración de dióxido de carbono en la atmósfera y en la temperatura promedio están fuertemente correlacionados.

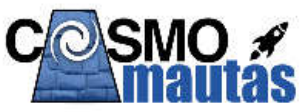

# 5. Galaxias y agujeros negros

Nuestro Sol y todas las estrellas que vemos brillar en el cielo nocturno son parte de nuestra galaxia, una gran estructura de cientos de miles de millones de estrellas, llamada la Vía Láctea. Así como la Tierra orbita alrededor del Sol, el Sol orbita alrededor del centro de la Vía Láctea, el cual se encuentra a unos 25 000 años luz de nuestro sistema solar. El Sol, que es una estrella ni muy cercana ni muy lejana al núcleo galáctico (se encuentra, aproximadamente, a mitad de camino desde el centro hacia el borde de la galaxia), tarda unos 250 millones de años en completar una vuelta.

Desde la tierra, la Vía Láctea puede verse a simple vista en una noche despejada, como una larga banda luminosa y blanca formada por estrellas que atraviesa el cielo. Por eso, se atribuye el nombre de "Vía láctea", el cual, al igual que la palabra "galaxia", se origina del griego *gala, galactos* y significa "Camino de leche". Para aprender más sobre la Vía Láctea la exploraremos en la actividad API 5 (capítulo 11).

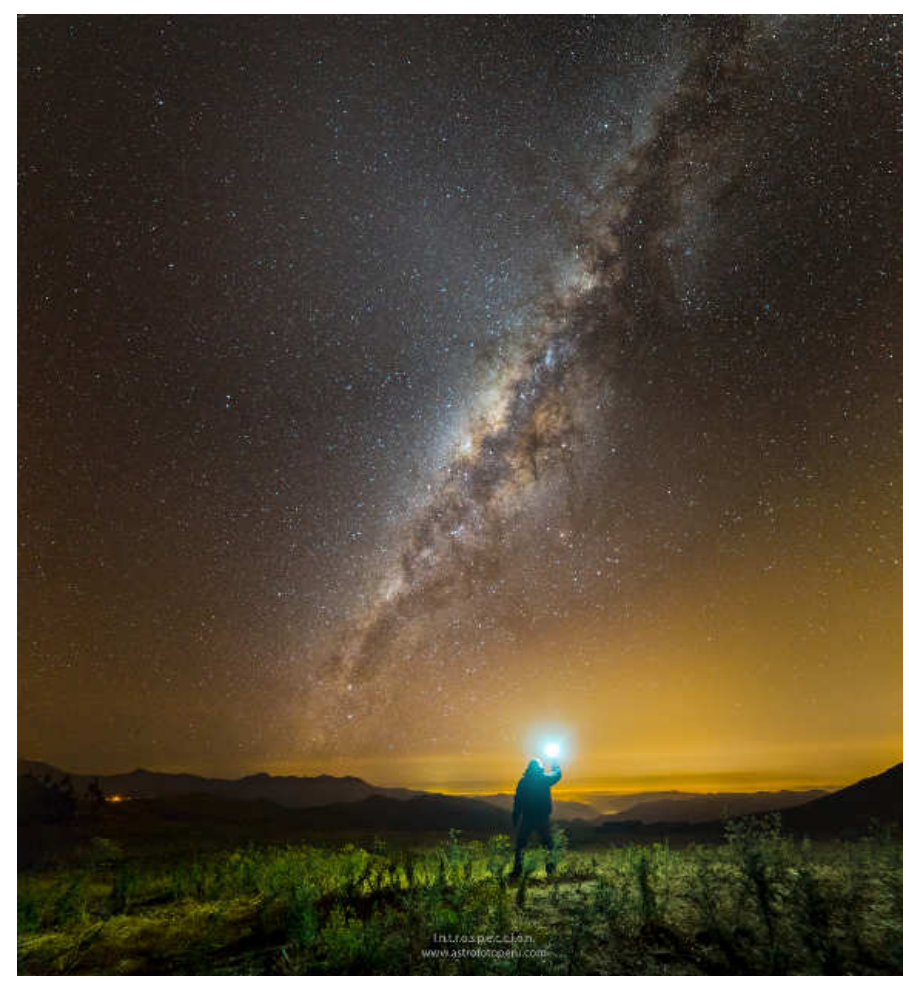

Figura 5.1: La Vía Láctea fotografiada desde Santiago de Tuna (Huarochirí, Lima, Perú) [Crédito: `astrofotoperu.com`]

La Vía Láctea es una de muchas galaxias en nuestro universo, el cual contiene más de dos billones de galaxias, según estimaciones recientes [**52**]. En el presente capítulo se dará a conocer las características básicas de las galaxias y su comportamiento dentro del universo. Además se hablará acerca de los enigmáticos agujeros negros y algunas de sus propiedades.

---

Imagen de encabezado: Galaxia Espiral NGC 1566. Crédito: Telescopio Hubble, NASA/ESA

## 5.1 ¿Qué es una galaxia?

Una galaxia es un complejo sistema conformado principalmente por estrellas, nubes de gas, polvo y materia oscura, el cual se mantiene unido por la gravedad. Además, se estima que muchas, o todas, las galaxias contienen agujeros negros supermasivos en sus centros. Las galaxias son consideradas los bloques de construcción fundamentales del universo.

Las distancias entre galaxias son tan grandes que estas prácticamente no se pueden observar desde la Tierra a simple vista, con unas pocas excepciones. Estas excepciones comprenden a las tres más cercanas a la Vía Láctea, las cuales, en noches muy despejadas y sin contaminación lumínica, pueden ser observadas a simple vista o con un telescopio pequeño. Dos de estas son galaxias enanas, llamadas las nubes de Magallanes, que podemos ubicar cerca a las constelaciones Mensa y Dorado y se encuentran a aproximadamente 160 000 años luz. En dirección a la constelación de Andrómeda también podemos observar a simple vista a la galaxia espiral más cercana, llamada Andrómeda, o “Messier 31”, la cual es visible desde el hemisferio sur durante los últimos meses del año.

A pesar de ser la galaxia espiral más cercana, Andrómeda se encuentra a 2.5 millones de años luz de la Tierra. Es decir que la luz que podemos ver en Andrómeda en realidad fue emitida y empezó su camino hacia nosotros hace 2.5 millones de años, lo que significa que al observar objetos celestes tan lejanos como esta galaxia, ¡estamos observando el pasado! Este efecto, que se origina gracias a que la luz tiene una velocidad constante, nos permite estudiar la historia del universo observando el universo profundo, usando telescopios como nuestras máquinas del tiempo.

**Historia 5.1.1: El gran debate**

Aunque ahora sabemos que nuestra Vía Láctea es tan solo una de muchas galaxias, esta es una noción relativamente nueva producto de un largo proceso de investigación científica. Un evento que marcó el inicio simbólico de la astronomía extragaláctica fue “el gran debate”, que se llevó a cabo en 1920 entre los astrónomos Harlow Shapley y Heber Curtis. En este debate, Shapley postulaba que la Vía Láctea comprendía todo el universo observable, y que los objetos que en aquel tiempo eran llamados ‘nebulosas’ (como Andrómeda y las nubes de Magallanes) eran solo objetos dentro de la Vía Láctea. Curtis, por encontrario, señalaba que estas ‘nebulosas’ eran galaxias externas a la nuestra, también denominadas ‘universos isla’. El debate fue resuelto en base a la evidencia encontrada por **Edwin Hubble** en 1925, quien, gracias a la ley propuesta por la astrónoma **Henrietta Swan Leavitt** sobre la relación lineal entre el periodo y la luminosidad de las estrellas variables Cefeidas, pudo medir la distancia a estas "nebulosas". De esta manera se demostró que la hipótesis de Curtis era correcta, estas ‘nebulosas’ son efectivamente galaxias a enormes distancias de la nuestra, naciendo así la astronomía extragaláctica.

### 5.1.1 Estructura y dinámica de una galaxia

Se observa que las estrellas y el resto de la materia visible en una galaxia giran en órbitas circulares o elípticas alrededor del centro galáctico por efecto de la gravedad, de la misma forma que la gravedad mantiene a los planetas en órbita alrededor del Sol. Pero la materia visible en una galaxia es tan solo una parte minúscula del contenido galáctico total, como veremos a continuación.

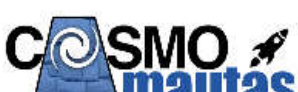

Según la ley de gravitación de Newton, las estrellas más lejanas al centro de la galaxia deberían orbitar más despacio que las más cercanas, y por lo tanto deberían tardar mucho más tiempo en dar una vuelta completa a la galaxia. Esto sería equivalente a lo que observamos en los planetas del sistema solar, en el cual los planetas más alejados del Sol como Urano y Neptuno, tienen velocidad circulares mucho menores que los planetas más cercanos como Mercurio o Venus (ver capítulo 2).

Sin embargo, esto no ocurre en las galaxias. Las velocidades circulares de las estrellas no disminuyen a mayores distancias del centro galáctico. Esta observación se explica por el efecto gravitatorio de una materia invisible que se distribuye hasta radios muy lejanos del centro, la cual es denominada **materia oscura**. Además de ser uno de los elementos más enigmáticos de la física moderna, es extremadamente importante para entender la composición de una galaxia. En base a la medición de velocidades del gas y las estrellas, se estima que la materia oscura puede comprender más del 90 % de la materia en una galaxia. Esto significa que es la materia oscura es lo que en realidad sostiene y mantiene unida a una galaxia. Exploraremos más sobre la materia oscura en el capítulo 6.3.2.

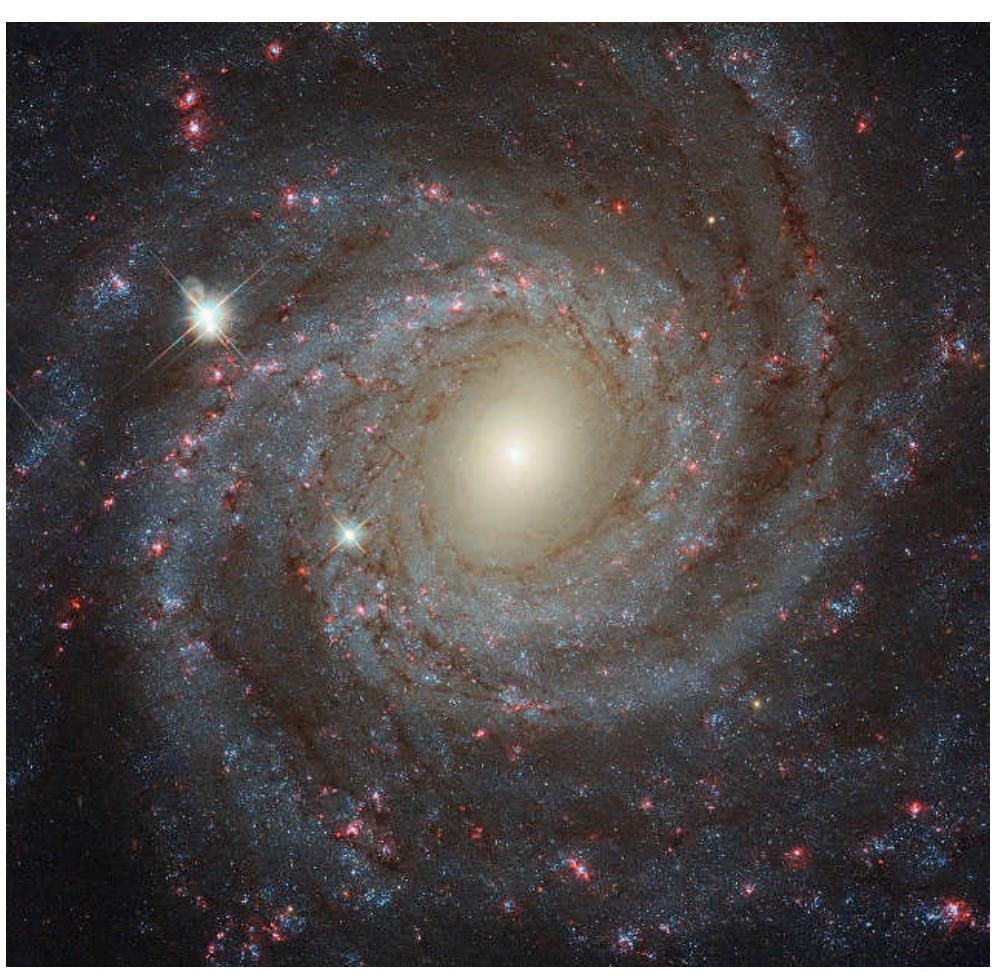

Figura 5.2: Galaxia espiral NGC 3344, ubicada a unos 20 millones de años luz de distancia [Crédito: NASA/ESA].

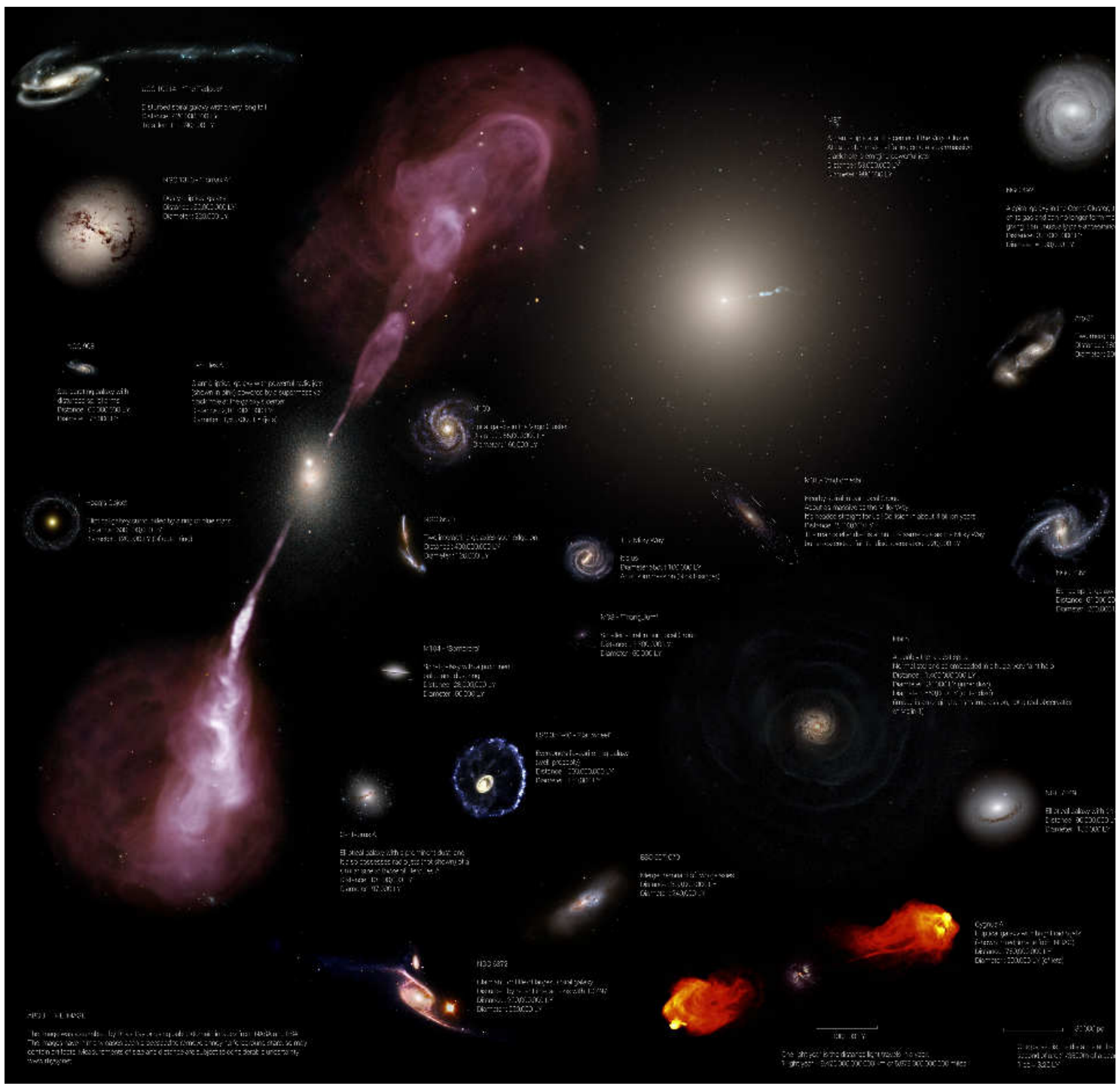

Figura 5.3: Galaxias a escala, que muestran la diversidad de formas, colores y tamaños en relación a la Vía Láctea (en el centro). [Crédito: Rhys Taylor, usando imágenes de NASA/ESA].

**Recurso TIC 5.1.1:**

Usando el programa Stellarium (`stellarium.org`), puede elegir la opción "galaxias" para ubicar algunas de las pocas galaxias cercanas que pueden ser observadas con ayuda de un telescopio pequeño.

### 5.1.2 Clasificación de galaxias

Observaciones profundas del cielo con telescopios de diferentes longitudes de onda han demostrado que el universo contiene galaxias de una gran diversidad de formas, tamaños, masas y luminosidades, como se puede ver en la Figura 5.3. Por ello, clasificar a las galaxias es importante para entender su formación y estructura, aunque no es siempre una tarea simple. El criterio de clasificación más común es la morfología, es decir, que las galaxias son clasificadas en base a su forma y apariencia. Las clases generales en las que se agrupan las galaxias según este criterio son las siguientes:

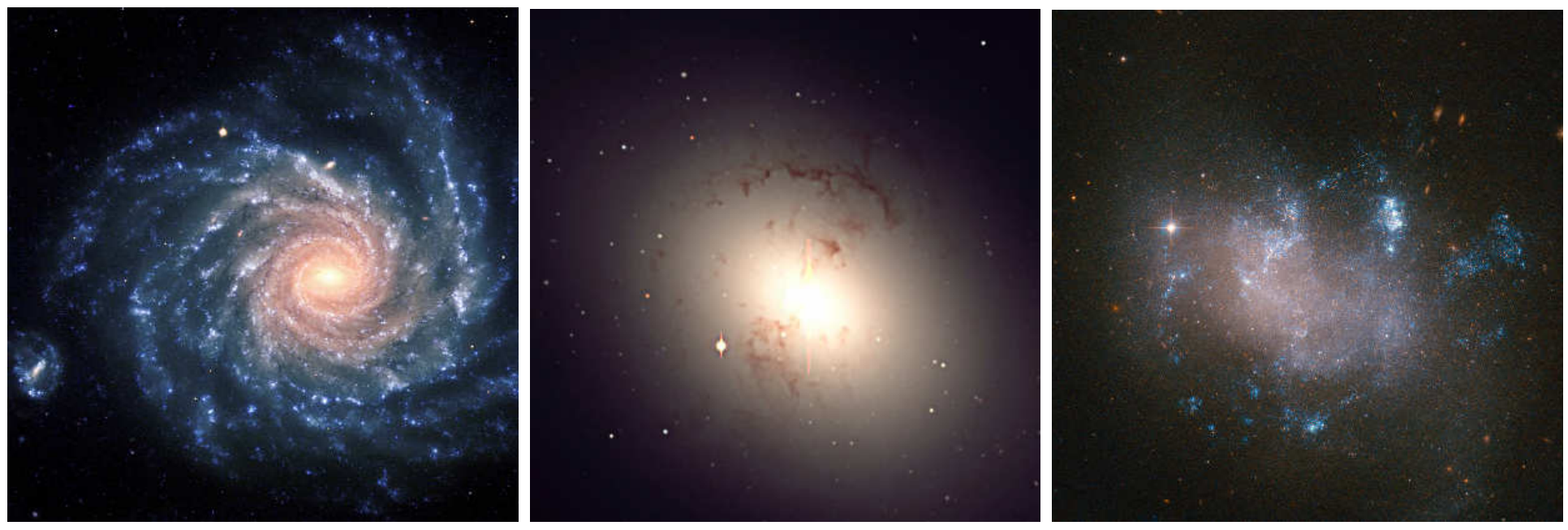

Figura 5.4: *(Izquierda)* Galaxia espiral NGC 1232. *(Centro)* Galaxia gigante elíptica NGC 1316. *(Derecha)* Galaxia irregular UGC 12682 [Crédito: ESO].

- **Galaxias espirales:** se caracterizan por tener un disco plano y brazos espirales, de los cuales obtienen su nombre. Las estrellas se concentran en el disco y orbitan el centro galáctico de manera ordenada. Ese movimiento giratorio es causante de su forma distintiva. Algunas de estas galaxias, como la Vía Láctea, tienen en su centro una estructura en forma de barra lineal, de cuyos extremos se extienden los brazos espirales. En el espectro visible, las galaxias espirales suelen verse de color blanco o azul, lo que indica que contienen estrellas jóvenes muy brillantes que definen la forma de los brazos espirales.
- **Galaxias elípticas:** se caracterizan porque sus estrellas no están concentradas en un disco, sino que se distribuyen en un volumen más grande, y se mueven en órbitas elípticas de orientación aleatoria, por lo que estas galaxias tienen una apariencia redonda y ovalada. Las galaxias elípticas pueden ser pequeñas, incluyendo las llamadas galaxias elípticas enanas, pero dentro de este tipo también se encuentran las galaxias más grandes conocidas del universo, denominadas galaxias elípticas gigantes. En el espectro visible, las galaxias elípticas suelen tener colores rojizos, lo que indica que están compuestas principalmente por estrellas viejas.
- **Galaxias irregulares:** las galaxias irregulares carecen de una forma distintiva, no son redondas ni similares a discos. A menudo adquieren esta apariencia por la influencia gravitacional de otras galaxias cercanas, o incluso porque se formaron como producto de una colisión de galaxias que aún está en curso.

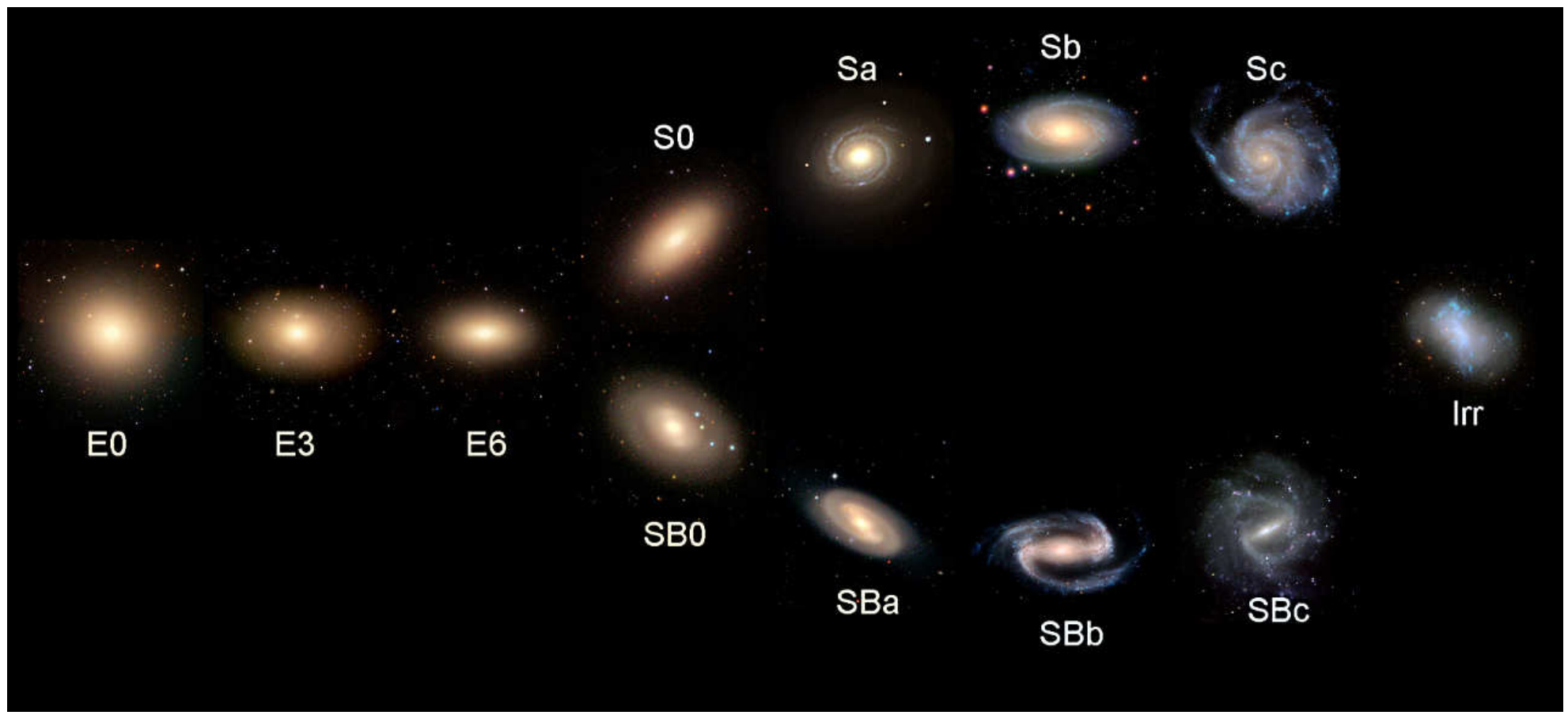

Figura 5.5: Secuencia de Hubble. [Crédito: Zooniverse]

La **secuencia de Hubble** es un esquema desarrollado por Edwin Hubble en 1926, en el que los diversos tipos de galaxias descritos anteriormente son ordenados en una secuencia. Las galaxias elípticas y lenticulares, conjuntamente llamadas galaxias de tipo "temprano", se caracterizan porque su distribución de estrellas no presenta mucha estructura, lo que les da una apariencia difusa. Las galaxias espirales, a veces llamadas de tipo "tardío", muestran una estructura clara, caracterizada por brazos espirales. Como se puede ver en la Figura 5.5, este grupo se divide en dos ramas, las espirales normales y las espirales barradas, las cuales tienen una barra central que conecta sus brazos. Además, Hubble también incluyó una clase de galaxias irregulares.

**Recurso TIC 5.1.2:**

Puede clasificar galaxias con sus alumnos y ayudar a los científicos a entender las formas de las galaxias en el proyecto de ciencia ciudadana "Galaxy Zoo". Puedes acceder a los datos y a la actividad también disponible en español aquí: `https://www.zooniverse.org/projects/zookeeper/galaxy-zoo/classify`.

**Luz 5.1.1: Radiación invisible - el polvo y el gas**

Las galaxias son ejemplos prominentes de emisión multi-onda en el universo, ya que su compleja composición resulta también en una compleja combinación de procesos físicos. Estos procesos incluyen, entre otros, a la física de las estrellas y la física del polvo y el gas en el medio interestelar. Estos procesos emiten radiación de diferentes energías que van más allá de la luz óptica, produciendo radiación "invisible", tanto en longitudes de onda de menor energía (infrarrojas y radio), como a mayor energía (ultra-violeta, rayos X). Un claro ejemplo de esto es la Vía Láctea, en la cual podemos reconocer estructuras de sombra producidas por el polvo, que impiden el paso de la luz óptica en esas regiones [Crédito A. Mellinger].

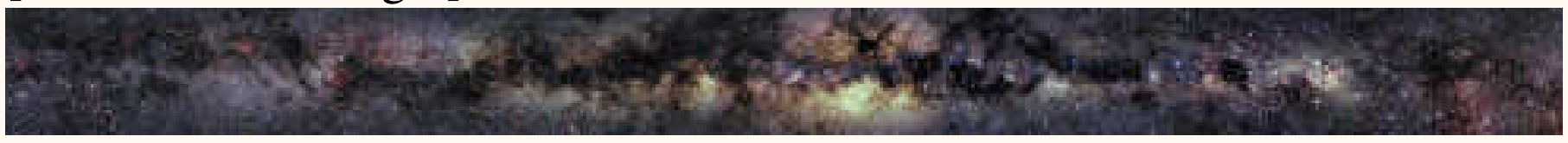

Si observamos nuestra vía láctea en ondas infrarrojas entonces se aprecia,

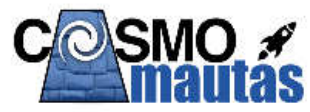

además de una parte de la radiación estelar, la emisión infrarroja brillante del polvo, que ha sido calentado por la luz de las estrellas, como demuestra esta imagen del telescopio IRAS [Crédito: NASA/NIVR/SERC].

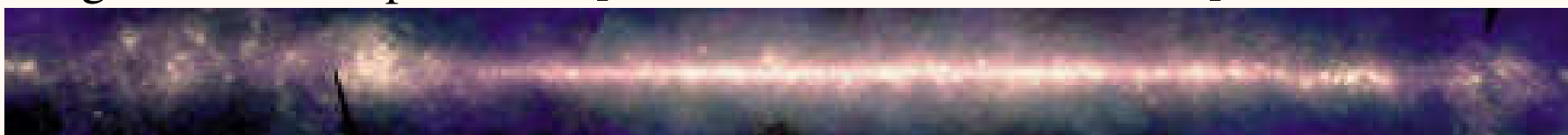

Finalmente también podemos observar la emisión del gas, en base a cómo elementos específicos interactúan con la radiación. En éste caso observamos cómo la radiación de hidrógeno molecular está distribuida en nuestra galaxia [Crédito: NASA].

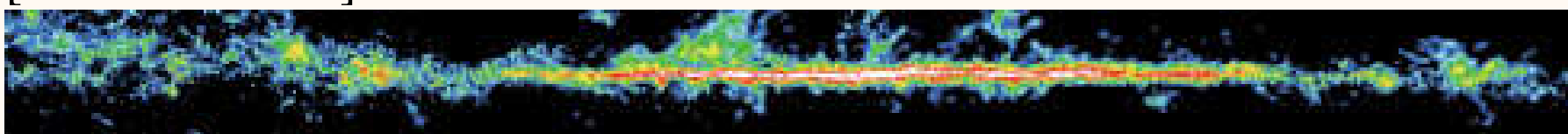

### 5.1.3 Componentes de una galaxia

A pesar de la gran diversidad de galaxias que observamos en el universo, éstas tienen varios componentes en común, los cuales se distinguen por su forma y por las poblaciones de estrellas que se encuentran en ellos. En general, los distintos tipos de estrellas tienden a estar distribuidos de manera diferente dentro de las galaxias. Por ejemplo, los brazos espirales en una galaxia espiral suelen contener una mayor densidad de estrellas jóvenes o recién nacidas, por lo que deducimos que dentro de ellos deben formarse estrellas nuevas con más frecuencia. Sabemos esto porque los brazos de las galaxias espirales suelen tener colores más azules que el resto de su disco. Recordemos que las estrellas azules, que son muy calientes, tienen un periodo de vida más corto que las estrellas rojas, que son más frías. Por lo tanto, las regiones de las galaxias que se ven más azules son donde las estrellas deben haberse formado más recientemente, mientras que en las regiones más rojas abundan las estrellas "viejas" y no ha habido formación estelar reciente.

En el caso de las **galaxias espirales**, como la Vía Láctea o Andrómeda, los componentes principales son los siguientes:

- **Disco:** El disco es una región aplanada en donde se encuentra la mayor parte de las estrellas de la galaxia, las cuales se distribuyen en brazos con forma de espiral. El disco de la Vía Láctea, por ejemplo, tiene 100 000 años luz de diámetro y 1 000 años luz de espesor. El disco contiene estrellas de distintas edades, así como polvo y grandes cantidades de gas, a partir del cual se forman estrellas nuevas. Aunque el gas y el polvo son como sombras en longitudes de onda óptica o "visible", estos son brillantes al observarlos en otras longitudes de onda (ver la caja de luz 5.1.2). Las estrellas dentro del disco orbitan alrededor del centro galáctico de manera muy ordenada, todas en la misma dirección y con órbitas muy semejantes a círculos.
- **Bulbo:** El bulbo es una estructura redonda que se concentra en la región central de la galaxia, y está compuesta principalmente de estrellas antiguas. La forma redondeada del bulbo se debe a que las estrellas que lo componen, a diferencia de aquellas en el disco, orbitan el núcleo galáctico en diferentes direcciones, y sus órbitas tienen inclinaciones muy variadas. El bulbo de la Vía Láctea tiene aproximadamente 10 000 años luz de diámetro.
- **Halo:** El halo es una estructura redondeada que engloba a la galaxia en su totalidad y se extiende más allá de los bordes del disco. Tiene una apariencia difusa, pues la concentración de estrellas en él es mucho menor que en el disco. Las estrellas del halo suelen ser muy antiguas, están mayormente concentradas en conjuntos llamados "cúmulos globulares", y sus órbitas también tienen inclinaciones variadas. Sin embargo, en base a las mediciones de las velocidades de las estrellas más lejanas, resulta evidente que el componente principal del halo es la materia oscura, la cual no podemos ver, pero podemos deducir su presencia debido a su influencia gravitacional. Se estima que el halo de la Vía Láctea tiene más de 130 000 años luz de diámetro.

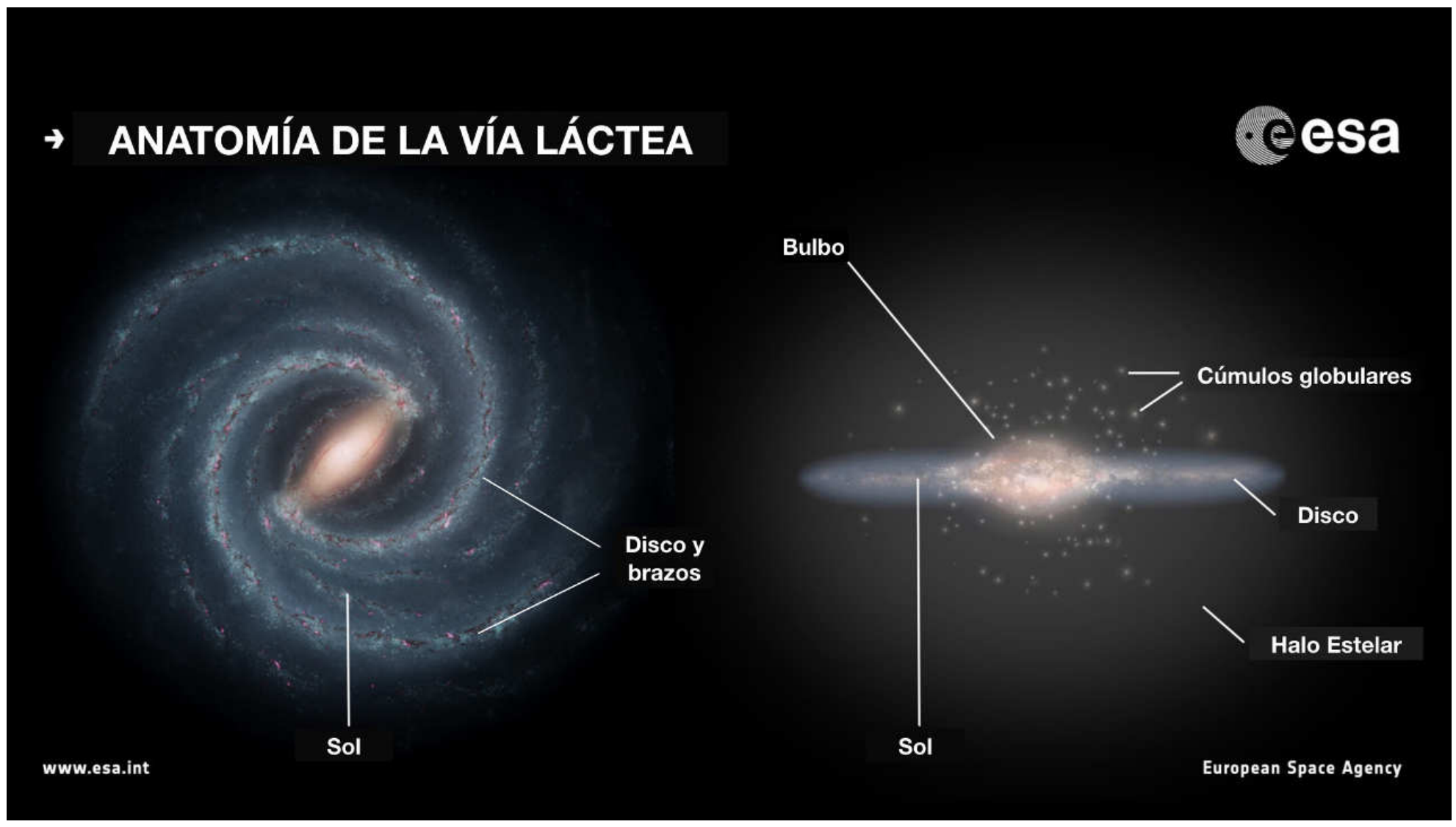

Figura 5.6: Representación artística de la de la Vía Láctea y esquema de sus componentes, como ejemplo de una galaxia espiral. [Crédito: adaptado de imagen de la Agencia Espacial Europea (ESA)].

**Unidades 5.1.1: Año luz (ly)**

Un año luz es la distancia que recorre la luz en el vacío en un año terrestre. La luz viaja a una velocidad de $c = 3 \times 10^8$ m/s, de tal forma que un año luz viene a ser 1 ly $= 9{,}46 \times 10^{15}$ m, o 63 241 AU.

Las **galaxias elípticas** no presentan componentes tan claramente definidos como los de las espirales. Estas galaxias carecen de un disco, por lo que todas sus estrellas se encuentran en un bulbo central y en el halo esferoidal que lo rodea, aunque puede ser difícil distinguir entre ambos pues las poblaciones de estrellas que los componen son muy similares. Estas son principalmente estrellas antiguas, que le dan a las galaxias colores rojizos o amarillos, con muy poco contenido de gas y polvo. Esto nos indica que en las galaxias elípticas se forman estrellas nuevas a un ritmo mucho menor que en las espirales. Además, a diferencia de las galaxias espirales, la dinámica de las estrellas en una galaxia elíptica está dominada por movimientos aleatoria y desordenados de las estrellas.

## 5.2 Familias y ciudades de galaxias

La mayoría de las galaxias no se encuentran aisladas en el universo, si no que se distribuyen en **grupos de galaxias**, formados por menos de 50 galaxias, o en **cúmulos de galaxias**, los cuales pueden contener de cientos a miles de galaxias.

De igual modo, los cúmulos forman conjuntos más grandes llamados supercúmu-los, los cuales a su vez forman estructuras en forma de filamentos. La Vía Láctea se encuentra dentro del Grupo Local, conformado por más de 50 galaxias, el cual forma parte del Supercúmulo de Virgo.

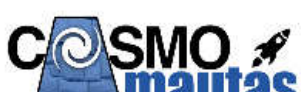

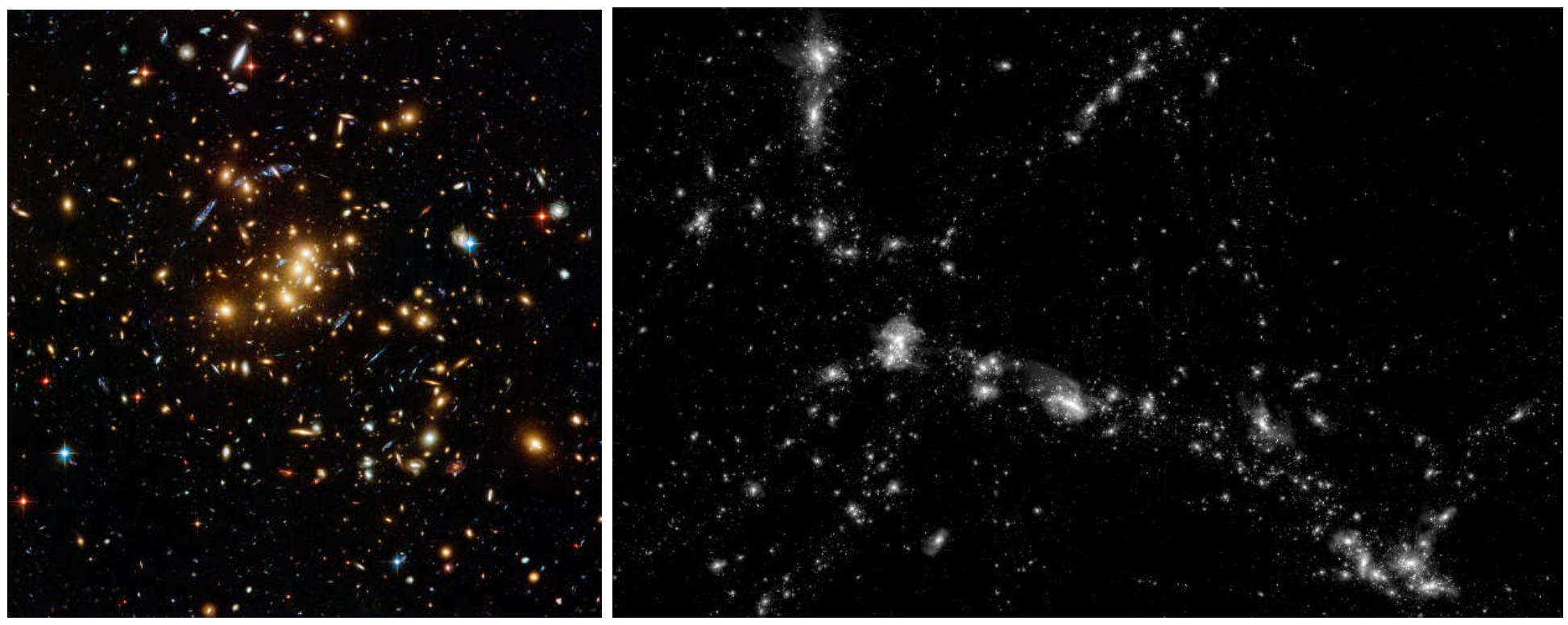

Figura 5.7: *(Izquierda)* Cúmulo de galaxias Cl 0024+17 (ZwCl 0024+1652) fotografiada por el *Telescopio Espacial Hubble* [Crédito: NASA/ESA]. *(Derecha)* Distribución de galaxias vistas a gran escala dentro de una simulación cosmológica (masa bariónica), distribuidas en cúmulos y filamentos [Crédito: Illustris TNG].

### 5.2.1 Interacciones entre galaxias

Las galaxias interactúan entre ellas también a través de la gravedad, cuando las distancias entre ellas son relativamente cortas. Las **interacciones menores** se dan entre una galaxia principal y otra más pequeña que orbite alrededor de esta, llamada galaxia satélite. Por ejemplo, la gravedad de la galaxia satélite puede atraer a uno de los brazos espirales de la galaxia principal, e incluso puede llegar a capturar algunas de las estrellas de la galaxia principal. Este proceso también puede ocurrir al revés, de modo que la galaxia principal le arrebate estrellas a la galaxia satélite.

Por otro lado, las **interacciones mayores**, como su propio nombre indica, son perturbaciones gravitatorias de mayor magnitud que las interacciones menores, y son sucesos frecuentes en la evolución de las galaxias. Un tipo de interacción mayor es la colisión de dos galaxias, proceso en el cual puede ocurrir uno de los siguientes escenarios:

1. Que las galaxias se atraviesen, de tal modo que cada una conserve la mayor parte de su material y su forma permanece prácticamente inalterada.
2. Que las galaxias prosigan su camino, pero que pierdan parte de su material galáctico durante la colisión, mientras que su estructura y forma cambien notablemente
3. Que las galaxias se fusionen, lo que suele suceder cuando dos galaxias chocan y no tienen suficiente cantidad de movimiento para continuar el viaje después de la colisión. En ese caso, el material de las dos galaxias se une poco a poco hasta formar una sola galaxia.

Este último tipo de colisión, por ejemplo, ocurrirá en el futuro entre nuestra galaxia y la galaxia Andrómeda, la galaxia espiral más cercana a la Vía Láctea. Mediciones de las velocidades en la galaxia de Andrómeda usando el corrimiento al azul de los espectros (debido al efecto Doppler), revelan que esta se está acercando a nuestra galaxia con una velocidad de 100-300 km/s respecto al Sol [**53**] y por lo tanto se estima que las galaxias se acercarán dentro de aproximadamente 4-6 mil millones de años. Cabe resaltar que esta colisión no tendrá un gran impacto en la vida humana. Debido a las grandes distancias entre las estrellas dentro de una galaxia, en una colisión entre galaxias la probabilidad que dos estrellas y sus sistemas planetarios coincidan y colisionen es casi inexistente. Además, se estima que en la escala de tiempo en la que ocurrirá la colisión con Andrómeda, la Tierra ya no será habitable debido a los cambios de luminosidad de nuestro Sol.

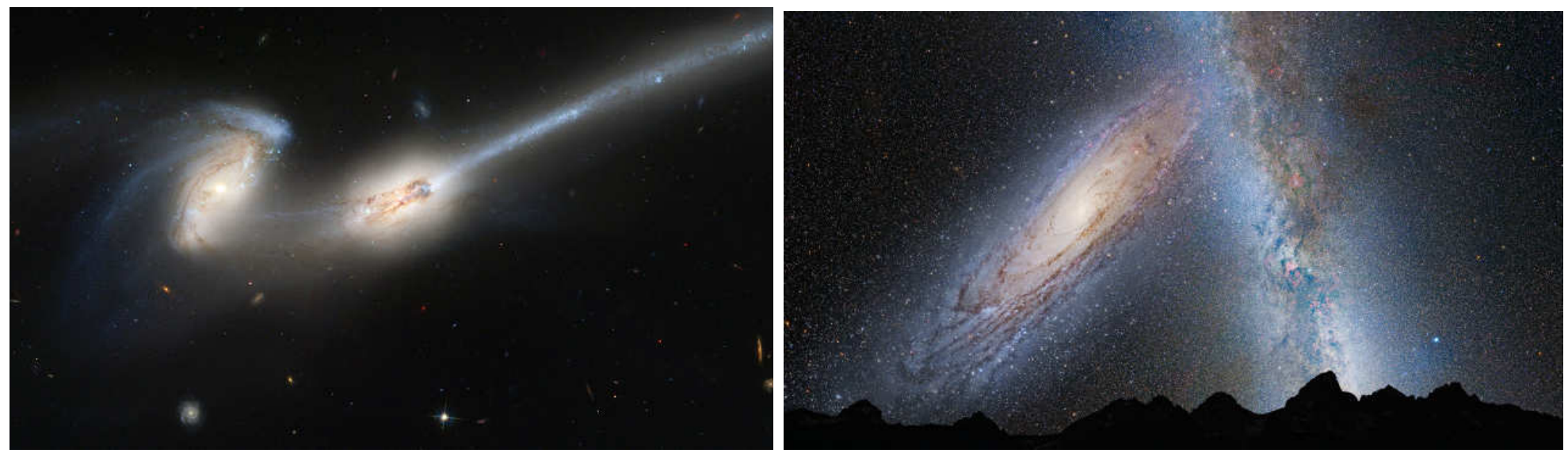

Figura 5.8: *(Izquierda)* Galaxia NGC4676, conocida como “galaxia los ratones” es producto de la colisión entre dos galaxias [Crédito: NASA/ESA]. *(Derecha)* Representación artística de cómo se vería el cielo nocturno en 4 mil millones de años producto de la colisión entre Andrómeda y la Vía Láctea [Crédito: NASA].

## 5.3 Agujeros negros

Anteriormente mencionamos que muchas galaxias contienen agujeros negros en su centro, pero ¿qué es un agujero negro? Un agujero negro es una región en el espacio donde la fuerza de atracción de la gravedad es tan fuerte que ningún tipo de materia o partícula, ni la luz, puede escapar. La gravedad es tan fuerte porque un agujero negro contiene una gran concentración de masa en un espacio relativamente diminuto, es decir, un agujero negro es un objeto de densidad extremadamente alta (ver caja 5.3). Por esta razón, los agujeros negros pueden tener cualquier tamaño, desde muy pequeños (del tamaño de átomos) hasta miles de kilómetros, lo importante es que una determinada masa esté contenida en un volumen suficientemente pequeño llamado singularidad, el mismo que está en el interior de una esfera determinada por el radio de Schwarzschild llamada horizonte de eventos, zona límite donde ni la luz no puede escapar(ver definición en la caja 5.3.2).

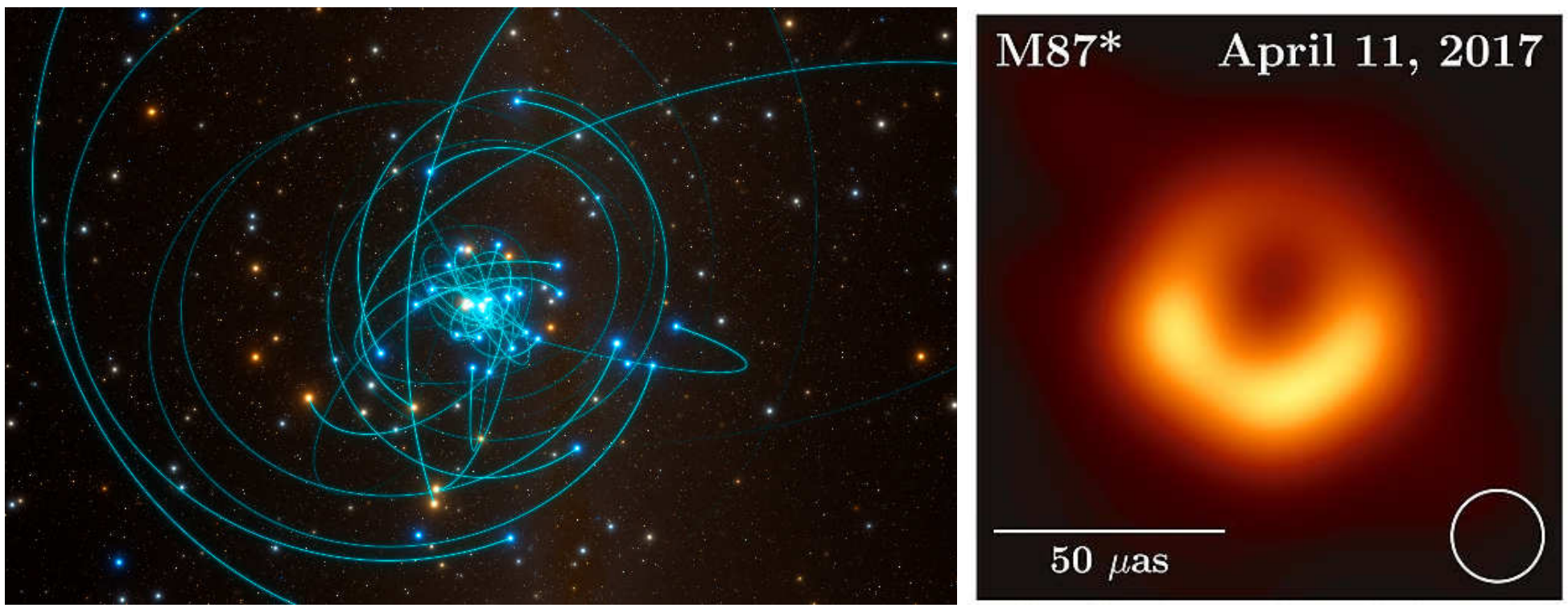

Figura 5.9: *(Izquierda)* Simulación de las órbitas que siguen las estrella en torno a Sagitario A*, el agujero negro supermasivo que vive en el corazón de nuestra galaxia [Crédito: ESO/L. Calçada/spaceengine.org]. *(Derecha)* Fotografía del disco de acreción del agujero negro en la galaxia M87, capturada por el telescopio de Horizonte de Eventos. Crédito: Colaboración EHT

**Gravedad 5.3.1: Velocidad de escape**

En física se denomina velocidad de escape a la velocidad debe tener un objeto para escapar del campo gravitatorio de otro. Por ejemplo, la velocidad inicial de un cohete para vencer la fuerza de la gravedad de la Tierra y escapar de ella es de 11 km/s, es importante mencionar que si deseamos salir del sistema solar, la velocidad debe ser mayor. La velocidad de despegue de un planeta o una estrella está directamente relacionada con el tamaño y la masa de éste. En particular, el cuadrado de la velocidad de escape es directamente proporcional a la masa del planeta (o de la estrella) e inversamente proporcional a su radio. Así, si conocemos dos de estos parámetros podemos calcular el tercero.

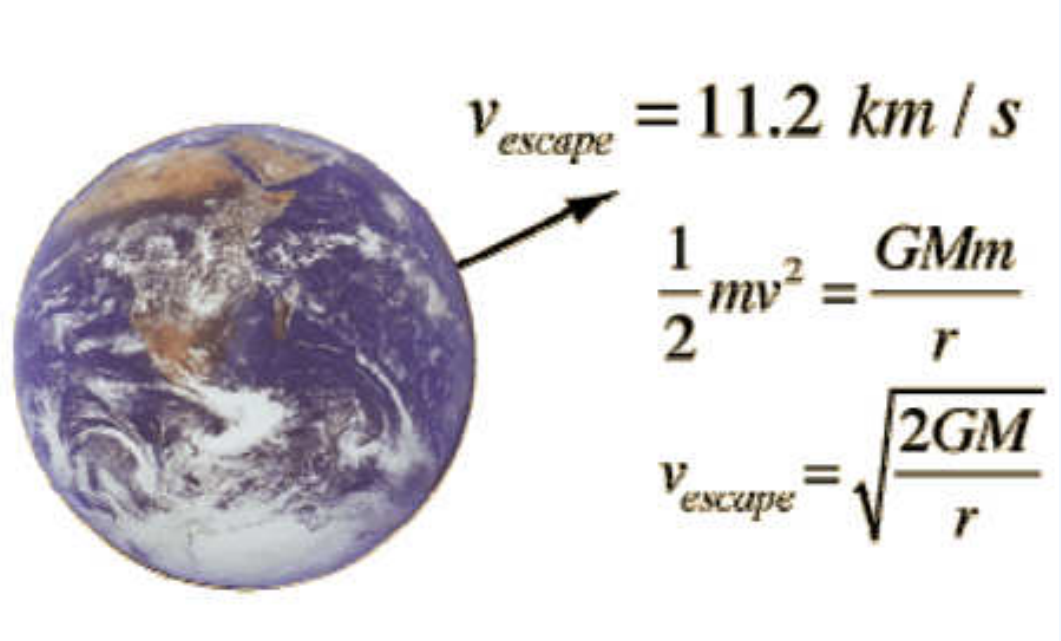

El concepto de agujero negro empezó como una idea puramente teórica, una predicción matemática de la Teoría de la Relatividad General de Einstein ya en 1916 (ver más sobre relatividad general en el Capítulo 6.1.2). Un largo proceso de investigación teórica por el físico Roger Penrose concluyó que la formación de estos objetos teóricos era viable en el universo observable, empezando así una larga aventura de busqueda y caracterización de estos objectos. Hoy en día tenemos una gran cantidad de observaciones astronómicas que muestran evidencia de que estos objetos realmente existen. Una de ellas es el seguimiento de estrellas en la región central de nuestra propia galaxia, cuyas órbitas demuestran que debe existir un objeto muy masivo en el centro y que este ocupa un volumen tan pequeño que solo puede tratarse de un agujero negro. La astrónoma Andrea Ghez y el astrónomo Reinhard Genzel, quienes lideraron investigaciones que confirmaron la presencia de este agujero negro en nuestra galaxia, denominado Sagitario A*, fueron recientemente galardonados con el premio Nobel 2020. Otra observación importante se realizó en el 2017, cuando se logró captar la primera imagen de la sombra de un agujero negro, aquel ubicado en el centro de la galaxia M87, usando el Telescopio de Horizonte de Sucesos [**54**]. Además, las altas luminosidades encontradas en los centros de muchas galaxias pueden ser explicadas por materia orbitando y siendo consumida por un agujero negro supermasivo (disco de acreción), y la emisión en ondas de radio, lo cual sugiere que la mayoría de galaxias, si no todas, contienen este tipo de agujeros negros en sus centros.

**Gravedad 5.3.2: El radio de Schwarzschild**

Recordemos que la densidad de la materia se define como la masa contenida en un determinado volumen, por lo que si se reduce el volumen de un objeto manteniendo constante su masa, su densidad será mayor:

$$\text{densidad} = \frac{\text{masa}}{\text{volumen}}$$

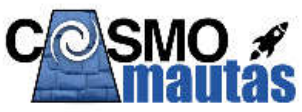

El **radio de Schwarzschild** de un objeto es el radio que éste debería tener para que la *velocidad de escape* (ver caja 5.3) desde su superficie sea igual a la velocidad de la luz. El tamaño de dicho radio depende únicamente de la masa del objeto, por lo que si se conoce uno de estos parámetros es posible calcular el otro. Un objeto podría convertirse en un agujero negro si toda su masa es comprimida hasta adquirir un radio menor a su radio de Schwarzschild. Ya que la velocidad de escape sería igual a la velocidad de la luz, y ningún objeto puede superar esta velocidad, entonces ningún objeto, ni siquiera la luz, es capaz de escapar de un agujero negro.
Veamos algunos ejemplos del radio de Schwarzschild de objetos conocidos. Para transformar a la Tierra en un agujero negro, habría que comprimir su tamaño actual (6357 km de radio) hasta 9 milímetros de radio, es decir toda la masa de la Tierra quedaría contenida en un volumen menor al de un grano frijol, maní o sacha inchi. Por otro lado, para transformar al Sol en un agujero negro habría que comprimirlo desde su tamaño actual (700 000 km de radio) hasta solo 3 km de radio. Es como si tratáramos de meter toda la masa del Sol en un astro del tamaño de un pueblo.

### 5.3.1 ¿Cómo funciona un agujero negro?

Aunque el término “agujero negro” puede sugerir la idea de un espacio vacío, y conlleva una fuerte connotación de misterio, es importante comprender que en realidad se trata de un objeto compacto y muy masivo, cuya atracción se debe a la fuerza de gravedad, la misma fuerza que mantiene a los planetas en órbita alrededor del Sol. Por ello, la atracción gravitacional que un agujero negro ejerce sobre los objetos ubicados a grandes distancias no es muy diferente de aquella que ejercen las estrellas, o cualquier otro objeto con masa. Por ejemplo, si el Sol llegara a convertirse en un agujero negro, la órbita de la Tierra no se vería afectada, ya que al no alterarse la masa, la atracción gravitatoria entre ambos objetos no cambiaría.

Aunque no nos es posible observar un agujero negro, podemos inferir su presencia estudiando su efecto gravitacional sobre su entorno inmediato, así como los procesos físicos que ocurren a consecuencia de este y que pueden llegar a ser muy luminosos. Aquí presentamos una descripción de los distintos componentes que pueden identificarse en un agujero negro y sus alrededores, los cuales están ilustrados en la Figura 5.10 (contenido y figura adaptados de ESO Astronomy).

Cuando la materia experimenta el potencial gravitatorio en la cercanía de un agujero negro, ésta empieza a orbitar en torno a él en un conjunto de círculos de materia difusa, acercándose a él en una espiral, con una velocidad cada vez mayor a medida que la materia se acerca al centro. En este proceso, la materia forma un disco en torno al agujero negro conocido como **disco de acreción**. Por la fuerza de la gravedad y la fricción, las temperaturas en el disco de acreción son extremadamente altas, produciendo altos niveles de energía luminosa. En el caso de muchos agujeros negros supermasivos ubicados en el centro de las galaxias, los discos de acreción pueden llegar a ser los fenómenos más luminosos del universo, emitiendo radiación de las más altas energías: óptica, ultra-violeta, y rayos X (ver caja 5.3.3).

En ocasiones, parte del material del disco es liberado en corrientes estrechas, largas y rápidas, llamadas **jets o chorros**, que pueden llegar a extenderse a escalas gigantescas y son muy luminosos en ondas de radio. Estos chorros emergen desde el agujero negro, generalmente en direcciones opuestas, en muchos casos perpendiculares al plano del disco de acreción. El origen físico de estos chorros, sin embargo, es aún una pregunta abierta en la física extragaláctica.

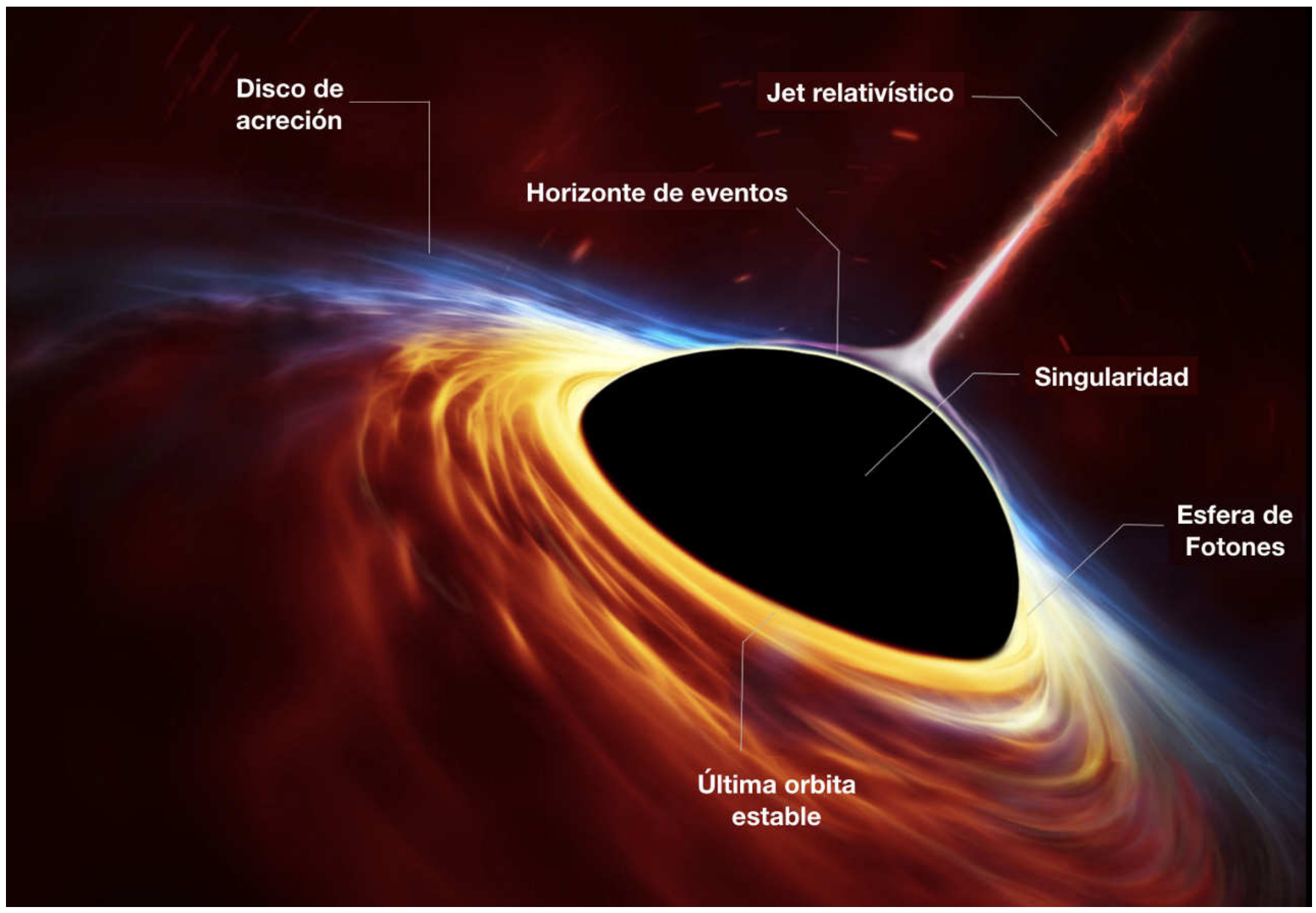

Figura 5.10: Anatomía de un agujero negro. Crédito: Imagen adaptada de ESO, ESA/Hubble, M. Kornmesser.

El borde interior de un disco de acreción es conocido como **la última órbita estable**, ya que a menores distancias del agujero negro la materia ya no puede orbitar de manera estable, por lo que caerá hacia el centro. A una distancia incluso más corta del centro se encuentra la esfera de fotones. La **esfera de fotones** es un lugar donde la gravedad es tan fuerte que la luz, que usualmente viaja en lineas rectas, empieza a viajar en círculos por la gravedad del agujero negro. Si uno viera directamente un agujero negro, vería la luz proveniente de la esfera de fotones como una delgada circunferencia luminosa, y en su interior no vería más que oscuridad, la llamada sombra del agujero negro. Esta 'sombra' es la que se llegó a observar en el agujero negro central de la galaxia M87 con el Telescopio de Horizonte de Sucesos.

Finalmente, el **horizonte de sucesos** es el "punto sin retorno" del agujero negro, del cual nada puede escapar, y toda materia o luz que llegue a cruzarlo caerá inevitablemente hacia el centro. No se trata de una barrera física, sino de una superficie imaginaria que rodea el agujero negro y marca la región en donde la velocidad de escape (ver caja 5.3) es igual a la velocidad de la luz. El radio del horizonte de sucesos que determina el tamaño del agujero negro es equivalente al radio de Schwarzschild y depende directamente de la masa del agujero negro (ver caja 5.3). Debido a la intensa gravedad, la materia se comprime hasta un volumen diminuto, llamado **singularidad** , que tiene esencialmente una densidad infinita. Es probable que las leyes de la física tal como las conocemos se rompan en la singularidad. Ya que ni siquiera la luz puede escapar tras haber cruzado el horizonte de sucesos, es imposible observar u obtener cualquier información de lo que ocurre en este punto. La investigación sobre lo que ocurre dentro de un agujero negro es por lo tanto materia del campo de la física teórica.

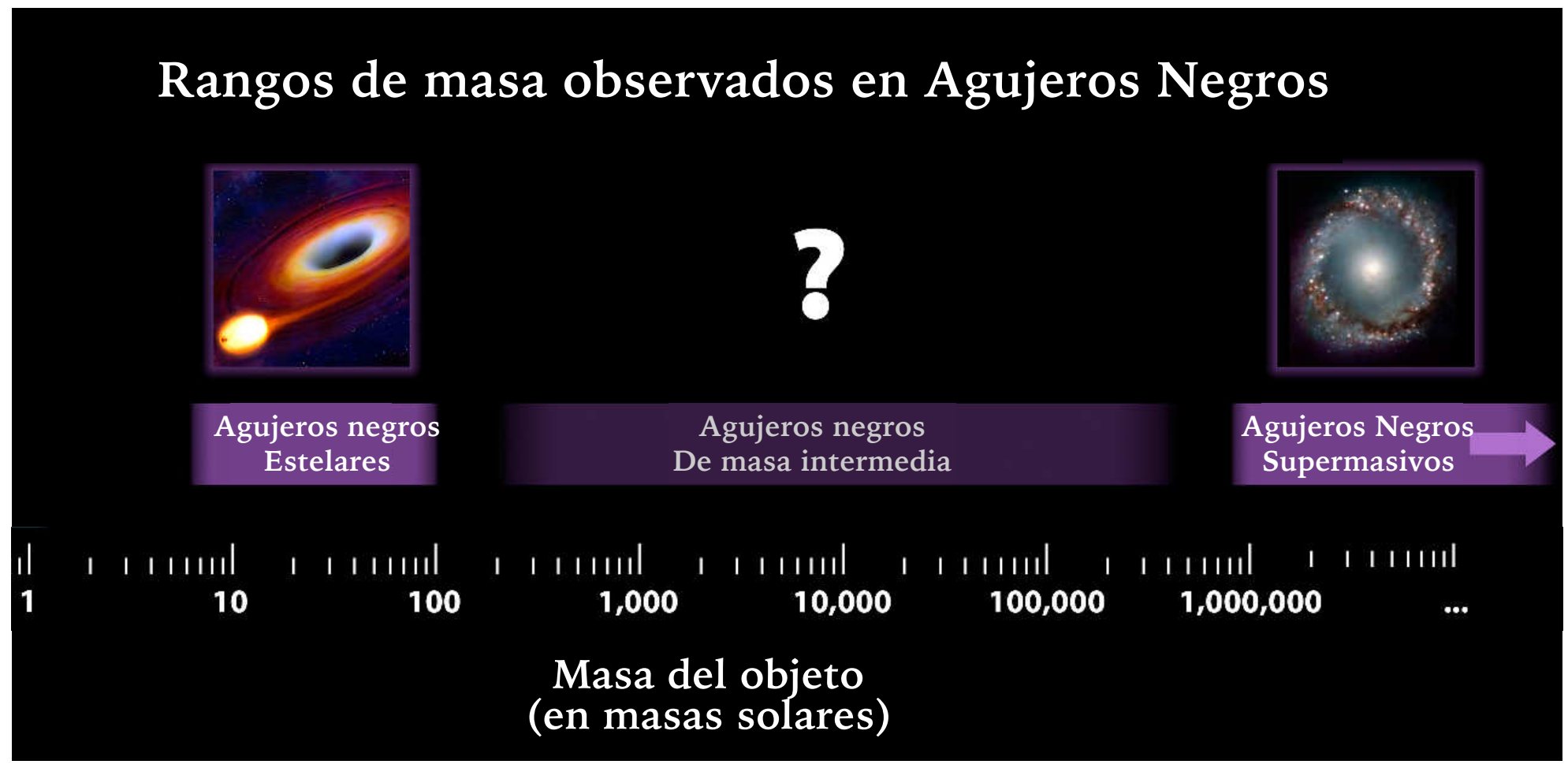

Figura 5.11: Tipos de agujeros negros observados, de acuerdo a su masa [Crédito: adaptado de NASA/JPL-Caltech].

### 5.3.2 Tipos de Agujeros negros

Tradicionalmente, los astrónomos han identificado dos clases básicas de agujeros negros, los que tienen masas de 5 a 20 veces la del Sol, que se denominan agujeros negros de masa estelar, y los que tienen masas de millones a miles de millones de veces la del sol, que son llamados agujeros negros supermasivos (Figura 5.10). La aparente ausencia de agujeros negros de masa intermedia es aún una incógnita en la investigación astronómica actual.

- Los **agujeros negros de masa estelar** se forman cuando una estrella masiva se queda sin combustible, por lo que su propia gravedad hace que colapse hasta un volumen diminuto (ver sección 3.6.2). Se encuentran dispersos por toda la galaxia, en las mismas regiones donde encontramos estrellas, ya que empezaron su vida como estrellas. Algunos de estos comenzaron siendo parte de un sistema estelar binario, y luego de convertirse en agujeros negros continúan orbitando su estrella acompañante. La forma en que el agujero negro afecta a su acompañante y su entorno puede servir como una pista para los astrónomos sobre su presencia.
- Los **agujeros negros supermasivos** se encuentran en el centro de todas las galaxias. Aunque hay diversas teorías, aun no se sabe exactamente cómo se forman, por lo que esta es un área activa de investigación en astronomía. Diversos estudios han demostrado que la masa del agujero negro está correlacionada con la masa de la galaxia, por lo que se cree que debe haber alguna conexión entre la formación del agujero negro y la galaxia.

### 5.3.3 Galaxias de núcleo activo o AGNs

Muchos agujeros negros supermasivos, que residen en los núcleos de las galaxias, son considerados activos, es decir que se encuentran acretando o "tragando" materia con una eficiencia muy alta, por lo tanto están en crecimiento. Estos **núcleos activos, o "AGN"** por sus siglas en inglés, pueden ser tan brillantes por lo que la región central es mucho más luminosa que el resto de la galaxia. La principal fuente de la luminosidad en estos núcleos activos es naturalmente el disco de acreción en torno al agujero negro supermasivo (ver sección 5.3.1).

Existe una gran variedad de AGNs, incluyendo quásares, radio-galaxias, y galaxias

Seyfert entre los más destacados. Los más luminosos entre estos, los quásares, recibieron su nombre (derivado del término “fuente cuasi-estelar”) por su intenso brillo, similar al de una estrella de nuestra Vía Láctea, a pesar de encontrarse a distancias extremadamente lejanas. En efecto, los quásares son las fuentes de luz constante más luminosas en todo el universo.

**Luz 5.3.1: Radiación “invisible” - ondas de radio y rayos-X**

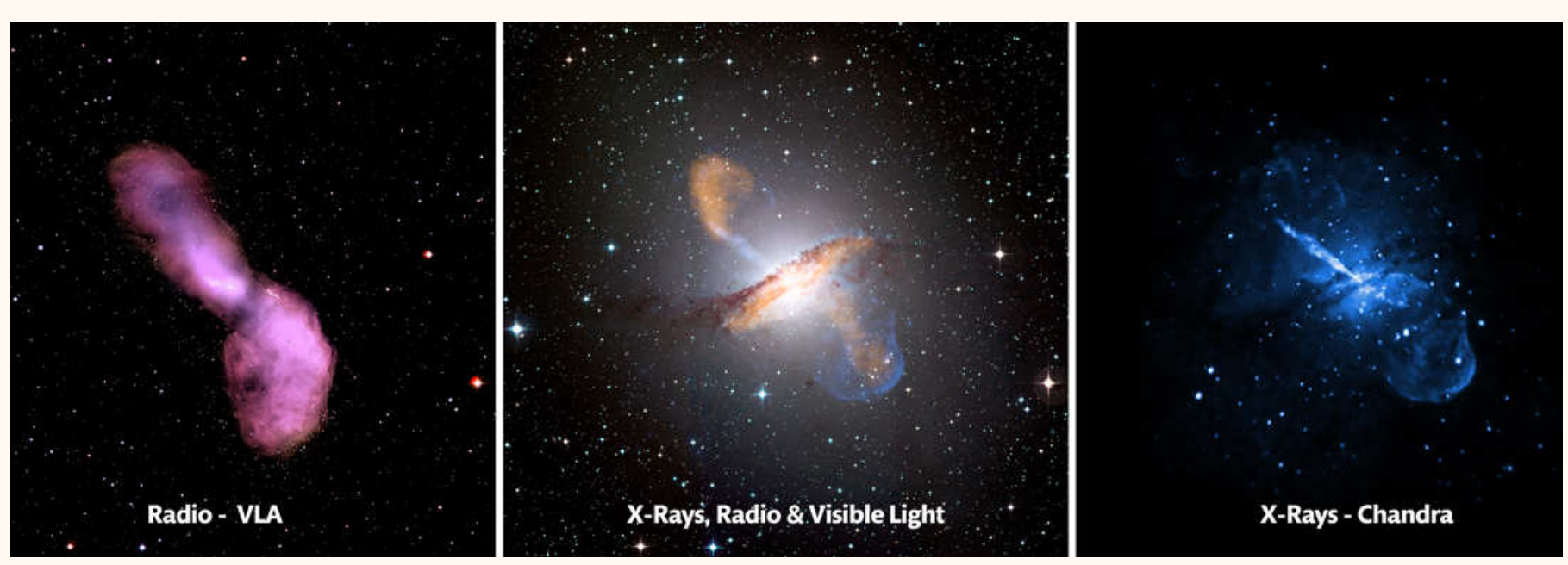

Los núcleos activos de galaxia, o AGNs, son también ejemplos prominentes de radiación “invisible”, adicional a aquella de la galaxia (caja 5.1.2). Un buen ejemplo de este fenómeno multi-onda es el AGN Centaurus A, cuyas fotografías podemos apreciar aquí usando un telescopio de radio (izquierda), un telescopio de rayos X (derecha), o una combinación de estos y un telescopio óptico (centro). Los chorros o jets son estructuras brillantes en ondas de radio que pueden abarcar escalas de tamaño muy superiores al tamaño de la propia galaxia. De forma similar las altas energías que se producen en las inmediaciones de un agujero negro también producen una potente emisión energética en rayos-X.

# 6. Cosmología

La Cosmología es el estudio del origen, naturaleza y evolución de nuestro universo. Pitágoras fue el primero en llamar a los cielos “cosmos”, lo que significa orden. Desde Pitágoras, que sugería que la Tierra, el Sol y otros cuerpos celestes son esferas perfectas y se mueven en círculos perfectos alrededor de un fuego invisible, se ha evolucionado mucho en el estudio del cosmos. Actualmente, **el modelo del *Big Bang*** es la teoría más ampliamente aceptada sobre el origen y la evolución de nuestro universo. Postula que hace aproximadamente 13 mil millones de años, la porción del universo que podemos ver hoy, era microscópica. Desde entonces, se ha expandido de este estado denso y caliente al cosmos vasto y mucho más frío que habitamos actualmente. Podemos ver las señales de esta materia densa y caliente como **la radiación de fondo de microondas cósmica**, ahora muy fría, que todavía impregna el universo y es visible para los detectores de microondas como un brillo uniforme en todo el cielo. En este capítulo exploraremos los últimos descubrimientos en la cosmología, principalmente en el contexto del modelo del Big Bang.

## 6.1 Pilares de la cosmología

La cosmología sienta sus bases en conceptos de física moderna, especialmente en la Teoría de Relatividad General y la física de partículas, junto con hallazgos observacionales que describiremos a continuación.

### 6.1.1 Las leyes del universo

Todas las interacciones en el universo están gobernadas por cuatro fuerzas fundamentales. A grandes escalas, es decir aquellas escalas que podemos percibir con nuestros sentidos (de milímetros a metros) y a escalas cósmicas (de miles de parsecs,

Ilustración a escala logarítmica del universo observable. Crédito: Imagen adaptada de Wikipedia/Pablo Carlos Budassi.

kpc), gobiernan la fuerza electromagnética y la fuerza de gravedad. Por otro lado, la física del reino microscópico del núcleo atómico está determinada por la fuerza fuerte y la fuerza débil.

**Física 6.1.1: Las 4 fuerzas fundamentales del universo**

Las 4 fuerzas fundamentales
Electromagnetismo
Mantiene unido el átomo y las moléculas entre sí.
Partícula Portadora: Fotón
Nuclear Débil
Provoca algunos tipos de desintegraciones radioactivas incluido el decaimiento Beta.
Partícula Portadora: Bosones W y Z
Gravedad
Afecta a todas las partículas y rige el movimiento de los astros.
Partícula Portadora: Gravitón* (hipotética)
Nuclear Fuerte
Mantiene unido protones y neutrones en el núcleo atómico.
Partícula Portadora: Gluón (Existen 8 tipos)
Carlos Pazos
mola SABER

Las cuatro fuerzas fundamentales del universo son:

- El electromagnetismo que controla la interacción entre el núcleo atómico y los electrones y es responsable por la electricidad, el magnetismo y los procesos químicos.
- La gravedad que media entre objetos con masa y energía, es responsable de la dinámica terrestre y de los astros.
- La fuerza nuclear débil que rige el comportamiento del núcleo atómico y es responsable por la radioactividad.
- La fuerza nuclear fuerte que mantiene unidos los neutrones y protones del núcleo.

### 6.1.2 Relatividad General: De Newton a Einstein

Como resulta evidente de los capítulos anteriores, la gravedad, formulada en la física Newtoniana como una fuerza de atracción, tiene un rol protagónico en la física del universo. Sin embargo, nuestro conocimiento sobre la gravedad cambió categóricamente en el siglo XX. En 1915, Albert Einstein publicó su **Teoría General de la Relatividad**, que propuso una nueva teoría de la gravedad. Esta teoría generaliza la ley de la gravitación universal formulada por Isaac Newton, encontrando una descripción que es válida tanto para cuerpos en movimiento uniforme o acelerado, como para cuerpos en reposo. La gravedad de Newton sólo es válida para cuerpos en reposo o que se mueven muy lentamente en comparación con la velocidad de la luz. Recordamos al lector que para la percepción humana en nuestro día a día, la física Newtoniana describe perfectamente el efecto de la gravedad, ya que las velocidades que experimentamos son siempre mucho menores a la velocidad de la luz. Sin embargo, esto cambia cuando estudiamos el Cosmos, ya que uno de los más importantes mensajeros sobre los eventos en el espacio exterior es precisamente la luz que viaja por millones de kilómetros para finalmente atravesar nuestra atmósfera y llegar a nosotros.

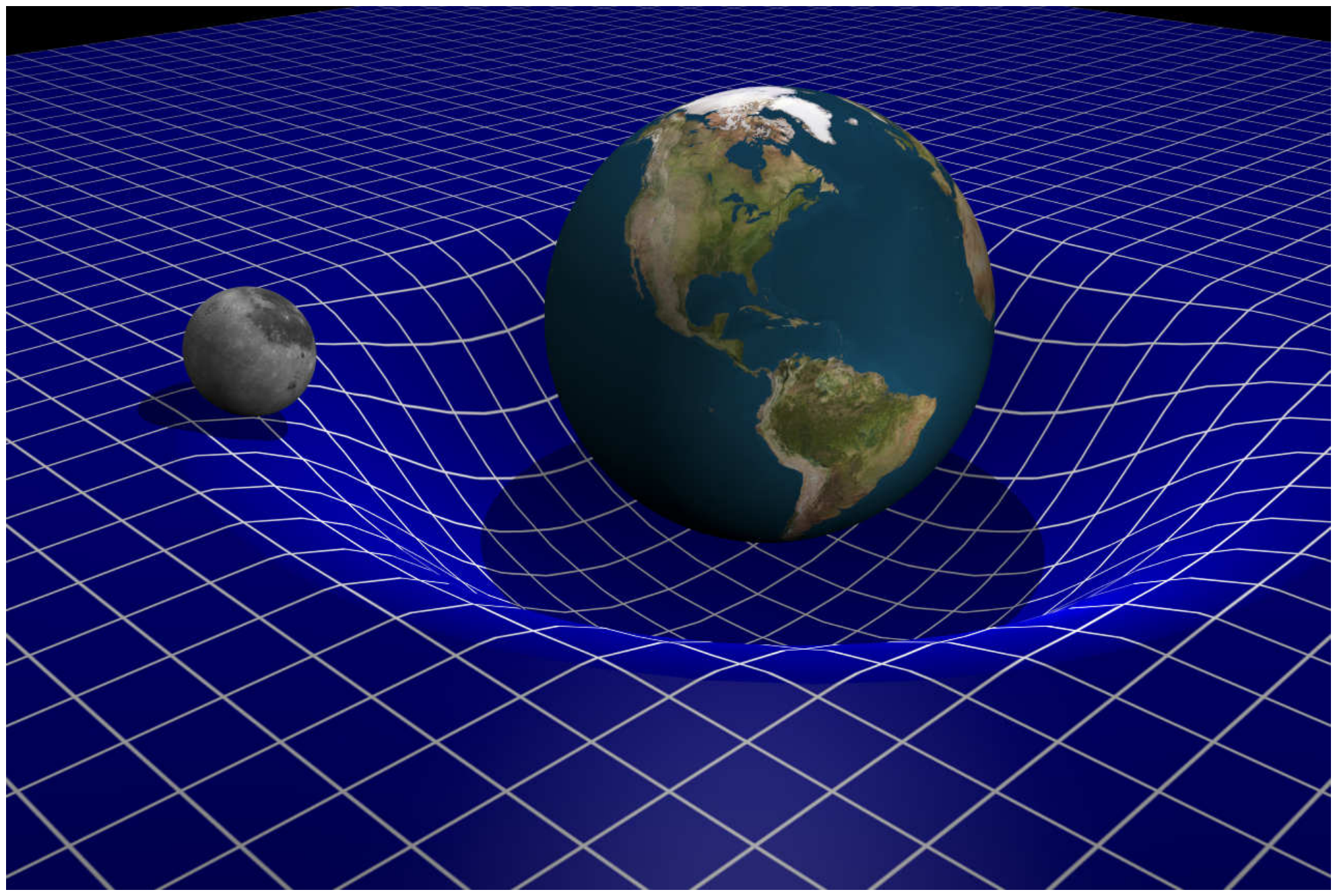

Figura 6.1: Curvatura del espacio-tiempo alrededor de un objeto masivo. En la teoría general de la relatividad de Einstein, el movimiento de la Luna alrededor de la Tierra, puede ser explicado como un simple movimiento uniforme sobre una superficie curva, en este caso las 4 dimensiones del espacio-tiempo. No obstante, en una imagen sólo podemos representar 3 dimensiones de modo ilustrativo [Crédito: `scienceatyourdoorstep.com`].

Un concepto clave de la relatividad general es que la gravedad ya no se describe mediante una fuerza y un "campo" gravitacional, sino que en la descripción de Einstein la gravedad es una distorsión del espacio-tiempo en sí. Para entender la gravedad en la relatividad general es conveniente imaginarse el efecto que tienen la masa (por ejemplo, un globo o una piedra pesada) cuando es colocada sobre una tela elástica estirada, que representa el espacio-tiempo. La deformación de la tela dependerá de la masa colocada, si la masa es pequeña, por ejemplo una pelota de tenis, la deformación será pequeña; y si la masa en mayor, como una piedra, la deformación será mayor. Esta deformación de la estructura del espacio-tiempo es lo que causa que las trayectorias de los objetos cambien cuando se encuentran cerca de un objeto masivo, lo cual observamos como "fuerza de gravedad". Así como en la superficie de la Tierra, las distancias más cortas entre dos puntos son círculos y no líneas rectas, de igual forma las trayectorias más directas entre objetos en el espacio-tiempo, son regidas por la curvatura del espacio-tiempo causadas por la masa y la energía de otros objetos.

El físico John Wheeler lo expresó bien cuando dijo: "La materia le dice al es-pacio cómo curvarse, y el espacio le dice a la materia cómo moverse". En un caso extremo de esta curvatura, los agujeros negros son objetos en el espacio tan masivos y densos que la luz no podía escapar de ellos. Estos fueron predichos por la Teoría de la Relatividad General de Einstein ya en 1916 y fueron recientemente observados directamente por la colaboración del Event Horizon Telescope en el 2019, que junto con otras observaciones indirectas de las últimas décadas, proveen fuerte evidencia de su existencia (ver más detalles sobre agujeros negros en el Capítulo 5.3). Otra predicción de la teoría de la relatividad general es la existencia de **ondas gravitacionales**, que son perturbaciones del espacio tiempo producidas por ejemplo, por la fusión de dos agujeros negros. La primera detección experimental de estas ondas por las colaboraciones LIGO, Virgo y GEO600 en el 2016, fue otra evidencia a favor de la teoría de relatividad general.

**Gravedad 6.1.1: Teoría científica**

¿Será lo mismo una teoría que una ley o un concepto?
Una **teoría** es una explicación extensamente fundamentada de un aspecto del mundo natural que puede incorporar leyes, hipótesis y hechos. La teoría de la gravitación, por ejemplo, explica por qué las manzanas caen de los árboles y los astronautas flotan en el espacio.
Una teoría no solo explica hechos conocidos; también permite a los científicos hacer predicciones de lo que deberían observar si una teoría es cierta. La nueva evidencia debe ser compatible con una teoría. Si no es así, la teoría se refina o se rechaza; es decir, son comprobables. Cuantas más observaciones predice y más hechos explica, más pruebas pasa y más sólida es la teoría.

### 6.1.3 Principio cosmológico: isotropía y homogeneidad

Después de la introducción de la Relatividad General, varios científicos, incluido Einstein, intentaron aplicar la nueva dinámica gravitacional al universo en su conjunto. En ese momento, esto requería una suposición sobre cómo se distribuía la materia en el universo. La suposición más simple que se puede hacer es que si se viera el contenido del universo con un filtro suficientemente amplio, aparecería aproximadamente igual en todas partes y en todas direcciones. Es decir, la materia en el universo es homogénea e isotrópica cuando se promedia a escalas muy grandes. A esto se le llama el **Principio Cosmológico**.

El ser **homogéneo** quiere decir que no se tiene un lugar central o preferido, así nos traslademos a cualquier punto, el universo en promedio se ve igual. La característica de ser **isotrópico** es similar a la anterior, sólo que esta vez en vez de existir una simetría traslacional, ahora es rotacional, es decir el universo se ve igual en todas las direcciones. Esta suposición se está probando continuamente a medida que observamos la distribución de las galaxias a escalas cada vez mayores usando modernos telescopios terrestres y espaciales. La figura 6.2 muestra cuán uniforme es la distribución de las galaxias medidas en una franja de 70° del cielo. Una de las principales razones por la que creemos que el principio cosmológico es válido es que podemos observar una radiación de microondas proveniente del Big Bang, con una misma temperatura muy uniforme en cualquier dirección del cielo. Este hecho apoya firmemente la idea de que la materia primordial del universo era muy hómogenea.

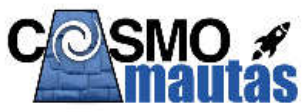

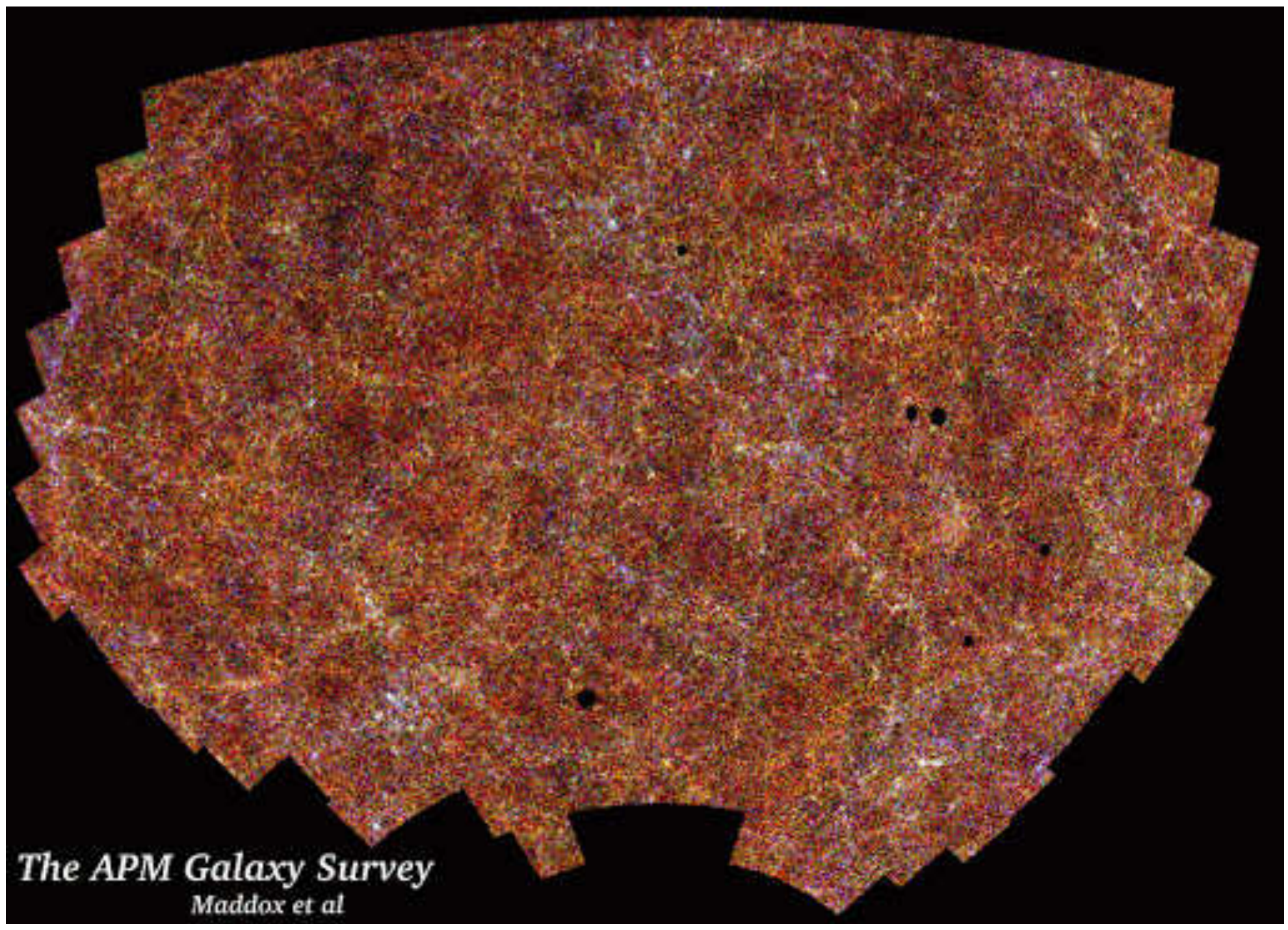

Figura 6.2: Distribución de las galaxias del estudio APM, que demuestra cuán uniforme están distribuidas las galaxias en una franja de 70° del cielo [Crédito: Maddox et al. 1990, APM Survey].

### 6.1.4 El universo observable

Como el lector ha notado en los capítulos anteriores, casi todo lo que sabemos sobre el universo más allá de la Tierra se ha aprendido “a distancia”. No podemos medir directamente una estrella o recoger un poco de su material para estudiarlo en un laboratorio. Solo tenemos como mensajera a la luz que ha viajado hacia nosotros desde distancias efectivamente ilimitadas. Debido a que la luz, o más precisamente, la radiación electromagnética, se produce e interactúa con la materia ordinaria, lo que al principio parece una desventaja insuperable se convierte en una fuente de gran información. Fue cuando comenzamos a comprender la naturaleza de la luz, cómo se produce y cómo interactúa con la materia, que marcó el final de la “Astronomía” como una simple catalogación de los cielos como lo había sido principalmente desde sus inicios, para pasar a ser el inicio del estudio moderno de la “Astrofísica” a finales del siglo XIX, cuando se dio inicio al uso de los espectros para averiguar la composición física de las estrellas.

El universo observable comprende entonces todo el espacio y la materia que podemos observar del universo desde nuestra posición. Cuánto podemos observar del universo está limitado por la mayor distancia que puede recorrer la luz, en base a su velocidad constante. La luz más antigua que llega hasta nosotros es en efecto la luz del fondo cósmico de microondas, definiendo al universo observable como una esfera en torno a nosotros de un radio limitado por el viaje de esta luz primordial. En especifico, la distancia física máxima que se calcula que podemos “observar” es de 14,0 mil millones de parsecs (aproximadamente 45,7 mil millones de años luz) [**55**].

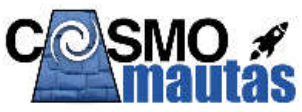

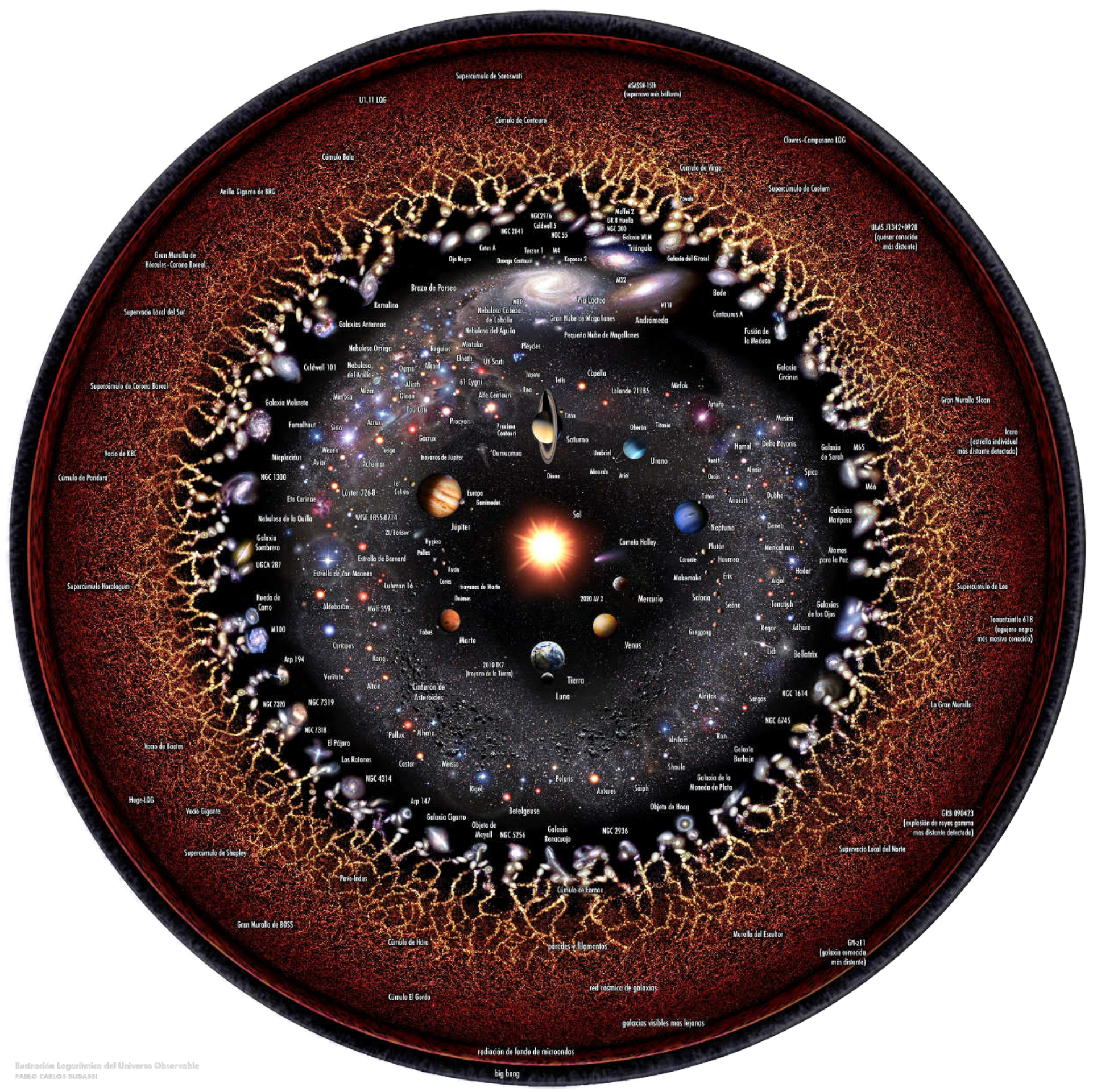

Figura 6.3: Ilustración a escala logarítmica del universo observable con el sistema solar en el centro, los planetas interiores, el cinturón de asteroides, los planetas exteriores, el cinturón de Kuiper, la nube de Oort, Alfa Centauri, el brazo de Perseus, la Vía Láctea, Andrómeda y las galaxias cercanas, la telaraña cósmica de cúmulos galácticos, la radiación de fondo de microondas y el plasma invisible del Big Bang en el borde. Crédito: Wikipedia/Pablo Carlos Budassi.

**Recurso TIC 6.1.1:**

Si desea explorar todas las escalas del universo, desde el tamaño de un electrón, hasta las distancias más extensas del universo, puede hacerlo de forma interactiva en `https://htwins.net/scale2/` [Crédito: Cary Huang/ Matthew Martori].

## 6.2 La expansión del universo y el *redshift*

Las observaciones recientes del universo comprueban que se está expandiendo en todas las direcciones. Así nos pongamos en cualquier punto del universo, la expansión se aprecia de igual forma a nuestro alrededor. Podemos aprender más sobre esto en la actividad 12. También se tiene que esa expansión es acelerada, es decir, la velocidad a la cual las galaxias se alejan de nosotros, es cada vez más rápida.

En 1929 Edwin Hubble descubrió una relación lineal entre la velocidad y la distancia a la que se encuentran distintas galaxias[**56**]. Esta relación es precisamente lo que predice la relatividad general para un universo homogéneo e isotrópico que comienza con un *Big Bang*. Por mucho tiempo se pensó que la expansión del universo debería frenarse hasta empezar a contraerse de nuevo, ya que esta sería la predicción de la teoría en un universo dominado únicamente por materia. No obstante, en 1998 se presentaron los datos experimentales que corroboraron la expansión acelerada del universo, lo cual mereció el Premio Nobel de Física en el 2011.

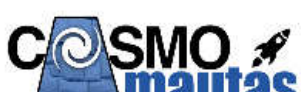

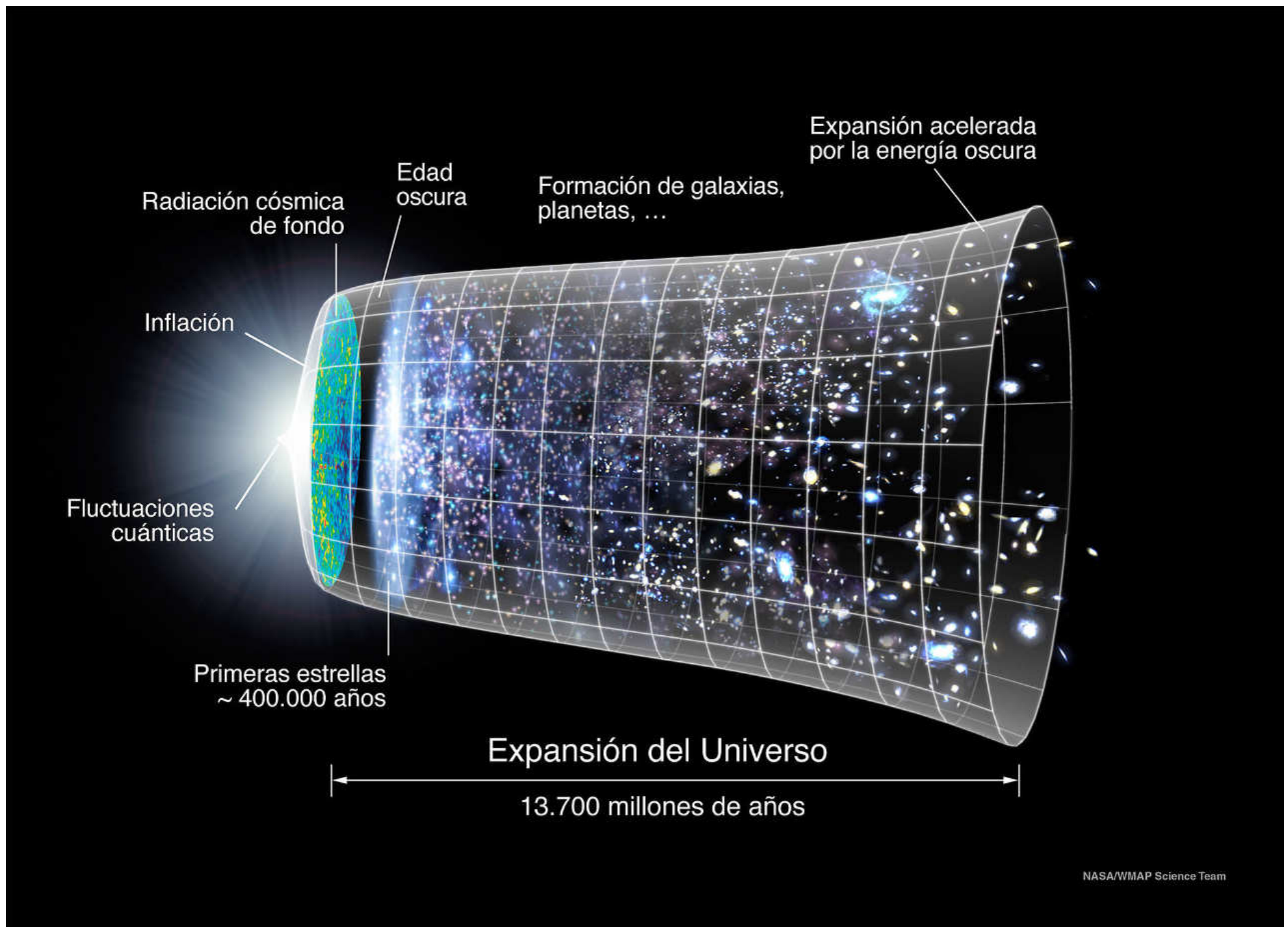

Figura 6.4: Linea de tiempo de la historia del universo desde el Big Bang. [Crédito: Wikipedia/NASA, Ryan Kaldari, adaptado al español: Luis Fernández García].

Un efecto de la expansión del universo es **el corrimiento al rojo cosmológico** (o **redshift** en inglés) que es el incremento en la longitud de onda de la luz a medida que viaja a través del espacio (en expansión). Al expandirse el universo, se expanden las medidas de distancia en sí mismas, por lo cual la longitud de onda de la luz aumenta con respecto a su valor original emitido en reposo. A mayor longitud de onda, la luz se mueve más hacia el color rojo y el infrarrojo, de allí el nombre de este efecto. Aunque el efecto del redshift cosmológico es similar al del efecto Doppler en la luz (ver Caja 4.2.2), el corrimiento al rojo cosmológico resulta de la expansión del espacio mismo y no del alejamiento a cierta velocidad de un cuerpo individual.

## 6.3 Receta para cocinar un universo: Lambda CDM

En este capítulo comenzaremos a conocer poco a poco la composición del universo, y cómo describimos sus componentes en el contexto de un universo en expansión, descrito en cuatro dimensiones (el espacio – 3D + tiempo – 1D) por la teoría de relatividad general de Einstein. El modelo estándar de la cosmología moderna se conoce como el modelo "Lambda CDM" ya que está conformado en un 95 % por dos componentes misteriosos: "Lambda" que representa una constante cosmológica ($\Lambda$) en las ecuaciones de Einstein y que explica la expansión acelerada del universo y “CDM”, que por sus siglas en inglés significa Materia Oscura Fría (Cold Dark Matter) y explica la velocidad de rotación de las galaxias y las estructuras que forman la red cósmica del universo. Sólo un 5 % del universo está formado por la materia que conocemos comúnmente en la Tierra y la que los cósmologos llaman **materia bariónica**. A continuación explicaremos un poco de qué se tratan cada uno de estos ingredientes cosmológicos.

**Recurso TIC 6.3.1:**

Pueden encontrar una simulación de los componentes del universo en: `https://www.youtube.com/watch?v=JAyrpJCC_dw`

### 6.3.1 Materia bariónica

Primero, cuando pensamos sobre la materia que nos rodea, podemos comenzar con las partículas fundamentales como el protón, el electrón y el neutrón (ver caja 3.1.1); luego, en los átomos que se forman debido a las distintas configuraciones de estas partículas, con lo cual ya podemos pensar en los elementos de la tabla periódica. Seguidamente podemos agrupar a los átomos y formar moléculas, con las cuales podemos explicar la composición de las cosas que rodean nuestro día a día, ya sea una mesa hecha de madera (Carbono) o el vidrio de una ventana (Silicio). Lo más interesante ocurre cuando elevamos nuestra mirada hacia el cielo, durante el día podemos observar al Sol, la estrella que mantiene unido a nuestro sistema solar. Durante la noche, nos podemos encontrar con un cielo estrellado que nos deja observar miles de puntos a la vez, muchos de estos pueden ser estrellas o galaxias.

Todo lo mencionado compone lo que conocemos como materia bariónica. La característica principal es que interactúa tanto con la gravedad (por lo que sentimos el peso en la Tierra, la cual gira alrededor del Sol, y el Sol alrededor del centro de la galaxia) y con la fuerza electromagnética, por lo que interactúa con la luz, siendo por lo tanto “visible”, si no en el rango óptico, en otras longitudes de onda.

### 6.3.2 Materia oscura

La mecánica de Newton nos dice que, por ejemplo, los planetas del sistema solar cuanto más alejados estén del Sol, más lento deberán trasladarse y al realizarse las mediciones de las velocidades de traslación de los planetas, vemos que todo marcha bien. Como lo hemos averiguado en la actividad 11, al aplicar la misma mecánica a nuestra galaxia resulta que los objetos celestes que se encuentran cerca al borde “visible” de la Vía Láctea no cumplen con lo predicho; es decir, se mueven más rápido de lo esperado. Esta observación indujo a los científicos a pensar que había algo más allá de lo visible.

**Historia 6.3.1: Vera Rubin y la materia oscura**

Durante los años 70s, la astrónoma Vera Rubin midió las velocidades de rotación de las estrellas en diferentes galaxias, entre ellas la galaxia espiral vecina Andrómeda. Con estas mediciones ella verificó que las estrellas en los bordes de las galaxias se movían a la misma velocidad que las estrellas en el centro, lo cual implicaba, usando cálculos simples basados en la fuerza de gravedad y la centrípeta (ver actividad 11), que la cantidad de materia era mayor en los bordes de la galaxia. Sin embargo, esta materia no era visible, al contrario, a esas distancias del centro había cada vez menos estrellas o nubes de gas. Con esta evidencia, Vera Rubin propuso la existencia de materia "invisible" permeando toda la galaxia, a la cual conocemos como **materia oscura**.

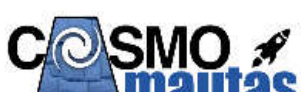

A diferencia de la materia bariónica, la característica principal de la materia oscura es que interactúa solo con la gravedad y no con la luz. Es por esto que no la podemos ver ni detectar con ninguno de nuestros telescopios, y cómo no interactúa electromagnéticamente, es muy difícil o imposible detectarla con sensores electrónicos o experimentos de laboratorio. La naturaleza de esta materia también es "oscura", ya que no se conoce aún qué tipo de partículas (si es que es una partícula) componen a la materia oscura o si se necesitan nuevas leyes de la física para explicar su naturaleza. La materia oscura es uno de los temas más intrigantes de la física moderna, y en particular de las ramas de la astrofísica y la física de partículas.

La materia oscura no solo es un componente importante en las galaxias, si no que es el componente mayoritario, correspondiendo aproximadamente al 90 % de la masa total dentro de una galaxia, mientras que la materia bariónica compone solo el 10 % restante.

### 6.3.3 Energía Oscura

Ahora, hablemos de otro ingrediente misterioso: la **energía oscura**. Como mencionamos anteriormente, Edwin Hubble comprobó que las galaxias se alejan de nosotros con una velocidad proporcional a su distancia, pero si el universo estuviera conformado sólo por materia, la fuerza atractiva de la gravedad debería frenar esta expansión y hacer que el universo colapsara sobre sí mismo en el futuro lejano. En los años 90s, diferentes grupos de astrónomos midieron el *redshift* de galaxias lejanas mediante supernovas tipo Ia que eran visibles a grandes distancias y pueden ser calibradas para obtener su magnitud absoluta. Luego, usando el brillo aparente de estas supernovas, pudieron calcular su distancia y con ello encontrar una relación entre el redshift y la distancia.Los astrónomos estaban pensando encontrar que la velocidad de expansión debería disminuir mientras más lejanas se encontraran esas galaxias, pero encontraron todo lo contrario. Las galaxias más distantes se alejan de nosotros aún más rápidamente. Ya que esto no puede ser causado por un efecto de la gravedad (que es atractiva solamente), por el momento, la única explicación que tenemos para este fenómeno, es asignarle una energía invisible al espacio, que acelera la expansión del universo y a la que usualmente se llama "energía oscura".

### 6.3.4 Distribución de los ingredientes

Ya que hemos hablado de los ingredientes para poder cocinar un universo, ahora tenemos que entender cuáles son las cantidades relativas contenidas en nuestro Universo y cómo podemos medirlas. Hacer un cálculo de la densidad total de materia en el universo basándonos en observaciones directas es una tarea prácticamente imposible, ya que esto requeriría observaciones infinitamente profundas de telescopios para detectar la materia bariónica. Incluso así, estas observaciones no detectarían la mayor parte de la materia, que es la materia oscura, ni medirían la energía oscura que gobierna la expansión acelerada del universo.

Afortunadamente, los modelos teóricos cosmológicos, construidos en base a los pilares mencionados anteriormente (ver sección 6.1), parecen brindar una descripción bastante detallada de lo que observamos, lo que nos permite usar estos modelos matemáticos y estas observaciones para medir indirectamente el contenido del universo. En particular, las observaciones principales para este objetivo son aquellas de la luz más antigua y lejana del universo, el Fondo Cósmico de Microondas (CMB, por sus siglas en inglés). Esta luz fue producida 380 000 años despúes del *Big Bang* y lleva en sí la información acerca de las fluctuaciones de materia y energía presentes en el universo de esa época, un universo que estaba formado por un plasma caliente de protones, neutrones, electrones y fotones fuertemente acoplados. El universo al expandirse, también se fue enfriando y cuando los núcleos atómicos atraparon a los electrones en átomos neutrales, los fotones (las partículas de la luz) fueron liberados, cargando en sus pequeñas fluctuaciones de temperatura toda la información de las interacciones físicas de ese instante. La intensidad de estas fluctuaciones, sus tamaños y la distribución de estas, combinados con la descripción matemática del universo en expansión y el modelo estándar de partículas elementales, nos permiten medir las cantidades precisas de materia y energía en el cosmos.

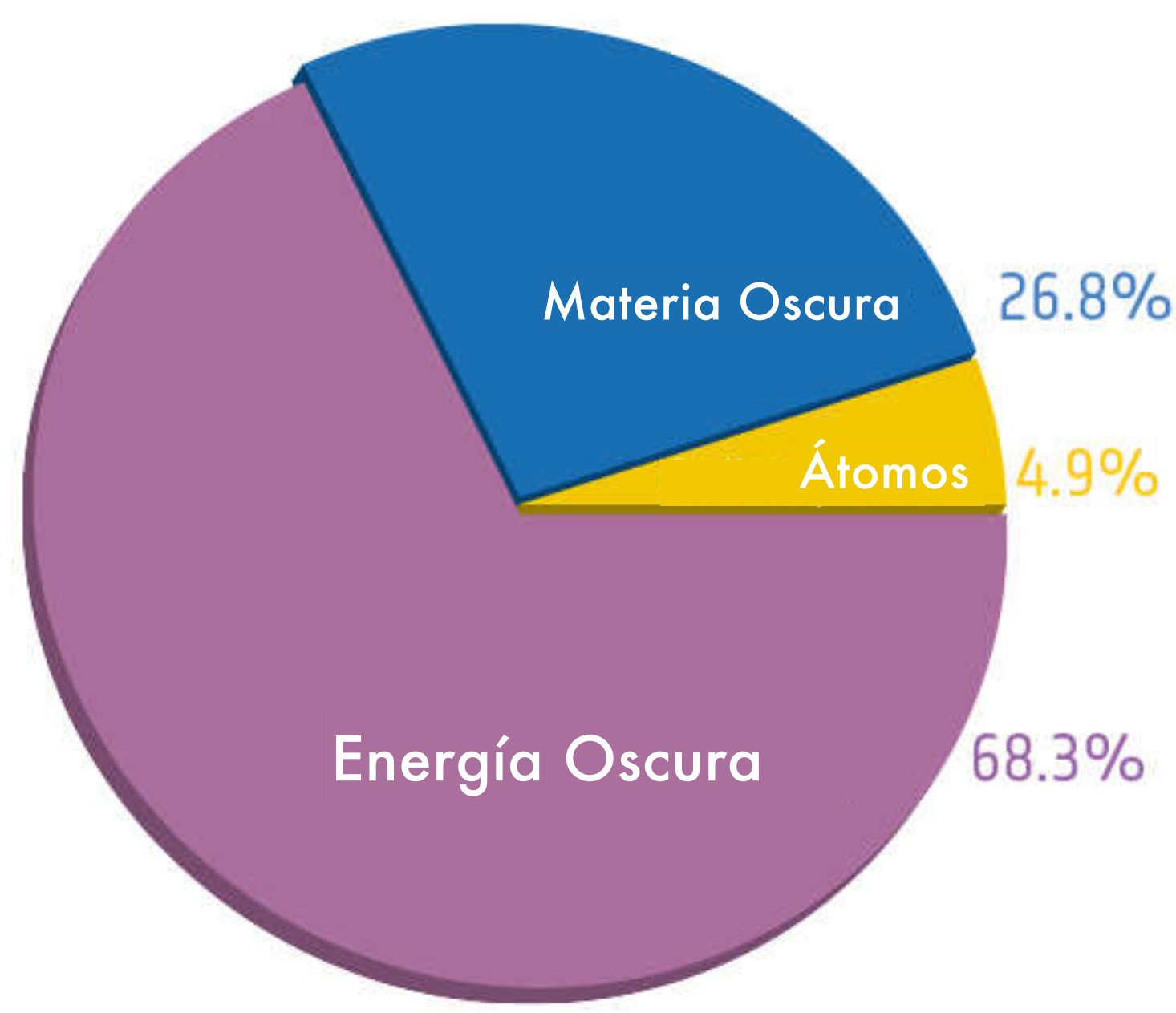

Figura 6.5: Composición del universo según las mediciones de las fluctuaciones de temperatura del fondo cósmico de microondas (CMB) por el observatorio Planck [Crédito: Imagen adaptada de ESA].

El *WMAP (Wilkinson Microwave Anisotropy Probe)* fue una sonda espacial de la NASA que tuvo como una de sus misiones medir las diferencias de temperatura que se observan en la radiación de fondo de microondas (CMB). Este experimento, efectuado entre el 2003 y el 2010, nos indicó que el universo está constituido un 71,4 % por energía oscura, un 24 % de materia oscura y 4,6 % de materia bariónica. Estas mediciones fueron mejoradas más recientemente, con datos publicados entre el 2013 y el 2018, por el observatorio espacial *Planck*, un satélite de la ESA (agencia espacial europea), que construyó un mapa más detallado del CMB. Estas nuevas mediciones nos brindaron la siguiente receta cósmica: 68.3 % de energía oscura, un 22.7 % de materia oscura y 4,9 % de materia bariónica. ¡Eso quiere decir que solo podemos ver (o detectar por medio de la luz) menos del 5 % de nuestro universo! Estas mediciones son en general bastante precisas, con márgenes de error en estos parámetros del 1 % aproximadamente, por lo que en la literatura se dice que actualmente nos encontramos en la era de “la cosmología de precisión”.

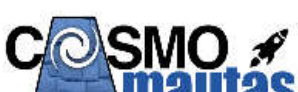

## 6.4 Viajemos desde el inicio del universo

### 6.4.1 Calendario cósmico: La edad del universo

Para poder comprender las etapas del universo, es útil compararlas con las escalas de tiempo de un calendario anual. Es decir, dividiremos la edad del universo en 12 meses. Sabemos que el universo tiene una edad de aproximadamente 13,8 mil millones de años, por lo que si lo dividimos en 12 meses, cada mes tendría una duración 1 000 millones de años, cada día 38 millones de años, cada hora 1,5 millones de años, cada minuto 26 mil años y cada segundo 500 años.

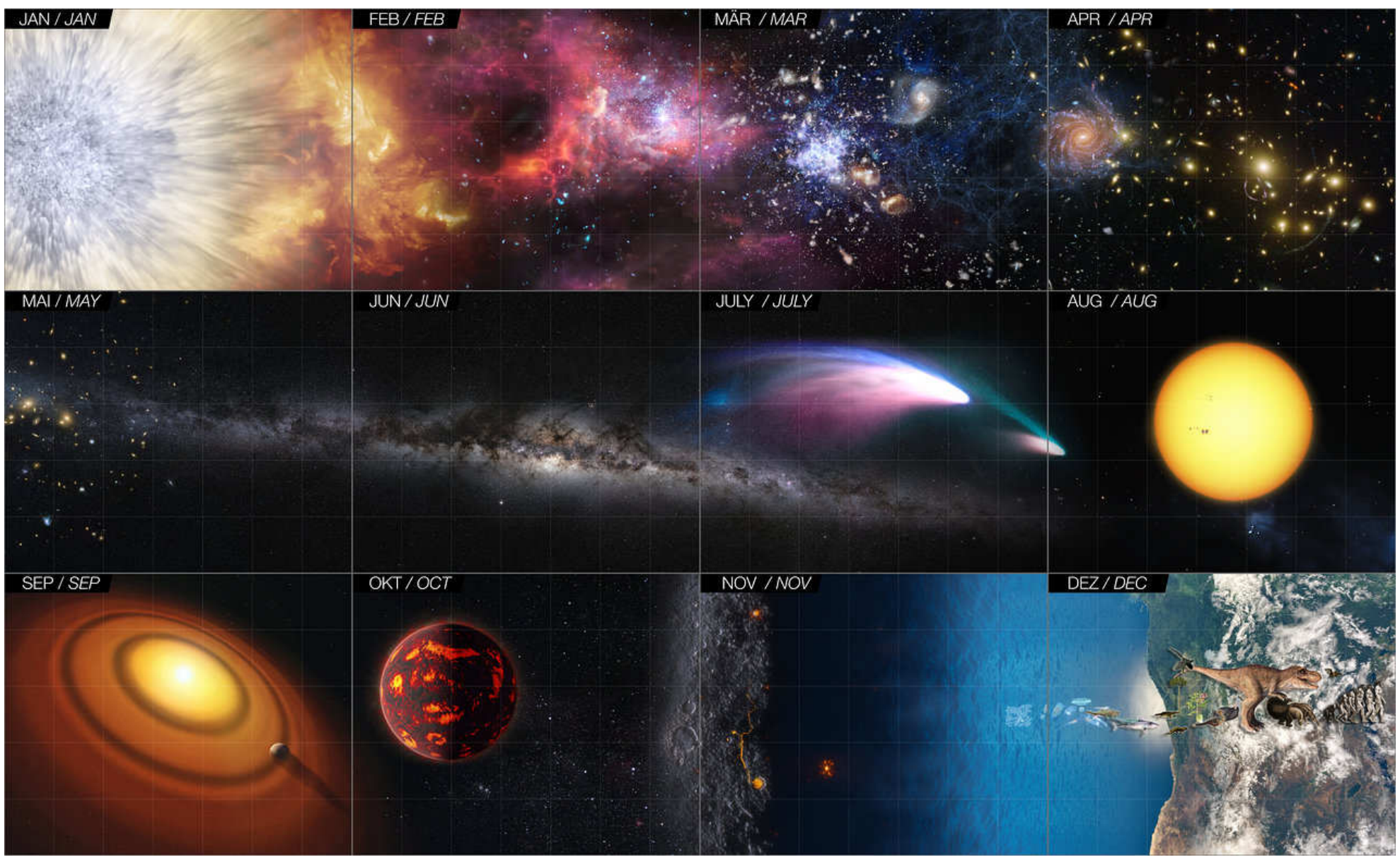

Figura 6.6: Calendario Cósmico [Crédito: ESO/M. Kornmesser].

### 6.4.2 Teoría del Big Bang

Comenzaremos haciéndonos las siguientes preguntas: ¿Cómo fue creado nuestro universo? ¿Existe el centro del universo? ¿Y qué será de él dentro de muchos años? El 1 de enero a las 00:00 horas ocurrió el Big Bang, y el universo estuvo sumergido en total oscuridad por 200 millones de años; es decir, entre 5 a 6 días de nuestro calendario cósmico. La hipótesis del Big Bang establece que toda la materia actual y pasada del universo nació al mismo tiempo, hace aproximadamente 13 800 millones de años (Figura 6.4). En este momento, toda la materia se compacta en un punto muy pequeño con una densidad infinita y un calor intenso formando lo que llamamos una *singularidad*. De repente, la singularidad comenzó a expandirse y éste fue el inicio del universo tal como lo conocemos. A partir del Big Bang, nos adentraremos en un viaje cósmico para descubrir cómo fueron apareciendo los distintos elementos que hoy en día podemos reconocer. ¡Todos a bordo!

### 6.4.3 Las primeras partículas

Luego del *Big Bang*, el universo era extremadamente caliente y denso. Mientras se iba enfriando, las condiciones eran óptimas para que los bloques de construcción de la materia empiecen a formarse: los quarks y electrones. Unas millonésimas de segundo después, se comenzaron a producir los protones y neutrones. Luego de minutos, estos comenzaron a combinarse y formar núcleos. Luego de 380 000 años los electrones comenzaron a ser atrapados en órbitas alrededor de estos núcleos para poder dar paso a los primeros átomos[**56**]. Según nuestro calendario cósmico, los átomos aparecieron luego de aproximadamente 14 minutos. ¡Aún seguimos en enero!

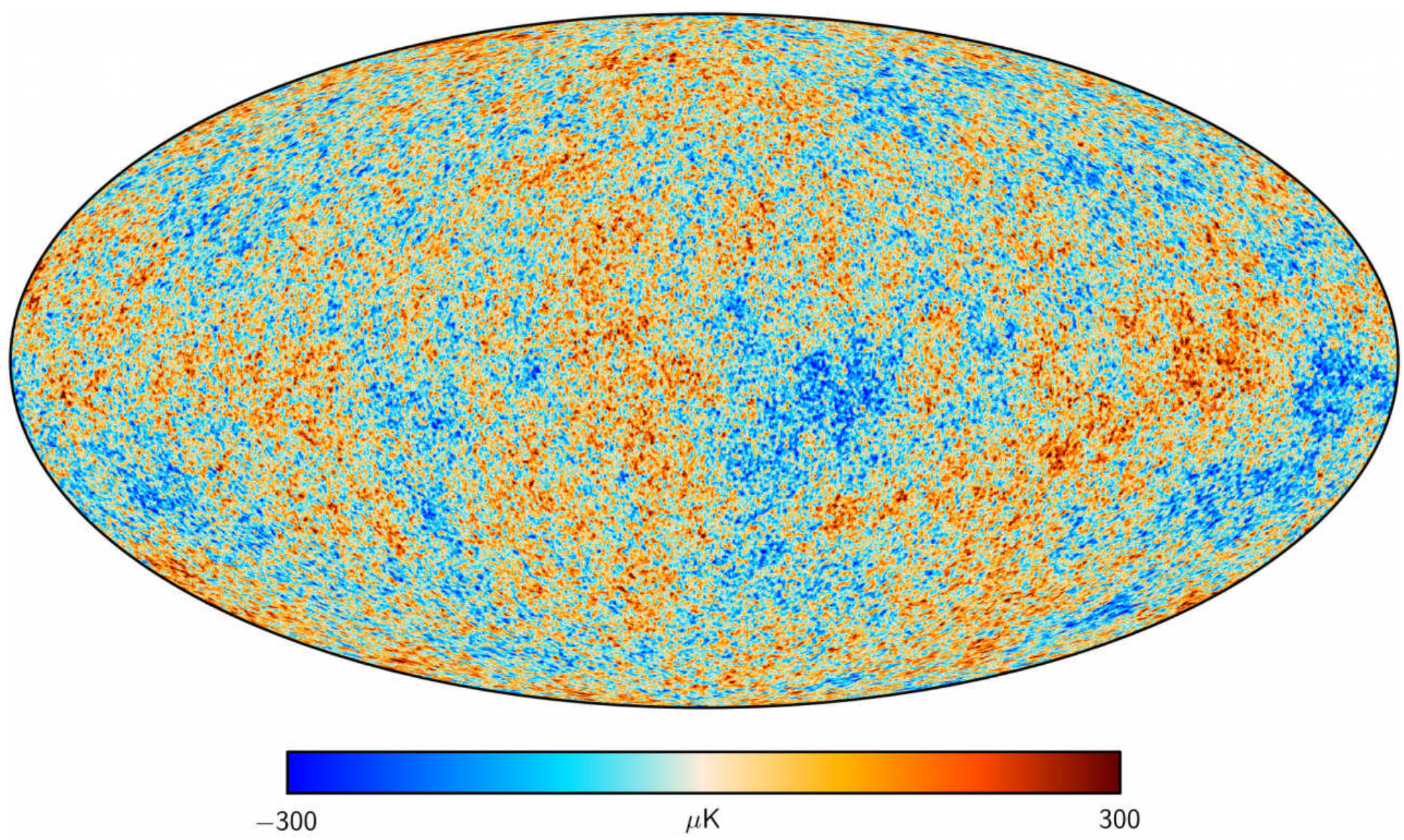

Figura 6.7: El Fondo de Radiación Cósmica de Microondas (CMB, por sus siglas en inglés) mapeado por la misión europea *Planck*. La temperatura actual del CMB es de aproximadamente 2.7 Kelvin (2.7 grados encima del cero absoluto). Esta imagen muestra las fluctuaciones sobre esa temperatura uniforme, que son del orden de micro-Kelvin. En azul se muestran las zonas más frías y en rojo, las zonas más calientes. El tamaño y la distribución específica de estas “manchas” de temperatura nos da información muy valiosa sobre los componentes primordiales del universo [Crédito: ESA].

### 6.4.4 La luz más antigua: Fondo de Radiación Cósmica

Cuando los electrones y los protones andaban libremente por el universo, llenándolo en su totalidad, los fotones no podían viajar largas distancias sin antes chocar e interactuar con una de estas partículas cargadas, por lo tanto el universo era opaco a la vista. No fue sino hasta que los electrones comenzaban a caer en las órbitas de los núcleos, formando átomos neutrales, que los fotones pudieron viajar libremente por el universo. Estos fotones nos llegan hasta hoy de todas las direcciones del cielo y conforman lo que se conoce como el Fondo de Radiación Cósmica o CMB. Luego de este proceso de recombinación de los electrones, el universo entró en lo que se denomina “la edad oscura”. Esto debido a que el universo aún no contenía estrellas ni galaxias que pudieran producir nueva luz, sino que estaba formado básicamente de sólo hidrógeno neutro y fotones libres a altas temperaturas.

### 6.4.5 Las primeras galaxias, las primeras estrellas

Una vez formados los átomos, estos fueron en principio hidrógeno, helio, y cantidades mínimas de litio, los cuales sabemos que son los componentes principales de las estrellas. La acumulación de estos elementos en protogalaxias de aproximadamente 30 a 100 años luz de largo conlleva a la formación de las primeras estrellas. Sin embargo, este proceso se dio luego de aproximadamente 150 a 250 millones de años [**56**]. Es decir, 6 días de nuestro calendario cósmico. A estas primeras estrellas se les denomina “Estrellas de primera generación”. Luego de mil millones de años, ya estamos comenzando el mes de febrero de nuestro calendario cósmico, estas protogalaxias comenzaron a unirse y formar las galaxias que hoy en día podemos ver en el cielo con nuestros telescopios.

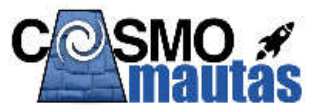

# II Aprendizaje por Indagación (API)

# 7. API 1: La Tierra y el Sol

## Orientándonos (8 min)

Para empezar nuestra exploración del universo tenemos que empezar por nuestro punto de partida: nuestro planeta Tierra. Observando muy atentamente a nuestro alrededor podemos averiguar algunas cosas sobre la forma, el tamaño y los movimientos que tiene nuestro planeta respecto a otros objetos celestes, como nuestra estrella, el Sol, y el resto del universo.

Por ejemplo, ¿qué sabemos respecto al Sol? Sabemos que aparece aproximadamente por el este y se esconde por el oeste. Eso podría hacer pensar que el Sol gira en torno a la Tierra, pero sabemos que no es así, que es el movimiento de la Tierra en torno a su eje lo que genera el "aparente" desplazamiento tanto del Sol y de todo el cielo.

¿Qué sabemos respecto a la Tierra? Todo lo que podemos ver a nuestro alrededor, es muy pequeño en comparación con el tamaño de la Tierra, es difícil sacar conclusiones generales sobre la forma de nuestro planeta. Podemos saber qué tan grande es nuestra casa, porque miramos alrededor y tenemos una idea del espacio que ocupa, o qué tan grande es nuestro barrio, porque podemos desplazarnos en un paseo por todos sus límites y cuantificar el área que este tiene. Saber el tamaño de toda la región o del país, y aún más, de toda la Tierra, ya no es posible de esta forma, pues son mucho más grandes, y por ello tenemos que usar otros métodos como satélites para tener una visión externa de nuestro planeta. Esto lo podemos lograr, por ejemplo, usando Google Maps.

Sin embargo, existen otras formas de averiguar el tamaño y la forma de nuestro planeta, ¡tan solo observando nuestro cielo! En esto consiste nuestro experimento de hoy.

**Unidades de aprendizaje incluidas en este API**
Matemática: Promedio, mediana, desviación estándar, margen de error
Matemática: Trigonometría (ángulos, circunferencia, sen, cos, tan)
Matemática: Trigonometría (ángulos internos, externos y alternos)
Matemática: Regla de tres
Geografía: Tierra y Sol
Ciencias: Testear modelos

## Materiales

- Mapa plano de la Tierra
- Globo terráqueo (opcional)
- Pegatipo o plastilina
- 4 palitos de diente
- Gnomon de 60 cm (puede ser reemplazado por cualquier palo que pueda ser posicionado verticalmente, como un recogedor)
- Papelógrafo y lápiz
- Regla
- Linterna

## Objetivos y preguntas (2 min)

El objetivo de este experimento es usar nuestras observaciones para responder dos preguntas:

A ¿Qué forma tiene la Tierra? Aunque ya sabemos la respuesta, ¡queremos comprobarlo!

B ¿Qué tan grande es la Tierra? ¿Cómo podemos medir la circunferencia o el radio de la Tierra?

## Hipótesis y diseño del experimento (20 min) Nivel 1

Ahora vamos a ir en grupos para discutir cómo podemos responder estas preguntas, observando a nuestro alrededor.

**A. ¿Qué forma tiene la Tierra?** Sigamos los siguientes puntos para la discusión:

- Vamos a testear dos hipótesis: 1) La Tierra es redonda y 2) la Tierra es plana.

- Discutamos ideas sobre cómo comprobar una de las dos hipótesis (nos ayudaremos con un globo terráqueo y un planisferio para representar dos modelos).

- Usando cuatro palitos de dientes y pegatipo, posicionemos dos palitos de forma perpendicular a la superficie del globo y dos palitos sobre el planisferio. Usemos una linterna para simular al Sol. Experimentemos iluminando los palitos con la linterna desde diferentes ángulos (de preferencia, ubicarse en una zona oscura que permita distinguir las sombres producidas). Luego, discutan las siguientes preguntas: ¿Qué diferencias se pueden observar entre las sombras de los palitos sobre el globo y las sombras sobre el planisferio? ¿Cómo nos puede ayudar la luz a comprobar que la Tierra es redonda?

Tener en consideración al representar con una linterna la luz del Sol: Por la inmensa distancia entre la Tierra y el Sol, la luz del Sol cae de forma paralela sobre todos los puntos de la Tierra. [Crédito: WYP Eratosthenes Project].

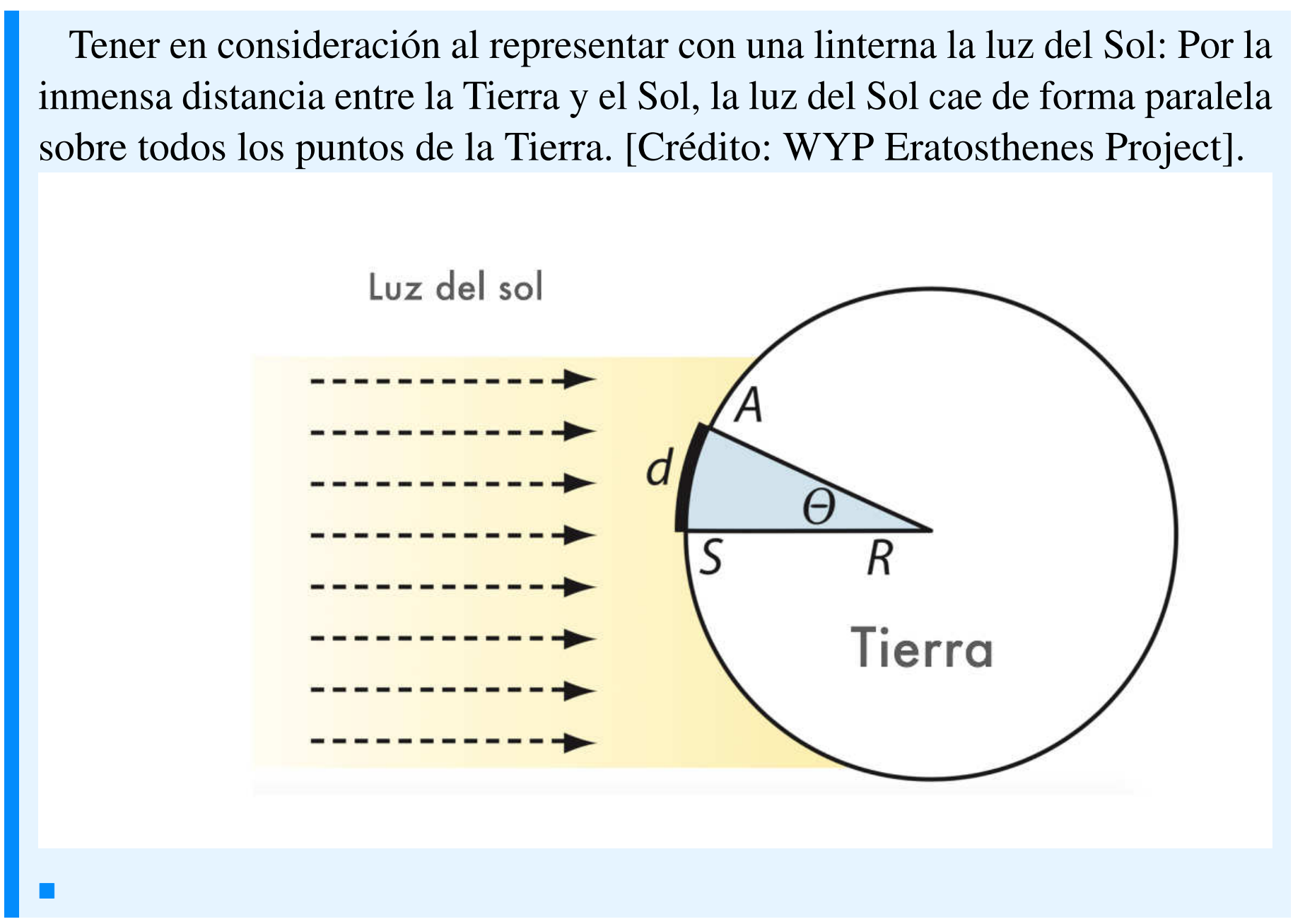

**B. ¿Qué tan grande es la Tierra? ¿Cómo podemos medir la circunferencia o el radio de la Tierra?** Ahora que hemos determinado que modelo representa mejor la forma de la Tierra, veamos como podemos medir su circunferencia. Sigan los siguientes puntos:

- Imaginemos a la Tierra (globo terráqueo ) y dos gnomon (palitos de diente) que son colocados de forma perpendicular a su superficie, tal como se muestra en la imagen.

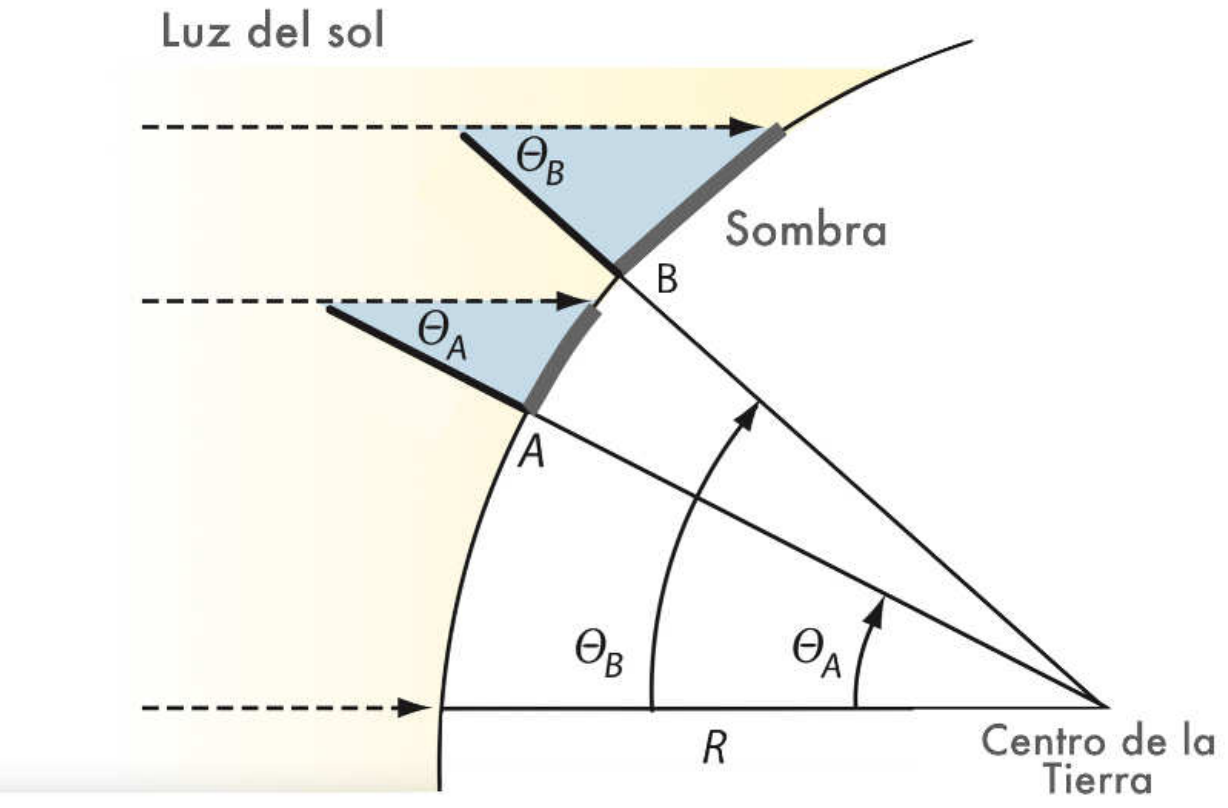

Figura 7.1: Dos gnomon colocados perpendicularmente sobre la Tierra.

- Al analizar la imagen, notamos que para medir la circunferencia, necesitamos repasar algunos conceptos geométricos simples: ángulo, radio, arco, circunferencia. Luego, respondamos las siguientes preguntas: ¿Cuántos grados hay en un círculo? ¿Qué es el arco de un círculo? ¿Cómo podemos calcular la circunferencia del círculo si sabemos cual es el arco de un respectivo ángulo? ¿Cómo podemos calcular el radio del círculo si sabemos cuál es la circunferencia?

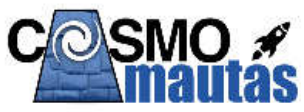

**Matemática 7.0.1: Radio de la circunferencia**

El radio ($r$), el ángulo ($\theta$), y la longitud de arco ($S$) en una circunferencia, están relacionados mediante la siguiente ecuación, por lo que, si conocemos dos de las cantidades, podemos calcular la tercera.

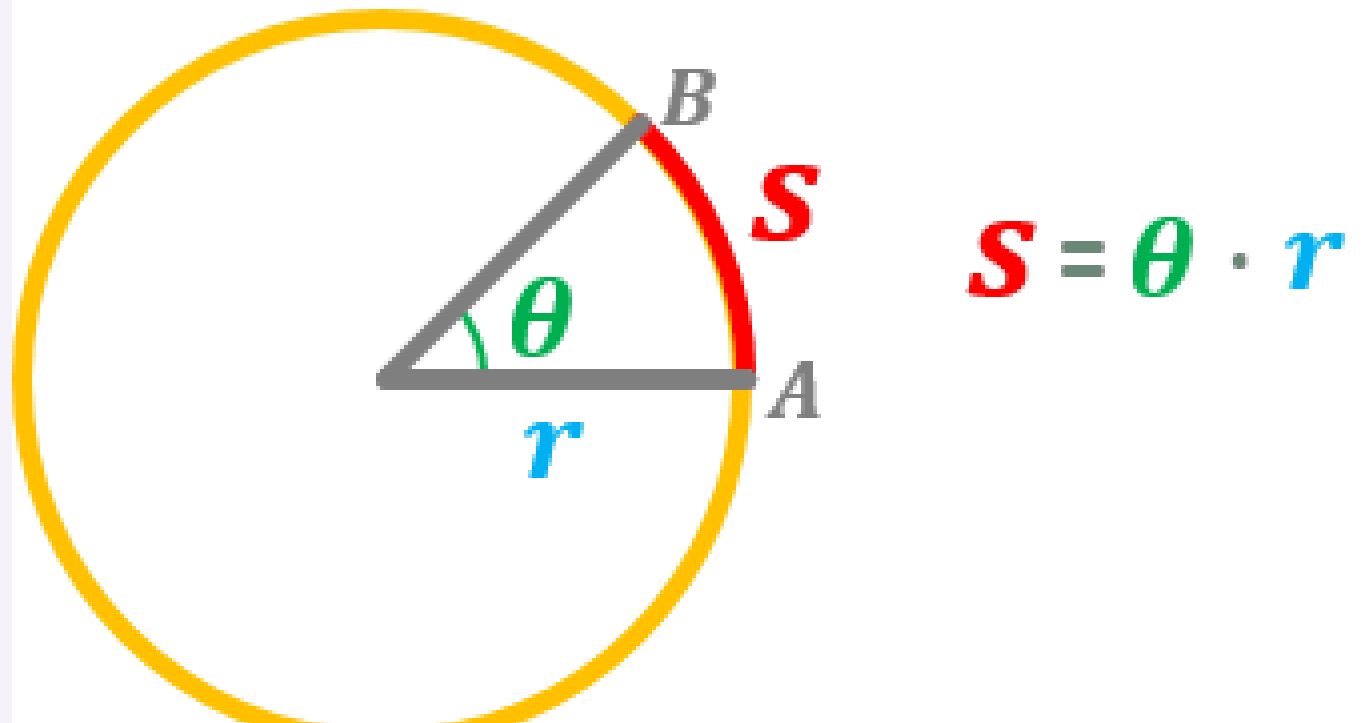

Es importante mencionar que:

- El ángulo debe estar expresado en radianes (rad).
- 1 vuelta $<> 360^\circ <> 2\pi$ rad
- Cuando el ángulo es muy pequeño, el arco $S$ se puede aproximar al segmento que une el punto $A$ y $B$.

- Ahora extrapolemos estos conceptos a la circunferencia de la Tierra. Usen el globo terráqueo para tener una imagen concreta del modelo y respondan las siguientes preguntas:
  - Si tenemos dos puntos sobre un meridiano del globo terráqueo, ¿a cuál de los conceptos geométricos equivale la distancia entre los dos puntos?
  - ¿Cómo podríamos medir el ángulo entre entre estos dos puntos A y B? Este ángulo se denomina distancia angular entre dos puntos.
  - Una vez que tenemos el ángulo y la distancia entre A y B, ¿cómo podemos calcular la circunferencia y el radio de la Tierra con una simple regla de tres?
  - Planifiquen juntos la medición y escriban el plan en un *Google Doc*.
- Ahora, luego de discutir las respuestas, haremos la medición.

## Experimento (35 min) Nivel 2

- Extendamos el papelógrafo sobre una superficie plana y al aire libre. Luego, coloquemos el palo o gnomon, de forma perpendicular (90°) a la Tierra. Si el suelo presenta desniveles, para encontrar el ángulo correcto, podemos utilizar un nivel o una plomada. El ángulo que forman la plomada y una línea vertical será siempre cero, por lo que la plomada caerá de forma perpendicular hacia la superficie.

- Midamos la altura H del gnomon, considerando la altura de la base, y lo anotamos.

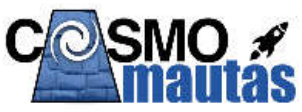

| Hora | Largo de la sombra (cm) | Largo del palo (cm) |
|---|---|---|
| 11:25 h. | | |
| 11:30 h. | | |
| 11:35 h. | | |
| 11:40 h. | | |
| 11:45 h. | | |
| 11:50 h. | | |
| 12:00 h. | | |
| 12:05 h. | | |
| 12:10 h. | | |

Cuadro 7.1: Medición del largo de la sombra del gnomón a diferentes momentos del día.

- Realizamos mediciones de la sombra del gnomon en las horas indicadas en el cuadro. Es importante marcar la longitud de la sombra en el papelógrafo y escribir la hora en la que se realizó cada medición al costado de la marca. Iniciamos a las 11:25 h., realizamos las marcas y anotaciones, y repetimos los pasos cada 5 minutos hasta completar todas las horas indicadas en el cuadro.

## Análisis e interpretación (20 min)

- Cada uno elija el valor más corto que ha encontrado para el largo de la sombra.

- Use el largo del palo, el largo de la sombra y calcule el ángulo $\Theta$, siguiendo el siguiente diagrama:

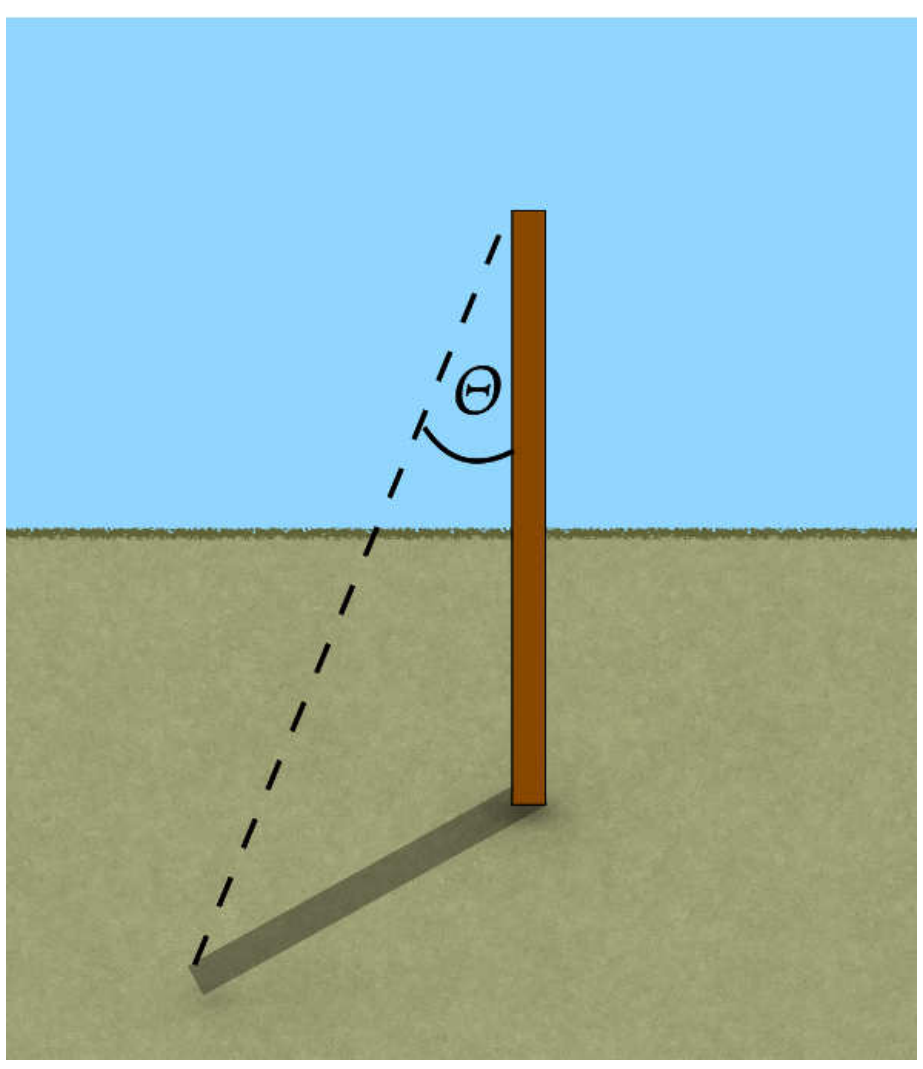

Al elegir la sombra más corta, nos aseguramos de que este ángulo sea aquel en el que caen los rayos solares en el momento en el que el Sol está en su punto más alto del día (llamado el zenith). En los días de los equinoccios, en el Ecuador el zenith está exactamente a 90 grados, y por lo tanto no hay sombra. Entonces, si midiéramos este ángulo en los equinoccios, este ángulo también equivaldría a la distancia angular entre nuestra posición y el Ecuador, es decir nuestra latitud [Crédito: adaptado de WYP2005 Eratosthenes project].

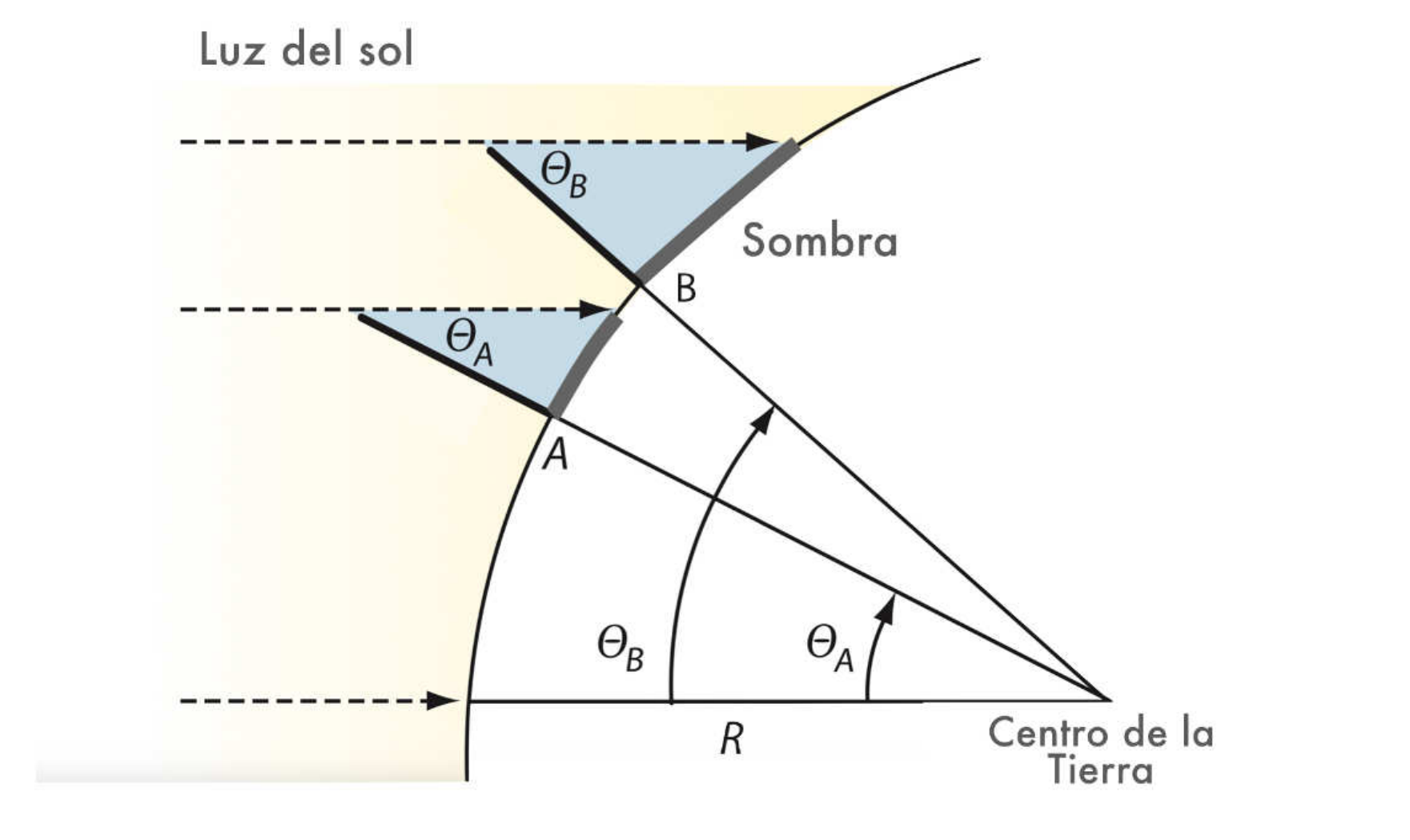

En el resto de los días para medir esta distancia angular requerimos comparar con las mediciones en otras latitudes. Para ello necesitamos trabajar con nuestros colegas de otras regiones.

### Matemática 7.0.2: Arco tangente

Si conocemos las longitudes de los catetos ($m$ y $H$) de un triángulo rectángulo, podemos hallar cualquiera de los ángulos agudos ($a$ y $b$) utilizando la función **tan** y **arctan**. Calculemos esta relación para hallar el ángulo $a$.

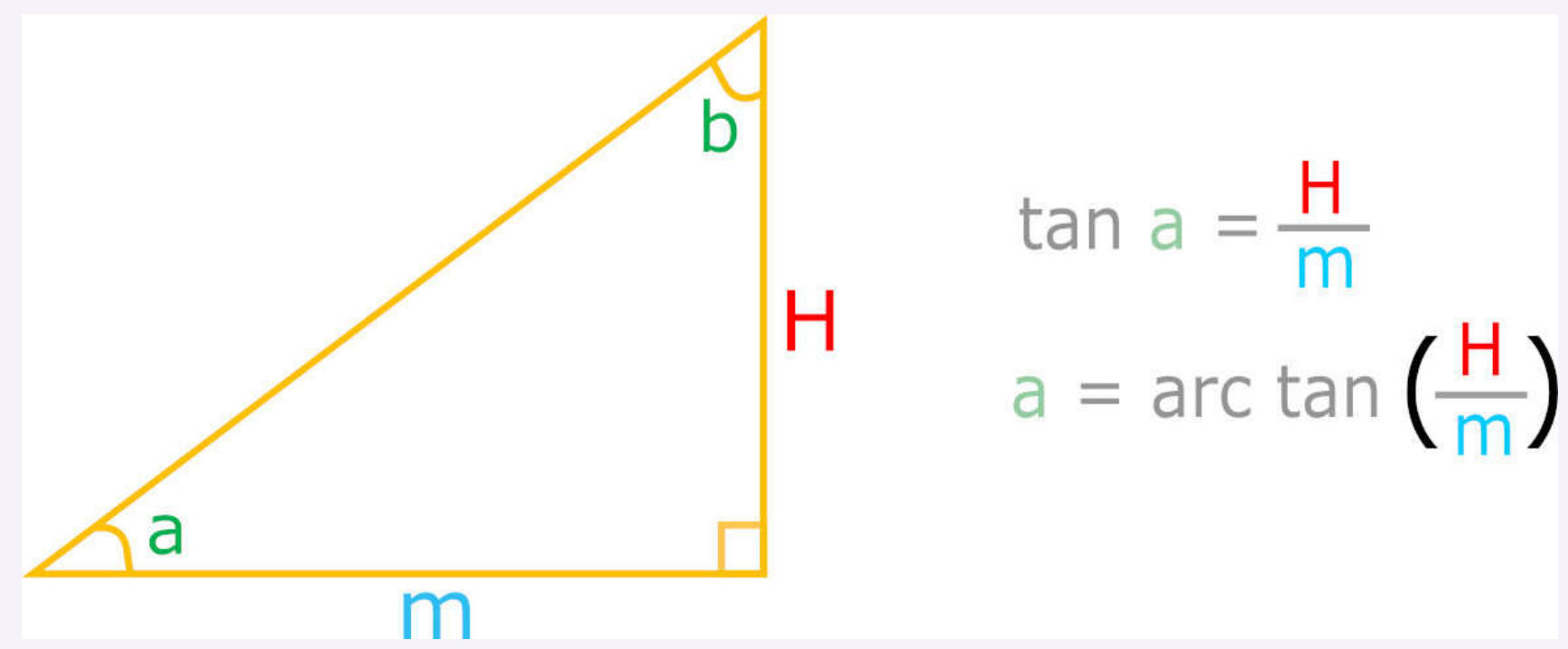

- Recolectemos los ángulos medidos por los compañeros de otras regiones del Perú.
- Ahora formaremos parejas con integrantes de diferentes regiones.
- Calculemos la diferencia entre el ángulo que medimos y el que midió nuestro colega de otras región. El valor resultante es la distancia angular entre las dos localidades, como lo muestra la siguiente imagen: (Crédito de la imagen: imagen adaptada de WYP2005 Eratosthenes project)

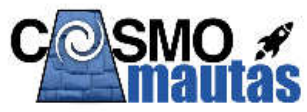

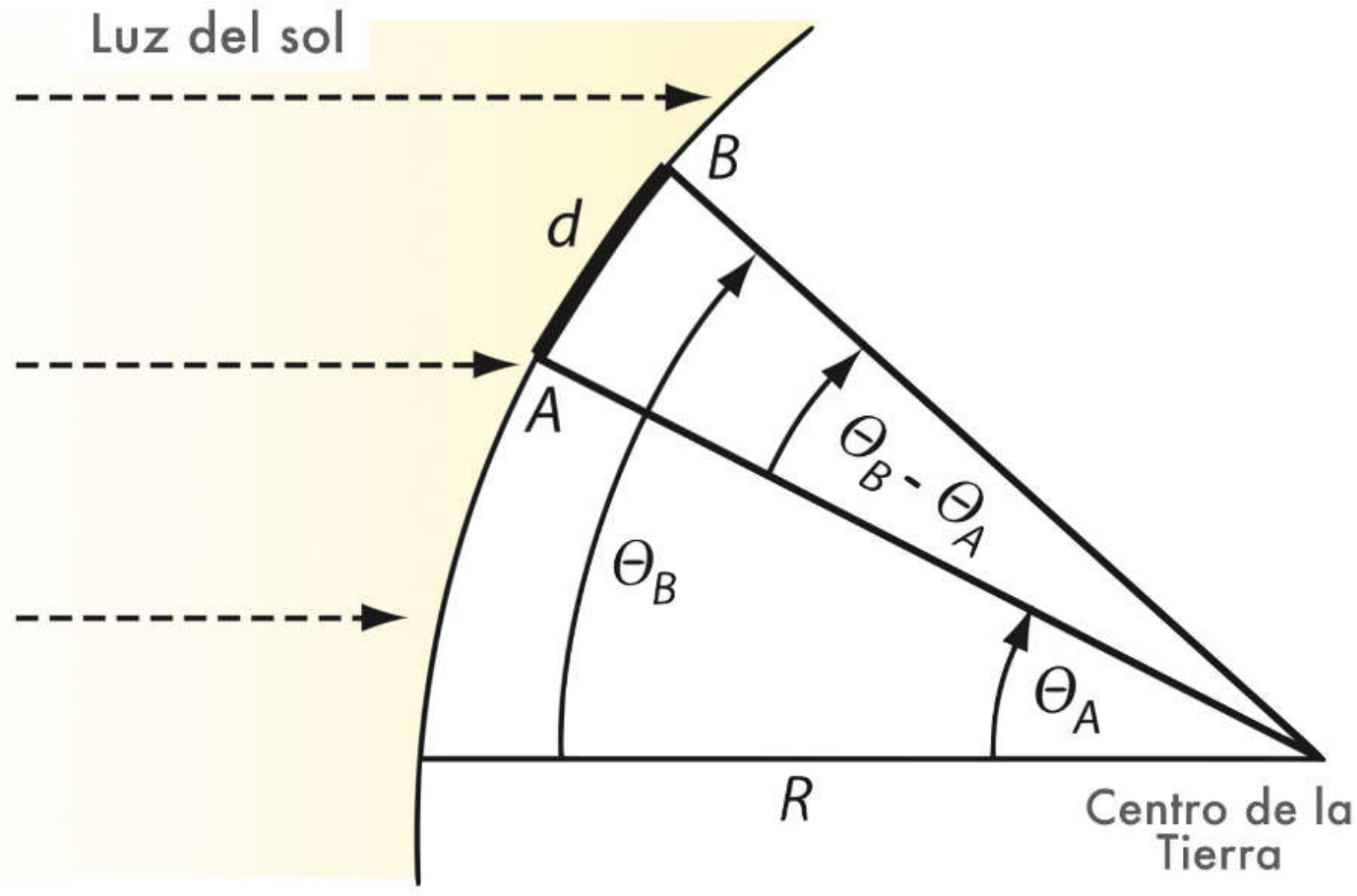

- Usando el arco (es decir la distancia lineal entre las dos localidades, en km), y el ángulo (es decir la distancia angular entre las escuelas), calcular la circunferencia de la Tierra usando la regla de tres. Reportar las diferentes mediciones en el *Google Docs* General.

- Comparar el valor resultante con las otras parejas del grupo, y discutir. Calcular la mediana y el promedio de las mediciones, y el margen de error.

## Conclusión y evaluación (10 min)

Al terminar, hablaremos sobre cómo Eratóstenes hizo la primera medición del tamaño de la Tierra en el año 300 a.C. Realizaremos una pequeña presentación sobre nuestras mediciones y aquellas hechas por satélites, y usaremos los datos recolectados en nuestro experimento para encontrar el resultado medio de CosmoAmautas, el margen de error de nuestra medición y compararlos con las mediciones satelitales e históricas.

## Reflexionando (5 min)

Discutir en grupo:

- ¿Qué fue lo que les impresionó más de este ejercicio?
- ¿Qué modificaciones le harían para implementarlo en su salón de clases?

## Referencias

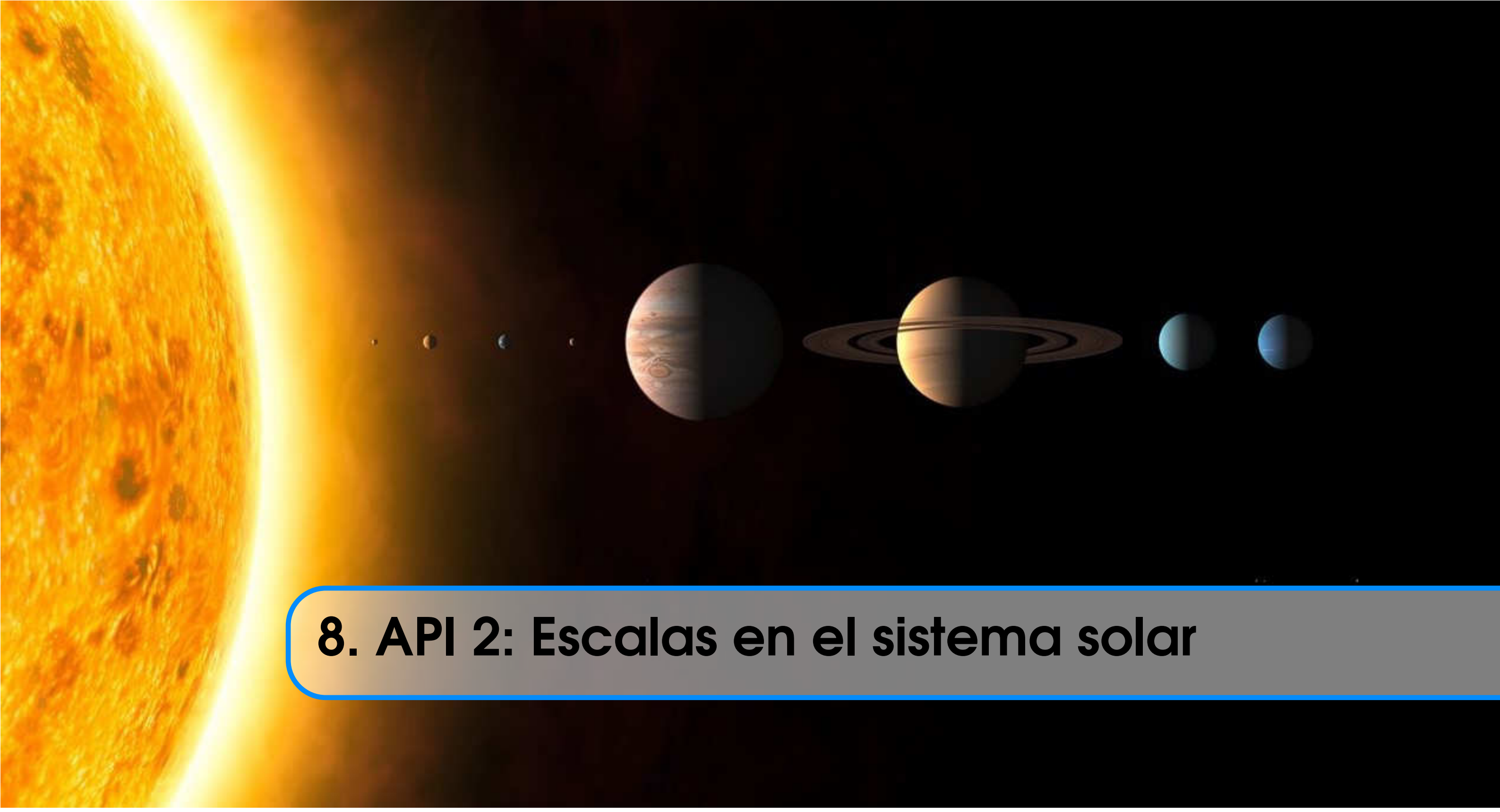

# 8. API 2: Escalas en el sistema solar

## Orientándonos (8 min)

Probablemente ya sabemos que Júpiter es el planeta más grande en nuestro sistema solar y que Neptuno es el más lejano, pero ¿realmente tenemos una idea de qué tan grande es Júpiter y qué tan lejos está Neptuno? Hoy vamos a explorar las escalas de masa y distancia en nuestro sistema solar.

**Unidades de aprendizaje incluidas en este API**
Matemática: regla de tres, fracciones

## Materiales

- Rollo de papel higiénico
- Lapicero
- 1/2 kg de arroz

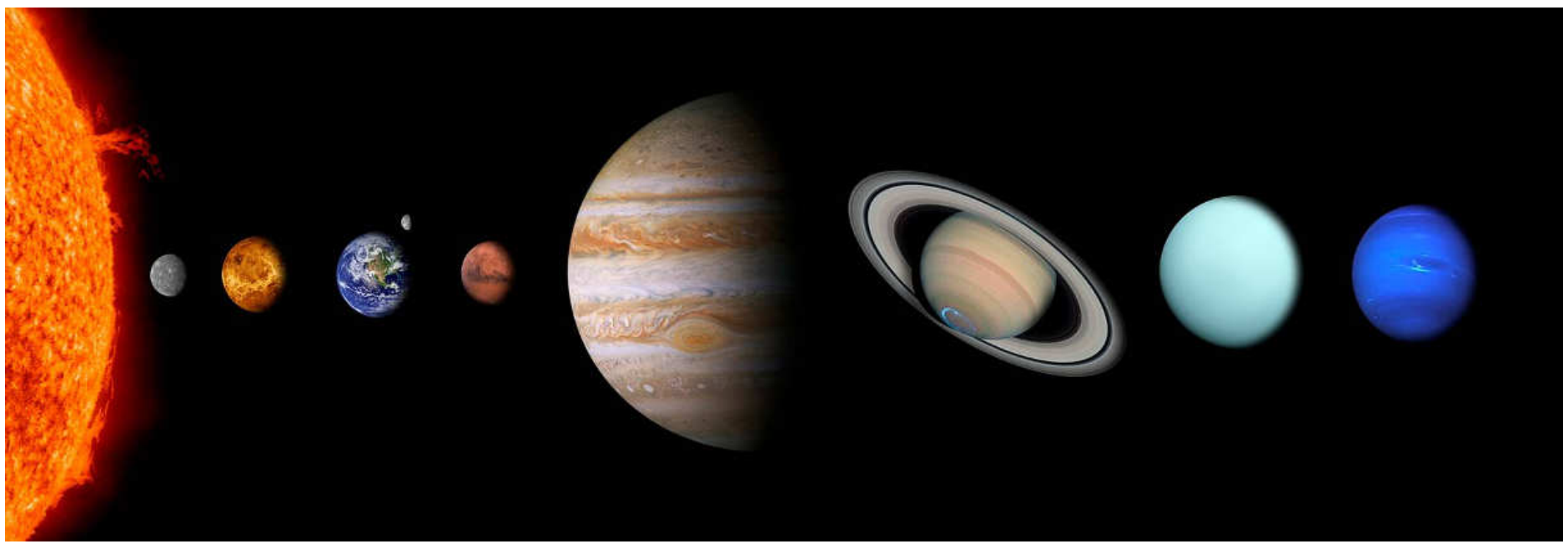

Figura 8.1: El sistema solar. De izquierda a derecha: Sol, Mercurio, Venus, Tierra, Marte, Júpiter, Saturno, Urano, Neptuno. Ni las distancias ni los tamaños están a escala.

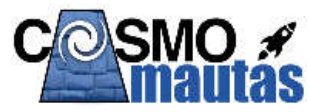

## Objetivos y preguntas (2 min)

Primero, hablemos sobre distancias. Una unidad de medida muy usada por astrónomos es la Unidad Astronómica (UA), que equivale a la distancia promedio entre la Tierra y el Sol. Antes de empezar este ejercicio, hagamos una apuesta:

**Pensamos, escribimos y discutimos:**

- ¿A cuántas unidades astronómicas cree que se encuentra Júpiter? ¿Y Neptuno?
- ¿Creen que los planetas están espaciados equitativamente?
- ¿Dónde cree que termina el sistema solar?

Ahora hablemos de masas:

- ¿A cuántas Tierras creen que equivale un Júpiter?
- ¿A cuántos Júpiteres creen que equivale un Sol?

Anoten sus apuestas en un Google Doc.

## Experimento y Análisis (30 min)

Ahora vamos a hacer una demostración para visualizar estas respuestas aproximadamente.

### Parte 1: Distancias Nivel 1

1. Estiren el rollo de papel higiénico hasta una longitud de mano a mano con los brazos estirados y cortar ese pedazo.
2. Con un lapicero, escribir “Sol” en un extremo y “Plutón” en el otro. El Sol va a ser nuestro punto de partida.
3. Doblar el papel por la mitad. ¿Qué planeta cree que corresponde a ese lugar (½ papel)?
4. Doblar el papel una vez más por la mitad hasta tener cuatro cuartos. ¿Qué planeta cree que corresponde al doblez (¾ papel) más cercano a Plutón?
5. Ahora, ¿qué planeta cree que corresponde en el doblez más cercano al Sol (¼ papel)?
   *¡Ya hemos cubierto tres cuartas partes del papel y todavía no hemos utilizado a todos los planetas gigantes!*
6. Doblar el papel entre el Sol y el primer ¼ . ¿Qué planeta cree que corresponde al nuevo doblez (⅛ papel) más cercano al Sol?
7. Doblar una vez más el papel entre el Sol y el primer ⅛. ¿Qué estructura corresponde en este lugar (1/16 papel)?
8. Doblar otra vez el papel entre el Sol y el doblez más cercano al Sol (1/16 papel). ¿Qué planeta corresponde a este lugar (1/32 papel)?
9. Doblar dos veces más el papel entre el Sol y el planeta que acabamos de nombrar (1/32 papel), de tal forma que tengamos cuarto cuartos entre ellos. ¿A qué planetas cree que corresponden estos tres nuevos dobleces (1/128, 1/64, 3/128 papel)?

*Fíjense en la gran distancia entre los planetas y lo enorme que es nuestro sistema solar. La mayor parte de él se encuentra vacío, ¡por eso lo llamamos **espacio**!*

**Pensamos, escribimos y discutimos:** Nivel 2

- Ahora, si les digo que 1 AU equivale a $1.5 \times 10^8$ km, ¿a qué distancia se encuentra Plutón? ¿y Júpiter?
- Si la distancia a Júpiter es 5.2 AU, ¿a cuánto se acerca la aproximación de la distancia de Júpiter en porcentaje?

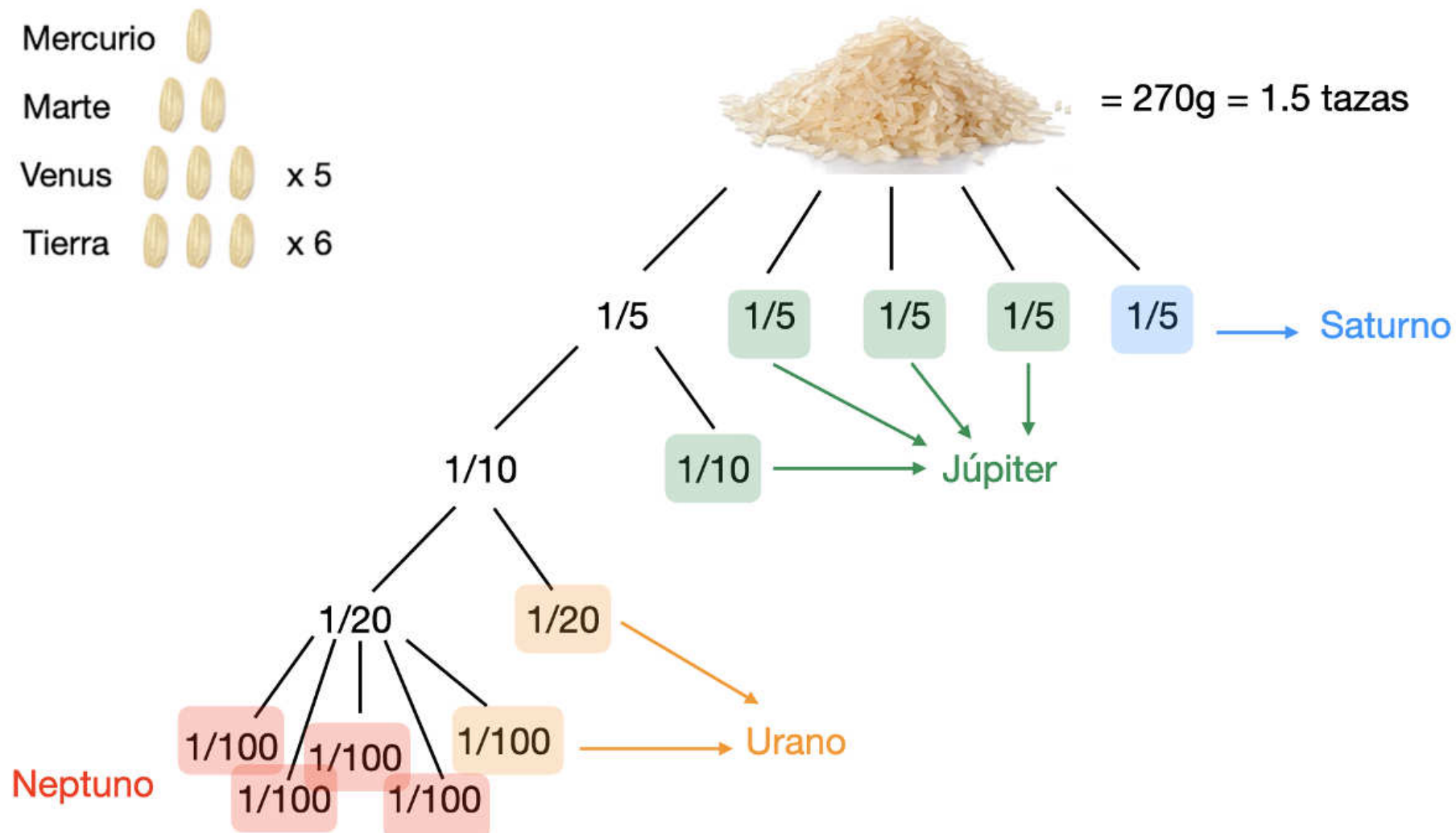

Figura 8.2: Fracciones de la bolsa de arroz que corresponden a la masa de cada planeta del sistema solar.

#### Parte 2: Masas Nivel 1

En esta demostración vamos a hacer "modelos" a escala para visualizar las masas de los planetas del sistema solar. Asegúrense de separar espacio en una mesa para poder dividir el arroz. Primero, escriban en un papel los nombres de los 8 planetas y corten los pedazos para tener etiquetas. Digamos que un grano de arroz equivale a la masa de Mercurio ($M_M = 3{,}3 \times 10^{23}$ kg). Para los planetas rocosos, vamos a contar granos de arroz uno por uno:

1. Un grano de arroz para Mercurio
2. Dos granos de arroz para Marte
3. Quince granos de arroz para Venus
4. Dieciocho granos de arroz para la Tierra
   *¡Ya terminamos con los planetas rocosos!* La masa de todos los planetas rocosos juntos equivale al $\sim 0{,}5\,\%$ de la masa total de los planetas del sistema solar. Ahora para los planetas gaseosos:
5. Con una taza de medida, asegúrense de tener 1 taza y media de arroz. Este volumen equivale a 270 g aproximadamente.
6. Dividir el arroz restante en 5 partes iguales. Una de esas partes equivale a la masa de Saturno.
7. Dividir uno de los quintos en dos partes.
8. Juntar tres partes del paso 5 con una parte del paso 6. Este montón equivale a la masa de Júpiter, el planeta más masivo del sistema solar.
9. Dividir el arroz restante en dos partes iguales.
10. Dividir una de las partes del paso 8 en 5 partes iguales.
11. Juntar una parte del paso 8 con una parte del paso 9. Este montón equivale a la masa de Urano.
12. Finalmente, juntar las 4 partes restantes del paso 9. Este montón equivale a Neptuno.
13. Ahora, juntar todo el arroz, y ¡a la olla!

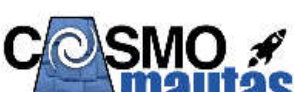

**Pensamos, escribimos y discutimos:** Nivel 2

- ¿Cómo podrían calcular aproximadamente cuántos granos de arroz hay en 1 kg? ¿Cuál sería la respuesta?
- Si les digo que la masa del Sol es $1.989 \times 10^{30}$ kg, ¿cuántos kilos de arroz se necesitan basados en su modelo de Mercurio?

## Conclusión y evaluación (10 min)

Al terminar, hablaremos sobre cómo Giovanni Cassini y Jean Richer midieron la distancia a Marte en 1673 usando el paralaje trigonométrico y cómo los astrónomos medimos distancias a estrellas hoy en día.

## Reflexionando (5 min)

Discutir en grupo:

- ¿Qué fue lo que más les impresionó de este ejercicio?
- ¿Qué modificaciones le harían para implementarlo en su salón de clases?

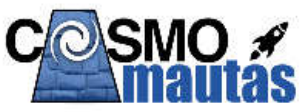

# 9. API 3: Explorando el diagrama HR

## Orientándonos (8 min)

En el cielo nocturno, a simple vista podemos ver cientos de puntos de luz, que son en su mayoría, estrellas. Las estrellas tienen una gran diversidad de masas, tamaños, temperaturas y colores. Existen desde estrellas tan "ligeras" que apenas son capaces de fusionar hidrógeno (8 % de la masa del Sol), hasta estrellas tan masivas en riesgo de que su propia radiación las haga explotar (R136a1, $\sim 300\,\mathrm{M}_\odot$). Hay estrellas tan pequeñas como Júpiter ($0{,}1\,\mathrm{R}_\odot$) y tan grandes como UY Scuti, cuyo radio excede la órbita de Júpiter ($1700\,\mathrm{R}_\odot$ o casi 8 AU). Las estrellas pueden tener temperaturas efectivas tan "frías" como 2400 K (300 grados menos que la temperatura de un filamento de tungsteno en un foco encendido) y todavía fusionar hidrógeno a una tasa bajísima, o tan calientes (estrellas de Wolf-Rayet, $\mathrm{T}_{eff} \sim 200,000\,\mathrm{K}$) que emiten la mayor parte de su luz en rayos X. Sin embargo, sin importar qué tan exótica sea una estrella, siempre la podemos ubicar en un diagrama de Hertzprung-Russell.

El Sol es una de miles de millones de estrellas en nuestra Vía Láctea y conocemos una gran diversidad de ellas gracias a observaciones con espectrógrafos, que son instrumentos instalados en algunos telescopios que separan la luz por longitudes de onda, similar a un arco iris. Cuando vemos un arco iris, la luz blanca del Sol pasa a través de gotas de agua suspendidas en la atmósfera. Al entrar la luz a la gota de agua, la velocidad de la luz se reduce dependiendo de su longitud de onda. Dentro de la gota de agua, la luz azul viaja más lentamente que la luz roja, en un fenómeno que se llama refracción, y de esta forma se separan los colores que componen la luz blanca y vemos un arco iris. Los astrónomos utilizamos espectrógrafos de una manera parecida para averiguar los gases que componen la atmósfera de las estrellas (sí, ¡las estrellas tienen atmósferas!). En lugar de una gota de agua, usamos un prisma o una rejilla de difracción para separar la luz en longitudes de onda. La intensidad que vemos por cada longitud de onda se llama espectro. El espectro de una estrella es como su huella digital: es una firma única que identifica a la estrella por la composición de su atmósfera.

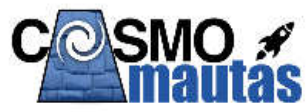

Hoy vamos a armar un espectrógrafo simple para ver el espectro del Sol y a explorar los espectros de otras estrellas para clasificarlas.

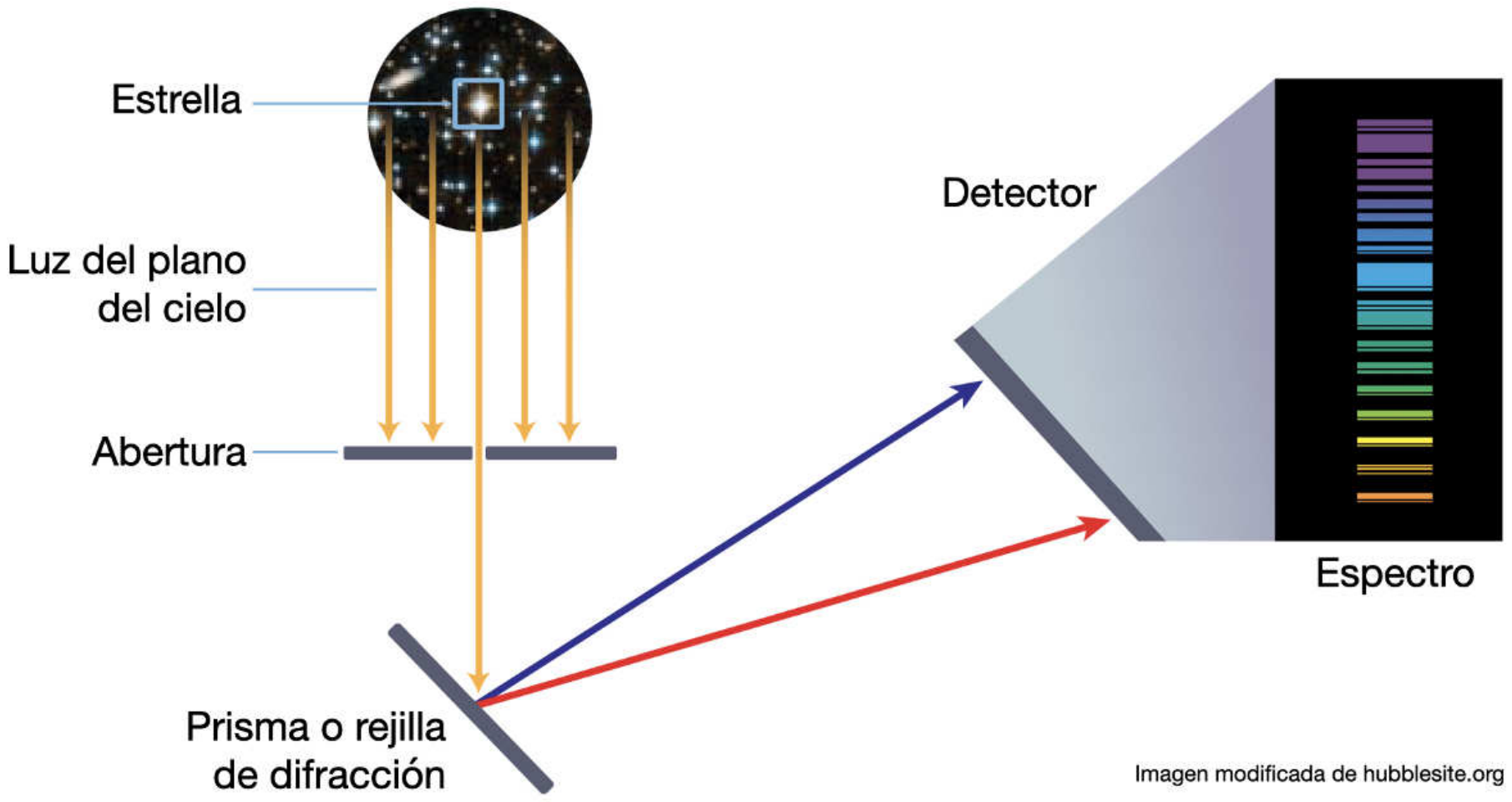

Figura 9.1: Esquema del funcionamiento de un espectrógrafo. La luz de las estrellas pasa por una apertura pequeña y luego por un prisma o rejilla de difracción que separa la luz en intensidades de acuerdo a su longitud de onda.

**Preparación:** Para este API, pedimos a los docentes, especialmente aquellos que no trabajan en matemáticas y física, que repasen conceptos de escalas logarítmicas, cuerpo negro, la ley de Planck, y la ley de Wien.

**Unidades de aprendizaje incluidas en este API**
Física: Ley de Wien, Ley de Planck, absorción atómica
Ciencias: Entender gráficos x e y, identificar tendencias

## Materiales

- CD
- Plantilla de espectrógrafo
- Papel celofán de colores
- 1 foco fluorescente
- Tijera, goma, lápiz, papel, calculadora
- Hojas impresas con los espectros estelares
- Hojas impresas con las curvas de cuerpo negro
- Hoja impresa con el diagrama de HR con datos de *Gaia*

## Explorando (20 min) Nivel 2

Armar el espectrógrafo según las instrucciones. Luego, observen con él la luz del Sol (hagan lo que hagan, ¡no miren directamente al Sol!). Prueben en apuntar el espectrógrafo a la pantalla de un televisor, luego cúbranlo con papel celofán de colores y vuelvan a observar; por último, enciendan el foco fluorescente y observen su luz con el espectrógrafo. ¿Qué diferencias observan?

Con un espectrógrafo profesional, el espectro del Sol se ve como en la Figura 9.2. Basados en esta Figura, discutan las siguientes preguntas en grupos:

**Construye tu propio espectrógrafo**

1. Pega esta plantilla sobre una cartulina de tamaño A4
2. Corta con tijeras por las líneas sólidas, incluyendo la línea **a,** y corta el rectángulo **b** y el círculo **c.**
3. Dobla por las líneas punteadas
4. Arma la caja juntando los lados del mismo número, e.g., 1 se une a 1.

a

b

5. Introduce un CD dentro del corte que hiciste en la línea **a** con el lado del “arco iris” mirando hacia arriba.
6. Mira hacia dentro de la caja a través del círculo y deberías poder ver la luz dividida en los colores del arco iris
7. Prueba mirar a diferentes tipos y colores de luz para ver qué cambia en tu espectrógrafo

c

Figura 9.2: El “código de barras” del Sol. Un espectro muy largo se ha cortado en pedazos pequeños y organizado en filas, una debajo de la otra [Crédito: NOAO/AURA/NSF/ESO].

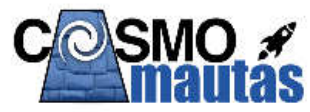

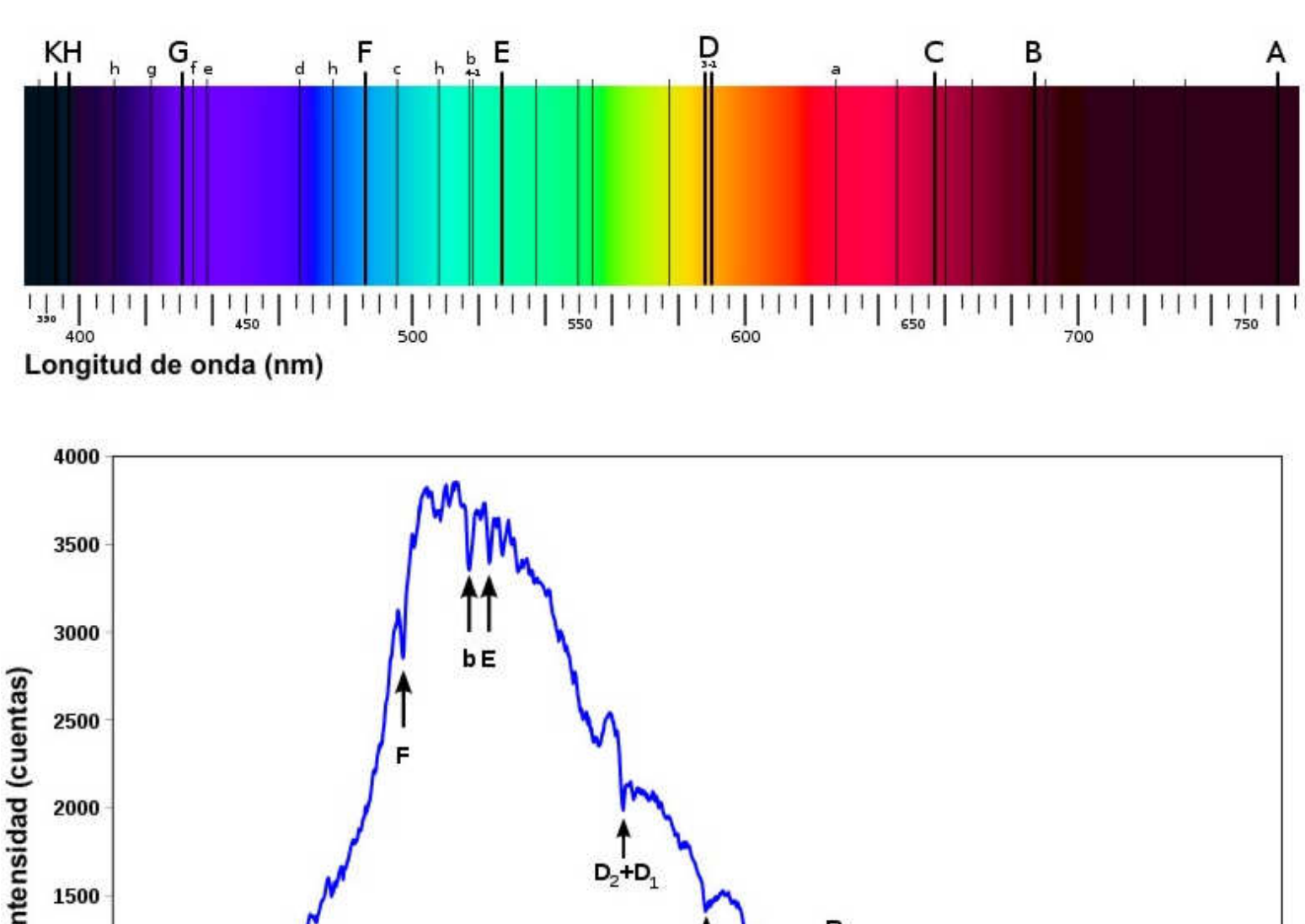

Figura 9.3: Espectro solar con líneas de absorción de Fraunhofer [Crédito: Wikimedia Creative Commons].

**Pensamos, escribimos y discutimos:**

- ¿En qué color emite más luz el Sol?
- ¿Cómo creen que se vería el espectro de una estrella más caliente o más fría? ¿De qué color sería?
- ¿A qué creen que corresponden esas zonas oscuras en el espectro?
- Mencionamos que las estrellas tienen atmósferas. ¿Cómo creen que son? ¿Cómo se parecen o diferencian de la atmósfera de la Tierra? ¿Qué moléculas hay en la atmósfera de la Tierra? ¿Esperaríamos moléculas complejas en una atmósfera mucho más caliente?

El espectro de la Figura 9.2 se ve hermoso, pero a los científicos nos gustan las gráficas, por eso podemos representar al espectro de la Figura 9.2 como la gráfica de la Figura 9.3.

## Experimento (15 min) Nivel 2

Ahora que hemos investigado el espectro del Sol, comparémoslo con el de otras estrellas. Para esta actividad, van a necesitar las dos hojas de espectros estelares y dos hojas de curvas de radiación de cuerpo negro. Corten las hojas para hacer tarjetas del mismo tamaño, cada una con un gráfico. En grupos, ordenen solo los espectros como mejor les parezca y discutan el orden que han logrado entre ustedes.

**Pensamos, escribimos y discutimos:**

- Ahora que han ordenado los espectros de manera visual, ¿creen que hay alguna variable física bajo esta organización? ¿Tal vez alguna que coincida con las curvas de radiación de cuerpo negro?

- ¿A qué creen que se deben las diferencias entre los espectros?

Cuando tengan ordenados sus espectros, compárenlos con las curvas de cuerpo negro para averiguar a qué temperatura están sus estrellas. Cada espectro corresponde a una estrella real, marcada en el diagrama de HR de *Gaia* (Figura 9.4). Anoten la temperatura de la estrella donde corresponda.

## Investigación y análisis (15 min) Nivel 3

Ahora vamos a entender los parámetros físicos que relacionan a estas estrellas en el diagrama HR:

- ¿Dónde se encuentran las estrellas más brillantes y las más débiles?
- ¿Dónde se encuentran las estrellas más calientes y las más frías?

Dibujemos flechas en el diagrama de HR para indicar las tendencias que hemos encontrado.

Sabiendo la temperatura de nuestras estrellas y utilizando la ley de Stefan-Boltzmann, calculen el flujo de la estrella más caliente ($F_c$) y de la más fría ($F_f$). La ley de Stefan-Boltzmann dice que:

$$F = \sigma T^4, \qquad \sigma = 5{,}678 \times 10^{-8} W m^{-2} K^{-4} \tag{9.1}$$

Finalmente, usando esta ecuación:

$$m_c - m_f = -2{,}5\, log_{10} \left( \frac{F_c}{F_f} \right) \tag{9.2}$$

Calculen la diferencia en brillo entre la estrella más caliente y la más fría ($m_c - m_f$) de los espectros con lo que hemos trabajado. ¿Cuántas veces más brillante es la estrella más caliente?

## Conclusión y evaluación (10 min)

Al terminar, hablaremos sobre cómo un grupo de astrónomas a comienzos del siglo XX definieron la clasificación estelar que se utiliza hoy en día.

## Reflexionando (5 min)

Discutir en grupo:

- ¿Qué fue lo que les impresionó más de este ejercicio?
- ¿Qué modificaciones le harían para implementarlo en su salón de clases?

## Referencias

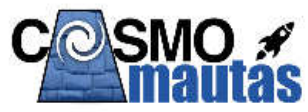

Figura 9.4: Diagrama de HR con datos de *Gaia*. En el eje $x$ usamos el color $B_p - R_p$ que equivale a la diferencia en intensidad de un objeto entre longitudes de onda azules y rojas. Esta cantidad se correlaciona directamente con la temperatura y el tipo espectral. En el eje $y$ usamos la magnitud absoluta en *Gaia G*, una banda visual equivalente al color blanco (centrada en verde, pero es combinación de todos los colores visibles). Esta cantidad se correlaciona con la luminosidad de una estrella.

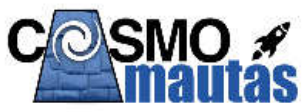

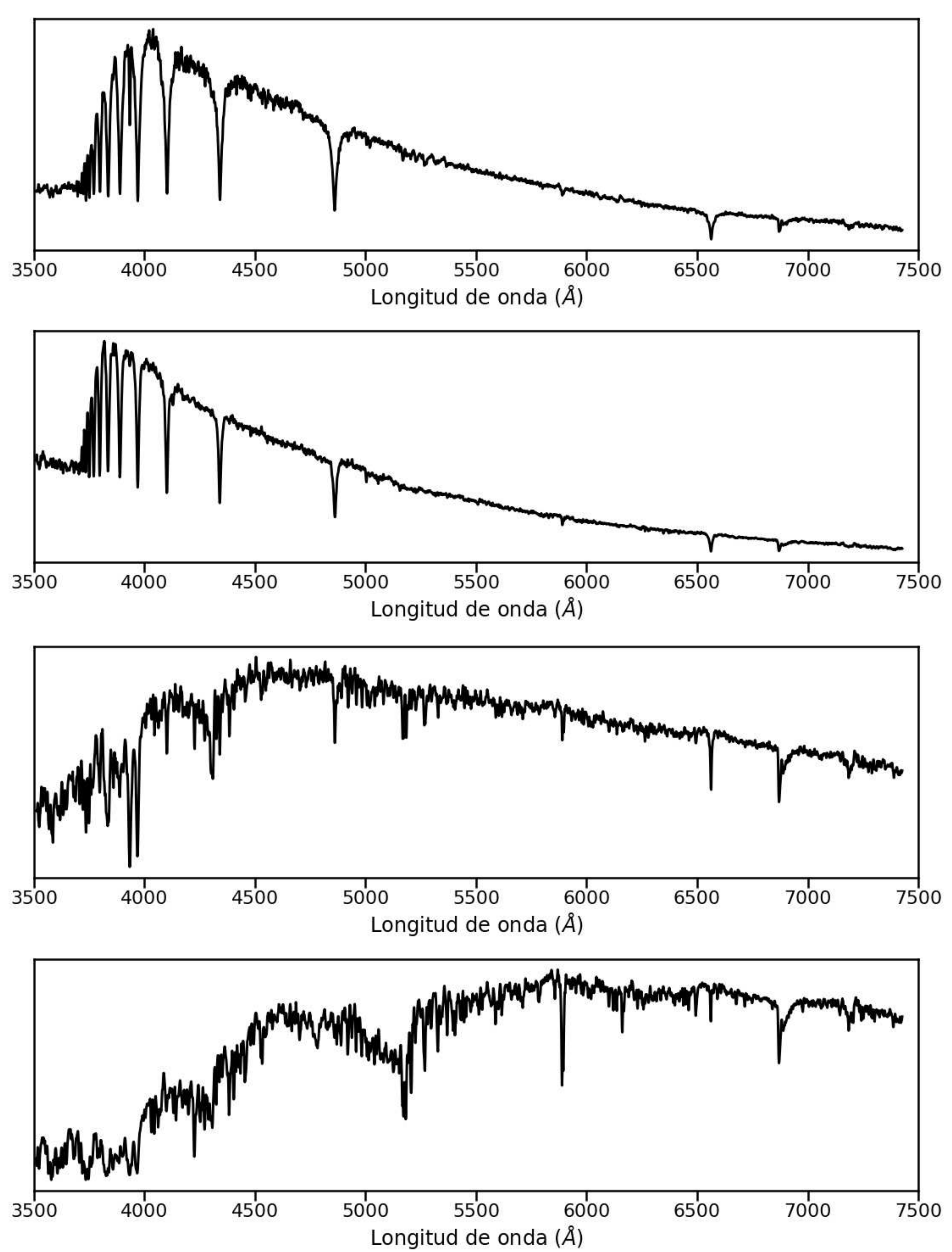

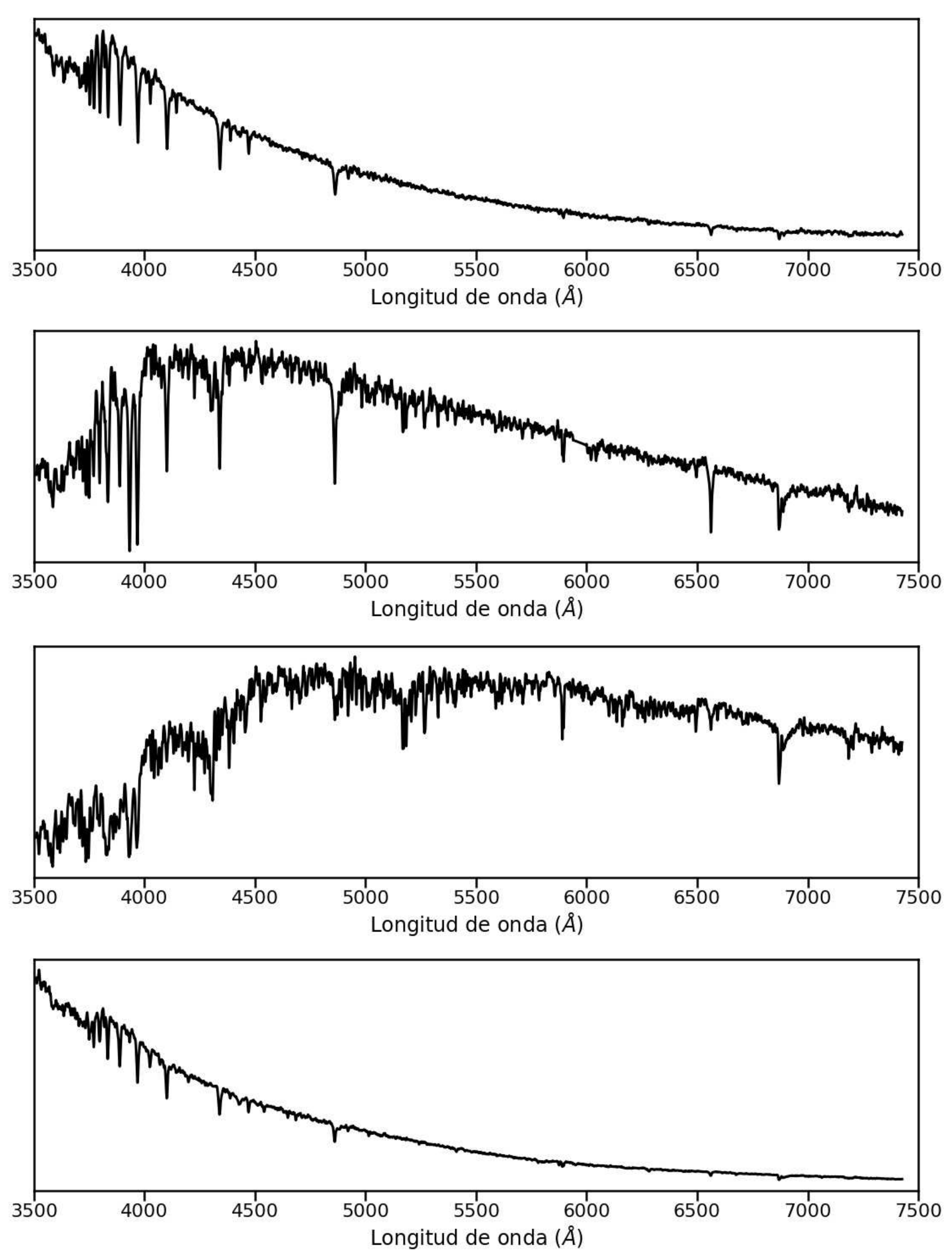

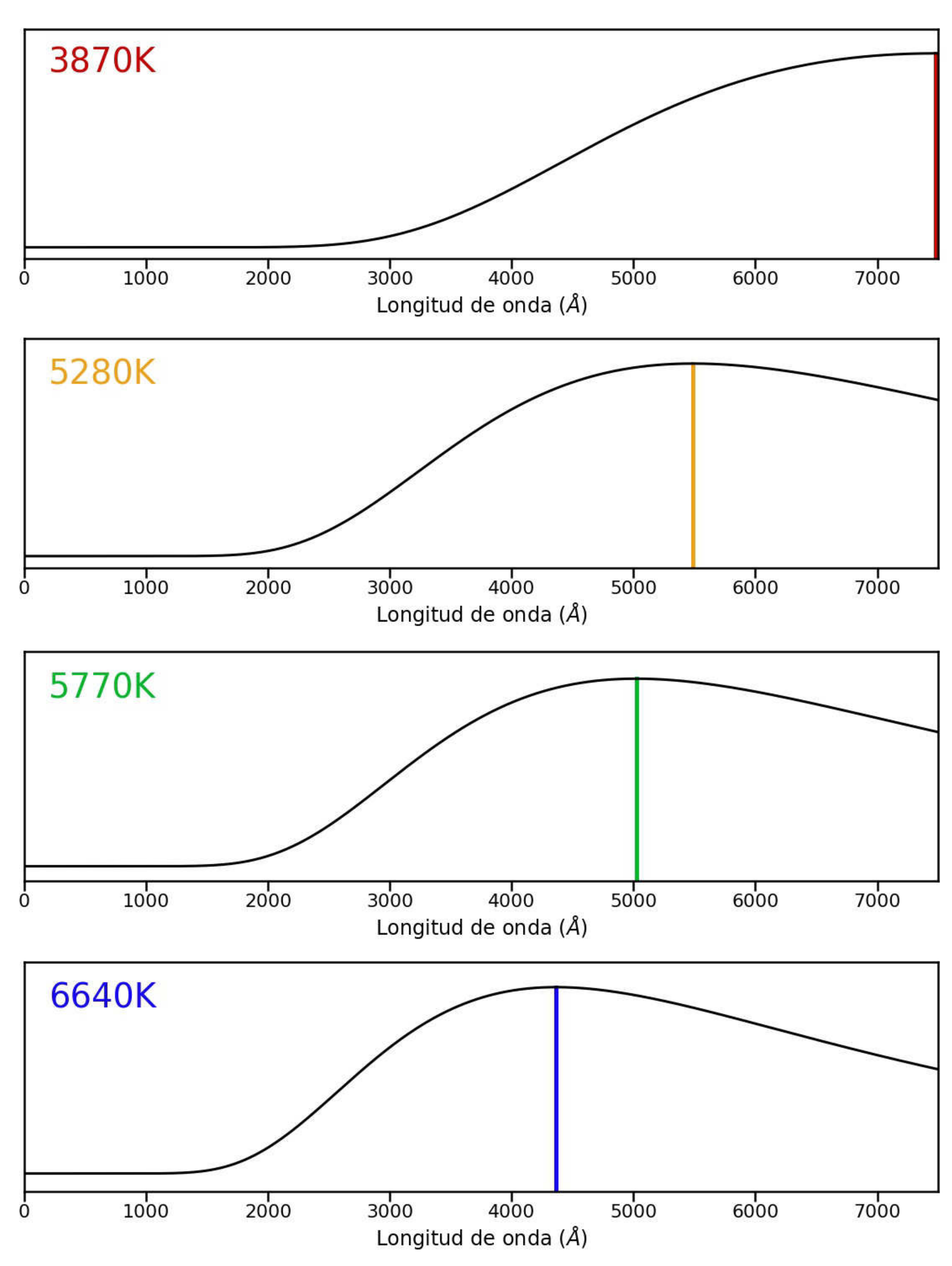

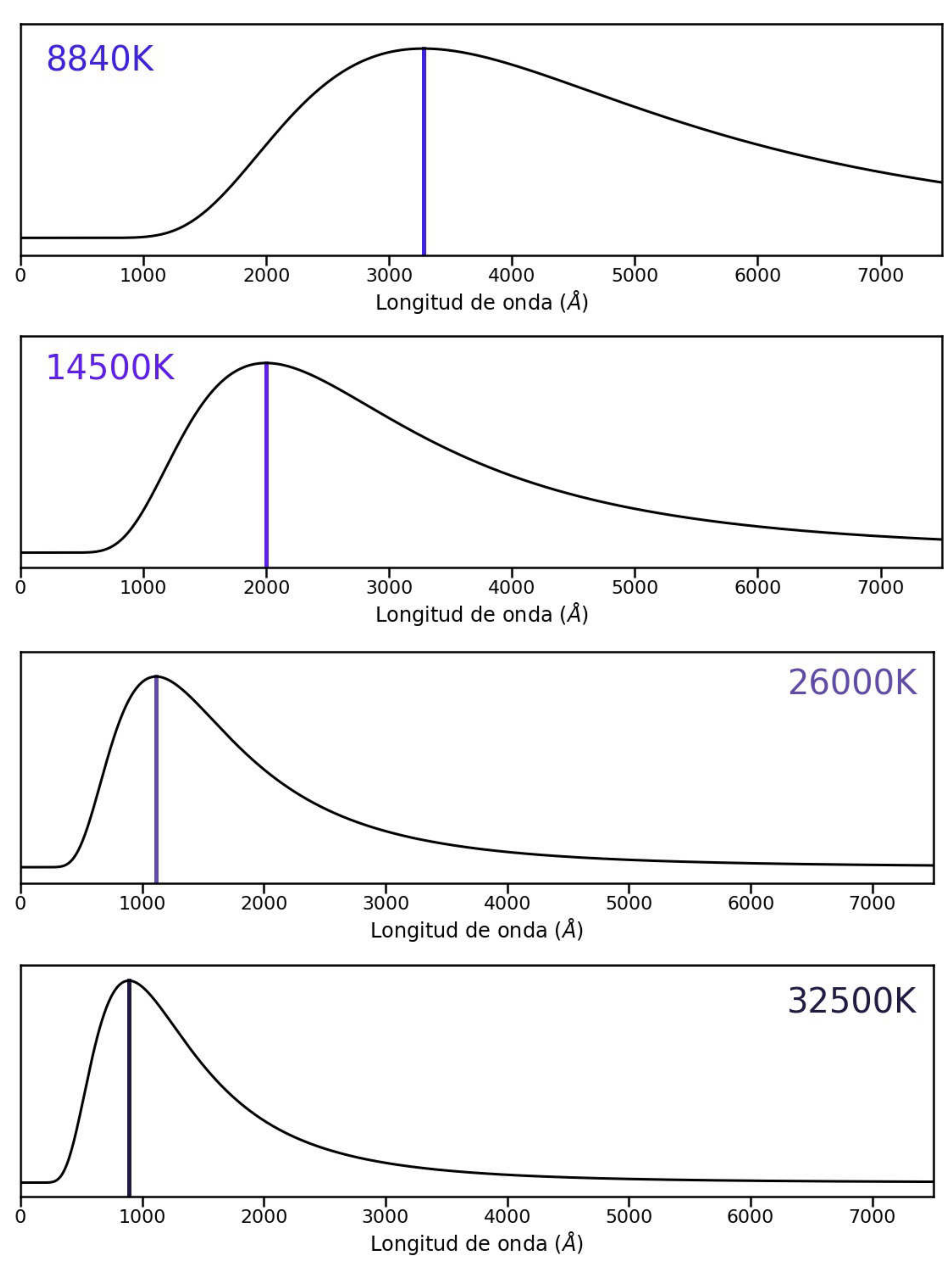

# 10. API 4: Exoplanetas

## Orientándonos (8 min)

Uno de los métodos más comunes para detectar exoplanetas es a través de tránsitos. En un tránsito, el planeta en órbita alrededor de su estrella cubre una parte de la superficie de la estrella y desde nuestro punto de vista en la Tierra, parece que se hubiera bloqueado un poco de luz proveniente de la estrella. Desde la Tierra podemos observar tránsitos de Mercurio y de Venus, por que estos planetas se encuentran entre la Tierra y el Sol. Para detectar el tránsito de un exoplaneta, debemos apuntar un telescopio hacia otro sistema estelar y tomar imágenes una tras otra durante varias horas. Después de calibrar las imágenes, aplicaríamos un proceso llamado **fotometría** donde contamos el número de fotones que hemos recolectado en cada imagen correspondientes a nuestra estrella. La suma total de fotones nos brinda el flujo de la estrella en un determinado momento. Cuando un planeta transita frente a una estrella desde nuestro punto de vista en la Tierra, el flujo total que hemos calculado con fotometría disminuye, y regresa a su valor base cuando el planeta termina de pasar frente a la estrella. El gráfico que describe el nivel de flujo con respecto del tiempo se conoce como **curva de luz** (ver Figura 10.2).

Si observamos a la estrella por suficiente tiempo, podemos incluso ver que el tránsito se repite periódicamente. Ese periodo equivale al periodo orbital del planeta, o la duración de su año. Para amplificar la señal y reducir el ruido de nuestra curva de luz, podemos "plegarla" (a lo largo de la dirección de tiempo) de tal forma que todos los tránsitos coinciden. A esta gráfica se le conoce como **curva de fase**. Dado que el planeta tiene una órbita periódica alrededor de su estrella, cada punto en la curva de fase corresponde al mismo punto en la órbita del planeta. Esta gráfica nos permite analizar las características de un tránsito y derivar parámetros físicos del planeta (ver Figura 10.3).

En el API de hoy, vamos a explorar el tránsito del planeta HAT-P-7b descubierto por *Kepler* para familiarizarnos con su geometría y así poder calcular algunos parámetros físicos del planeta y evaluar su habitabilidad.

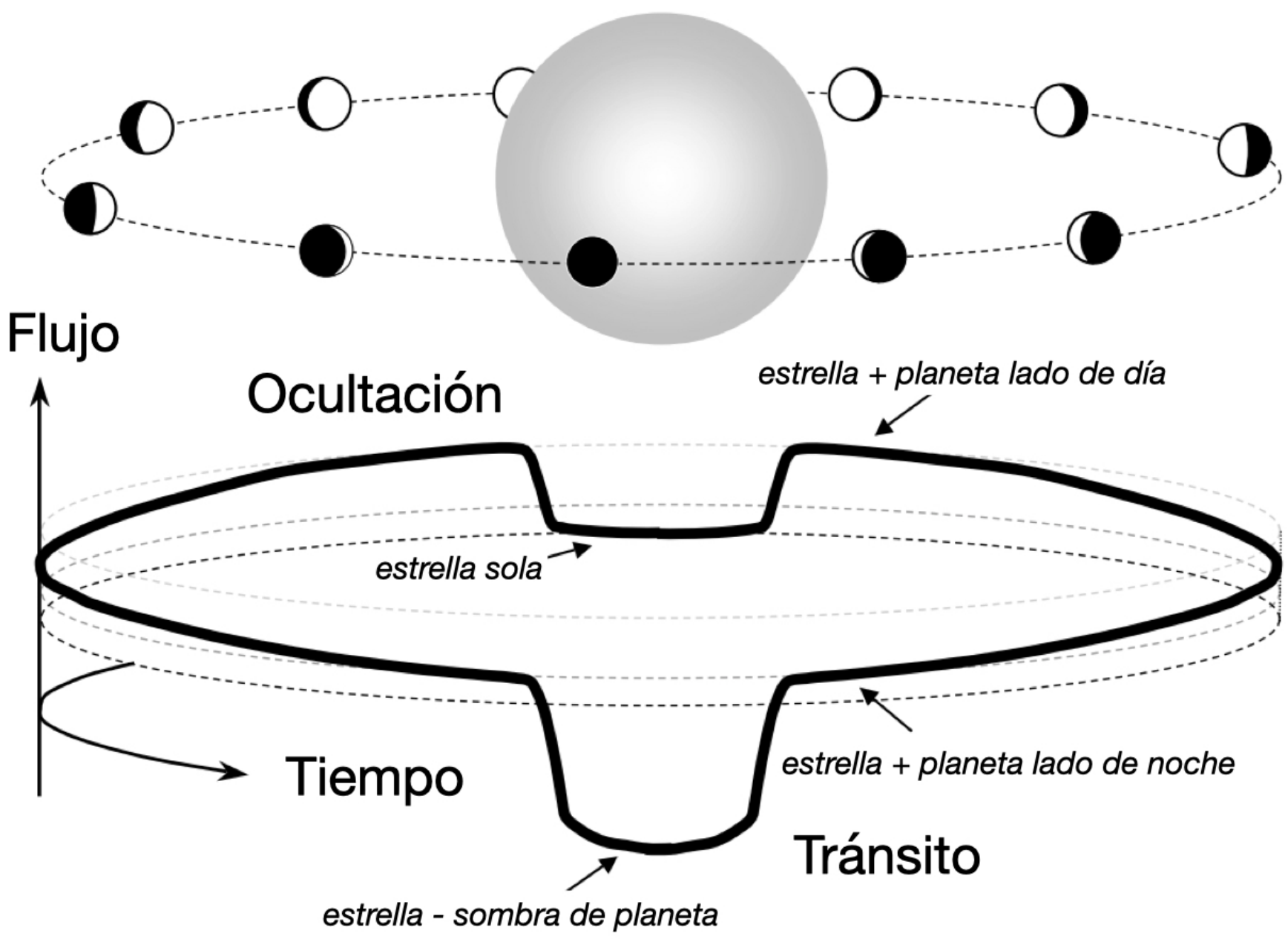

Figura 10.1: Esquema del tránsito y ocultación (eclipse) de un exoplaneta. Imagen modificada de Deming & Knutson (2020), *Nature Astronomy*, 4, 453.

**Unidades de aprendizaje incluidas en este API**
Matemática: geometría: áreas, regla de tres
Física: movimiento rectilíneo uniformemente variado (MRUV): velocidad, aceleración, período
Ciencias: ecuaciones, relaciones entre variables, y simplificaciones entre realidad y modelo

## Materiales

- Linterna
- Esferas de poliestireno de 3z de 1 1/2"
- Palito de brocheta
- Lápiz, papel, regla

## Hipótesis y diseño del experimento (10 min) Nivel 1

Usaremos una esfera de poliestireno de 3"para representar un "planeta", inserta un palito de brocheta hasta la mitad de la esfera para que se pueda sostener. En un papel, dibuja un círculo que abarque todo el papel y con un pedazo de cinta adhesiva, pégalo en la pared. Esta va a ser nuestra “estrella”. Enfoca la linterna hacia el papel con una mano y con la otra (¡o con un ayudante!) mueve el planeta de forma horizontal frente a la estrella.

**Pensamos, escribimos y discutimos:**

- Desplacemos el planeta acercándolo y alejándolo de la estrella, ¿qué notamos? Tomemos en cuenta que una diferencia entre este modelo y la situación real es que la luz proviene de la estrella (no de la linterna espacial) y que la distancia entre planeta y estrella es muchísimo más pequeña que la distancia entre nuestro punto de observación en la Tierra y cualquier estrella. **En la práctica, no podemos determinar la separación entre planeta y estrella a partir de un tránsito.**

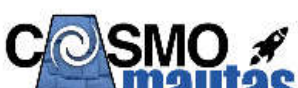

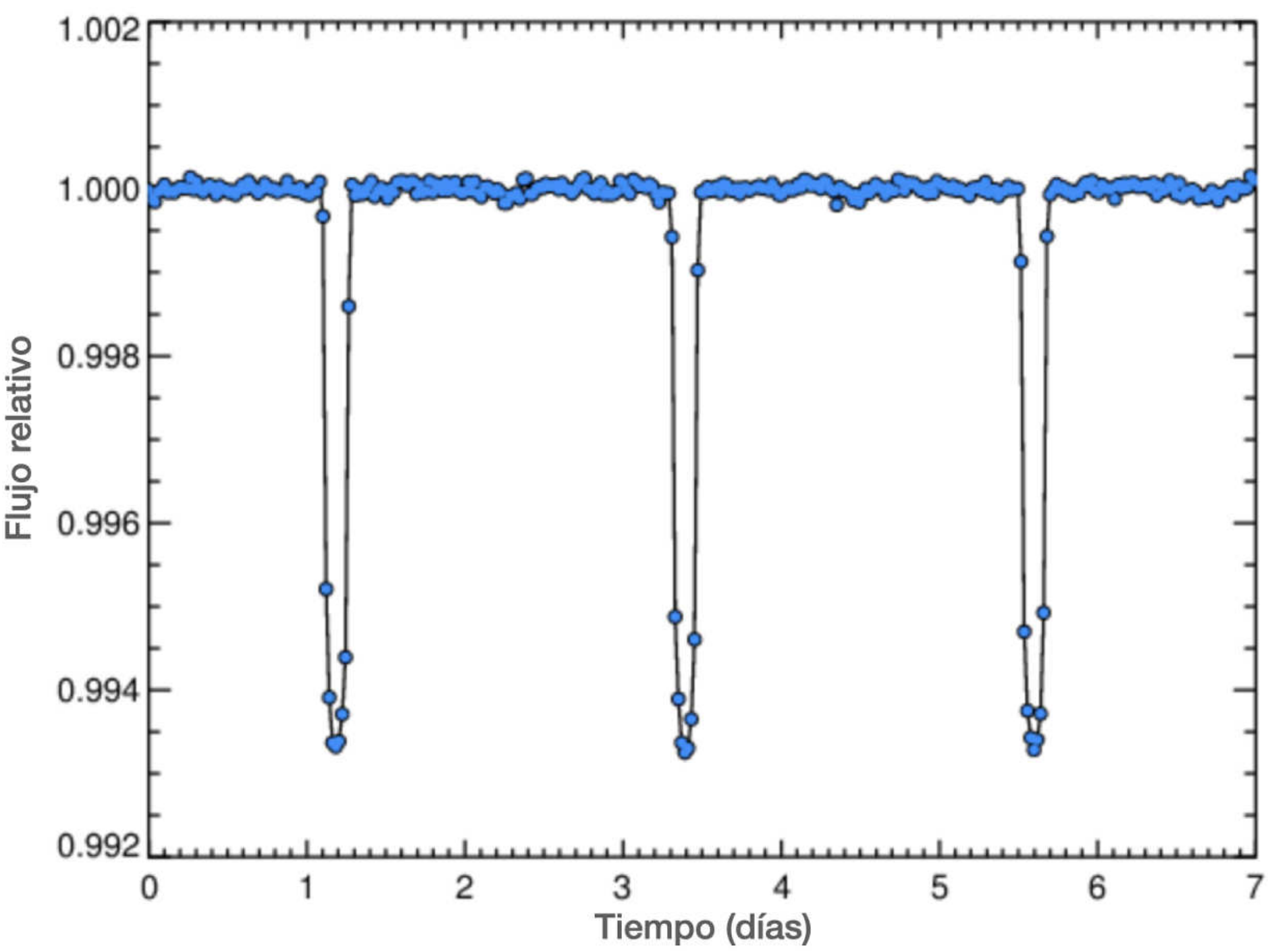

Figura 10.2: Curva de luz de HAT-P-7. Datos de Borucki et al. 2009, *Science*, 325, 7. Figura modificada de A. Vanderburg.

- Ahora, ubiquemos el planeta entre la linterna y la estrellas. Luego, movamos el planeta a lo largo de una línea vertical, de arriba hacia abajo y de abajo hacia arriba. ¿Hay alguna altura en la que ya no se pueda ver la sombra del planeta?

Reemplazamos la esfera de 3"por la de 1 1/2"para representar un planeta más pequeño y repitan el experimento.

- Con los dos planetas a la misma distancia de la estrella, ¿qué diferencias existen entre las sombras que proyectan?

## Investigación y análisis (30 min) Nivel 3

Aquí les presentamos la curva de luz y la curva de fase (Figura 10.3) de la estrella HAT-P-7, alrededor de la cual existe un planeta. Empecemos por encontrar el periodo del planeta a partir de la curva de luz (Figura 10.2). Para esto, van a necesitar una regla.

- ¿Cuál es el periodo del planeta HAT-P-7 b en unidades terrestres (días, años)?
- ¿Cómo se compara el año de HAT-P-7 b con el año terrestre?

Utilizando el lado largo de un papel, dibujen la curva de la Figura 10.2 repetida varias veces sin levantar el lapicero. Es como dibujar una línea recta y de vez en cuando una forma de “U”. No tiene que ser igual a la Figura; se pueden tomar una licencia artística y exagerar la “U” para que sea más fácil de dibujar. Ahora que tienen varias “U” dibujadas, doblen el papel como un acordeón haciendo que las “U” coincidan.

Usando el periodo que han encontrado, los astrónomos han sobrepuesto o “plegado” (de la misma forma que su acordeón de papel) un tránsito sobre el otro para mejorar la

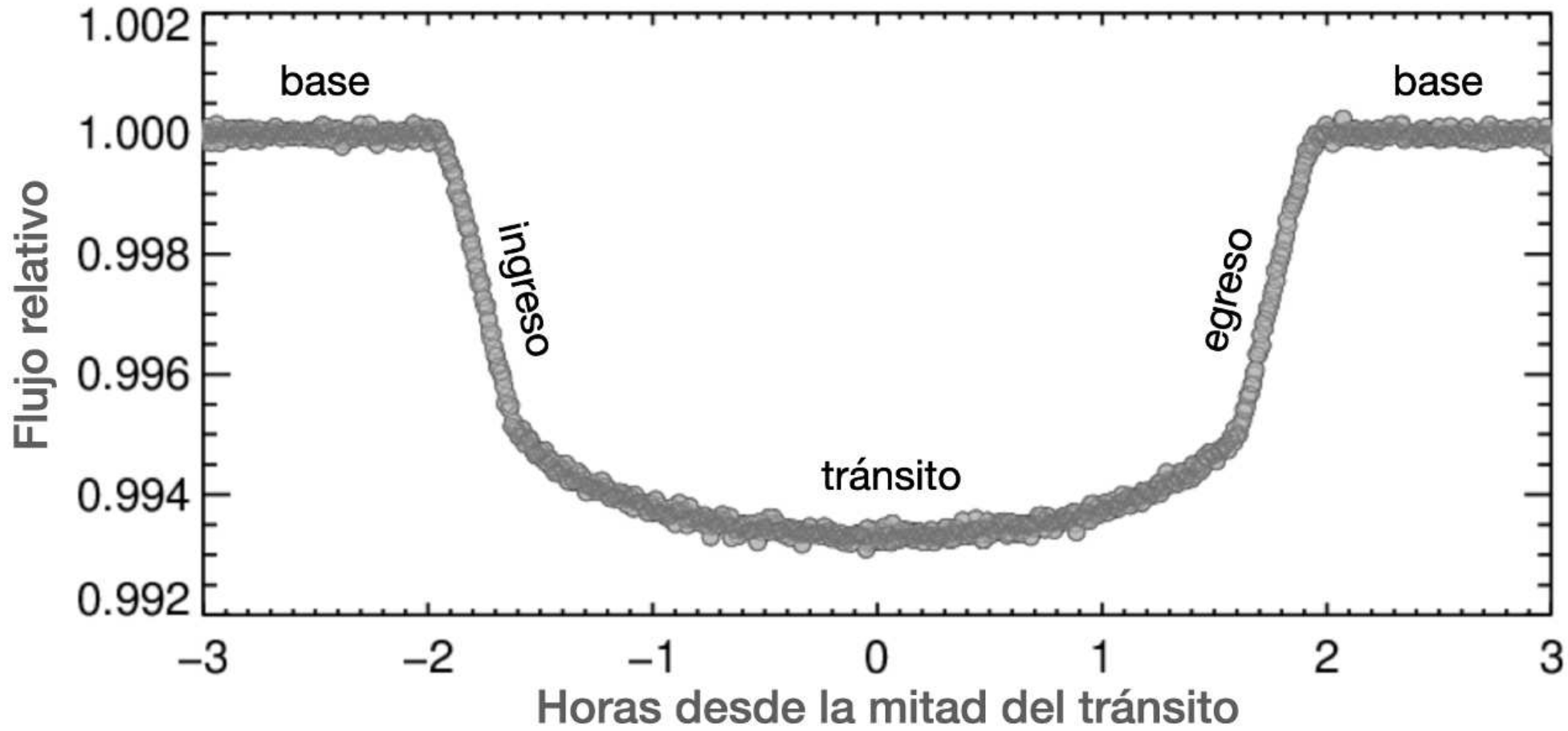

Figura 10.3: Curva de fase de HAT-P-7. Datos de Borucki et al. 2009, *Science*, 325, 7. Figura modificada de A. Vanderburg.

señal y reducir el ruido de los datos, haciendo sus mediciones más precisas. Como resultado, han obtenido una curva de fase (Figura 10.3). Basándose en esta figura, responder las siguientes preguntas:

**Pensamos, escribimos y discutimos:**

- ¿Cuánto tiempo dura el tránsito?
- ¿Cuánto tiempo dura el ingreso y egreso del planeta?

HAT-P-7 es una estrella de tipo espectral F8 que se encuentra a una distancia de 344 pc de la Tierra en la constelación de Cygnus. Tiene una masa de 1.5 $M_\odot$, un radio de 1.84 $R_\odot$ y una temperatura de 6440 K. Esto quiere decir que es una estrella un poco más masiva, un poco más caliente y un poco más grande que nuestro Sol. Cuando su planeta transita en frente de ella, bloquea una parte de la luz. Fijándonos en la curva de fase,

**Pensamos, escribimos y discutimos:**

- ¿Cuánto flujo estelar es bloqueado por el planeta a la mitad del tránsito? Vamos a llamar a esta cantidad $\Delta F$.
- Visualmente, a qué parte de la órbita corresponde este punto? Fíjense otra vez en la figura 10.1
- ¿Qué pasaría si el planeta fuera más grande? ¿O más chico?
- ¿Cómo podríamos determinar el radio del planeta?

¡Ahora ya tenemos todos los elementos necesarios para calcular el radio del planeta usando proporciones!

*Si el flujo de la estrella fuera 1, cuando el planeta transita en frente, decrece una cantidad* $\Delta F$*?. Dado que el área de la estrella es* $\pi R_\star^2$*, ¿cuál es el radio del planeta?*

Combinando estudios de velocidad radial de la estrella HAT-P-7 con el periodo hallado a través del tránsito, los astrónomos encontraron que la separación entre HAT-P-7b y su estrella es de 0.038 UA. ¡Esto es 10 veces más cerca de lo que Mercurio está del Sol!

**Pensamos, escribimos y discutimos:**

- ¿Cuál es el radio de HAT-P-7 b en unidades del radio de Júpiter? $1\,R_J = 0{,}1\,R_\odot$.
- ¿Cómo imaginas al planeta HAT-P-7 b en términos de temperatura? ¿Creen que puede existir agua líquida en su superficie?

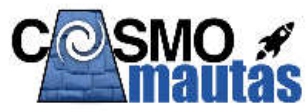

## Conclusión y evaluación (10 min)

Al terminar, hablaremos sobre cómo los astrónomos obtienen datos y calculan fotometría para obtener la curva de luz con la que trabajamos hoy.

## Reflexionando (5 min)

Discutir en grupo:

- ¿Qué fue lo que les impresionó más de este ejercicio?
- ¿Qué modificaciones le harían para implementarlo en su salón de clases?

## Referencias

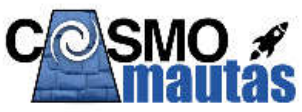

# 11. API 5: La Vía Láctea

## Orientándonos (5 min)

Todo lo que hemos estudiado en las sesiones anteriores, el sistema solar, estrellas, exoplanetas, son objetos que han sido observados todos dentro de nuestra galaxia, la Vía Láctea. Desde la Tierra, podemos observar a la Vía Láctea como una larga banda luminosa que atraviesa el cielo, prácticamente partiéndolo en dos, producto de una inmensa densidad de estrellas. Además de la luz, se puede reconocer sombras superpuestas, producto de polvo interestelar que impide el paso de luz de las estrellas.

Desde inicios del siglo XX se sabe que nuestra galaxia es tan solo una de muchas. En base a observaciones, se ha calculado que hay billones de otras galaxias en el universo. A pesar de que la Vía Láctea es nuestra galaxia, donde se encuentra el sistema solar y el planeta en el que vivimos, sabemos en realidad relativamente poco sobre ella. Resulta que sabemos mucho más sobre otras galaxias vecinas como Andrómeda o las nubes de Magallanes, o incluso sobre galaxias muy lejanas. La razón es porque, al encontrarnos dentro de la galaxia, nuestra perspectiva no nos permite tener una imagen completa de ella. Es como cuando nos encontramos en un punto muy alto de la cordillera, y desde lejos podemos ver la forma y tamaños de los pueblos del valle. Sin embargo, si estamos en una calle dentro del pueblo, sin poder movernos, es práctica-

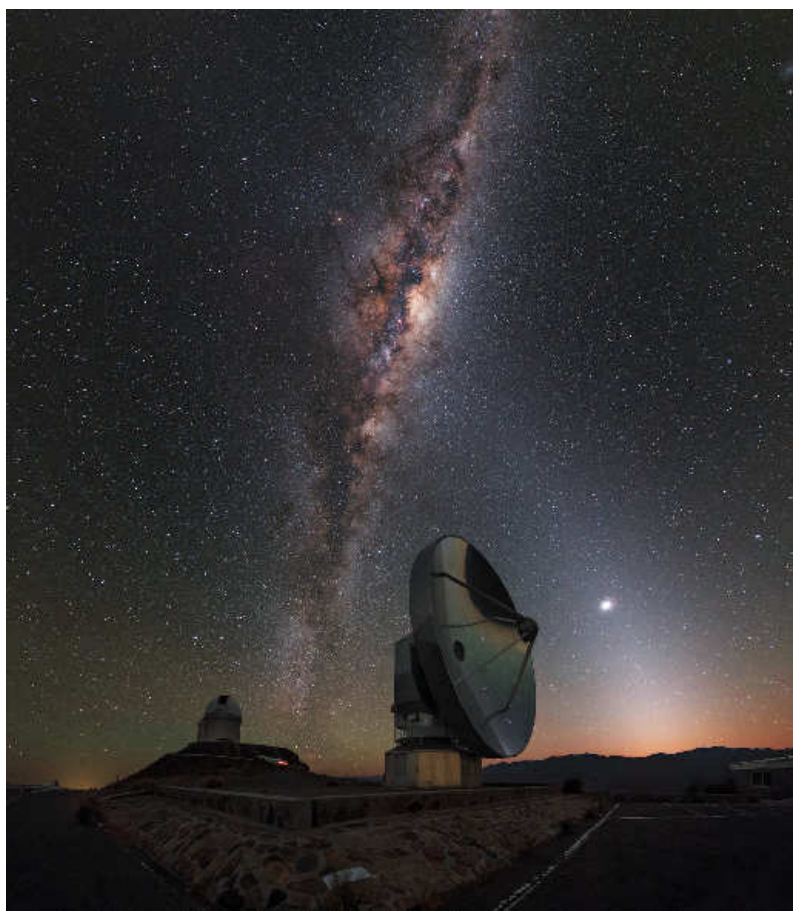

Figura 11.1: Fotografía de la Vía Láctea vista desde el observatorio La Silla en el desierto de Atacama, Chile. Crédito: ESO

Imagen de encabezado: Extracto de la vista del cielo del satélite Europeo Gaia. Este es el mapa más grande, preciso y profundo de nuestra Vía Láctea (aproximadamente 1.8 miles de millones de estrellas). Crédito: Gaia DR3, ESA mente imposible averiguar la forma del mismo. En esta actividad, vamos a investigar qué forma tiene nuestra galaxia, cómo se mueve, cómo se ve nuestro vecindario extragaláctico y finalmente averiguar de cuánta materia está compuesta nuestra galaxia. Empezamos observando a la Vía Láctea una noche desde el cielo peruano, gracias a un video de Astrofoto Perú: `https://www.youtube.com/watch?v=0UvzinzYAls`

**Unidades de aprendizaje incluidas en este API**
Física: Velocidades positivas y negativas
Física: El efecto Doppler
Matemáticas: Regla de tres
Física: La ley de gravedad de Newton
Matemáticas: Dibujar un gráfico con ejes x, y

## Objetivos y preguntas

El objetivo de este experimento es usar lo que observamos en el cielo nocturno para responder las siguientes preguntas:

A ¿Qué forma tiene nuestra galaxia, y cómo se mueve?
B ¿Cómo es nuestro vecindario extragaláctico?
C ¿Cuánta materia contiene nuestra Vía Láctea? (opcional)

## Materiales

- Un CD o un disco cualquiera.
- Dos palitos de helado
- Plumones rojos y azules
- Hojas de fotografías de galaxias: impresión artística de la Vía Láctea, tarjetas de la Gran Nube de Magallanes, Andrómeda, M87 y GN11 (adjuntas al final de la actividad).
- Cartón para la base de rotación
- Gancho para hacer girar el modelo

## A. ¿Qué forma tiene nuestra galaxia, y cómo se mueve? Nivel 1

### Hipótesis y diseño del experimento (20 min)

- En base a nuestra observación de la Vía Láctea (similar a la Figura 11.1), propongamos ideas sobre cómo podríamos empezar a investigarla.

- Gracias a observaciones de otras galaxias, sabemos que la mayoría de galaxias regulares pueden tener forma de un huevo o pelota (galaxias elípticas) o de disco (galaxias espirales). Basándonos en cómo vemos a la Vía Láctea desde la Tierra, y usando los modelos que tenemos aquí, imaginémonos que somos una partícula dentro del huevo o del CD (disco), y discutamos. ¿Nuestra galaxia tiene forma elíptica o espiral?

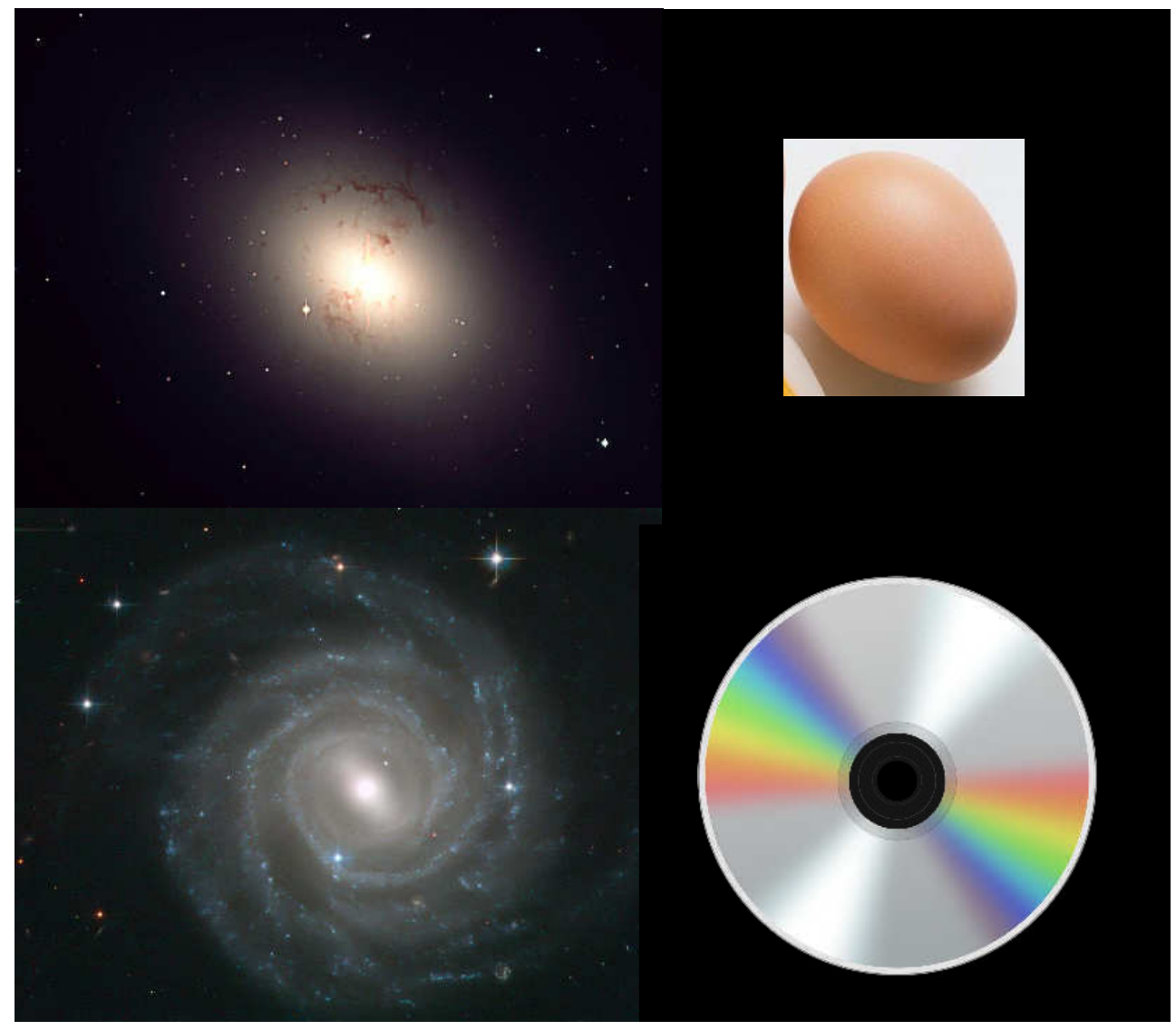

- Aunque no podemos observar a nuestra galaxia desde fuera, desde nuestra perspectiva sí podemos observar a las estrellas vecinas. ¿Creen que podemos entender el movimiento y estructura de toda la galaxia tan solo observando el movimiento de las otras estrellas?

El efecto Doppler (ver Caja 4.2.2) nos permite detectar si una estrella se acerca o si se aleja respecto al lugar en el que estamos. Si la estrella se acerca, su onda de luz se 'comprime' y la luz se ve mas azul, mientras que si se aleja, su onda de luz se 'extiende' y su luz se ve más roja.

- Ahora queremos entender su movimiento. Si nuestra galaxia es un disco que vemos de lado, podemos usar el CD como modelo 3D para entender mejor nuestra perspectiva y como sería la rotación del mismo.

- Para entender mejor esa rotación, construiremos un modelo simple de la vista de nuestra Vía Láctea en dos dimensiones usando un palito de helado y estrellas o puntos negros dibujadas sobre el, como vemos en la imagen.

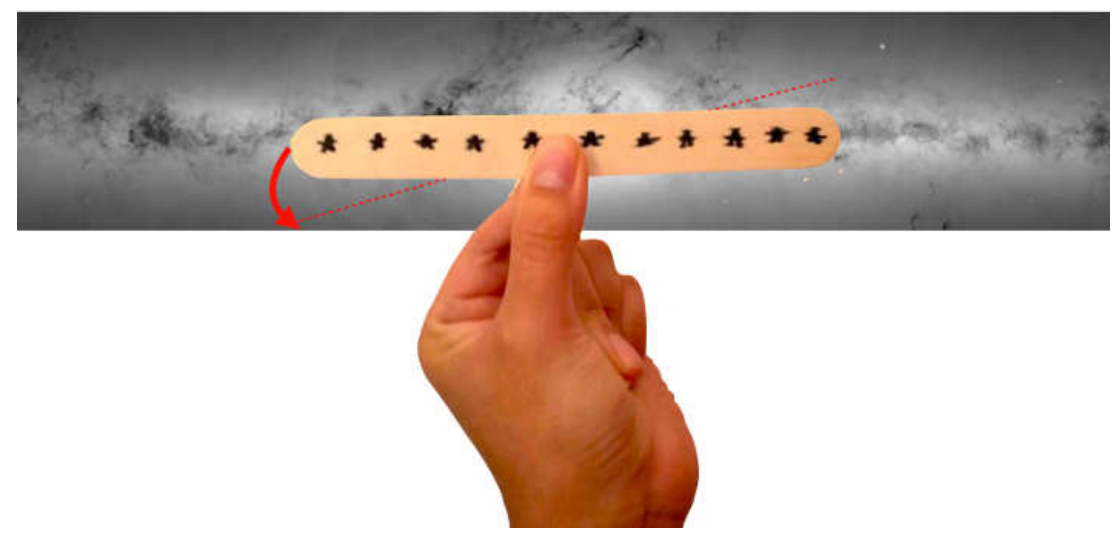

- Sujetemos el palito (horizontal) y hagámoslo girar muy lentamente respecto a su centro (solo unos pocos grados *en dirección anti-horaria*, visto desde arriba). Ahora observemos a las estrellas dibujadas en el palito, ¿en qué dirección respecto a nosotros se mueven las estrellas cuando el palito rota, las estrellas se alejan o se acercan? Marquemos con un circulo azul a las estrellas que tienen

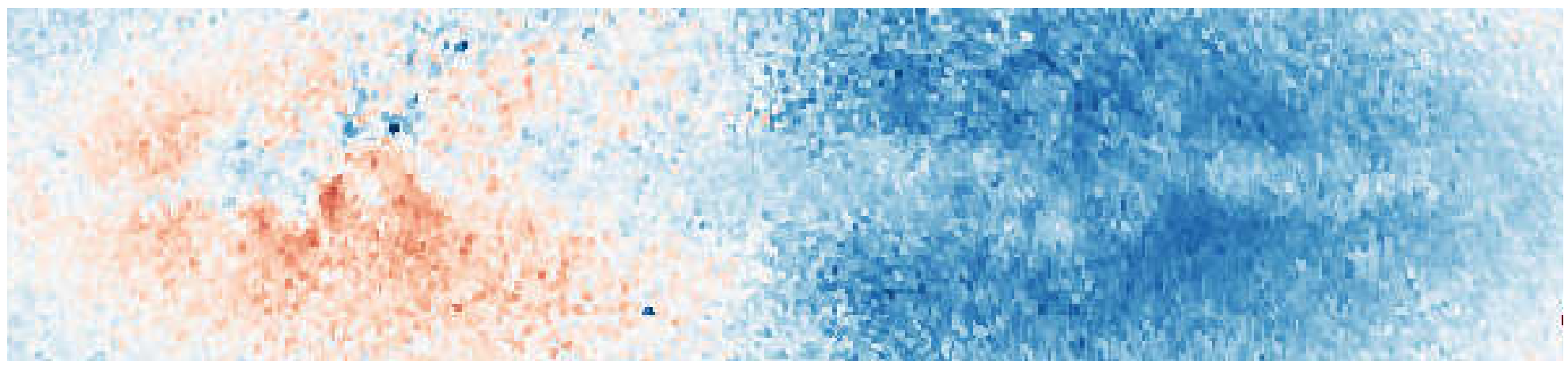

Figura 11.2: En esta imagen se muestra una parte (la región interna de la Galaxia) del mapa de las velocidades radiales de las estrellas, vistas desde la posición del Sol. El centro de la imagen corresponde al centro galáctico. Las estrellas que se acercan a nuestra perspectiva están codificadas con el color azul y las que se alejan con rojo. Crédito: Imagen modificada de ESA/Gaia/DPAC, CC BY-SA 3.0 IGO

velocidades negativas (aquellas que se acercan) y con un circulo rojo a las estrellas con velocidades positivas (aquellas que se alejan). ¿Cómo están distribuidas las velocidades en el palito?

- Ahora con un segundo palito, hagamos lo mismo pero girándolo en *dirección horaria* (visto desde arriba).

### Experimento (8 min)

Además de tomar fotografías profundas de la Vía Láctea, como la que vemos en la imagen del encabezado del capítulo, el telescopio Gaia también ha recolectado los espectros de todas las estrellas. Usando estos datos y el efecto Doppler se han medido las velocidades radiales (respecto al Sol) de las estrellas, como podemos apreciar en la Figura 12.2. De forma similar a como diseñamos en nuestros modelos, las estrellas de velocidad negativas (que se acercan) son mostradas como puntos azules y las estrellas de velocidades positivas (que se alejan) son mostradas como puntos rojos.

- Discutamos el patrón en las velocidades. ¿Cómo se comparan los datos observativos con el modelo de palito de helado que construimos anteriormente? Podemos concluir en base a estos datos si la rotación de nuestra galaxia es horaria o anti-horaria (vista desde arriba)? En base a nuestro modelo y los datos ¡hemos averiguado en qué dirección rota nuestra galaxia!

### B. ¿Cuáles son las distancias a otras galaxias? (20 min) Nivel 1

Ahora que confirmamos con los datos que nuestra Vía Láctea es un disco que gira, haremos una maqueta de la Vía Láctea, de su vecindario y de otras galaxias interesantes.

- Cortar un cuadrado de cartón (20*20cm) como base.

- Identifiquemos las hojas que contienen la impresión artística de nuestra galaxia (como creemos que se vería desde afuera) y las otras galaxias de referencia. Peguemos las hojas sobre cartulinas, para obtener tarjetas más estables.

- Recortar la imagen de nuestra Vía Láctea (ahora pegada a la cartulina) donde está indicado por lineas punteadas blancas.

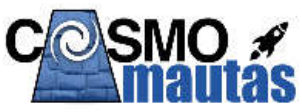

- Ahora construiremos la maqueta de nuestra galaxia. Con una aguja o un lapicero perforemos el centro del modelo. Centrémoslo en el cartón y perforemos también el cartón. Ahora insertemos el gancho adjunto.

- Tenemos nuestro modelo y lo podemos hacer rotar en la dirección que hemos aprendido más arriba. Identifiquemos dónde nos encontramos nosotros.

Hagamos ahora un modelo de nuestro vecindario extragaláctico y galaxias aún más lejanas. En nuestra maqueta la Vía Láctea mide aproximadamente 20 cm de diámetro. En realidad el diámetro de la Vía Láctea es de aproximadamente 120 000 años luz.

- Usemos la regla de tres para calcular ¿a cuántos años luz equivale 1 cm en la escala de nuestro modelo?

$$
\begin{aligned}
20\,\text{cm} &\rightarrow 120000\,\text{años luz} \\
1\,\text{cm} &\rightarrow X \\
X &= \frac{120000\,\text{años luz} \times 1\,\text{cm}}{20\,\text{cm}}
\end{aligned}
$$

$$
\begin{aligned}
1\,\text{cm} &\rightarrow \underline{\hspace{3cm}}\text{años luz} \\
1\,\text{m} &\rightarrow \underline{\hspace{3cm}}\text{años luz} \\
1\,\text{km} &\rightarrow \underline{\hspace{3cm}}\text{años luz}
\end{aligned}
$$

- Ahora leamos las distancias de las galaxias de las tarjetas y hagamos los cálculos de la tabla usando la regla de tres. ¿A qué distancias tendríamos que colocar a las otras galaxias para hacer una maqueta a escala. Situemos las tarjetas de las galaxias a la distancia que corresponden (1 metro equivale a la distancia del pecho a la mano).

| Galaxia | Distancia real (años luz) | Distancia en maqueta (cm, m, o km) |
|---|---|---|
| Gran nube de Magallanes | | |
| Andrómeda | | |
| M87 | | |
| GN-z11 | | |

- Considerando que la distancia de la Tierra al Sol (1 UA) es de 8.3 minutos luz, en la escala de nuestra maqueta esta distancia equivale a $2{,}63 \times 10^{-11}$ metros, ¡es decir más pequeño que un átomo! Reflexionemos entonces sobre lo que nuestra maqueta nos enseña sobre el espacio extragaláctico.

## Conclusión y evaluación (10 min)

Repasaremos todo lo que hemos descubierto, les contaremos brevemente la historia del estudio de la Vía Láctea, observaciones en diversas longitudes de onda, mencionando también el secreto que tiene escondido en su centro. Finalmente les contaremos sobre *Gaia* y las preguntas abiertas que aún quedan sobre la Vía Láctea.

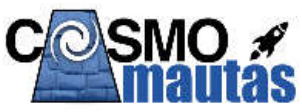

Discutir en grupo:

- ¿Qué fue lo que les impresionó más de este ejercicio?

- ¿Qué modificaciones le harían para implementarlo en sus respectivos salón de clases?

## Referencias

- Imágenes de *Gaia* DR3 (mapa) y *Gaia* DR2 (velocidades radiales) [Crédito ESA].
- Datos de la curva de rotación de la Vía Láctea son de una medición reciente usando *Gaia* DR2, *HST* y APOGEE de Eilers et al. 2018.

---

En las siguientes páginas mostramos el modelo de nuestra Vía Láctea y las tarjetas de galaxias selectas y su distancia. [Crédito: Wikipedia/ESO/VMC Survey/David Dayag/Hubble (NASA/ESA)/P. Oesch, G. Brammer]

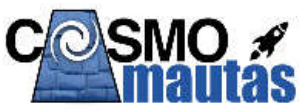

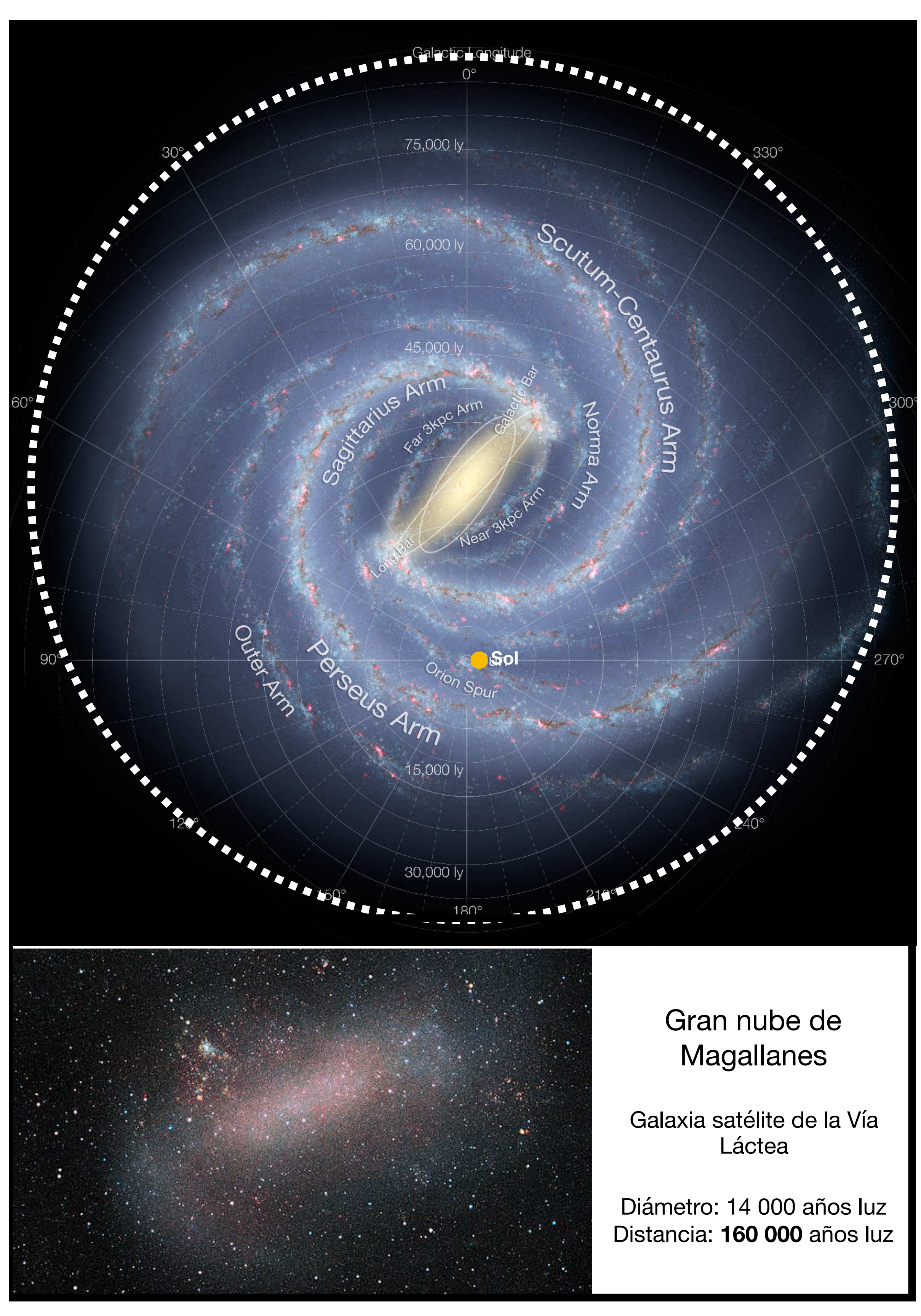

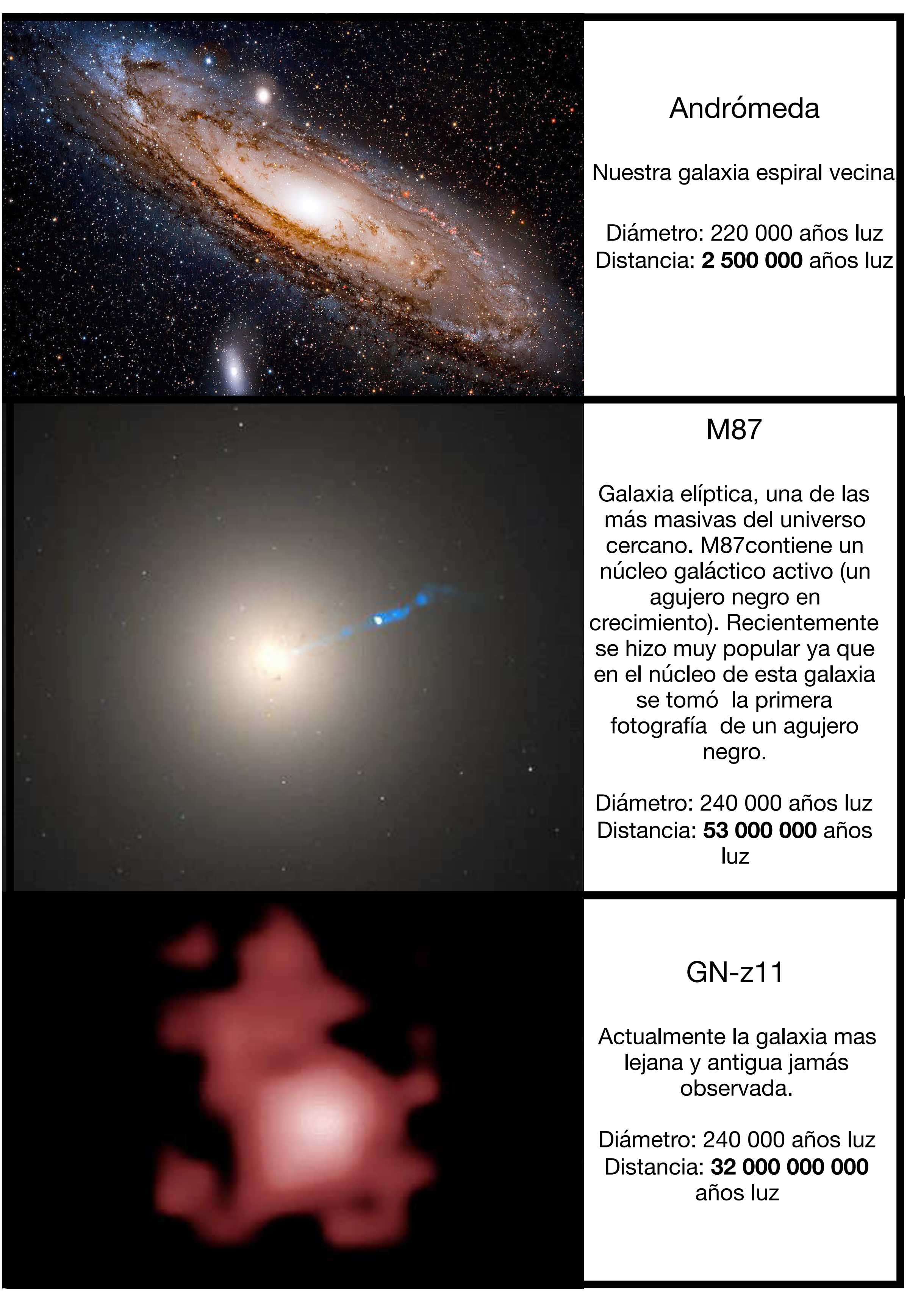
Andrómeda
Nuestra galaxia espiral vecina
Diámetro: 220 000 años luz
Distancia: **2 500 000** años luz
M87
Galaxia elíptica, una de las más masivas del universo cercano. M87contiene un núcleo galáctico activo (un agujero negro en crecimiento). Recientemente se hizo muy popular ya que en el núcleo de esta galaxia se tomó la primera fotografía de un agujero negro.
Diámetro: 240 000 años luz
Distancia: **53 000 000** años luz
GN-z11
Actualmente la galaxia mas lejana y antigua jamás observada.
Diámetro: 240 000 años luz
Distancia: **32 000 000 000** años luz

## C. ¿Cuánto pesa nuestra Vía Láctea? (opcional) Nivel 3

### Hipótesis

Ahora sabemos que nuestra Vía Láctea tiene forma de disco, el cual vemos de lado. En base a lo que aprendimos con los planetas del sistema solar, sabemos que la **velocidad orbital** de una estrella orbitando lejos de la galaxia sería proporcional a la raíz cuadrada de la masa total de la galaxia en torno a la que gira, dividida entre la distancia de la estrella hacia el centro. Esto se puede calcular usando la ley de gravedad de Newton y la fuerza centrípeta.

$$F_{centripeta} = F_{gravedad}$$
$$m\frac{v^2}{R} = Gm\frac{M}{R^2}$$
$$\frac{v^2}{R} = G\frac{M}{R^2}$$
$$v^2 = G\frac{M}{R}$$
$$v = \sqrt{G\frac{M}{R}}$$

donde M es la masa de la Galaxia, la constante gravitacional $G$, y R la distancia (radio) hacia el centro galáctico. Tomando en cuenta las unidades que deseamos, esto se simplifica a:

$$v = 2\times 10^{-3}\sqrt{\frac{M}{R}}$$

Podemos calcular esta cantidad usando la constante gravitacional $G$, el valor de una masa solar en kg, y el valor de un kpc en km.

- MODELO 1 : ¿Cómo serían las velocidades de las estrellas si la masa de la galaxia estuviera toda concentrada solo en el centro, como los planetas que orbitan alrededor del Sol?

| Distancia (kpc) | Masa central ($M_\odot$) | Velocidad (km/s) |
|---|---|---|
| 2.5 | $4\times 10^{10}$ | |
| 5 | $4\times 10^{10}$ | |
| 7.5 | $4\times 10^{10}$ | |
| 10 | $4\times 10^{10}$ | |
| 15 | $4\times 10^{10}$ | |
| 25 | $4\times 10^{10}$ | |

- Dibujemos la distribución de velocidades (curva de rotación) en la Figura 1.

- ¿Cómo cambiar la velocidad en función a la distancia del centro? ¿Qué pasa si una estrella está en la parte central? ¿Qué pasa si está lejos?

### Experimento

- Ahora trabajemos con los datos. Usando el telescopio *Gaia* se han medido la distribución de velocidades de estrellas. Leyendo esta tabla, grafiquemos en la Figura 11.4 la distribución real de las velocidades en función de su distancia al

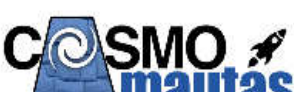

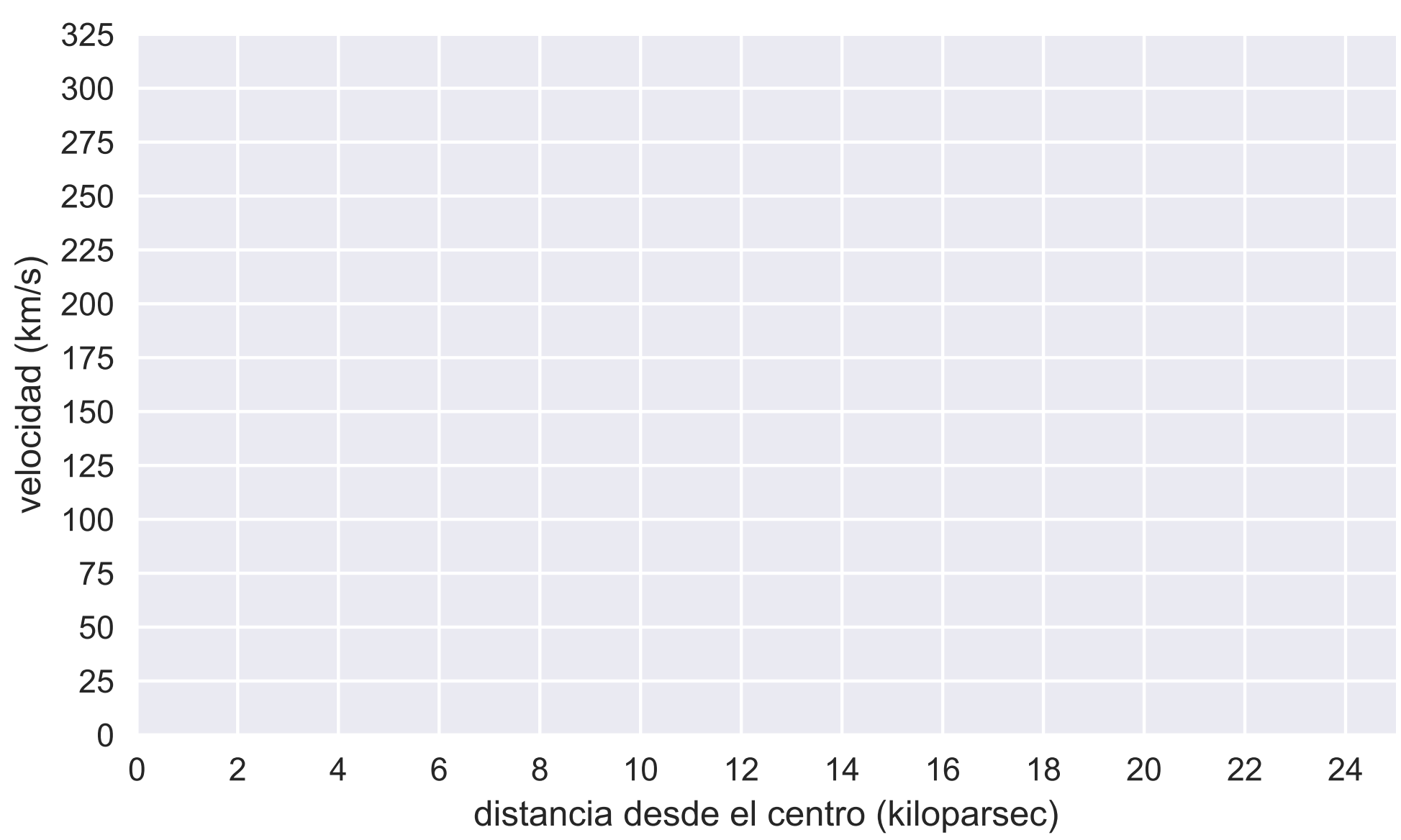

Figura 11.3: Gráfico de las velocidades orbitales de las estrellas en función a su distancia del centro galáctico.

centro. Usemos un color diferente al que usaste para el modelo 1.

| Distancia ( kiloparsec) | Velocidad ( km/s) |
|---|---|
| 0 | 0 |
| 1 | 240 |
| 2.5 | 210 |
| 5.3 | 226 |
| 9.3 | 226.2 |
| 13.2 | 220.92 |
| 17.3 | 216.14 |
| 21.2 | 195.3 |
| 24.8 | 198.2 |

- En base a este gráfico, ¿cuál es la velocidad con la que el Sol (y por ende nosotros), nos movemos alrededor del centro galáctico? Tengan en cuenta que la distancia desde el Sol al centro galáctico es 8 kpc.

### Análisis e interpretación (20 min)

- Comparemos el modelo 1 con las mediciones de *Gaia*. ¿Es nuestro modelo correcto?

- Describamos las diferencias entre nuestro modelo y las mediciones.

- ¿Qué nos dicen estas diferencias sobre cuánta masa hay en la parte más central del disco? ¿Y la parte más lejana del disco?

- Calculemos, en base a la ecuación 1 y usando la velocidad observada a 25 kpc,

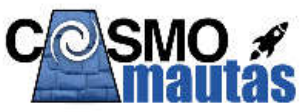

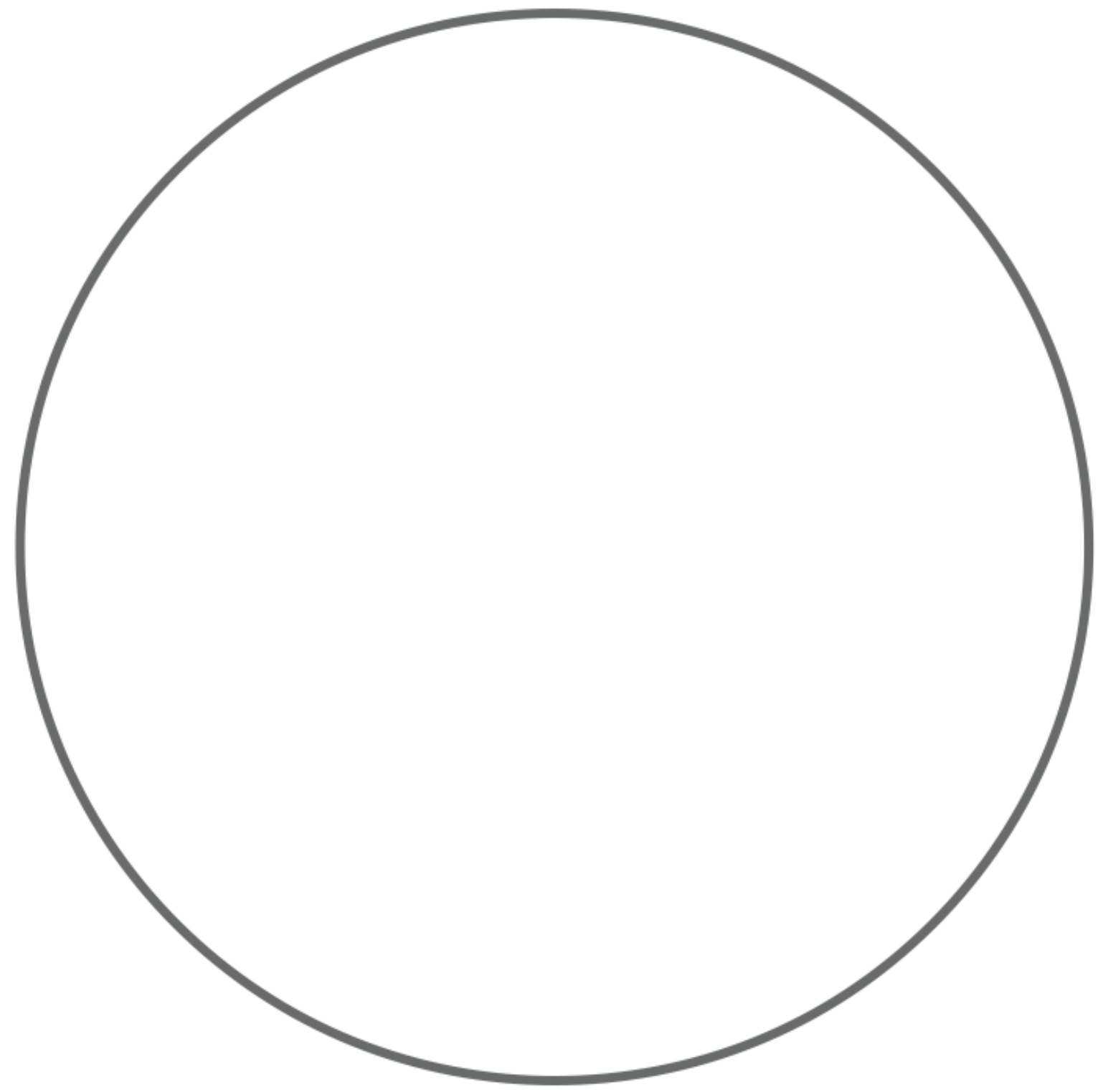

Figura 11.4: Gráfico circular (o de torta) de las porciones equivalentes a la masa total de las estrellas, la masa del gas y del polvo.

¿cuál es la masa total de la galaxia?

- Dibujemos en el siguiente gráfico circular los porcentajes de la masa total que ocupan las estrellas ($6 \times 10^{10}$ masas solares), y la masa del gas ($8 \times 10^{9}$ masas solares) y polvo ($8 \times 10^{8}$ masas solares) observados.

- Como podemos apreciar, hay una gran porcentaje de la masa total de la galaxia que no podemos ver o explicar, pero que es evidente por la curva de velocidades de las estrellas en la Galaxias. ¡Acabamos de descubrir la materia oscura!

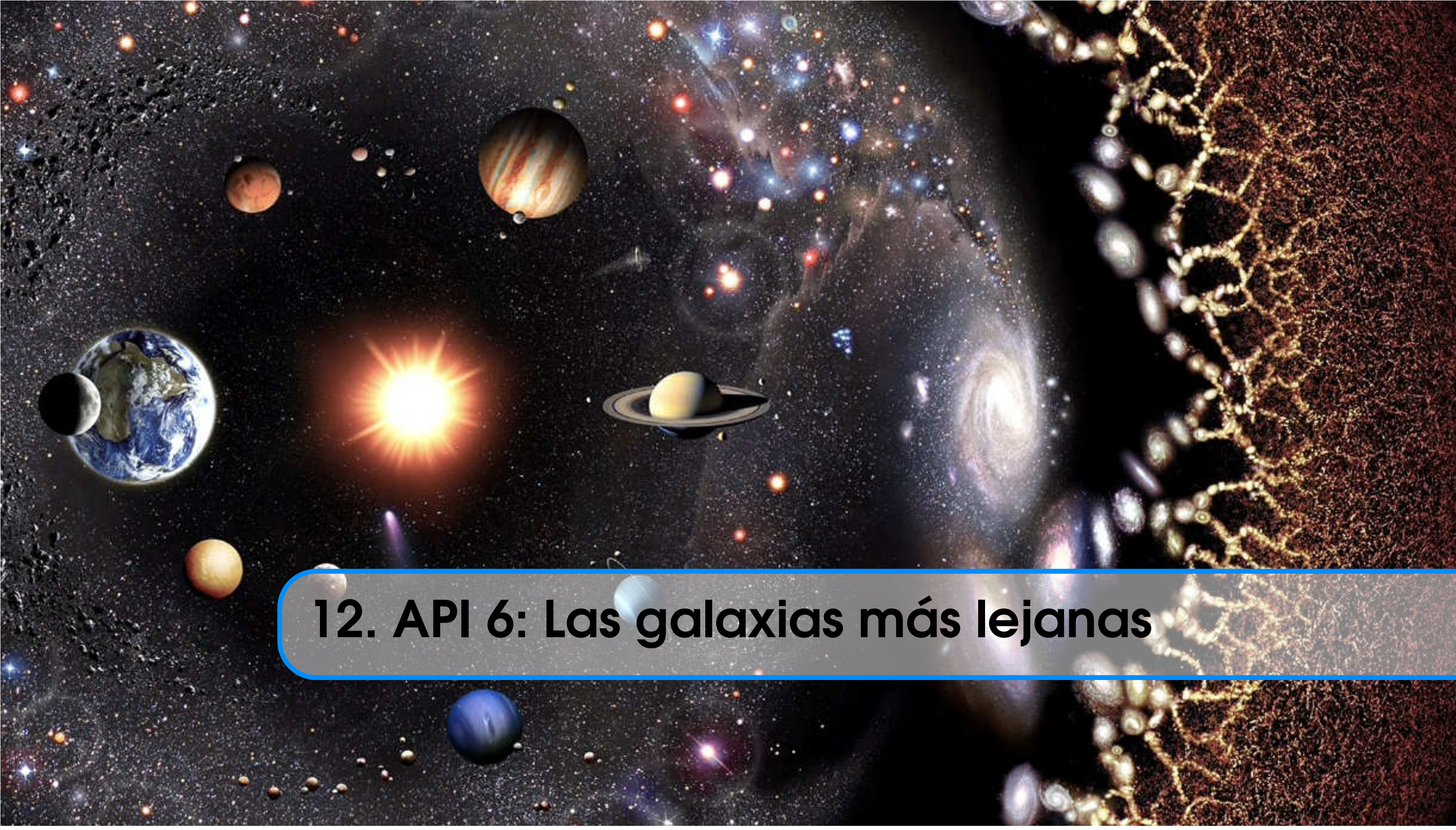

# 12. API 6: Las galaxias más lejanas

## Orientándonos (5 min)

Imagínense que en una noche despejada se encuentran observando un cielo lleno de estrellas. Si usan un telescopio, además de ver muchas estrellas centelleantes también podrán ver muchas galaxias. Sin embargo, incluso con los telescopios más potentes que tenemos sobre la Tierra, encontraremos aún muchos espacios profundamente oscuros y aparentemente vacíos de objetos celestes.

El Campo Ultra-profundo del telescopio *Hubble* (HUDF, por sus siglas en inglés) es justamente la fotografía de uno de esos espacios aparentemente vacíos. Para obtener esta imagen, los astrónomos enfocaron al telescopio *Hubble*, el telescopio espacial óptico más potente que tenemos, en un área muy pequeña seleccionada por ser completamente oscura. Para lograr una imagen tan profunda tuvieron que integrar, es decir recolectar luz, por casi 40 horas. Imagínense la sorpresa que se llevaron cuando al ver la fotografía de ese espacio "vacío" encontraron esta gran cantidad de objetos.

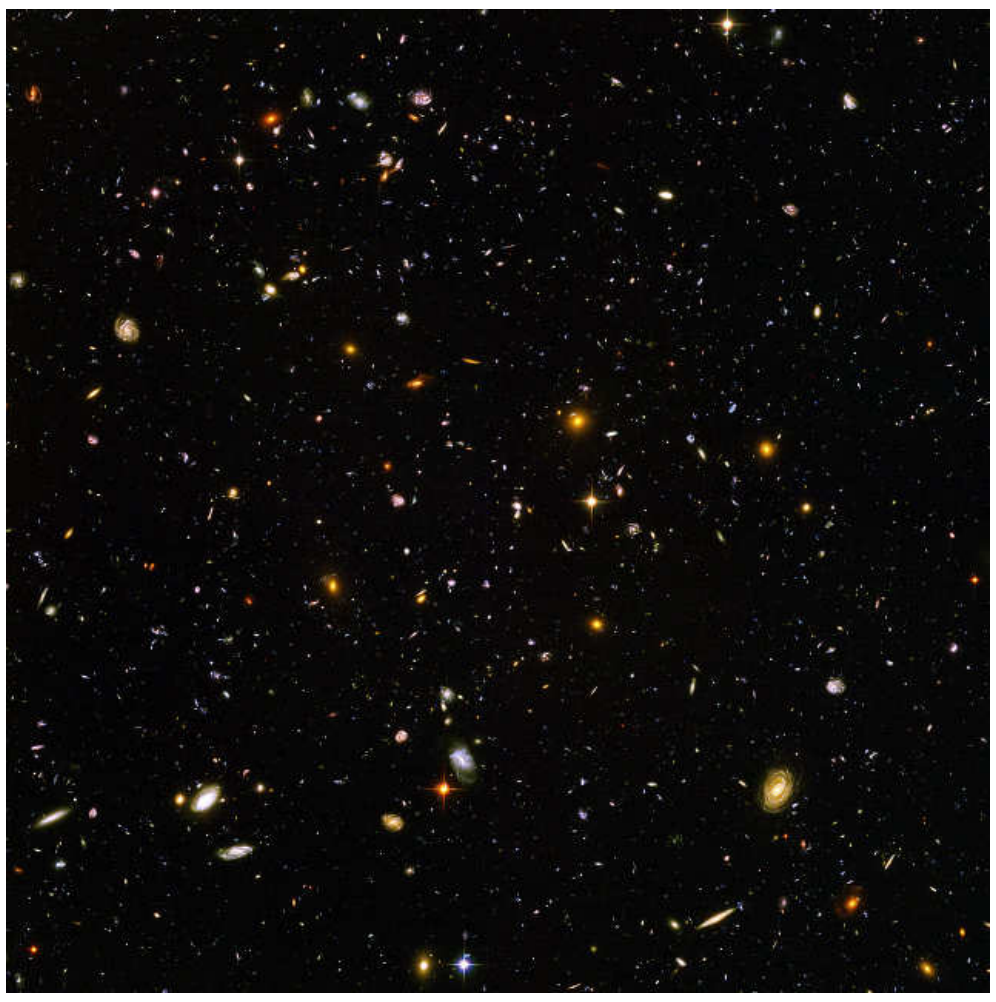

Figura 12.1: Campo Ultra Profundo de *Hubble* [Crédito: NASA].

¿Qué creen que son esos objetos? En efecto, es aún más impresionante pensar que ninguno de estos objetos es una estrella de nuestra Vía Láctea, sino que se trata de miles de galaxias, muchas de las cuales son tan lejanas que su luz empezó el viaje hacia nosotros en las épocas cercanas al inicio del universo. Gracias a esta fotografía sabemos que en cualquier área del cielo, si es que observamos durante suficiente tiempo, encontraremos una densidad parecida de galaxias, y gracias a esta podemos calcular que el universo está compuesto de miles de millones de galaxias. Estas galaxias son de todo tamaño y forma, unas más pequeñas y otras mucho más grandes que nuestra Vía Láctea, unas menos y otras mucho más luminosas que la nuestra. Si consideramos que cada galaxia lleva en su centro un agujero negro supermasivo, podemos entonces imaginar también cuántos agujeros negros hay escondidos cada vez que miremos el cielo. El universo profundo esconde muchísimos secretos, y en esta actividad vamos a explorar algunos de ellos.

**Unidades de aprendizaje incluidas en este API**
Física: Velocidad
Matemáticas: leer un gráfico con eje x e y
Matemáticas: Identificar correlaciones
Física: Efecto Doppler

## Materiales

- 1 Elástico blanco de 20 cm.
- 1 Globo blanco
- Notas adhesivas
- Marcador dos colores, rojo y negro
- Limpiatipo
- Foto del Fondo Cósmico de Microondas (CMB)
- Fotos pequeñas de galaxias
- Poster del HUDF
- Cartón para base de maqueta
- Palitos de brocheta
- Cinta adhesiva

## Objetivos y preguntas

Ahora bien, aunque vemos esta gran densidad de galaxias en el plano del cielo, naturalmente esta es solo una fotografía en dos dimensiones. Sin embargo, estas galaxias también están distribuidas diversamente en una tercera dimensión: aunque todas son galaxias bastante lejanas, algunas se encuentran más cerca a nosotros que otras. ¿Cómo podemos averiguar las distancias a estas galaxias?

El objetivo de este experimento es explorar esta fotografía y con ella tratar de entender distancias y escalas en el universo. Estas son las preguntas que exploraremos:

A ¿Cómo se miden distancias en el universo?
B ¿Cómo podemos calcular las distancias a estas galaxias lejanas?

## Hipótesis y diseño del experimento (20 min)

**A. Midiendo distancias sin movernos** Nivel 2

Antes de explorar las grandes distancias y profundidades que nos ofrece el campo profundo del telescopio *Hubble*, es importante entender cómo medimos distancias en astronomía. Vamos a empezar haciendo un modelo de medición a escalas pequeñas, dentro de nuestras casas. El principal desafío para las mediciones en el espacio es que no podemos movernos desde nuestro punto de observación. Observen alrededor y encuentren un ángulo en donde puedan identificar tres puntos a diferentes distancias desde donde se encuentran.

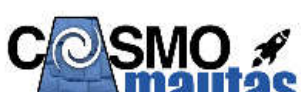

- ¿Cómo podríamos estimar las distancias hasta estos tres puntos sin tener que movernos?
- Coloquen las notas adhesivas grandes en cada una de esas tres distancias y regresen a su posición inicial. ¿Cómo cambian los tamaños aparentes de las notas adhesivas dependiendo de la distancia a la que lo colocaron?
- Ahora, ¿qué pasa si usan notas de diferentes tamaños? ¿Todavía pueden obtener información sobre estas distancias?

De forma similar al tamaño de las notas, el cambio de la intensidad de la luz también nos indica si un objeto está más lejos o más cerca que otro. Sin embargo, tal como con las notas adhesivas, esto solo es útil si es que la intensidad intrínseca de los objetos que usamos para comparar son idénticas y las conocemos a detalle. Los objetos luminosos de los cuales conocemos la intensidad a detalle y podemos utilizarlos como herramientas de medición se llaman **candelas estándar**.

- En base a lo que hemos aprendido en la sesión de estrellas, ¿creen que podemos usar estrellas como candelas estándar?

### B. Midiendo distancias de galaxias Nivel 3

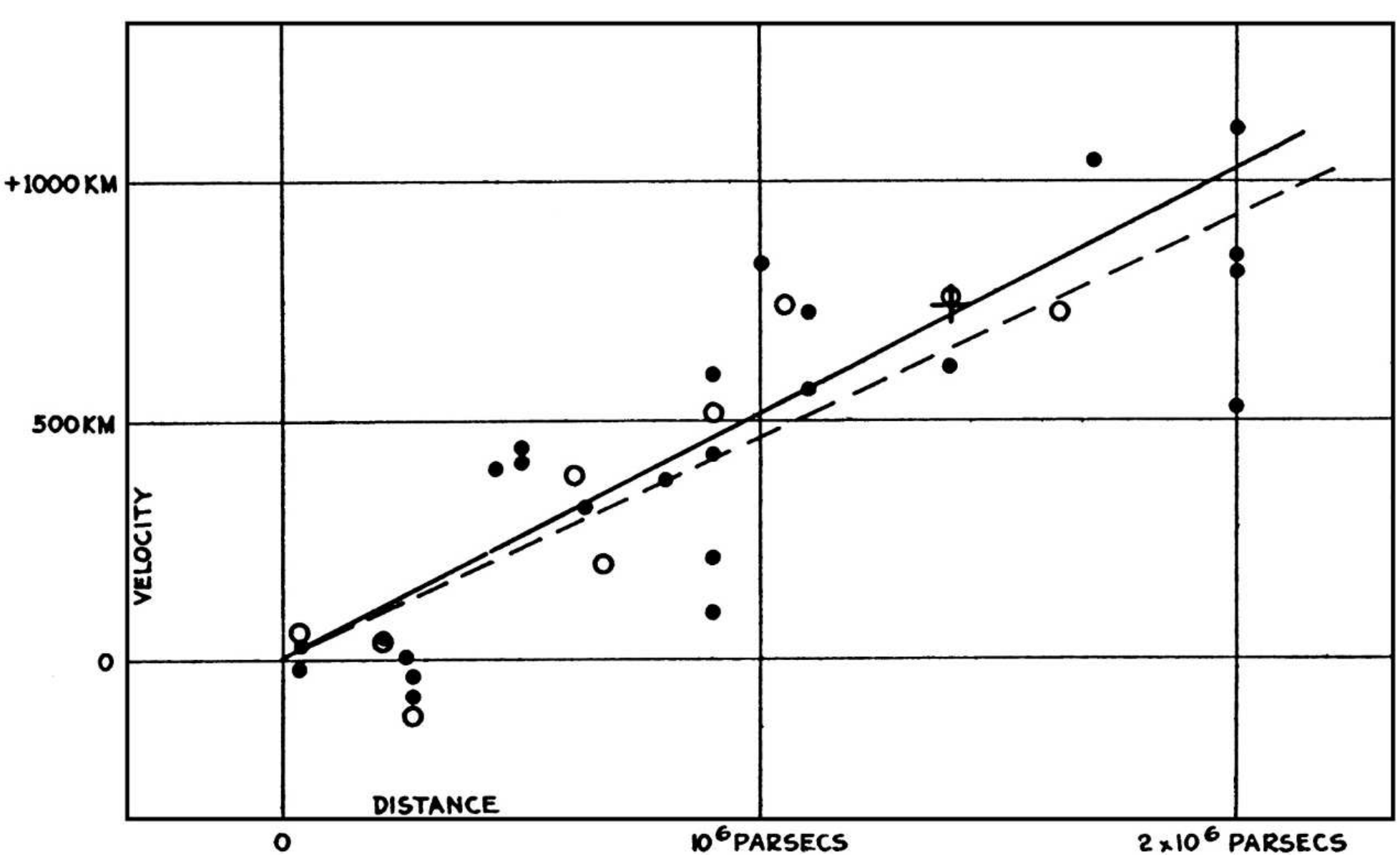

Figura 12.2: Medición de E. Hubble. Cada punto representa una galaxia, y se grafica la velocidad estimada (eje y) en función a la distancia a la que se encuentra (eje x).

Las cefeidas son un tipo particular de estrellas variables, es decir que varían su luminosidad en un respectivo periodo de tiempo. Gracias a la astrónoma Henrietta Levitt, quien descubrió en 1912 una relación exacta entre el periodo de variación y la luminosidad de las cefeidas, estas estrellas son hasta el día de hoy una gran herramienta para medir distancias. Este descubrimiento permitió que en 1925 Edwin Hubble pueda medir la distancia hacia la galaxia Andrómeda, confirmando por primera vez que Andrómeda era un objeto fuera de la Vía Láctea: una galaxia en sí misma. Este descubrimiento fue fundamental, ya que marcó el inicio de la astronomía extragaláctica.

Pero E. Hubble no se detuvo allí, sino que al localizar cefeidas en otras galaxias un poco más lejanas, también pudo calcular las distancias a diferentes galaxias. Además de las distancias, Hubble midió las velocidades de estas galaxias, usando el efecto Doppler en los espectros obtenidos de ellas. Éstas son las mediciones que Hubble reportó en 1929:

- ¿Qué relación encuentran entre la distancia de una galaxia y su velocidad? ¿Cómo es la velocidad de una galaxia cercana y cómo se compara con la velocidad de una galaxia lejana?
- **Modelos:** Ahora tratemos de entender esto usando un modelo. Usemos un elástico que vamos a alinear junto a una regla y dibujemos cuatros puntos sobre él con el plumón: un punto rojo a 0 cm y los otros puntos negros, el primero a $x_0 = 5$ cm, el segundo a $x_0 = 10$ cm y el tercero a $x_0 = 15$ cm. Con una mano sujetamos el elástico desde el primer punto rojo.

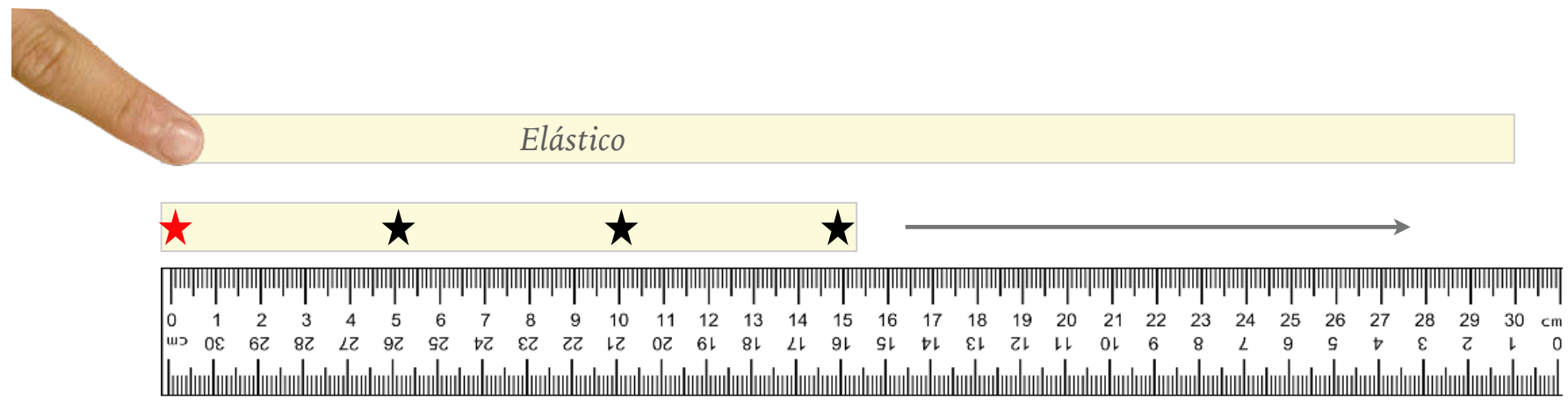

- Durante un segundo estiremos el elástico hasta que el tercer punto llegue a 30 cm. Lean las posiciones finales x1 de los puntos y llenen la tabla.

| Posición inicial $x_0$ [cm] | Posición final $x_1$ [cm] | Distancia $x_1 - x_0$ [cm] | Tiempo [s] | Velocidad [cm/s] |
|---|---|---|---|---|
| 5 | | | 1 | |
| 10 | | | 1 | |
| 15 | | | 1 | |

- ¿Cuánta distancia ha recorrido cada punto en un segundo? Escriban la distancia recorrida en cm y la velocidad promedio en cm/segundo.
- ¿Cómo se comparan la velocidad del punto más cercano y la del punto más lejano?
- En la superficie del globo desinflado, dibujen un punto rojo y muchos puntos negros. Al inflarlo pueden ver el mismo efecto del elástico (2D), esta vez en tres dimensiones.
- Discutan cómo estas dos observaciones (elástico y globo) podrían estar relacionadas a la medición de Hubble. ¿Qué nos dice esta comparación acerca del universo?

## Experimento (25 min) Nivel 2

Gracias a la medición de Hubble tenemos un modelo sobre cómo medir las distancias a galaxias lejanas a partir de la 'velocidad aparente' con las que estas se alejan de nosotros. Pero notemos que las galaxias en sino tienen una velocidad, si no que se alejan de nosotros a raíz de la expansión del universo. Esta expansión del universo afecta los espectros de las galaxias, corriendo los espectros a mayores longitud de onda (por ejemplo de la longitud de onda azul a roja), de forma análoga a como lo aprendimos por el efecto Doppler . Esto se denomina “corrimiento al rojo cosmológico” o ‘*redshift*”, por su nombre en inglés, y está esquematizado en la Figura 12.3. En base a observaciones de los espectros podemos calcular el corrimiento al rojo de las galaxias comparando las longitudes de onda esperadas en reposo ($\lambda_0$) con las observadas de la galaxia alejándose ($\lambda_1$) y calculando la diferencia $\Delta\lambda = \lambda_1 - \lambda_0$ de la siguiente forma:

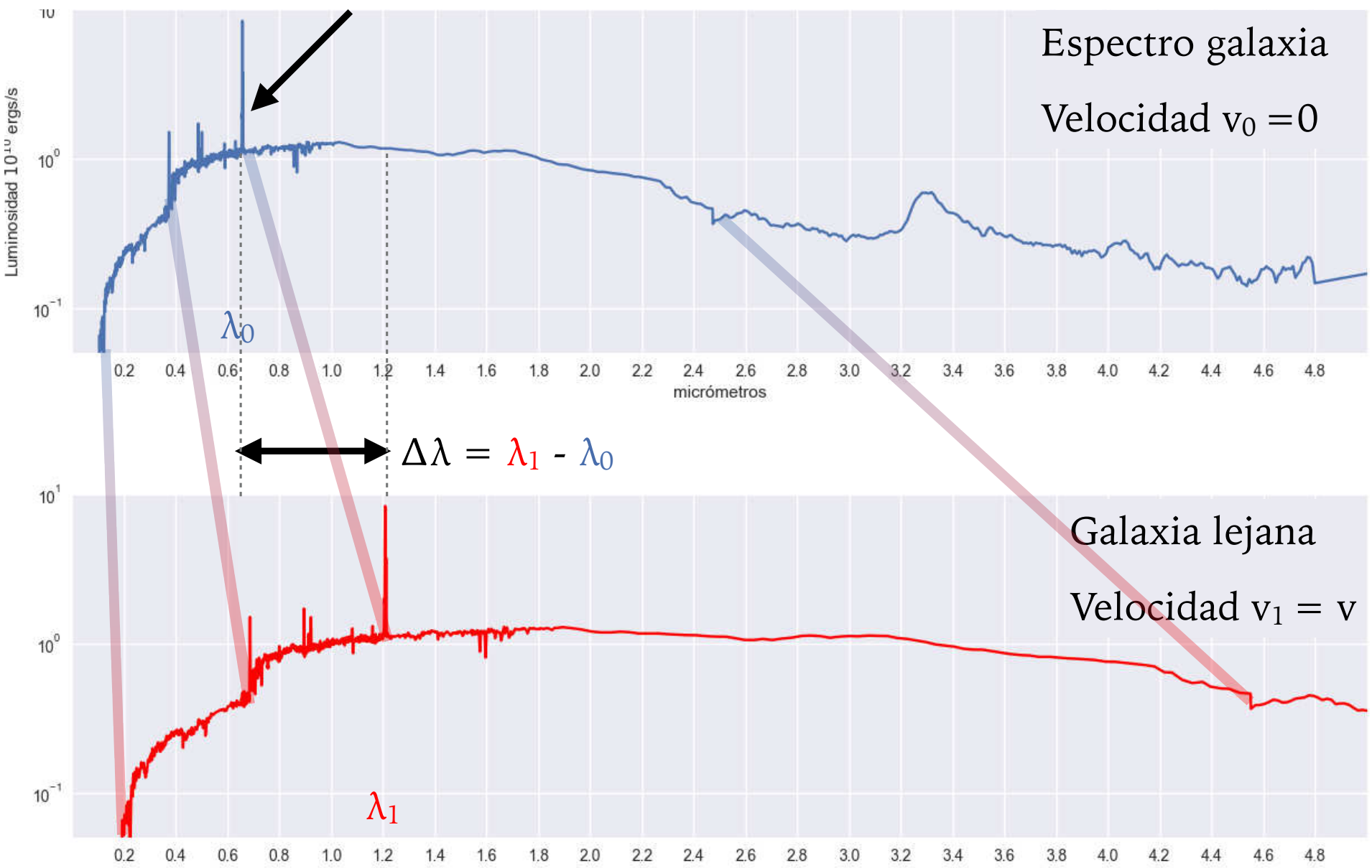

Figura 12.3: Efecto de corrimiento al rojo cosmológico en el espectro de una galaxia. Debido a la expansión del universo, objetos más distantes se alejan de nosotros con una mayor velocidad, y esto viene reflejado en el corrimiento al rojo del espectro.

$$\frac{\Delta\lambda}{\lambda_0} = \frac{v}{c} = \frac{\text{velocidad de la galaxia}}{\text{velocidad de la luz}}$$

$$\frac{\Delta\lambda}{\lambda_0} = z = \text{redshift}$$

Usando este método hemos calculado el redshift $z$ de 15 galaxias en el campo ultra profundo del *Hubble* que pueden encontrar en la figura 12.4

La distancia se calcula de la velocidad de acuerdo a la relación D = v/H0. La constante de proporcionalidad H0 es llamada la constante de Hubble, y tiene un valor estimado de 70 (km/s)/mpc. Esta relación, entre otras, es lo que vamos a utilizar para medir las distancias de las galaxias lejanas en el HUDF que encuentran en la regla cosmológica adjunta.

**Maqueta del universo lejano**

Ahora vamos a construir una maqueta en 3D del campo ultra profundo del *Hubble*, usando como base de nuestra maqueta un cartón:

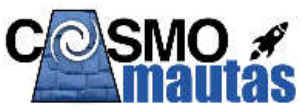

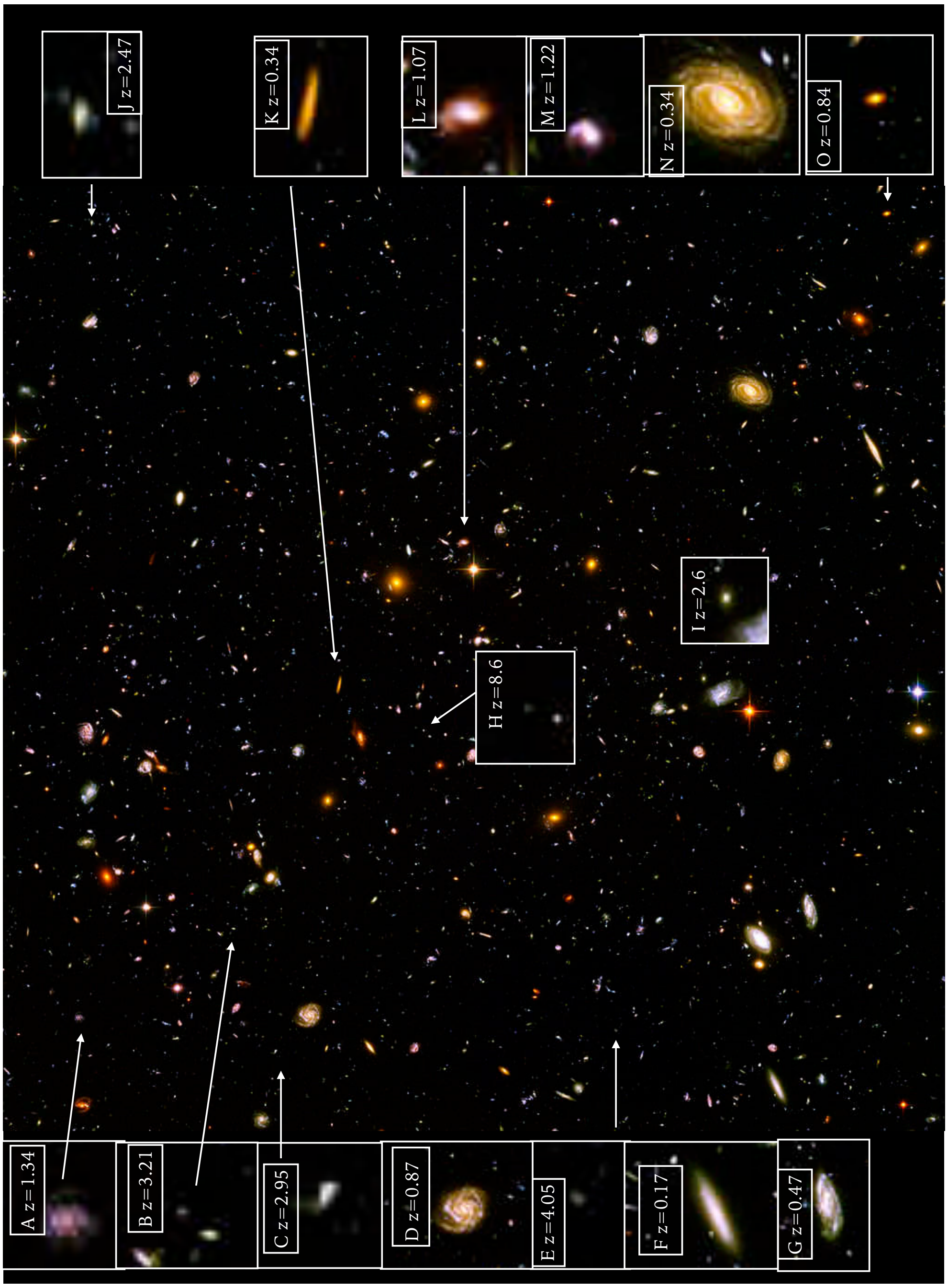

Figura 12.4: Galaxias del campo profundo del *Hubble* con respectivos “redshifts”. Éstos han sido estimados usando fotometría y no espectros. Los redshifts han sido extraídos de `http://ahah.asu.edu/clickonHUDF/udfmain_high.html`. [Crédito: adaptado de NASA/ESA].

- Peguemos o engrapemos la regla cosmológica adjunta al final de la actividad a lo largo del lado más largo del cartón.
- Recortemos las dos imágenes del CMB y del fondo de galaxias del campo profundo de Hubble. Luego, con la ayuda del limpiatipo, coloquemos cada imagen de forma vertical sobre el cartón, considerando su ubicación según la regla cosmológica.
- Ahora, recortemos las imágenes pequeñas de las galaxias y con la cinta adhesiva peguemos las galaxias en los palos de madera, en forma de carteles.
- En base a las mediciones del redshift, colocar los carteles en las respectivas posiciones correspondientes a cada galaxia (fijar cada palito con un poco de limpiatipo). De esta forma, observando la maqueta desde el extremo mas angosto de la base de cartón podemos ver el campo profundo del Hubble tal como en la imagen 2D, pero desde el extremo largo podemos apreciar su profundidad.
- Leyendo la escala de edad del universo, pueden darse cuenta de que usando estas pocas galaxias hemos reconstruido una gran parte de la historia del Universo!
- Escojan a su galaxia favorita y presenten a sus compañeros la distancia a la que se encuentra y cuál era la edad del universo cuando la luz de esta galaxia inició su viaje hasta nosotros.

## Conclusión y evaluación (10 min)

Veremos juntos un video explorando el espacio del HUDF. Discutiremos un poco más sobre la expansión del universo, y sobre los recientes debates sobre la medición de la variable de Hubble.

## Referencias

En las siguientes páginas mostramos los componentes de nuestra maqueta: una imagen del fondo cósmico de microondas, una imagen del HUDF, una escala de tiempo cósmico calculada para un universo ΛCDM con parametros cosmológicos estimados por el telescopio Planck (adaptada de Philipenko 2021) [Crédito: Planck/ESA, Hubble/NASA/ESA, NAOJ]

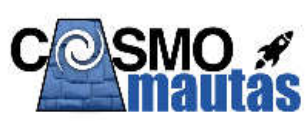

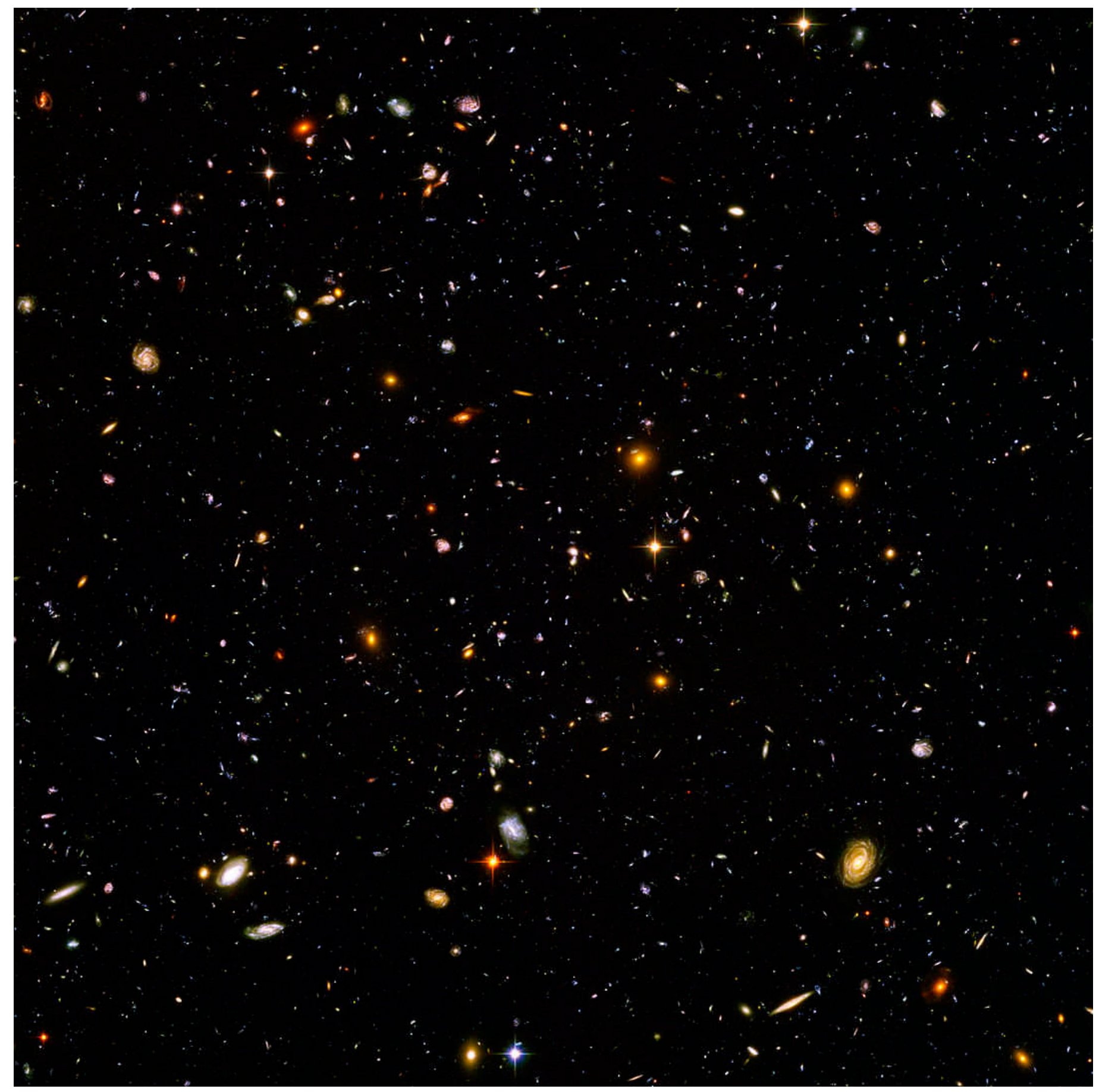

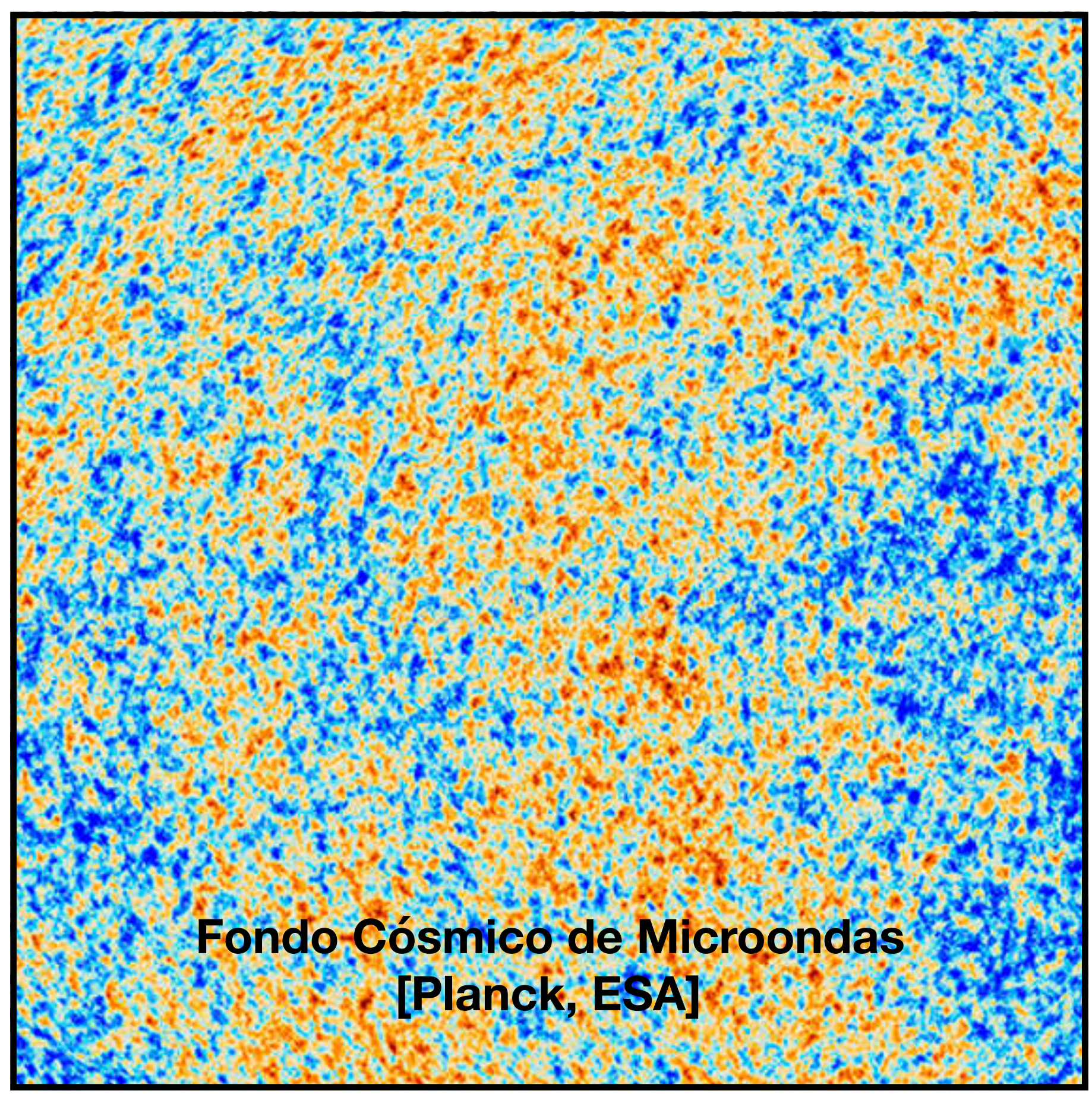

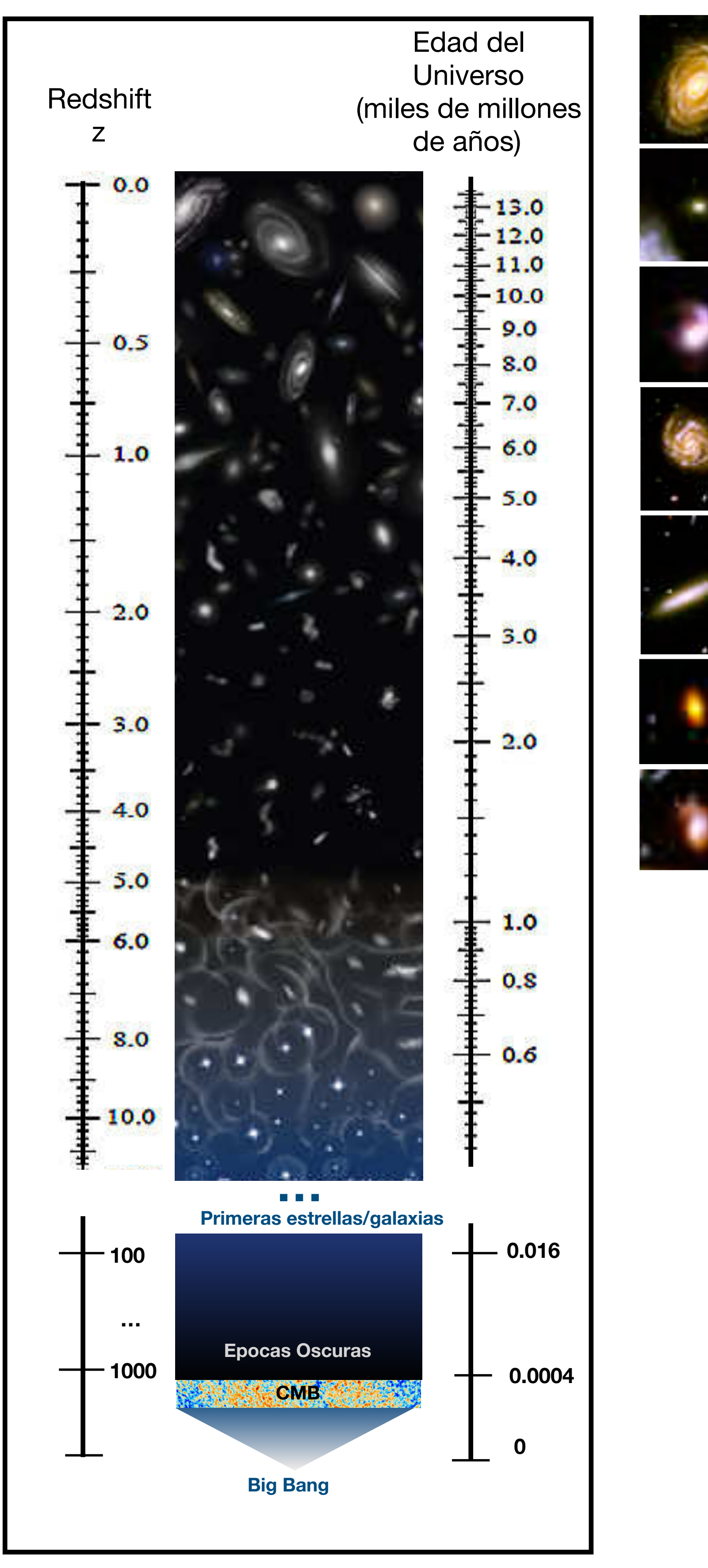

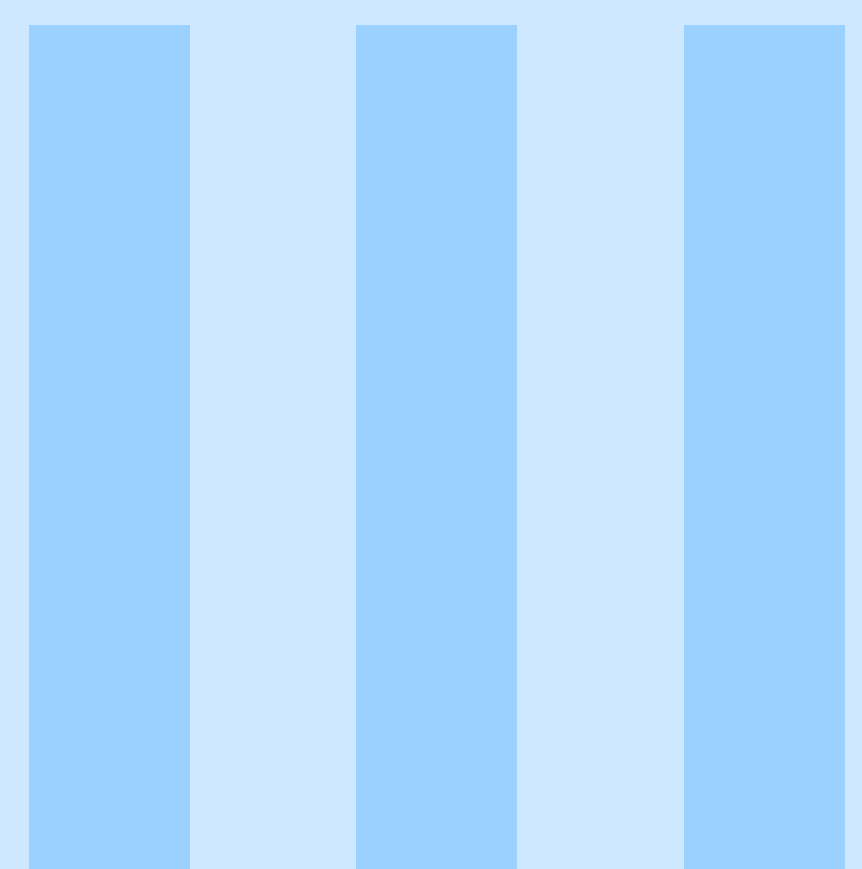

# Bibliografía

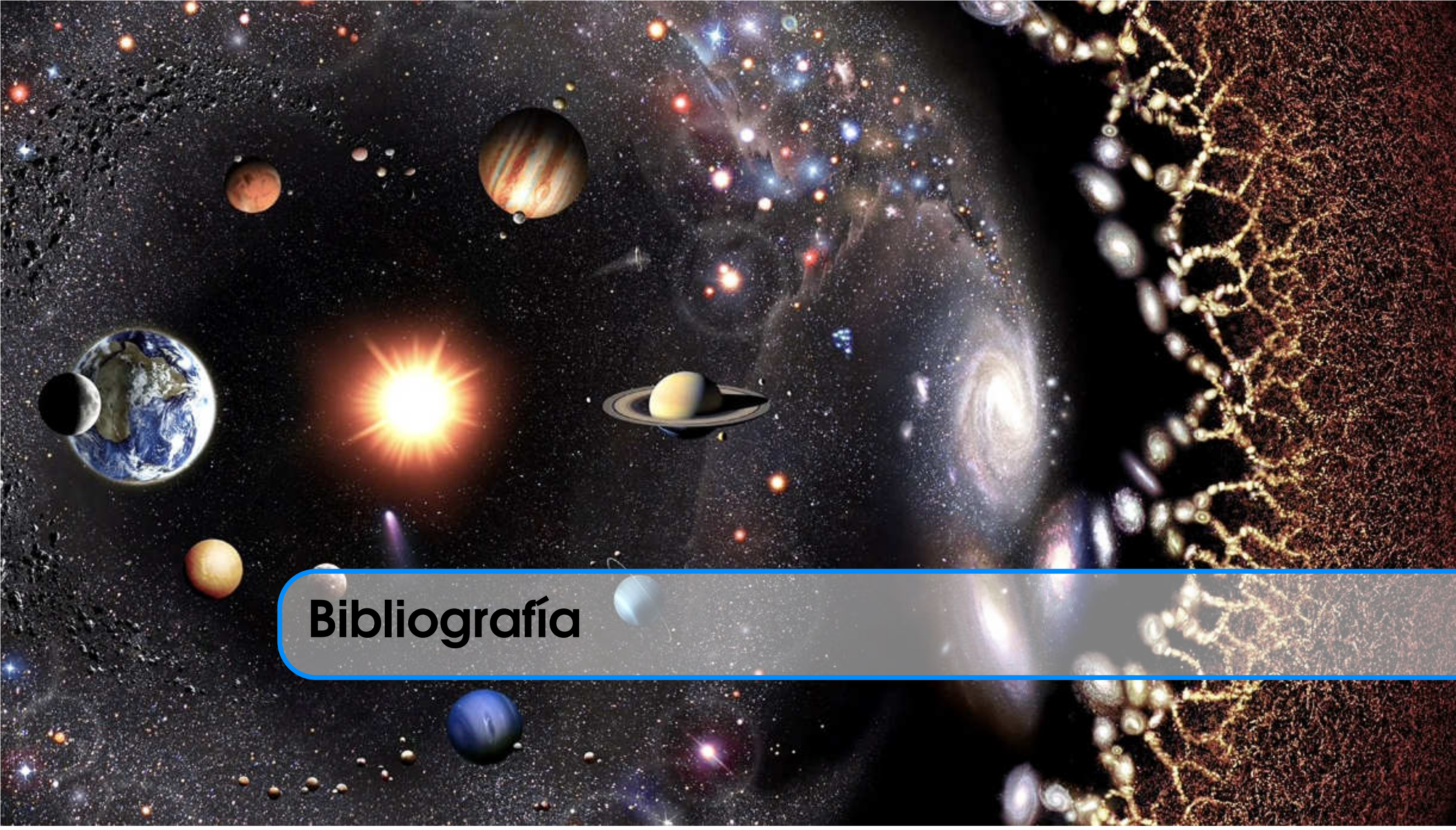

# Bibliografía

## Capítulo 1: La Tierra, el Sol y la Luna

## Capítulo 2: El sistema solar

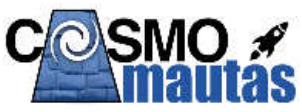

### Capítulo 3: Estrellas

### Capítulo 4: Exoplanetas

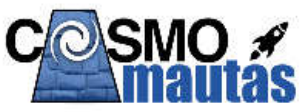

## Capítulo 5: Galaxias y Agujeros Negros

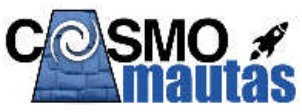

## Capítulo 6: Cosmología

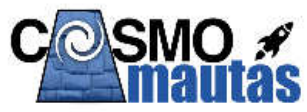

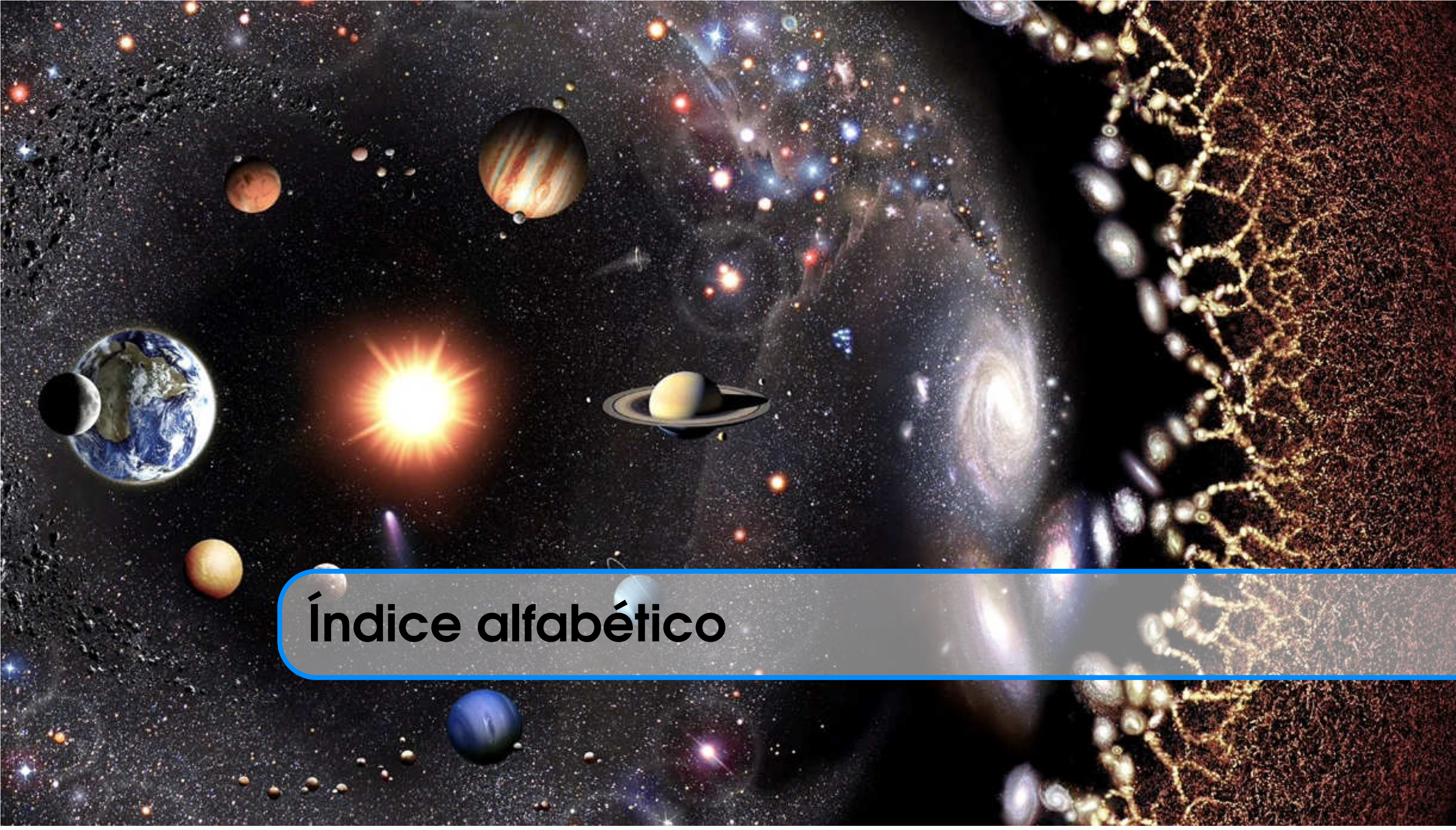

# Índice alfabético

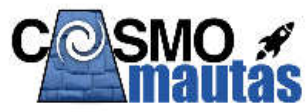

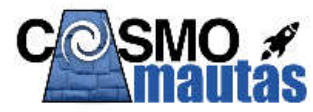